\documentclass[a4paper,11pt]{article}
\usepackage{jheppub} % for details on the use of the package, please
\usepackage[T1]{fontenc} % if needed

\title{\boldmath A New Method for Indirect Mass Measurements using the Integral Charge Asymmetry at the LHC}

\author[1]{G. Steve Muanza,\note{Corresponding author.}}
\author{and Thomas Serre}

\affiliation{Aix Marseille Universit\'e, CNRS/IN2P3,\\ CPPM UMR 7346, Marseille, France}

\emailAdd{muanza@in2p3.fr}

\abstract{Processes producing a charged final state at the LHC have a positive or null integral charge asymmetry. We propose a novel method 
for an indirect measurement of the mass of these final states based upon the process integral charge asymmetry.
We present this method in three stages. Firstly, the theoretical prediction of the integral charge asymmetry and its related uncertainties 
are studied through parton level cross sections calculations. Secondly, the experimental extraction of the
integral charge asymmetry of a given signal, in the presence of some background, is performed using particle 
level simulations. Process dependent templates enable to convert the measured integral charge asymmetry 
into an estimated mass of the charged final state.
Thirdly, a combination of the experimental and the theoretical uncertainties determines the full uncertainty of the indirect mass measurement.
\par\noindent
This new method applies to all charged current processes at the LHC.
In this article, we demonstrate its effectiveness at extracting the mass of the W boson, as a first step, and the sum of the masses of a chargino and a neutralino  
in case these supersymmetric particles are produced by pair, as a second step.}

\begin{document} 
\maketitle
\flushbottom
%\newpage

%%%%%%%%%%%%%%%%%%%%%%%%%%%%%%%%%%%%%%%%%%%%%%%%%%%%%%%%%%%%%%%%%%%%%%%%%%%%%%%%%%%%%%%%%%%%%%%%
\vspace*{5mm}
\section{Introduction} 
\par
Contrarily to most of the previous high energy particle colliders, the LHC is a charge asymmetric machine.
For charged final states \footnote{We defined these as event topologies containing an odd number of high $p_{T}$ charged and isolated leptons within the fiducial volume of the detector.}, denoted $FS^{\pm}$, the integral charge asymmetry, denoted $A_C$, is defined by

\begin{equation}
A_{C}=\frac{N(FS^{+})-N(FS^{-})}{N(FS^{+})+N(FS^{-})}
\end{equation}
\noindent
where $N(FS^{+})$ and $N(FS^{-})$ represent respectively the number of events bearing a positive and a negative charge in the FS.

For a $FS^{\pm}$ produced at the LHC in $p+p$ collisions, this quantity
is positive or null, whilst it is always compatible with zero for a $FS^{\pm}$ produced at the TEVATRON in $p+\bar{p}$ collisions.  
\par\noindent
To illustrate the $A_{C}$ observable, let's consider the Drell-Yan production of $W^{\pm}$ bosons in $p+p$ collisions. 
It is obvious for this simple $2\to 2$ s-channel process that more $W^+$ than $W^-$ are produced. Indeed, denoting $y_{W}$ the rapidity of the W boson, the 
corresponding range of the Bj\"orken x's: $x_{1,2}=\frac{M_{W^{\pm}}}{\sqrt{s}}\times e^{\pm y_{W}}$, probes the 
charge asymmetric valence parton densities within the proton. This results in having more $U+\bar{D}\to W^{+}$ than $\bar{U}+D\to W^{-}$ 
configurations in the initial state (IS). Here U and D collectively and respectively represent the up and the down quarks.

\noindent
In the latter case the dominant contribution to $A_{C}$ comes from the difference in rate between the $u+\bar{d}$ and the $d+\bar{u}$ quark currents in the IS. Using the usual notation $f(x,Q^{2})$ for the parton density functions (PDF) and within the leading order (LO) approximation, this can be expressed as: 

\begin{equation}
A_{C}\approx 
\frac{u(x_{1,2},M_{W}^{2})\bar{d}(x_{2,1},M_{W}^{2})-\bar{u}(x_{1,2},M_{W}^{2})d(x_{2,1},M_{W}^{2})}
{u(x_{1,2},M_{W}^{2})\bar{d}(x_{2,1},M_{W}^{2})+\bar{u}(x_{1,2},M_{W}^{2})d(x_{2,1},M_{W}^{2})}
\label{ACFLV}
\end{equation}

\noindent
where the squared four-momentum transfer $Q^{2}$ is set to $M_{W}^{2}$. 

\par\noindent
From equation \ref{ACFLV}, we can see that the $Q^{2}$ evolution of the parton density functions (PDFs) governs the $Q^{2}$ evolution of $A_{C}$. The former are known, up-to the NNLO in QCD, as solutions of the DGLAP equations \cite{Cafarella:2005zj}. One could therefore think of using an analytical functional form to relate $A_{C}$ to the squared mass of the s-channel propagator, here $M_{W}^{2}$. However there are additional contributions to the $W^{\pm}$ inclusive production. At the Born level, some come from other flavour combinations in the IS of the s-channel, and some come from the u-channel and the t-channel. On top of this, there are higher order corrections. These extra contributions render the analytical expression of the $Q^{2}$ dependence of $A_{C}$
much more complicated. Therefore we choose to build process-dependent numerical mass template curves for $A_{C}$ by varying M$_{FS^{\pm}}$. These mass templates
constitute inclusive and flexible tools into which all the above-mentioned contributions to $A_{C}$ can be incorporated, they can very easily be built within restricted domain
of the signal phase space imposed by kinematic cuts.

\par
The $A_{C}$ for the $W^{\pm}\to\ell^{\pm}\nu$ production at the LHC is large enough to be measured and it has relatively
small systematic uncertainties since it's a ratio of cross sections. The differential charge
asymmetry of this process in $p+p$ collisions have indeed been measured by the ATLAS \cite{Aad:2011yna}, the CMS \cite{Chatrchyan:2012xt} \cite{Chatrchyan:2013mza} 
and the LHCb \cite{LHCb:2011xha} experiments \cite{ATLAS:2011pha} for the first times in their 2011 datasets.

\par
In this article we exploit the $A_{C}$ to set a new type of constraint on the mass of the charged $FS^{\pm}$ as initially proposed
in \cite{Djouadi:1998di}\cite{Muanza:1st-note}.
\par
We'll separate the study into two parts. The first one, in section 2, is dedicated to present in full length the method of
indirect mass measurement that we propose on a known Standard Model (SM) process. We choose the 
$W^{\pm}\to\ell^{\pm}+\rlap{\kern0.25em/}E_{T}$ inclusive production at the LHC to serve as a test bench.
\par\noindent
In the second part, in section 3, we shall repeat the method on a "Beyond the Standard Model" (BSM) process. We choose a SUSY
search process of high interest, namely
\begin{equation}
\tilde\chi^{\pm}_{1}+\tilde\chi^{0}_{2}\to 3\ell^{\pm}+\rlap{\kern0.25em/}E_{T}.
\end{equation}
\par\noindent
For both the SM and the BSM processes, we obviously tag the sign of the FS by choosing a decay into one (or three) charged
lepton(s) for which the sign is experimentally easily accessible.
\par\noindent
It's obvious that for these two physics cases other mass reconstruction methods exist. These standard mass reconstruction techniques are all based on the kinematics
of the FS. For the $W^{\pm}\to\ell^{\pm}+\rlap{\kern0.25em/}E_{T}$ process mass templates based upon the transverse mass allow to extract
$M_{W^{\pm}}$ with an excellent precision that the new technique proposed here cannot match. In constrast, for the $\tilde\chi^{\pm}_{1}+\tilde\chi^{0}_{2}\to 3\ell^{\pm}+\rlap{\kern0.25em/}E_{T}$ process, even if astute extensions of the transverse mass enable to acurrately measure some mass differences, no standard techniques is able to measure accurately the mass of the charged FS: $M_{FS^{\pm}}=M_{\tilde\chi^{\pm}_{1}}+M_{\tilde\chi^{0}_{2}}$.
\par\noindent
Therefore this new mass reconstruction technique should not be viewed as an alternative to the standard techniques but rather as an unmined complement to them.
In a few cases, especially where many FS particles escape detection, this new technique can be more accurate than the standard ones. It also has the advantage of being
almost model independent.
\par
For each signal process we sub-divide the method into four steps that are described in four sub-sections. In the first
sub-sections~\ref{sec:Part1-parton-LVL} and~\ref{sec:next-parton-LVL}, we start by deriving the theoretical $A_{C}$ template curves at the parton level. 

\par\noindent
In the second sub-sections~\ref{sec:Part1-particle-LVL} and~\ref{sec:next-particle-LVL}, we place ourselves in the situation of an experimental
measurement of the $A_{C}$ of the signal in the presence of some background. For that we generate samples of Monte Carlo (MC) events that we reconstruct using a fast simulation of the response of the ATLAS detector. 
This enables to account for the bias of the signal $A_{C}$ induced by the event selection. In addition we can quantify the bias of $A_{C}$ due to the residual contribution of some background processes passing this event selection.
\par\noindent
Then,  in the third sub-sections~\ref{sec:Part1-mass-constraint} and~\ref{sec:next-mass-constraint}, we convert the measured $A_{C}$ into an estimated $M_{FS}$ using fitted experimental $A_C$ template curves that account for all the experimental uncertainties.
\par\noindent
In the fourth sub-sections~\ref{sec:Part1-MRST2007} and~\ref{sec:Part2-MRST2007}, we combine the theoretical and the experimental uncertainties on the signal $A_{C}$ to derive the full uncertainty of the indirect mass measurement.
The conclusions are presented in section~\ref{Concl} and the prospects in section~\ref{Prospects}.
\par\noindent
Note that we'll always express the integral charge asymmetry in $\%$ and the mass of the charged final state in $GeV$ throughout this article. The uncertainty on the integral charge asymmetry $\delta A_{C}$ will also be expressed in $\%$ but will always represent an absolute uncertainty as opposed to a relative uncertainty with respect to $A_{C}$.

\newpage

%%%%%%%%%%%%%%%%%%%%%%%%%%%%%%%%%%%%%%%%%%%%%%%%%%%%%%%%%%%%%%%%%%%%%%%%%%%%%%%%%%%%%%%%%%%%%%% 
%%% PART I:  W -> l + mET
%%%%%%%%%%%%%%%%%%%%%%%%%%%%%%%%%%%%%%%%%%%%%%%%%%%%%%%%%%%%%%%%%%%%%%%%%%%%%%%%%%%%%%%%%%%%%%%
\vspace*{5mm}
\label{sec:Part1-Intro} 
\section{Inclusive Production of $W^{\pm}\to\ell^{\pm}\nu$}
\vspace*{2.5mm}

\subsection{Theoretical Prediction of $A_{C}(W^{\pm}\to\ell^{\pm}\nu)$}
\label{sec:Part1-parton-LVL}
\par
In this section we calculate separately the cross sections of the "signed processes", i.e. the cross sections of the positive and negative FS: 
$\sigma^{+}=\sigma(p+p\to W^{+}\to\ell^{+}\nu)$ and $\sigma^{-}=\sigma(p+p\to W^{-}\to\ell^{-}\bar{\nu})$. The process integral 
charge asymmetry therefore writes:
\begin{equation}
A_{C} = \frac{\sigma^{+}-\sigma^{-}}{\sigma^{+}+\sigma^{-}}
\end{equation} 

\vspace*{1.5mm}
\subsubsection{Sources of Theoretical Uncertainties on $A_{C}$}
\vspace*{0.5mm}

Since these cross sections integration are numerical rather than analytical, they each have an associated statistical uncertainty $\delta\sigma^{\pm}_{Stat}$ 
due to the finite sampling of the process phase space. Even though these are relatively small we explicitely include them
and we calculate the resulting statistical uncertainty on the process integral charge asymmetry: $\delta (A_{C})_{Stat}$
for which we treat $\delta\sigma^{+}_{Stat}$ and  $\delta\sigma^{-}_{Stat}$ as uncorrelated uncertainties. Hence:
\begin{equation}
\delta (A_{C})_{Stat}=\frac{2}{(\sigma^{+}+\sigma^{-})^{2}}\sqrt{(\sigma^{-}\cdot\delta\sigma^{+}_{Stat})^{2}+(\sigma^{+}\cdot\delta\sigma^{-}_{Stat})^{2}}
\end{equation} 

\noindent
For each cross section calculation we choose the central Parton Density Function (PDF) from a PDF set (or just the single PDF when there's
no associated uncertainty set). Whenever we use a PDF set, it contains $2N_{PDF}$ uncertainty PDFs on top of the central
PDF fit, the PDF uncertainty is calculated as proposed in \cite{Campbell:2006wx}:

\begin{equation}
\begin{cases}
\delta (A_{C})_{PDF}^{Up}=\sqrt{\sum_{i=1}^{N_{PDF}} (Max[A_{C}(i)^{up}-A_{C}(0),A_{C}(i)^{down}-A_{C}(0),0])^{2}}
\\
\delta (A_{C})_{PDF}^{Down}=\sqrt{\sum_{i=1}^{N_{PDF}} (Max[A_{C}(0)-A_{C}(i)^{up},A_{C}(0)-A_{C}(i)^{down},0])^{2}}
\end{cases}
\end{equation}
where $A_{C}(0)$, $A_{C}(i)^{up}$, and $A_{C}(i)^{down}$ represent the integral charge asymmetries calculated with $\sigma_{0}$, $\sigma_{i}^{up}$, and 
$\sigma_{i}^{down}$, respectively. $\sigma_{0}$ represents the cross section calculated with the central PDF fit. $\sigma_{i}^{up}$ represent the $N_{PDF}$ upward uncertainty
PDFs such that generally $\sigma_{i}^{up} > \sigma_{0}$, and $\sigma_{i}^{down}$ represent the $N_{PDF}$ downward uncertainty PDFs such that generally 
$\sigma_{i}^{down} < \sigma_{0}$. 
\par\noindent
We choose the QCD renormalization and factorization scales: $\mu_{R}=\mu_{F}=\mu_{0}$ to be equal, and we choose a
process dependent dynamical option to adjust the value of $\mu_{0}$ to the actual kinematics event by event.
The scale uncertainty is evaluated using the usual factors 1/2 and 2 to calculate variations with respect to the central value $\mu_{0}$: 
\begin{equation}
\begin{cases}
\delta (A_{C})_{Scale}^{Up}=A_{C}(\mu_{0}/2)-A_{C}(\mu_{0}) \\  
\delta (A_{C})_{Scale}^{Down}=A_{C}(2\mu_{0})-A_{C}(\mu_{0})
\end{cases}
\end{equation}
The total theoretical uncertainty is defined as the sum in quadrature of the 3 sources:
\begin{equation}
\begin{cases}
\delta (A_{C})_{Total}^{Up}=\sqrt{[\delta (A_{C})_{PDF}^{Up}]^{2}+[\delta (A_{C})_{Scale}^{Up}]^{2}+[\delta (A_{C})_{Stat}]^{2}}\\
\delta (A_{C})_{Total}^{Down}=\sqrt{[\delta (A_{C})_{PDF}^{Down}]^{2}+[\delta (A_{C})_{Scale}^{Down}]^{2}+[\delta (A_{C})_{Stat}]^{2}}
\end{cases}
\end{equation}

\vspace*{1.5mm}
\subsubsection{Setup and Tools for the Computation of $A_{C}$}
\vspace*{0.5mm}

We calculate the $\sigma^{+}=\sigma(p+p\to W^{+}\to\ell^{+}\nu)$ and $\sigma^{-}=\sigma(p+p\to W^{-}\to\ell^{-}\bar{\nu})$ cross sections and their uncertainties at $\sqrt{s}=$7 TeV using MCFM v5.8 \cite{Campbell:1999ah}\cite{Campbell:2000bg}\cite{Campbell:2002tg}. We include both the $W^{\pm}+0Lp$ and the $W^{\pm}+1Lp$ matrix elements (ME) in the calculation in order to have a better representation of the $W^{\pm}$ inclusive production  (the notation "Lp" stands for "light parton", i.e. u/d/s quarks or gluons). We set the QCD scales as $\mu_{R}=\mu_{F}=\mu_{0}=\sqrt{M^{2}(W^{\pm})+p_{T}^{2}(W^{\pm})}$
and we run the calculation at the QCD leading order (LO) and next-to-leading order (NLO). For both the phase space pre-sampling and the actual cross section integration, we run 10 times 20,000 sweeps of VEGAS \cite{Lepage:1980dq}. We impose the following parton level cuts: $M(\ell^{\pm}\nu) > 10$ \rm\ GeV, $|\eta(\ell^{\pm})| < 2.4$ and
$p_{T}(\ell^{\pm}) > 20$ \rm\ GeV. We artificially vary the input mass of the $W^{\pm}$ boson and we repeat the computations for the 3 following couples of respective LO and NLO PDFs: MRST2007lomod \cite{Sherstnev:2007nd} - MRST2004nlo \cite{Martin:2004ir}, CTEQ6L1 \cite{Pumplin:2002vw} - CTEQ6.6 \cite{Nadolsky:2008zw}, and MSTW2008lo68cl - MSTW2008nlo68cl \cite{MSTW2008} which are interfaced to MCFM through LHAPDF v5.7.1 \cite{Whalley:2005nh}. 
As the LO is sufficient to present the method in detail, we'll restrict ourselves to LO MEs and LO PDFs throughout the article for the sake of simplicity. We shall however provide the theoretical $A_{C}$ mass templates up to the NLO for the W process. And we recommend to establish them using the best theoretical calculations available for
any use in a real data analysis, including at the minimum the QCD NLO corrections. 
\par\noindent
The MRST2007lomod is chosen as the default PDF throughout this article. The two other LO PDFs serve for comparison of the central value and the uncertainty of $A_{C}$ with respect to MRST2007lomod. In that regard, MSTW2008lo68cl is especially useful to estimate the
impact of the $\delta (A_{C})_{PDF}$.

%%%%%%%%%%%%%%%%%%%%%%%%%%%%%%%%%%%%%%%%%%%%%%%%%%%%%%%%%%%%%%%%%%%%%%%%%%%%%%%%%%%%%%%%%%%%%%%%%%%%%%%%%%

\vspace*{1.5mm}
\subsubsection{\label{I-parton-LVL:Fit_Functions} Modeling of the Theoretical $A_{C}(W^{\pm}\to e^{\pm}\nu_{e})$ Template Curves}
\vspace*{0.5mm}

\noindent
The theoretical MRST2007lomod and MRST2004nlo raw template curves are obtained by sampling $A_{C}^{Raw}$ at different values of $M_{W^{\pm}}$. The
corresponding theoretical uncertainties are also calculated: $A_{C}^{Raw}\pm\delta A_{C}^{Raw}$. This discrete sampling is then transformed 
into a continuous template curve through a fit using a functional form $A_{C}^{Fit}=f(M_{W^{\pm}})$ which is constrained by the theoretical uncertainties.
\par\noindent
We have considered three different types of functional forms for these fits with $f$ being either a:
\begin{enumerate}
\item polynomial of logarithms: $f(x) = \sum\limits_{i=0}^{N_{FP}} A_{i}\times \lbrace Log(x)\rbrace ^{i}$
\item polynomial of logarithms of logarithms: $f(x) = \sum\limits_{i=0}^{N_{FP}} A_{i}\times \lbrace Log\lbrack Log(x)\rbrack\rbrace ^{i}$
\item series of Laguerre polynomials: $f(x) = \sum\limits_{i=0}^{N_{FP}} A_{i}\times L_{n}(x)$ where $L_{n}(x) = \frac{e^{x}}{n!}\frac{d^{n}}{dx^{n}}(e^{-x}x^{n}).$
\end{enumerate}
\par\noindent
The types of functional forms that we're considering are not arbitrary, they are all related to parametrizations of solutions of the DGLAP equations for the evolution of
the PDFs. The polynomial of logarithms of logarithms is inspired by an expansion of the PDF in series of $Log\lbrack Log(Q^{2})\rbrack$ as suggested in \cite{Cafarella:2005zj}. The polynomial of logarithms was just the simplest approximation of the aforementioned series that we first considered. And the expansion of the PDF in series of Laguerre polynomials is proposed in \cite{Schoeffel:1998tz}.
\par\noindent
In the Appendix \ref{AppendixA}, we give a numerical example of the evolution of the $u(x,Q^{2})$,  $\bar{u}(x,Q^{2})$, $d(x,Q^{2})$, $\bar{d}(x,Q^{2})$ proton density functions calculated with QCDNUM \cite{Botje:2010ay} and the \\MSTW2008nlo68cl PDF. We also provide a few toy models to justify the main properties of the functional forms used for $A_{C}^{Fit}$.
\par\noindent
Ultimately, the model of the theoretical template curve uses the functional form $f$ for the $A_{C}^{Fit}$ central values and re-calculate their uncertainty $\delta A_{C}^{Fit}$
by accounting for the correlations between the uncertainties of the fit parameters:
\begin{equation}
(\delta A_{C}^{Fit})^{2} = (\delta f)^{2} = \sum\limits_{i=0}^{N_{FP}}\sum\limits_{j>i}^{N_{FP}} \left(\frac{\partial f}{\partial A_{i}}\right)^{2}\cdot VAR(A_{i}) + 2\cdot\frac{\partial f}{\partial A_{i}}\cdot\frac{\partial f}{\partial A_{j}}\cdot COVAR(A_{i},A_{j})
\label{Fit_Correl_Uncert}
\end{equation}
\noindent
The diagonal and off-diagonal elements of the fit uncertainty matrix are denoted $VAR(A_{i})$ and $COVAR(A_{i},A_{j})$, they correspond
to the usual variances of the parameters and the covariances amongst them, respectively.
\par\noindent
The number of fit parameters $N_{FP}$ is taken as the minimum integer necessary to get a good $\chi^{2}/N_{dof}$ for the fit and it is adjustable for each $A_{C}$ template curve. 
\par\noindent
Comparing the three types of polynomials cited above as functional forms to fit all the $A_{C}$ template curves of sub-sections~\ref{sec:Part1-parton-LVL} and~\ref{sec:next-parton-LVL}, we find that the polynomials of logarithms of logarithms of $Q$ give the best fits. They are henceforth chosen as the default functional form
to model the $Q$ evolution of $A_{C}$ throughout this article.

%%%%%%%%%%%%%%%%%%%%%%%%%%%%%%%%%%%%%%%%%%%%%%%%%%%%%%%%%%%%%%%%%%%%%%%%%%%%%%%%%%%%%%%%%%%%%%%%%%%%%%%%%%

\vspace*{1.5mm}
\subsubsection{\label{I-parton-LVL:MRST} $A_{C}(W^{\pm}\to e^{\pm}\nu_{e})$ Template Curves for MRST}
\vspace*{0.5mm}

\begin{table}[h]
\begin{center}
\begin{tabular}{|c|c|c|c|c|c|}
\hline\hline
$\rm M_{W^{\pm}}$ & $A_{C}$        & $\delta (A_{C})_{Stat}$ & $\delta (A_{C})_{Scale}$ & $\delta (A_{C})_{PDF}$ & $\delta (A_{C})_{Total}$\\ 
(\rm\ GeV)                   & ($\%$)           & ($\%$)	                    & ($\%$)                             & ($\%$)                             & ($\%$)		   \\
\hline
20.1                               &   LO:    2.20    & $\pm 0.24$  	               & $^{+0.47}_{+0.10}$        &  0.00		                   & $^{+0.52}_{-0.26}$   \\
                                      &   NLO:  2.09   & $\pm 0.11$                     & $ ^{+0.04}_{-0.14}$        & 0.00                                  & $^{+0.12}_{-0.18}$ \\
\hline
40.2                               &   LO:    6.77    & $\pm$0.12  	               & $^{+0.02}_{-0.11}$         &  0.00		                   & $^{+0.12}_{-0.16}$   \\
                                      &   NLO:  8.05   & $\pm  0.07$                    & $ ^{-0.18}_{-0.06}$         & 0.00                                 & $^{+0.19}_{-0.09}$ \\
\hline
\underline{80.4}         &   LO:    20.18  & $\pm$0.06  	            & $^{+0.05}_{-0.03}$         &  0.00		                & $^{+0.08}_{-0.07}$   \\
                                      &   NLO:  21.49 & $\pm  0.03$                    & $^{-0.08}_{-0.00}$          &  0.00                                & $^{+0.09}_{-0.03}$ \\
\hline
160.8                             &   LO:    29.39  & $\pm$0.05  	            & $^{+0.00}_{+0.03}$         &  0.00		                & $^{+0.05}_{-0.06}$   \\
                                      &   NLO: 30.55  & $\pm  0.03$                   & $^{ -0.02}_{ -0.01}$         &  0.00                                & $^{+0.04}_{-0.03}$ \\
\hline
321.6                             &   LO:    35.92  & $\pm$0.05  	            & $^{-0.11}_{+0.10}$         &  0.00		                & $^{+0.11}_{-0.11}$   \\
                                      &   NLO:  36.90 & $\pm  0.03$                   & $^{-0.05}_{-0.04}$          &  0.00                                & $^{+0.06}_{-0.05}$ \\
\hline
643.2                             &   LO:    43.99  & $\pm$0.05  	            & $^{-0.14}_{+0.13}$         &  0.00		                & $^{+0.15}_{-0.14}$   \\
                                      &   NLO:  45.11 & $\pm  0.03$                   & $^{-0.05}_{-0.05}$          &  0.00                                & $^{+0.06}_{-0.06}$ \\
\hline
1286.4	                         &   LO:     52.36 & $\pm$0.06  	            & $^{+0.03}_{-0.02}$         &  0.00		               & $^{+0.07}_{-0.07}$   \\
                                      &   NLO:  55.33 & $\pm  0.04$                    & $^{+0.01}_{-0.02}$         &  0.00                               & $^{+0.04}_{-0.04}$ \\
\hline\hline
\end{tabular}       
\end{center}
\caption{\label{I-parton-LVL:Tab_MRST} The MRST $A_{C}$ table with the breakdown of the different sources of theoretical uncertainty. The MRST2007lomod
PDF is used for the LO and the MRST2004nlo for the NLO.}
\end{table}

\begin{figure}[h]
\begin{center}
\includegraphics[scale=0.35]{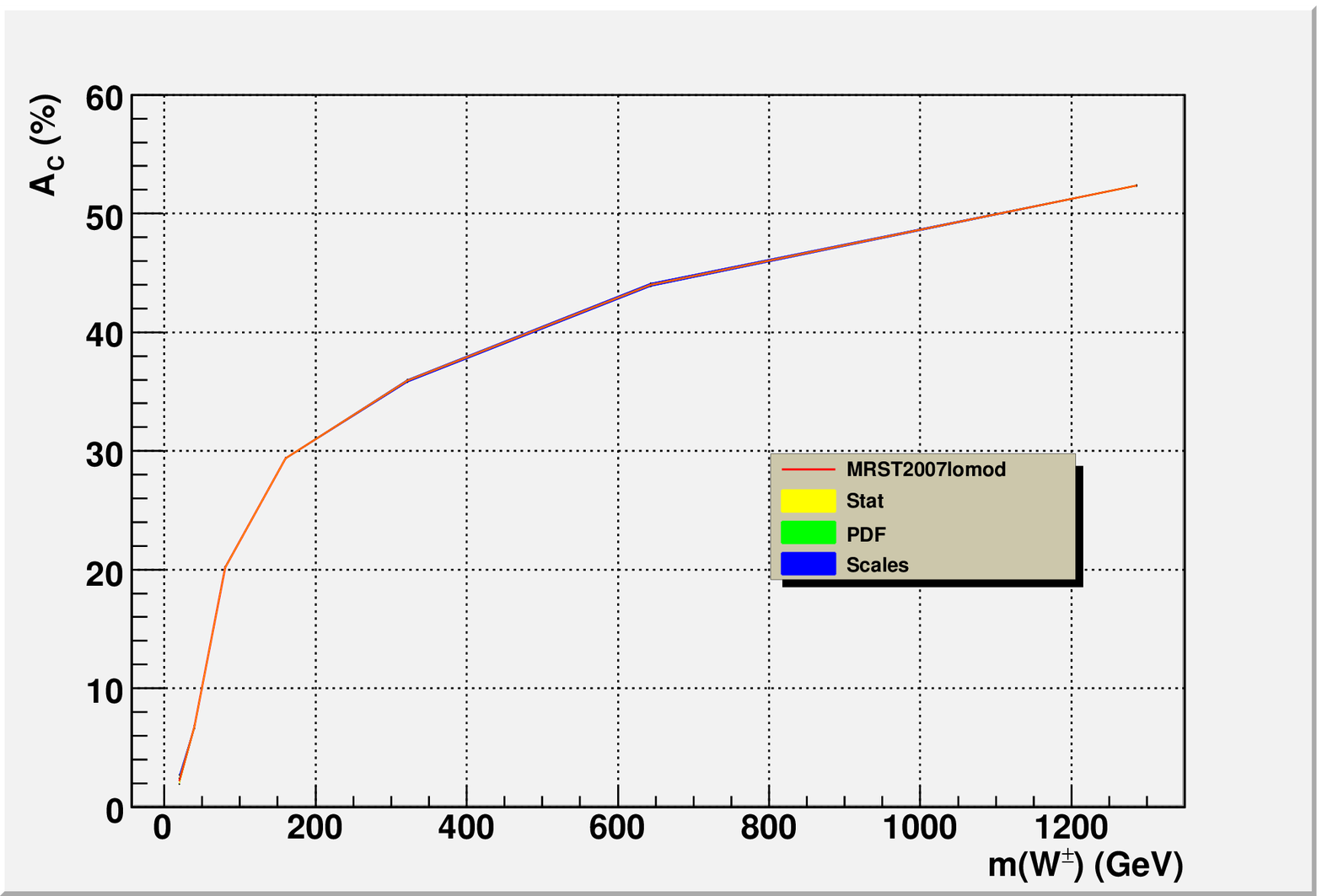}
\includegraphics[scale=0.35]{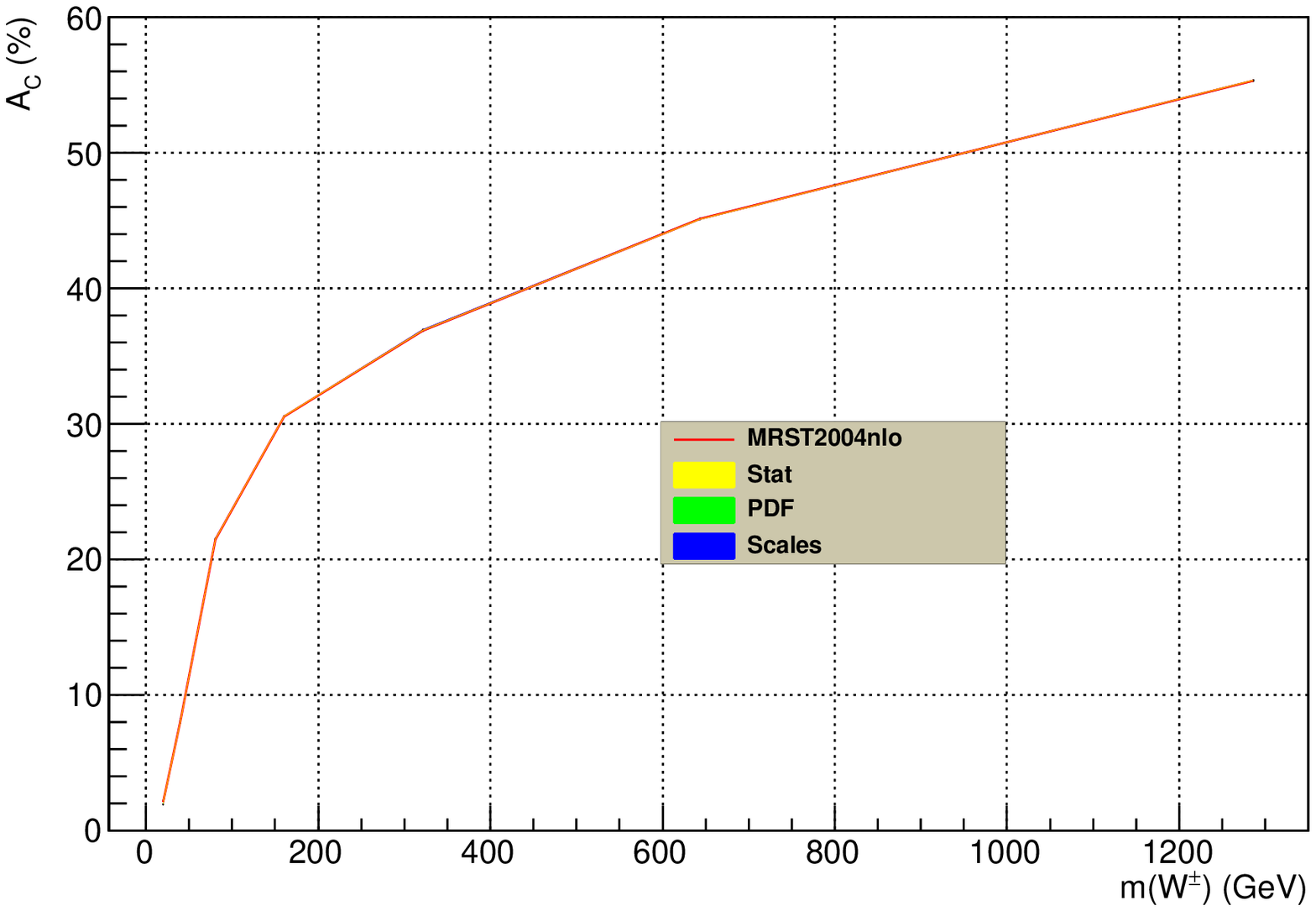}\\
\includegraphics[scale=0.35]{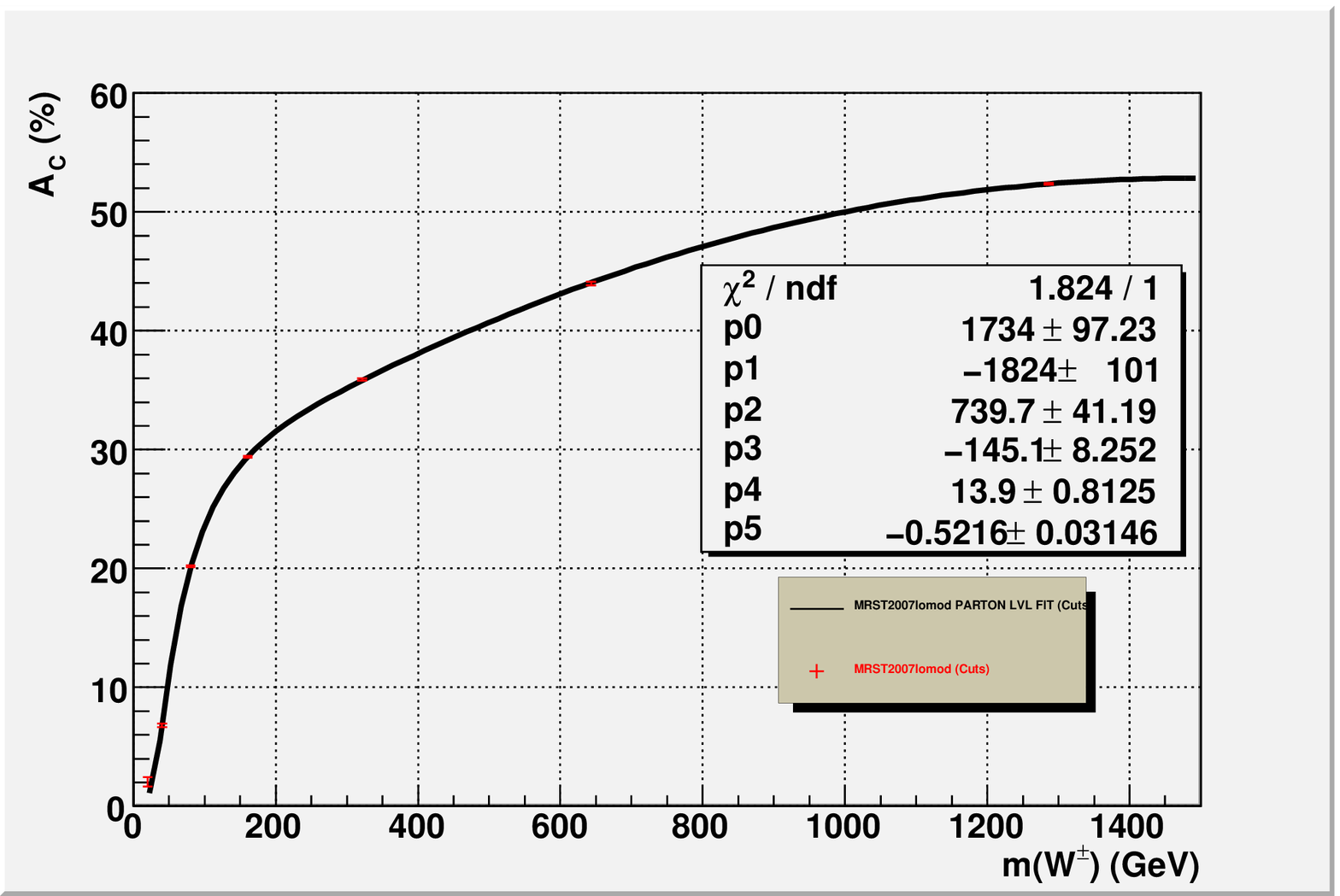}
\includegraphics[scale=0.35]{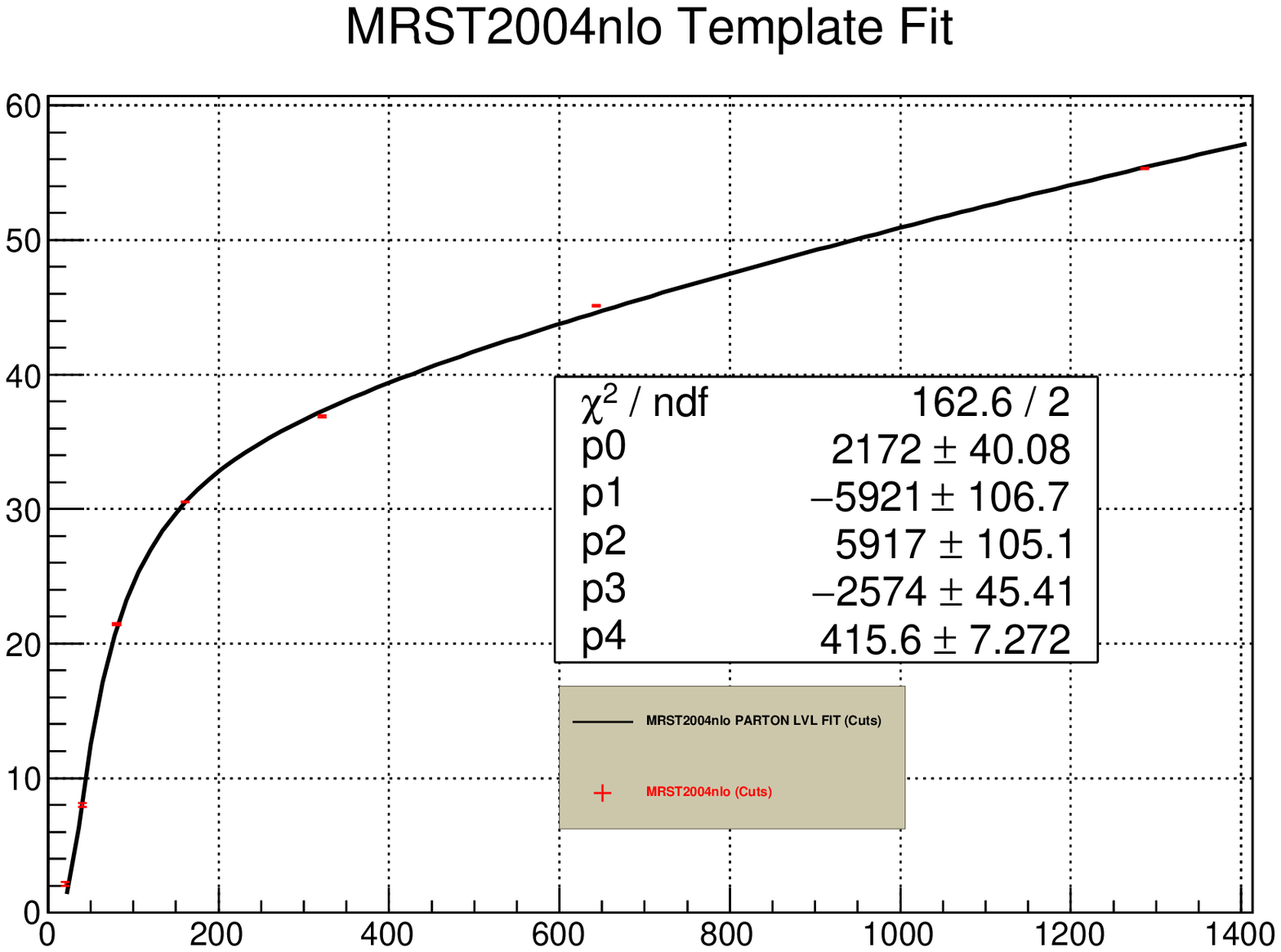}\\
\includegraphics[scale=0.35]{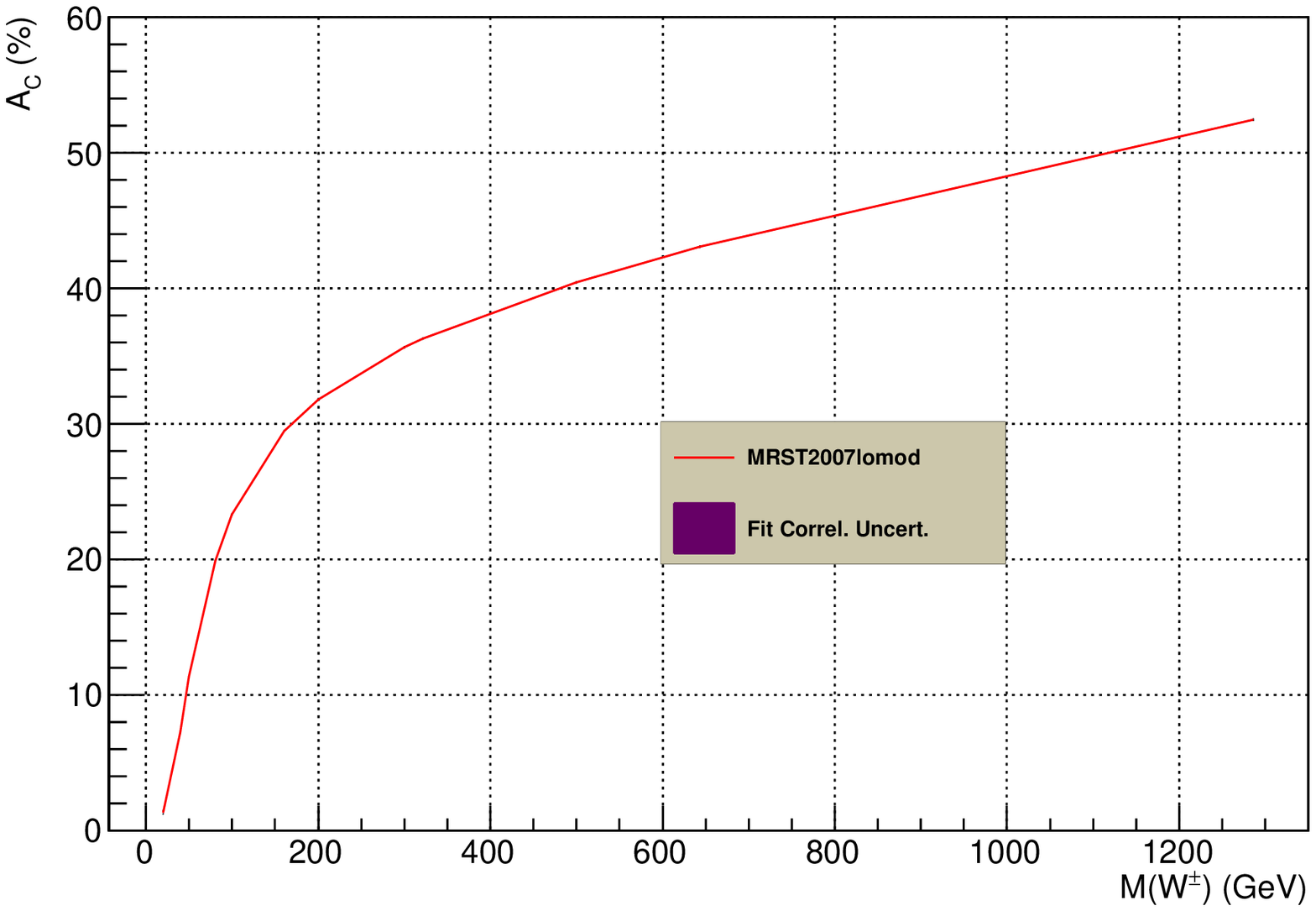}
\includegraphics[scale=0.35]{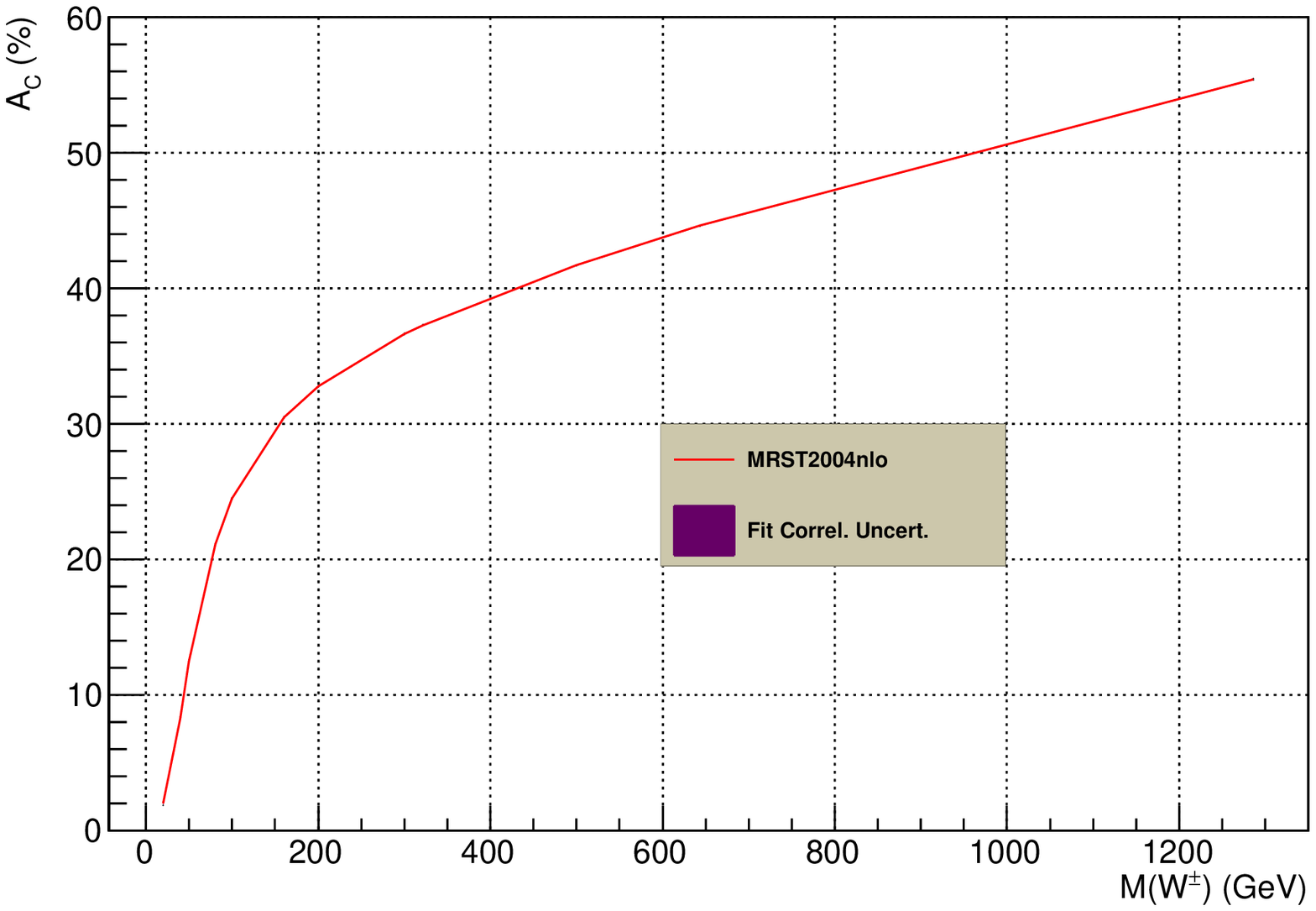}
\end{center}
\caption{\label{I-parton-LVL:Fig1} The theoretical  MRST $A_{C}$ template curves at LO with MRST2007lomod on the left-hand side (LHS)
and NLO with the MRST2004nlo on the right-hand side (RHS).
The raw curve with its uncertainty bands,  the corresponding fitted curve  and the fitted curve with the correlations 
between the fit parameters uncertainties are displayed on the top, the middle and the bottom rows, respectively.}
\end{figure}

\noindent
The theoretical  MRST2007lomod and MRST2004nlo $A_{C}$ template curves are obtained from the signed cross sections used for table \ref{I-parton-LVL:Tab_MRST}.
Since there is no MRST2007lomod PDF uncertainty set, we simply set $\delta (A_{C})_{PDF} = 0$. 
In this case, $\delta_{Total}^{Theory} A_{C} =\sqrt{\delta_{Stat}^{2} A_{C} +\delta_{Scale}^{2} A_{C}}$. 
Figure \ref{I-parton-LVL:Fig1} displays the fit to the $A_C$ template curve using a polynomial of $Log\left(Log(Q)\right)$. In the case of the MRST2007lomod PDF, it is sufficient to limit the polynomial to the degree $N_{FP}=5$ to fit the $A_{C}$ template curve
in the following (default) range: $M_{W^{\pm}}\in [15,1500]\rm\ GeV$. 

\begin{table}[h]
\begin{center}
\begin{tabular}{|c|c|c|}
\hline\hline
$\rm M_{W^{\pm}}$ & $A_{C}^{Fit}$ & $\delta A^{Fit}_{C}$\\ 
(\rm\ GeV)                   & ($\%$)               &  ($\%$)		   \\
\hline
20.1                               &  LO: 1.35            & $\pm 0.10$  	    \\
                                      &  NLO:  2.00        & $\pm 0.12$  	    \\
\hline
40.2                               &  LO: 7.27            & $\pm 0.07$  	    \\
                                      &  NLO: 8.31         & $\pm 0.08$  	    \\
\hline
\underline{80.4}         &   LO: 19.93         & $\pm 0.05$  	    \\
                                      &  NLO:  21.12      & $\pm 0.05$  	    \\
\hline
160.8                             &   LO: 29.46        & $\pm 0.04$  	    \\
                                      &  NLO:  30.49     & $\pm 0.04$  	    \\
\hline
321.6                              &   LO: 36.29        & $\pm 0.04$  	    \\
                                      &  NLO:  37.29      & $\pm 0.04$  	    \\
\hline
643.2                              &  LO: 43.07         & $\pm 0.05$  	    \\
                                      &  NLO: 44.61       & $\pm 0.04$  	    \\
\hline
1286.4	                         &  LO:  52.43         & $\pm 0.06$  	    \\
                                      &  NLO: 55.40       & $\pm 0.04$  	    \\
\hline\hline
\end{tabular}       
\end{center}
\caption{\label{I-parton-LVL:Tab_Fit_MRST} The MRST $A_{C}^{Fit}$ table with $\delta A_{C}^{Fit}$ calculated using Eq. \ref{Fit_Correl_Uncert}. The MRST2007lomod PDF is used at LO and the MRST2004nlo one is used at NLO.}
\end{table}

%%%%%%%%%%%%%%%%%%%%%%%%%%%%%%%%%%%%%%%%%%%%%%%%%%%%%%%%%%%%%%%%%%%%%%%%%%%%%%%%%%%%%%%%%%%%%%%%%%%%%%%%%%
\newpage
\vspace*{1.5mm}
\subsubsection{\label{I-parton-LVL:CTEQ} $A_{C}(W^{\pm}\to e^{\pm}\nu_{e})$ Template Curves for CTEQ6}
\vspace*{0.5mm}

\par
The theoretical  CTEQ6L1 and CTEQ6.1 $A_{C}$ template curves are obtained from the signed cross sections used for table \ref{I-parton-LVL:Tab_CTEQ}.

\begin{figure}[h]
\begin{center}
\includegraphics[scale=0.35]{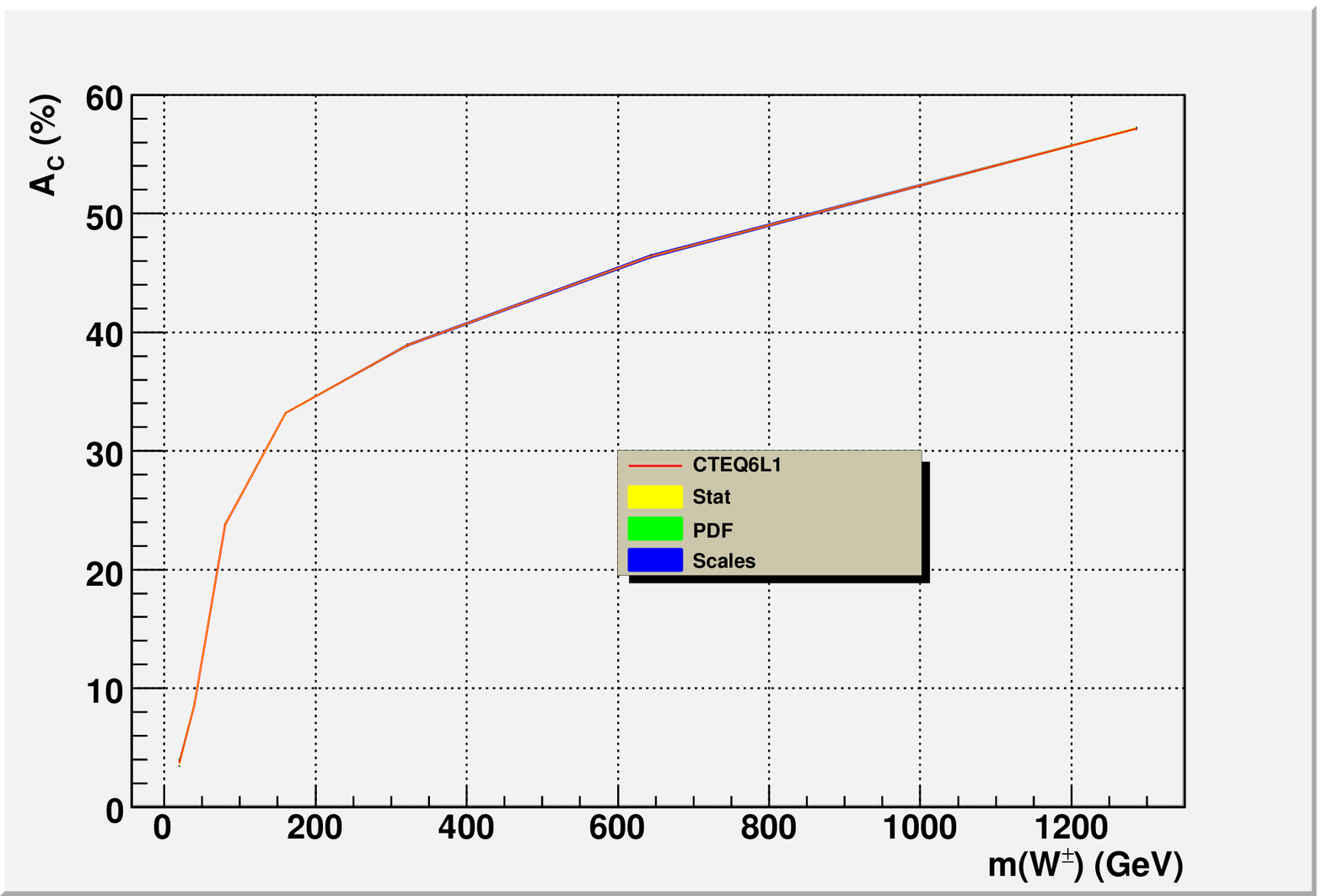}
\includegraphics[scale=0.35]{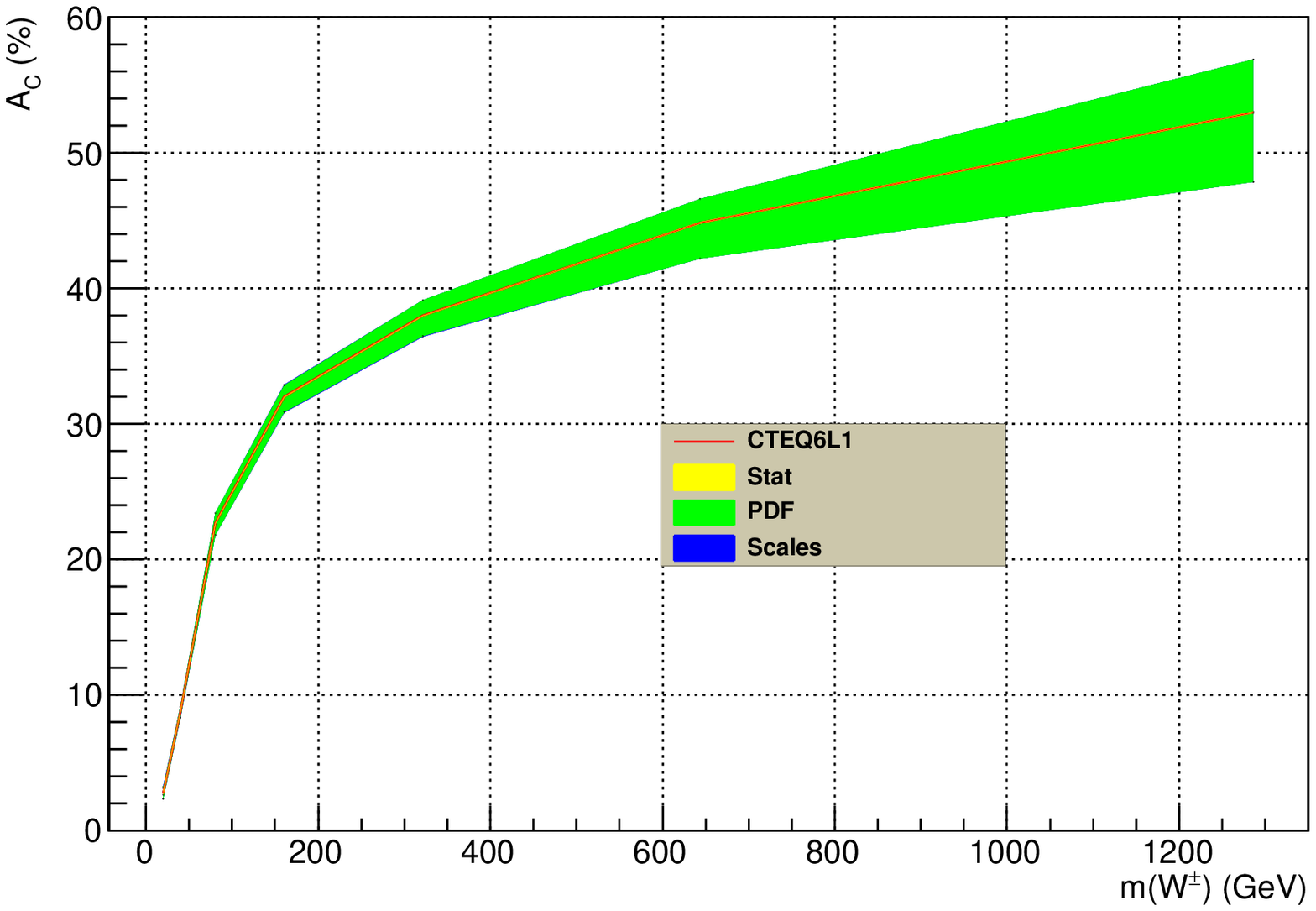}\\
\includegraphics[scale=0.35]{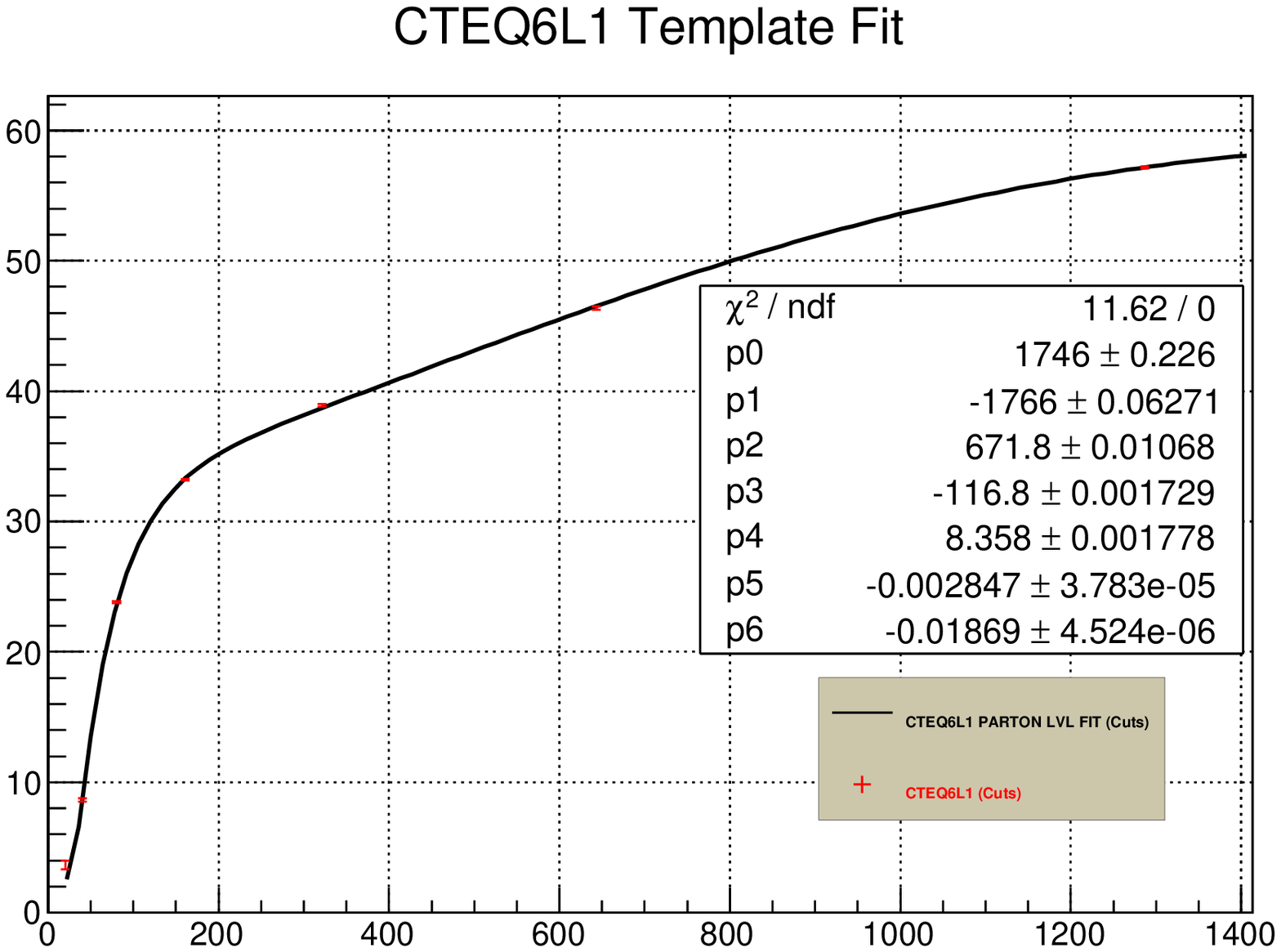}
\includegraphics[scale=0.35]{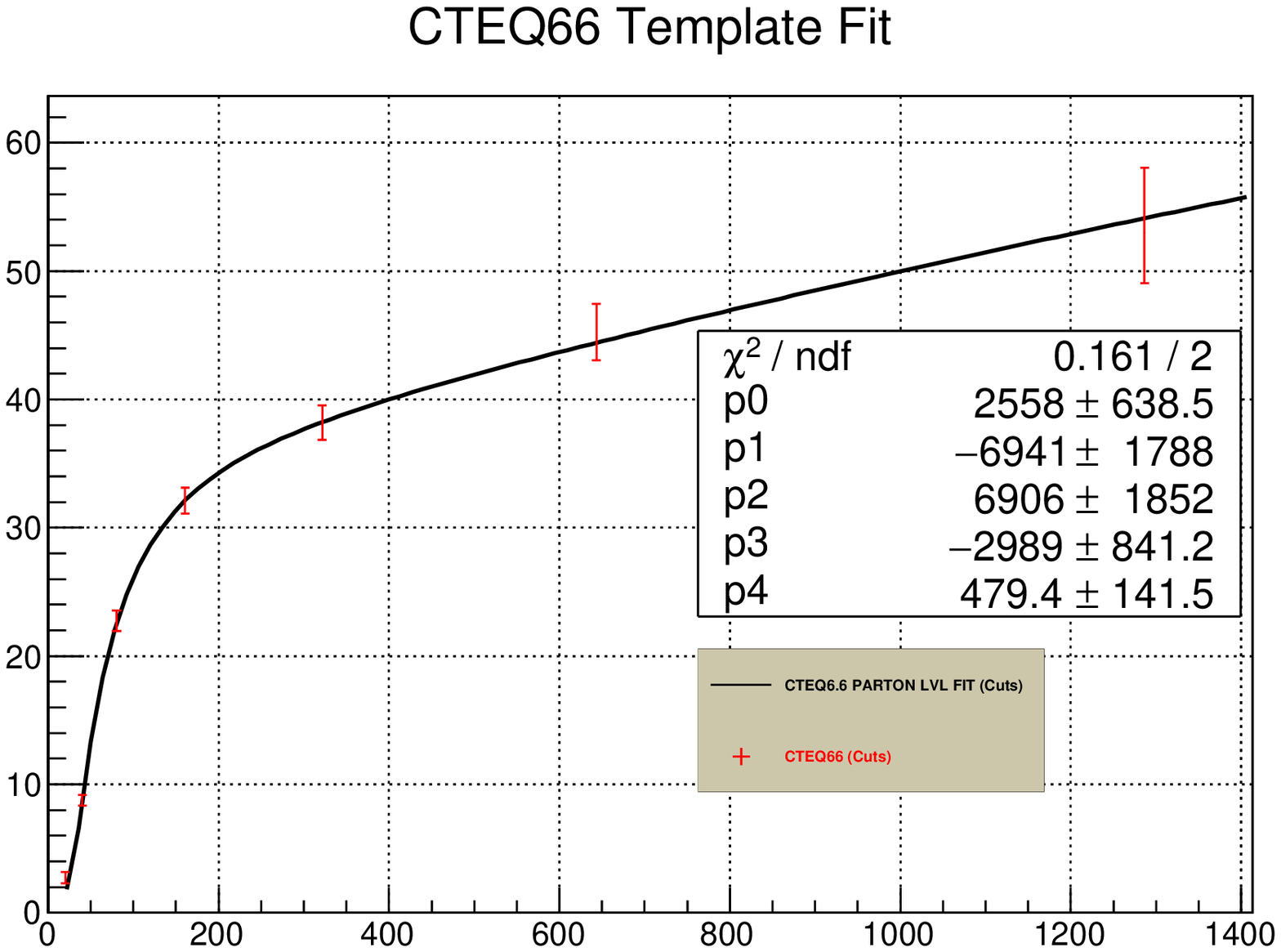}\\
\includegraphics[scale=0.35]{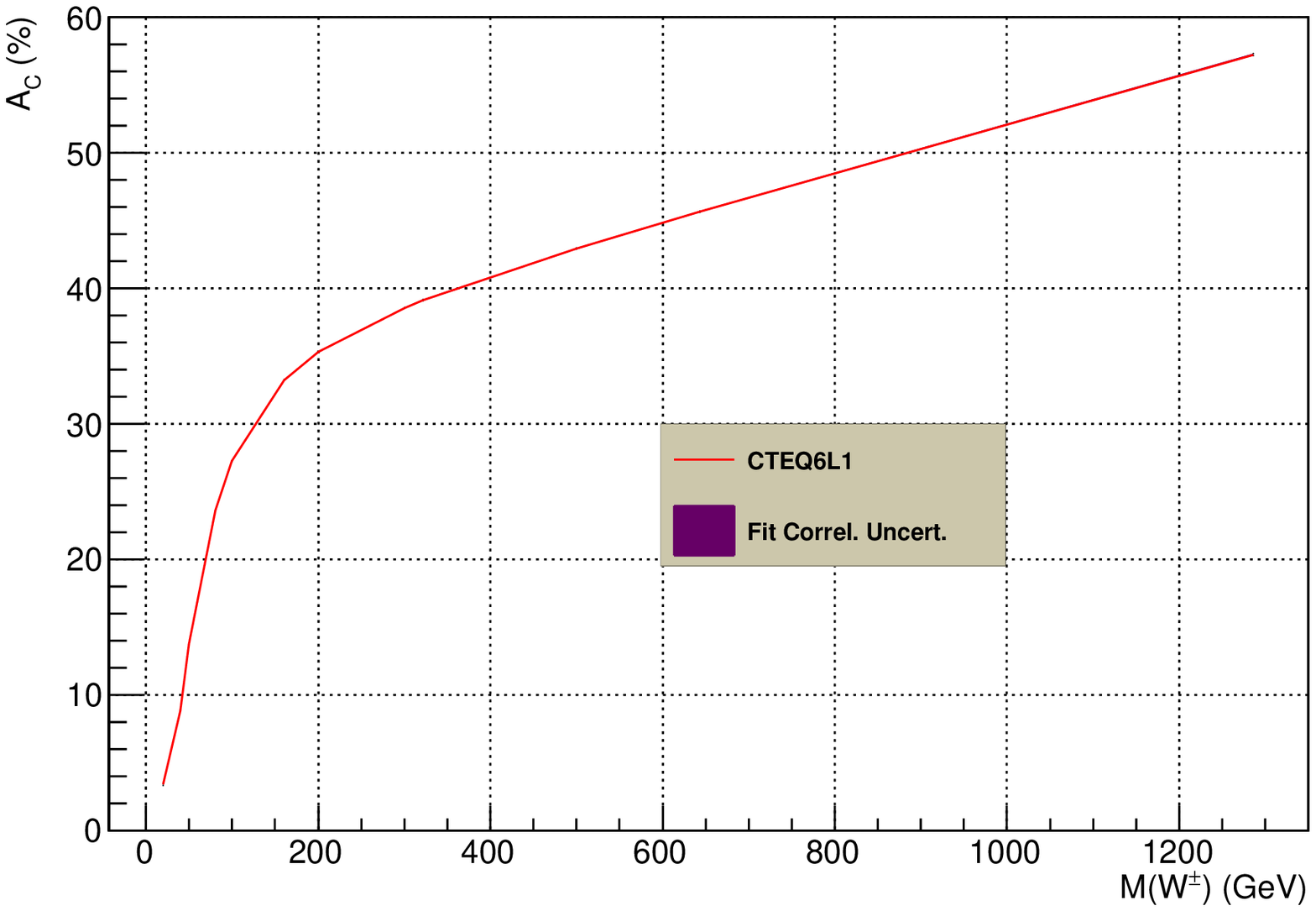}
\includegraphics[scale=0.35]{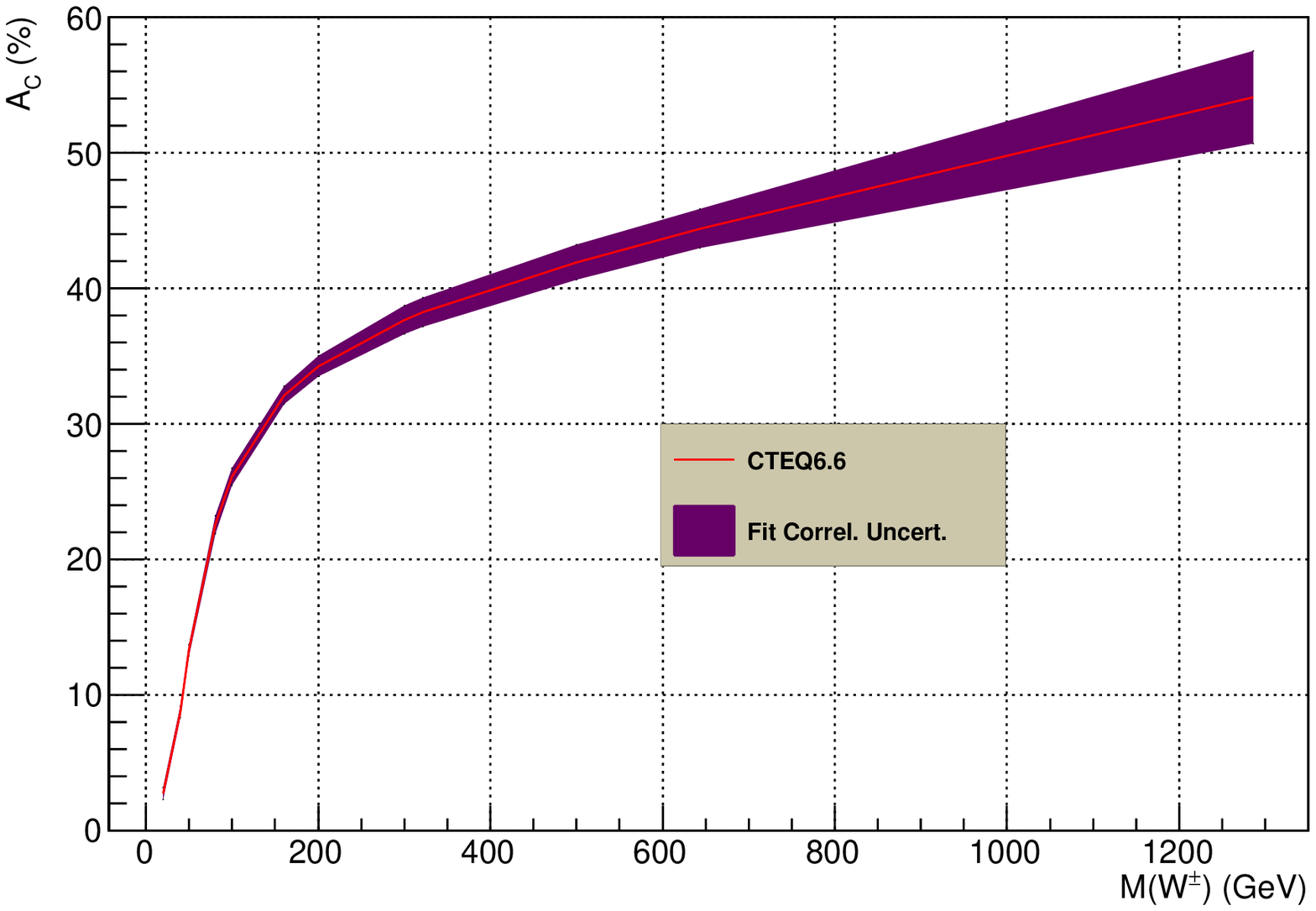}
\end{center}
\caption{\label{I-parton-LVL:Fig2} The theoretical CTEQ6 $A_{C}$ template curves at LO with CTEQ6L1 (LHS)
and NLO with the CTEQ6.6 (RHS). The raw curve with its uncertainty bands,  the corresponding fitted curve  and
the fitted curve with the correlations between the fit parameters uncertainties are displayed on the top, the middle and
the bottom rows, respectively.}
\end{figure}

\begin{table}[h]
\begin{center}
\begin{tabular}{|c|c|c|c|c|c|}
\hline\hline
$\rm M_{W^{\pm}}$   & $A_{C}$         & $\delta (A_{C})_{Stat}$ & $\delta (A_{C})_{Scale}$ & $\delta (A_{C})_{PDF}$ & $\delta (A_{C})_{Total}$\\ 
(\rm\ GeV)                     & ($\%$)            & ($\%$)	                       & ($\%$)                             & ($\%$)                             & ($\%$)		\\
\hline
20.1                                 &   LO:    3.70     & $\pm 0.24$  	                  & $^{-0.27}_{+0.11}$        &  0.00		                      & $^{+0.36}_{-0.26}$  \\
                                        &   NLO:  2.76    & $\pm 0.11$                     & $^{-0.24}_{-0.13}$         & $^{+0.37}_{-0.39}$        & $^{+0.45}_{-0.43}$ \\
\hline
40.2                                 &   LO:    8.65    & $\pm 0.12$  	                  & $^{ -0.02}_{-0.00}$         &  0.00		                       & $^{+0.12}_{-0.12}$  \\
                                        &   NLO: 8.75    & $\pm 0.07$                      & $^{+0.09}_{-0.09}$         & $^{+0.38}_{-0.41}$       & $^{+0.40}_{-0.43}$ \\
\hline
\underline{80.4}           &    LO:    23.81  & $\pm 0.06$  	               & $^{+0.07}_{-0.06}$          &  0.00		                    & $^{+0.09}_{-0.08}$  \\
                                        &   NLO:  22.67 & $\pm 0.03$                      & $^{+0.14}_{-0.20}$         & $^{+0.74}_{-0.85}$       & $^{+0.75}_{-0.87}$ \\
\hline
160.8                               &   LO:    33.21  & $\pm 0.05$  	                & $^{+0.01}_{-0.00}$          &  0.00		                     & $^{+0.05}_{-0.05}$  \\
                                        &   NLO:  31.99 & $\pm 0.02$                      & $^{+0.23}_{-0.24}$          & $^{+0.86}_{-1.11}$       & $^{+0.89}_{-1.14}$ \\
\hline
321.6                               &   LO:    38.90  & $\pm 0.05$  	               & $^{ -0.09}_{+0.07}$           &  0.00		                      & $^{+0.10}_{-0.09}$  \\
                                        &   NLO:  37.99 & $\pm 0.03$                     & $^{+0.18}_{ -0.18}$           & $^{+1.11}_{-1.52}$       & $^{+1.12}_{-1.53}$ \\
\hline
643.2                               &   LO:    46.38  & $\pm 0.05$  	               & $^{-0.14}_{0.13}$              &  0.00		                        & $^{+0.15}_{-0.14}$  \\
                                        &   NLO:  44.83 & $\pm 0.03$                     & $^{+0.06}_{-0.09}$            & $^{+1.76}_{-2.64}$       & $^{+1.76}_{-2.64}$ \\
\hline
1286.4	                           &   LO:    57.17  & $\pm 0.06$  	              & $^{ -0.06}_{+0.06}$           &  0.00		                     & $^{+0.08}_{-0.08}$  \\
                                        &   NLO:  52.97 & $\pm 0.04$                     & $^{+0.05}_{+0.04}$           & $^{+3.90}_{- 5.10}$      & $^{+3.90}_{-5.10}$ \\
\hline\hline
\end{tabular}       
\end{center}
\caption{\label{I-parton-LVL:Tab_CTEQ} The CTEQ6 $A_{C}$ table with the breakdown of the different sources of theoretical uncertainty. The CTEQ6L1 PDF is used at LO and the CTEQ6.6 one is used at NLO.}
\end{table}

\begin{table}[h]
\begin{center}
\begin{tabular}{|c|c|c|}
\hline\hline
$\rm M_{W^{\pm}}$ & $A_{C}^{Fit}$ & $\delta A^{Fit}_{C}$\\ 
(\rm\ GeV)                   & ($\%$)  &  ($\%$)		   \\
\hline
20.1                                &  LO:  3.40  & $\pm 0.09$  	    \\
                                       &  NLO: 2.76    & $\pm 0.44$  	    \\
\hline
40.2                                &   LO:  8.85     & $\pm 0.06$  	    \\
                                       &  NLO:  8.76   & $\pm 0.42$  	    \\
\hline
\underline{80.4}          & LO:   23.59   & $\pm 0.04$  	    \\
                                      &  NLO: 22.57  & $\pm 0.64$  	    \\
\hline
160.8                             &  LO:   33.24   & $\pm 0.04$  	    \\
                                      &  NLO:  32.11 & $\pm 0.66$  	    \\
\hline
321.6                            & LO:   39.11    & $\pm 0.04$  	    \\
                                     &  NLO:  38.23  & $\pm 1.08$  	    \\
\hline
643.2                           &  LO:  45.67   & $\pm 0.05$    \\
                                    &  NLO:  44.41 & $\pm 1.43$  	    \\
\hline
1286.4	                      & LO:   57.24    & $\pm 0.07$	    \\
                                   &  NLO:  54.11 & $\pm 3.42 $  	    \\
\hline\hline
\end{tabular}       
\end{center}
\caption{\label{I-parton-LVL:Tab_Fit_CTEQ} The CTEQ6 $A_{C}^{Fit}$ table with $\delta A_{C}^{Fit}$ calculated using Eq. \ref{Fit_Correl_Uncert}. The CTEQ6L1 PDF is used at LO and the CTEQ6.6 one is used at NLO.}
\end{table}

%%%%%%%%%%%%%%%%%%%%%%%%%%%%%%%%%%%%%%%%%%%%%%%%%%%%%%%%%%%%%%%%%%%%%%%%%%%%%%%%%%%%%%%%%%%%%%%%%%%%%%%%%%

\vspace*{1.5mm}
\subsubsection{\label{I-parton-LVL:MSTW} $A_{C}(W^{\pm}\to e^{\pm}\nu_{e})$ Template Curves for MSTW2008}
\vspace*{0.5mm}

\par
The theoretical  MSTW2008lo68cl and MSTW2008nlo68cl $A_{C}$ template curves are obtained from the signed cross sections used for table \ref{I-parton-LVL:Tab_MSTW}.

\begin{figure}[h]
\begin{center}
\includegraphics[scale=0.35]{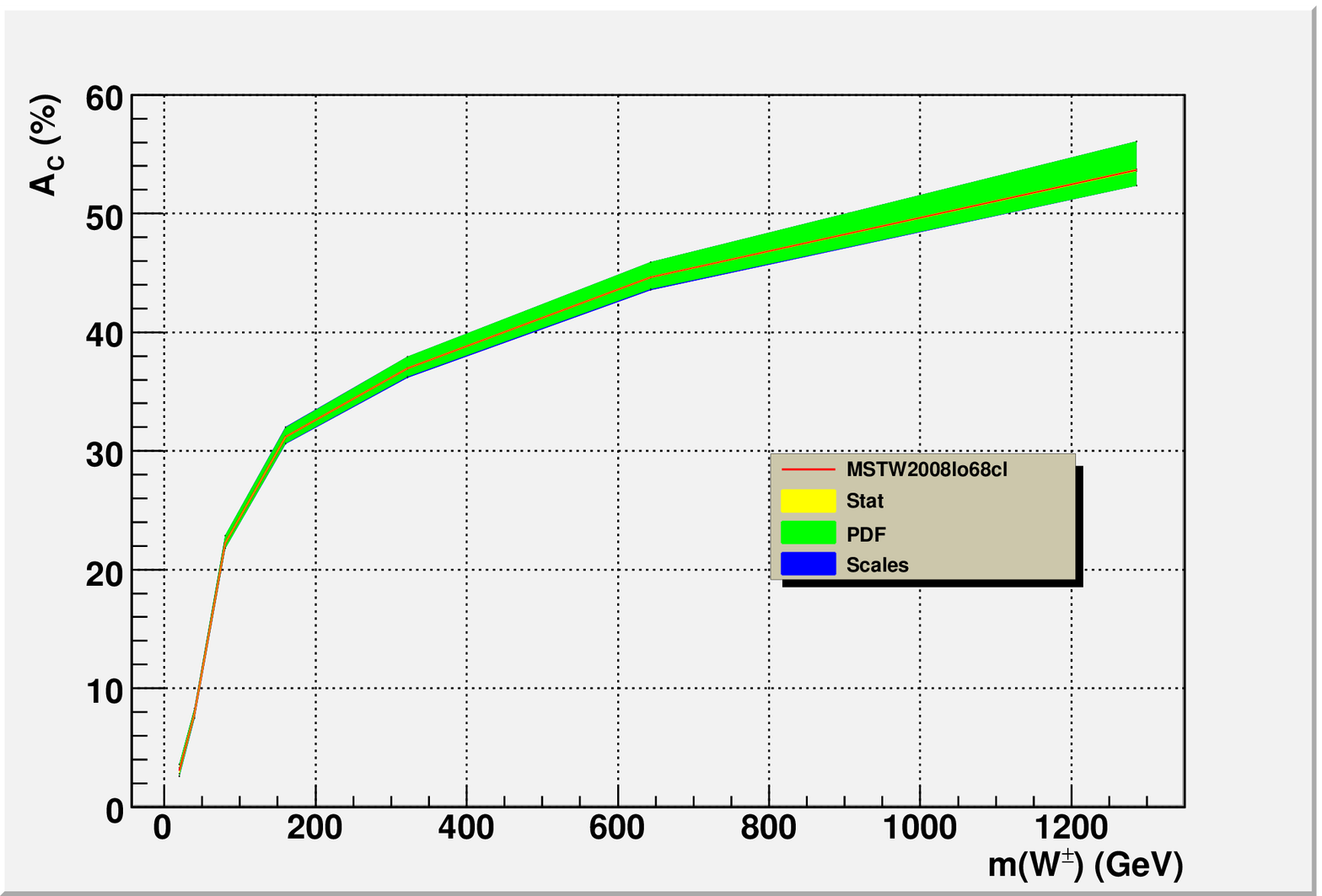}
\includegraphics[scale=0.35]{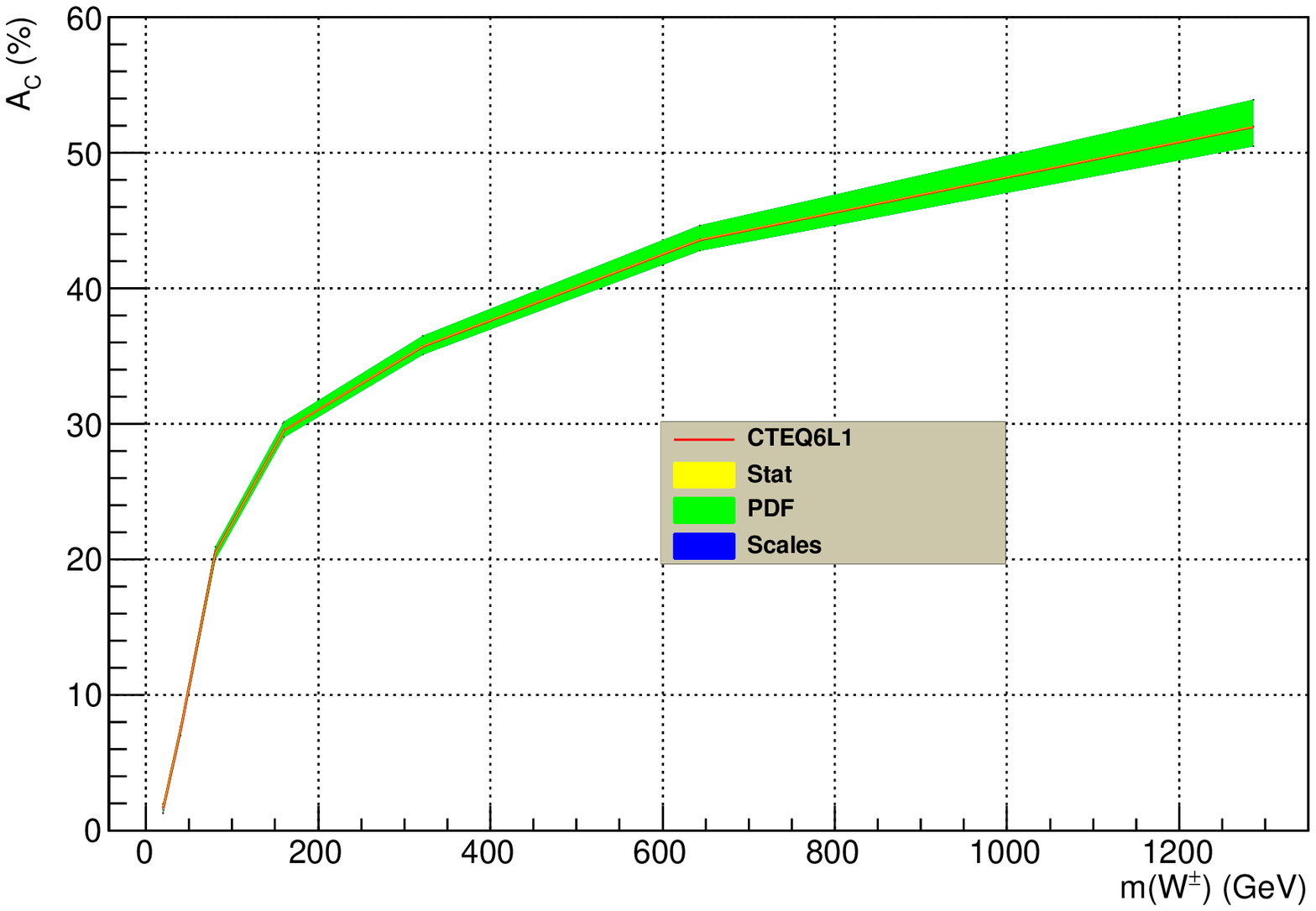}\\
\includegraphics[scale=0.35]{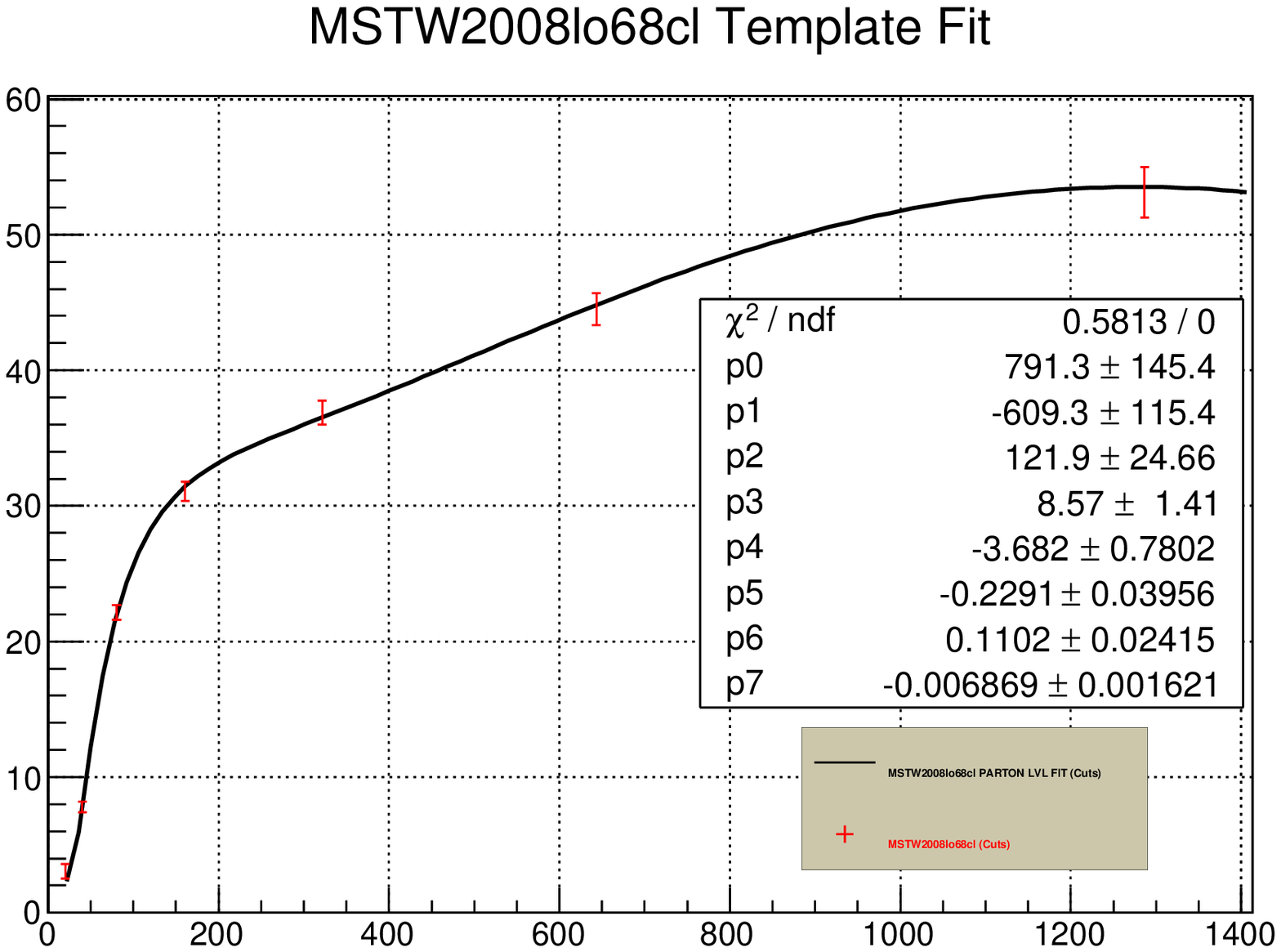}
\includegraphics[scale=0.35]{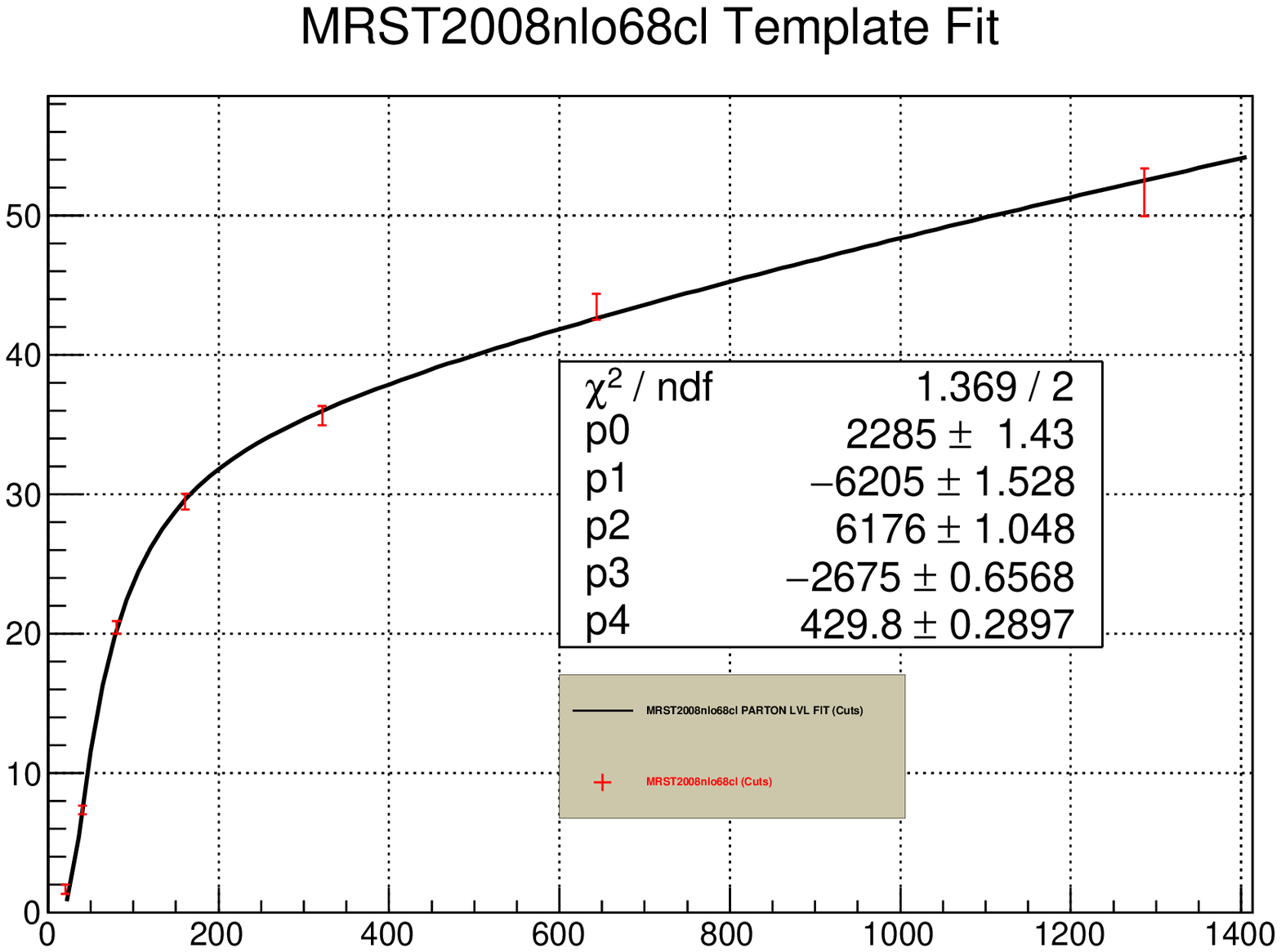}\\
\includegraphics[scale=0.35]{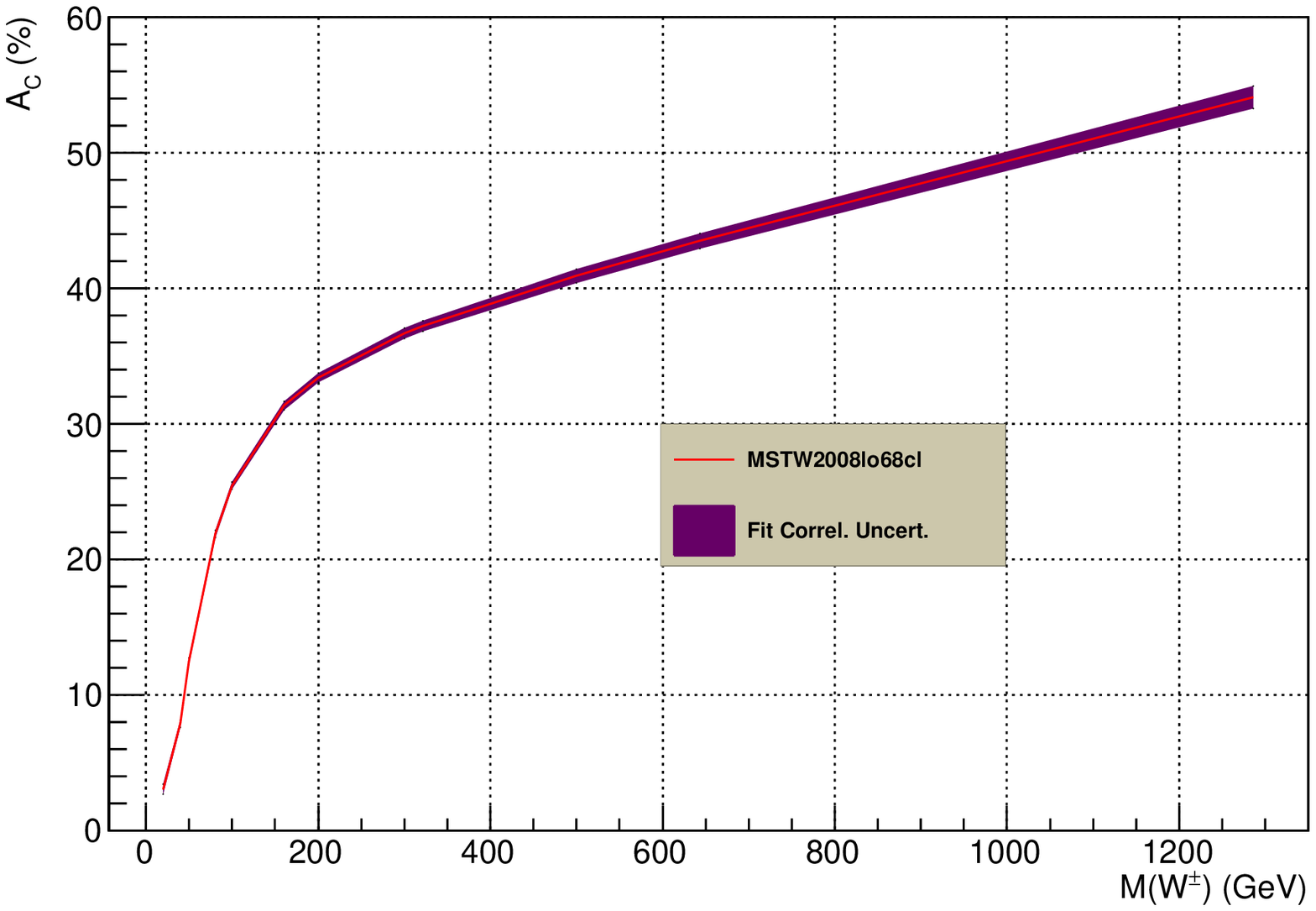}
\includegraphics[scale=0.35]{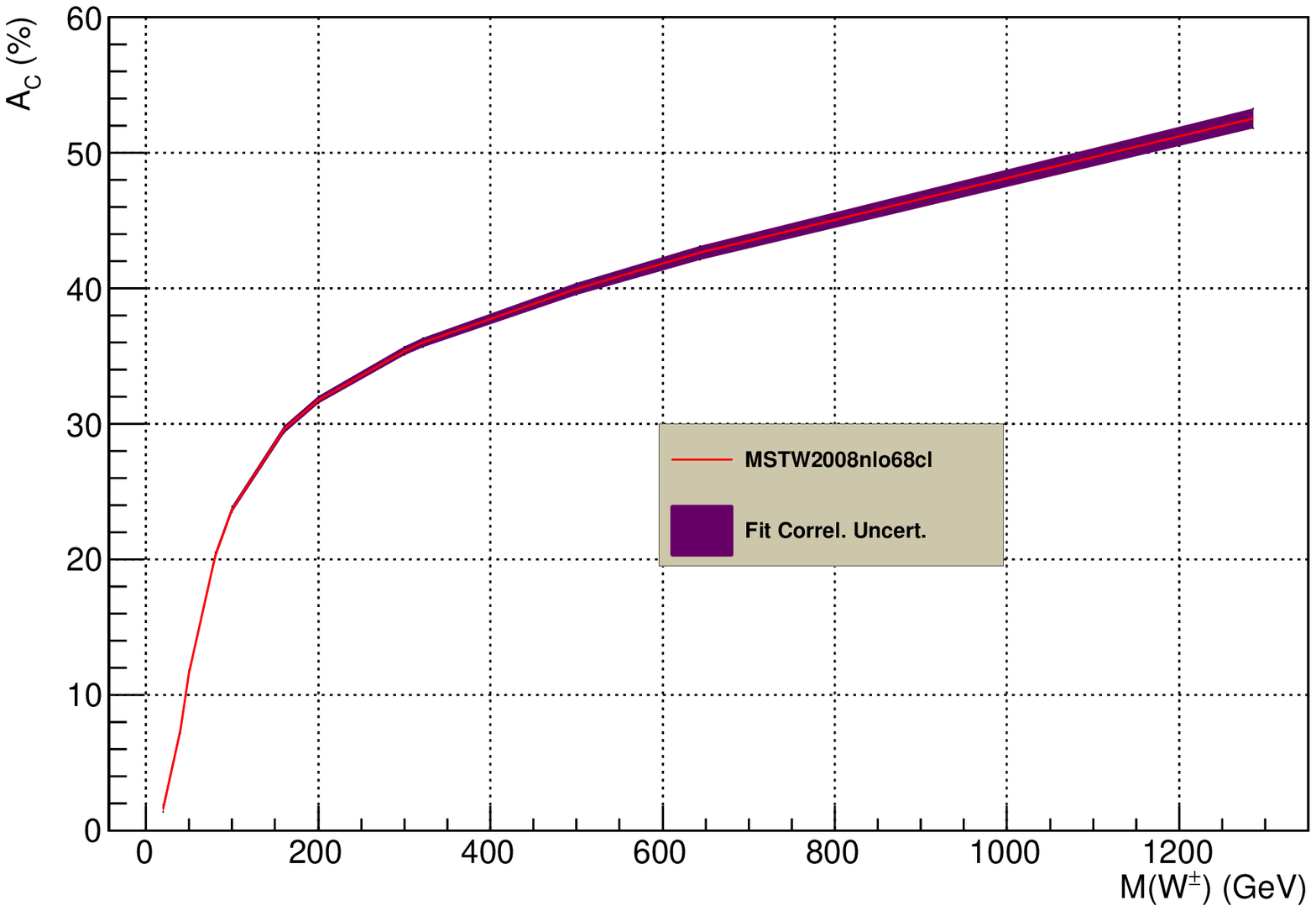}
\end{center}
\caption{\label{I-parton-LVL:Fig3} The theoretical MSTW2008 $A_{C}$ template curves at LO with MSTW2008lo68cl (LHS)
and NLO with the MSTW2008nlo68cl (RHS). The raw curve with its uncertainty bands and the corresponding fitted curve are displayed on the LHS and on the RHS, respectively.}
\end{figure}

\begin{table}[h]
\begin{center}
\begin{tabular}{|c|c|c|c|c|c|}
\hline\hline
$\rm M_{W^{\pm}}$  & $A_{C}$          & $\delta (A_{C})_{Stat}$ & $\delta (A_{C})_{Scale}$ & $\delta (A_{C})_{PDF}$ & $\delta (A_{C})_{Total}$\\ 
(\rm\ GeV)                    & ($\%$)             & ($\%$)	                       & ($\%$)                              & ($\%$)                            & ($\%$)		\\
\hline
20.1                                 &   LO:     3.07    & $\pm $0.24  	                  & $^{-0.21}_{+0.14}$         &  $^{+0.46}_{-0.40}$      & $^{+0.56}_{-0.49}$\\
                                        &   NLO:  1.64    & $\pm  0.12$                    & $ ^{-0.08}_{-0.17}$         & $ ^{+0.29}_{-0.31}$      & $^{+0.32}_{-0.37}$ \\
\hline
40.2                                &   LO:  7.85       & $\pm$0.12  	                  & $^{+0.10}_{+0.07}$        &  $^{+0.43}_{-0.33}$      & $^{+0.46}_{-0.36}$\\
                                       &   NLO:   7.35   & $\pm  0.07$                     & $ ^{+0.05}_{-0.06}$       & $^{+0.30}_{-0.33}$       & $^{+0.31}_{-0.34}$ \\
\hline
\underline{80.4}         &   LO:  22.24      & $\pm$0.06  	               & $^{+0.15}_{+0.13}$        &  $^{+0.64}_{-0.42}$      & $^{+0.66}_{-0.44}$\\
                                      &   NLO:  20.47  & $\pm  0.03$                     & $ ^{-0.06}_{-0.01}$         & $^{+0.48}_{-0.46}$        & $^{+0.48}_{-0.46}$ \\
\hline
160.8                            &   LO:  31.19      & $\pm$0.05  	               & $^{+0.21}_{+0.19}$        &  $^{+0.78}_{-0.53}$       & $^{+0.81}_{-0.57}$\\
                                     &   NLO:  29.52   & $\pm  0.03$                    & $ ^{-0.10}_{+0.02}$        &  $^{+0.62}_{-0.51}$       & $^{+0.63}_{-0.51}$ \\
\hline
321.6                            &   LO:  36.96     & $\pm$0.05  	               & $^{+0.16}_{+0.33}$        &  $^{+0.96}_{-0.70}$      & $^{+0.97}_{-0.77}$\\
                                     &   NLO:  35.73  & $\pm 0.03$                      & $^{-0.05}_{-0.05}$          & $ ^{+0.76}_{-0.59}$      & $^{+0.76}_{-0.59}$ \\
\hline
643.2                            &   LO:  44.63     & $\pm$0.06  	               & $^{+0.17}_{+0.41}$         &  $^{+1.28}_{-0.96}$     & $^{+1.29}_{-1.05}$\\
                                     &   NLO: 43.58   & $\pm  0.03$                      & $^{ -0.08}_{ -0.03}$         &  $^{+1.05}_{-0.78}$     & $^{+1.05}_{-0.78}$ \\
\hline
1286.4	                        &   LO:  53.66     & $\pm$0.07  	              & $^{+0.31}_{+0.33}$         &  $^{+2.39}_{-1.28}$     & $^{+2.42}_{-1.32}$\\
                                     &   NLO:  51.92  & $\pm 0.04$                       & $^{+0.03}_{+0.02}$         &  $^{+1.99}_{-1.45}$     & $^{+1.99}_{-1.45}$\\
\hline\hline
\end{tabular}       
\end{center}
\caption{\label{I-parton-LVL:Tab_MSTW} The MSTW2008lo68cl $A_{C}$ table with the breakdown of the different sources of theoretical uncertainty. The MSTW2008lo68cl PDF is used at LO and the MSTW2008nlo68cl one is used at NLO.}
\end{table}

\newpage
In this case, the PDF uncertainty is provided and it turns out to be the dominant source of theoretical uncertainty on $A_{C}$.

\begin{table}[h]
\begin{center}
\begin{tabular}{|c|c|c|}
\hline\hline
$\rm M_{W^{\pm}}$ & $A_{C}^{Fit}$ & $\delta A^{Fit}_{C}$\\ 
(\rm\ GeV)                   & ($\%$)  &  ($\%$)		   \\
\hline
20.1                               &  LO:  3.05   & $\pm 0.38$  	    \\
                                      &  NLO: 1.63    & $\pm 0.26$  	    \\
\hline
40.2                               & LO:   7.90     & $\pm 0.26$  	    \\
                                      &  NLO:  7.39   & $\pm 0.21$  	    \\
\hline
\underline{80.4}          & LO:    21.89  & $\pm 0.27$  	    \\
                                      &  NLO:  20.30   & $\pm 0.22$  	    \\
\hline
160.8                              &  LO:   31.35   & $\pm 0.31$  	    \\
                                      &  NLO:  29.59   & $\pm 0.26$  	    \\
\hline
321.6                              &  LO:   37.22   & $\pm 0.40$  	    \\
                                      &  NLO:  35.99   & $\pm 0.34$  	    \\
\hline
643.2                              & LO:   43.49   & $\pm 0.57$  	    \\
                                      &  NLO: 42.61    & $\pm 0.51$  	    \\
\hline
1286.4	                           &   LO:   54.08  & $\pm 0.83$  	    \\
                                      &  NLO:  52.53   & $\pm 0.74$  	    \\
\hline\hline
\end{tabular}       
\end{center}
\caption{\label{I-parton-LVL:Tab_Fit_MSTW} The MSTW2008lo68cl $A_{C}^{Fit}$ table with $\delta A_{C}^{Fit}$ calculated using equation \ref{Fit_Correl_Uncert}.
The MSTW2008lo68cl PDF is used at LO and the MSTW2008nlo68cl one is used at NLO.}
\end{table}

%%%%%%%%%%%%%%%%%%%%%%%%%%%%%%%%%%%%%%%%%%%%%%%%%%%%%%%%%%%%%%%%%%%%%%%%%%%%%%%%%%%%%%%%%%%%%%%%%%%%%%%%%%

\vspace*{1.5mm}
\subsubsection{\label{I-parton-LVL:Comp} Comparing the Different $A_{C}$ Template Curves}
\begin{figure}[h]
\begin{center}
\includegraphics[scale=0.35]{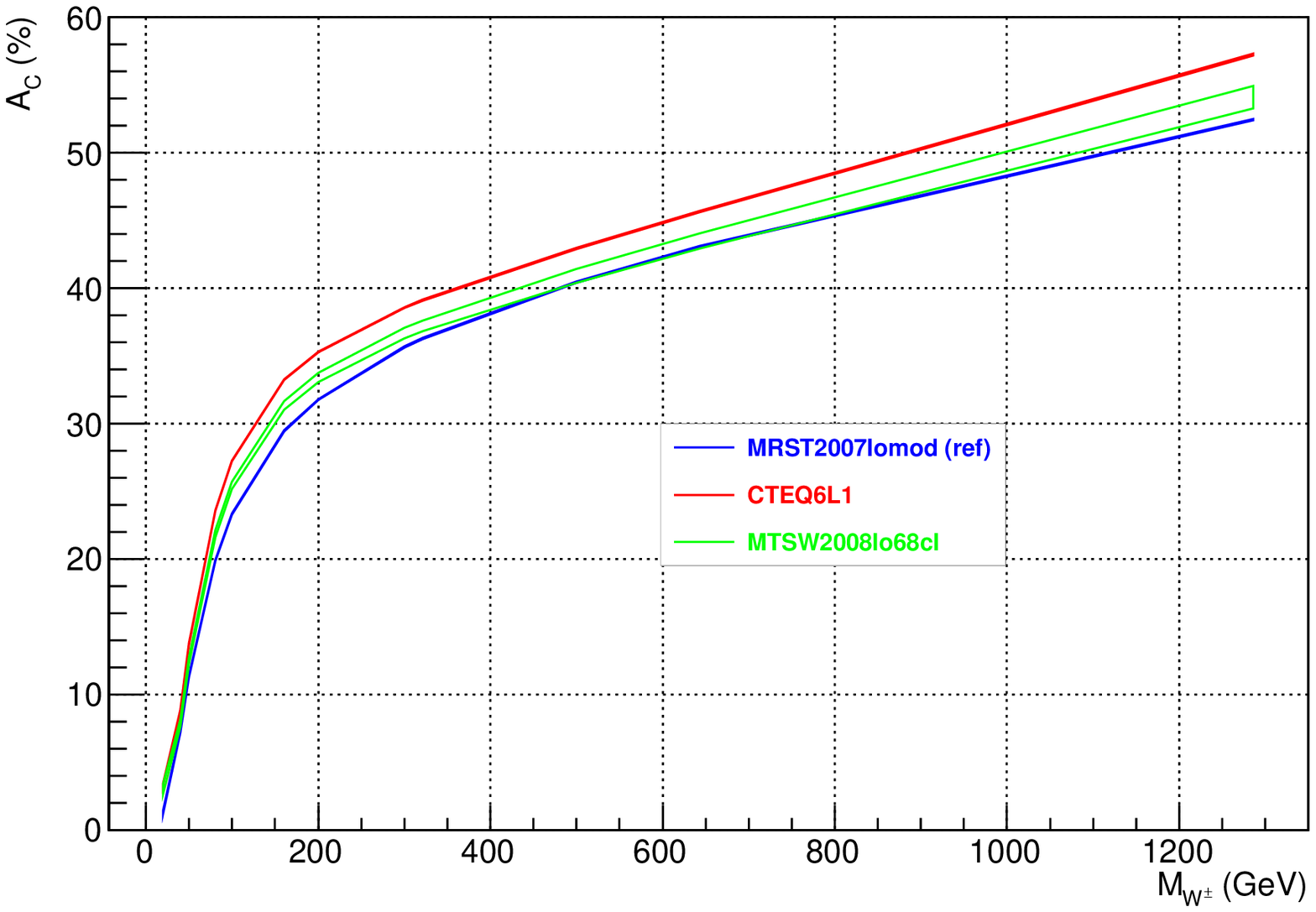}
\includegraphics[scale=0.35]{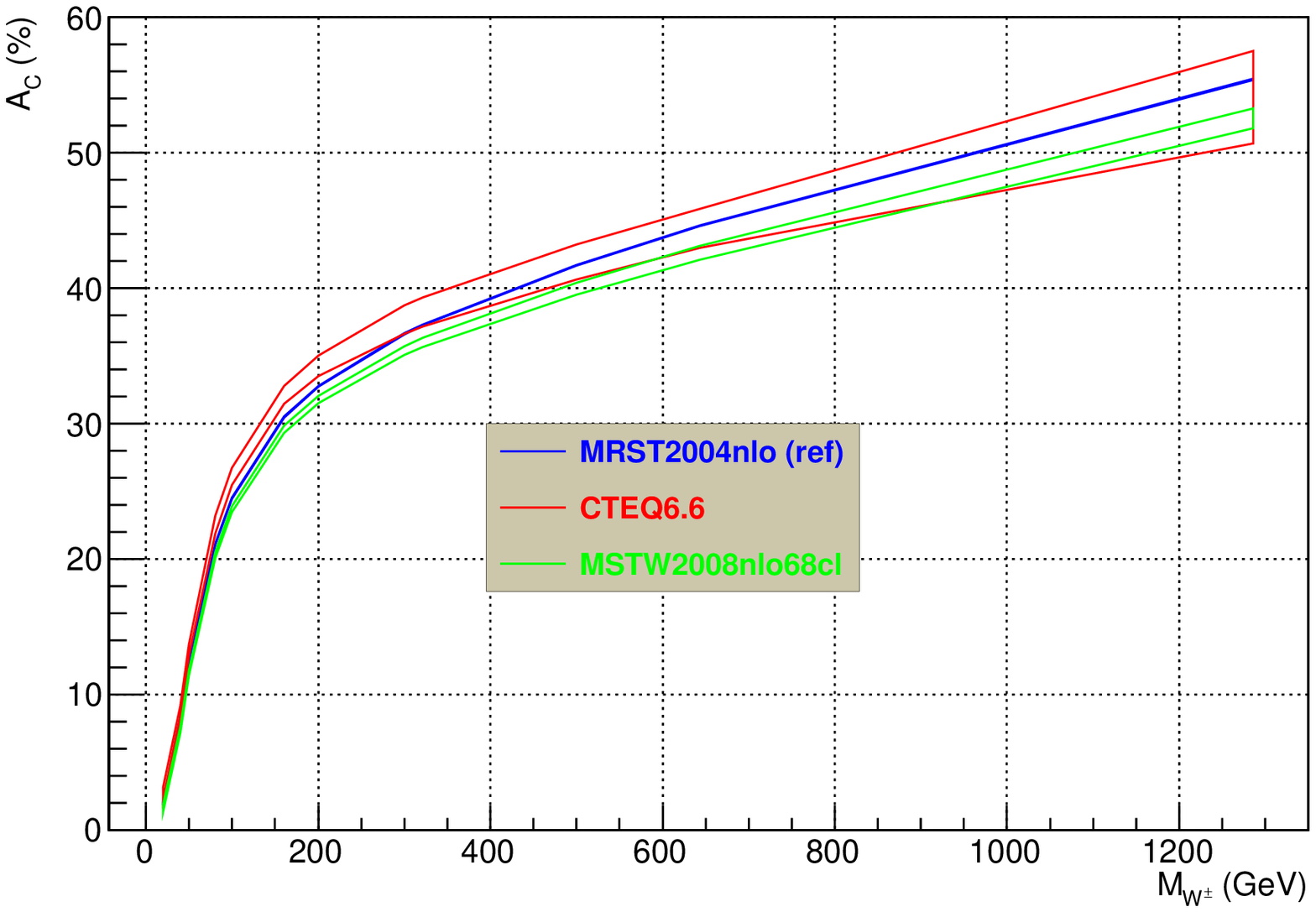}\\
\includegraphics[scale=0.35]{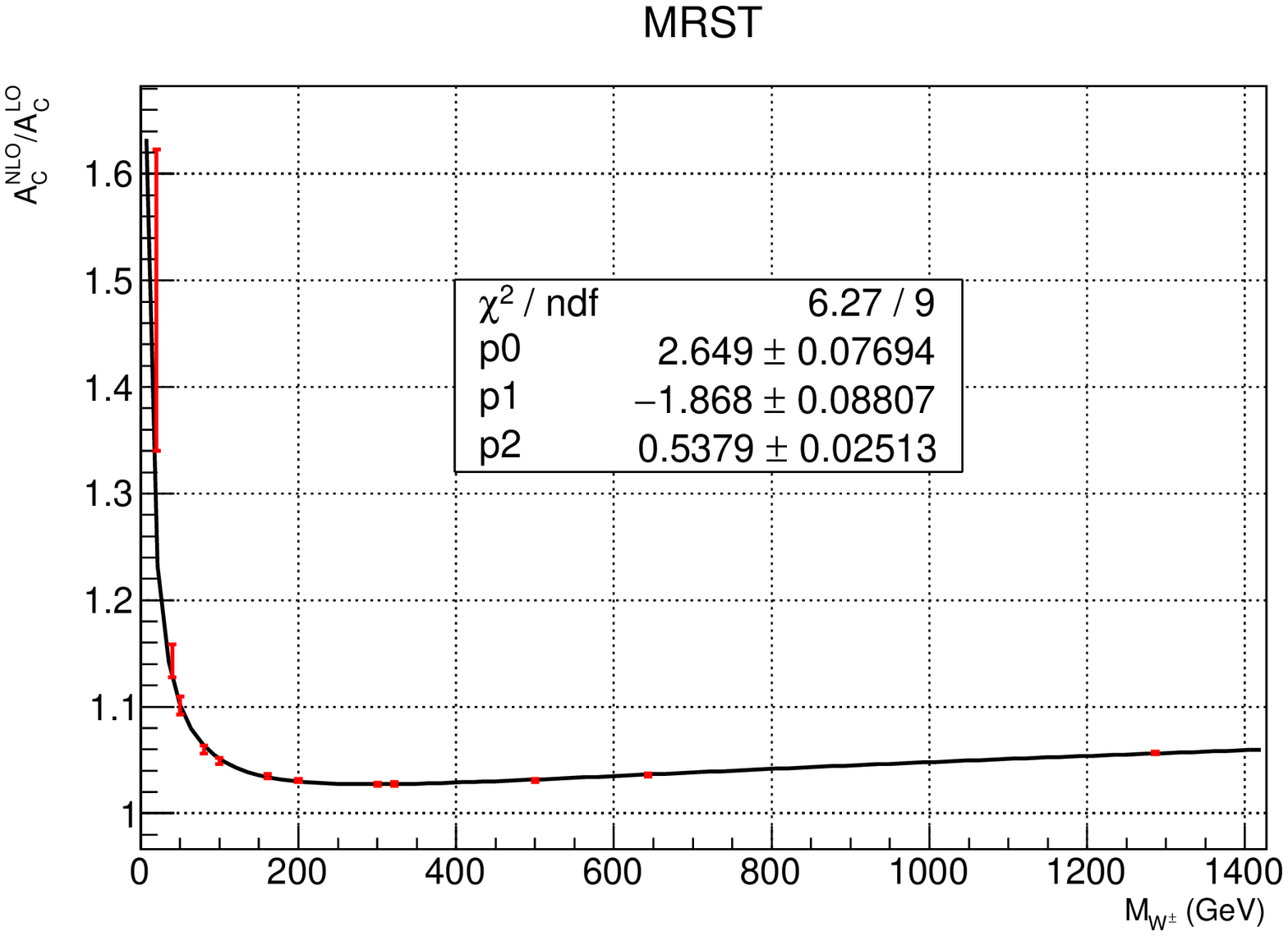}
\includegraphics[scale=0.35]{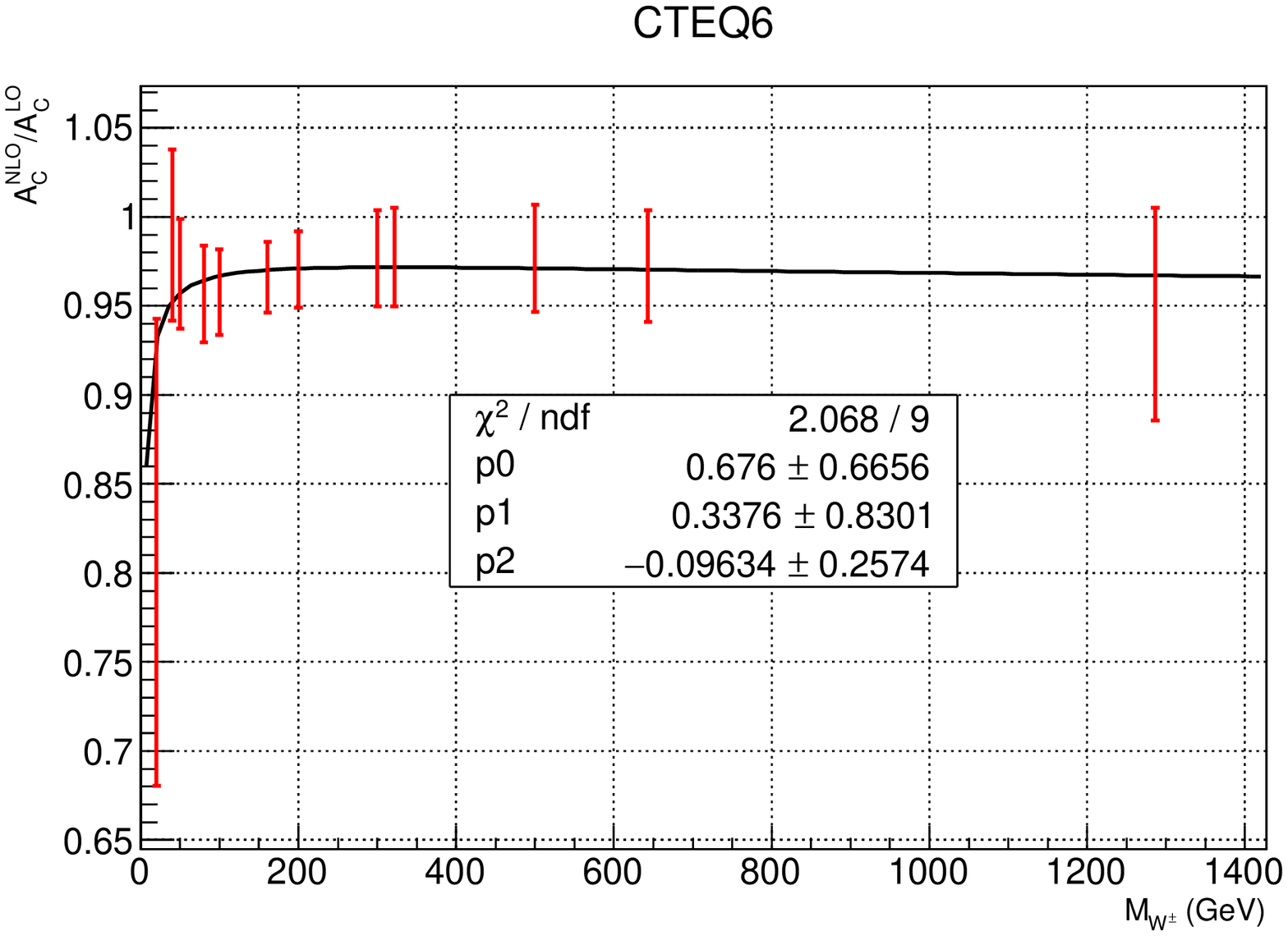}\\
\includegraphics[scale=0.35]{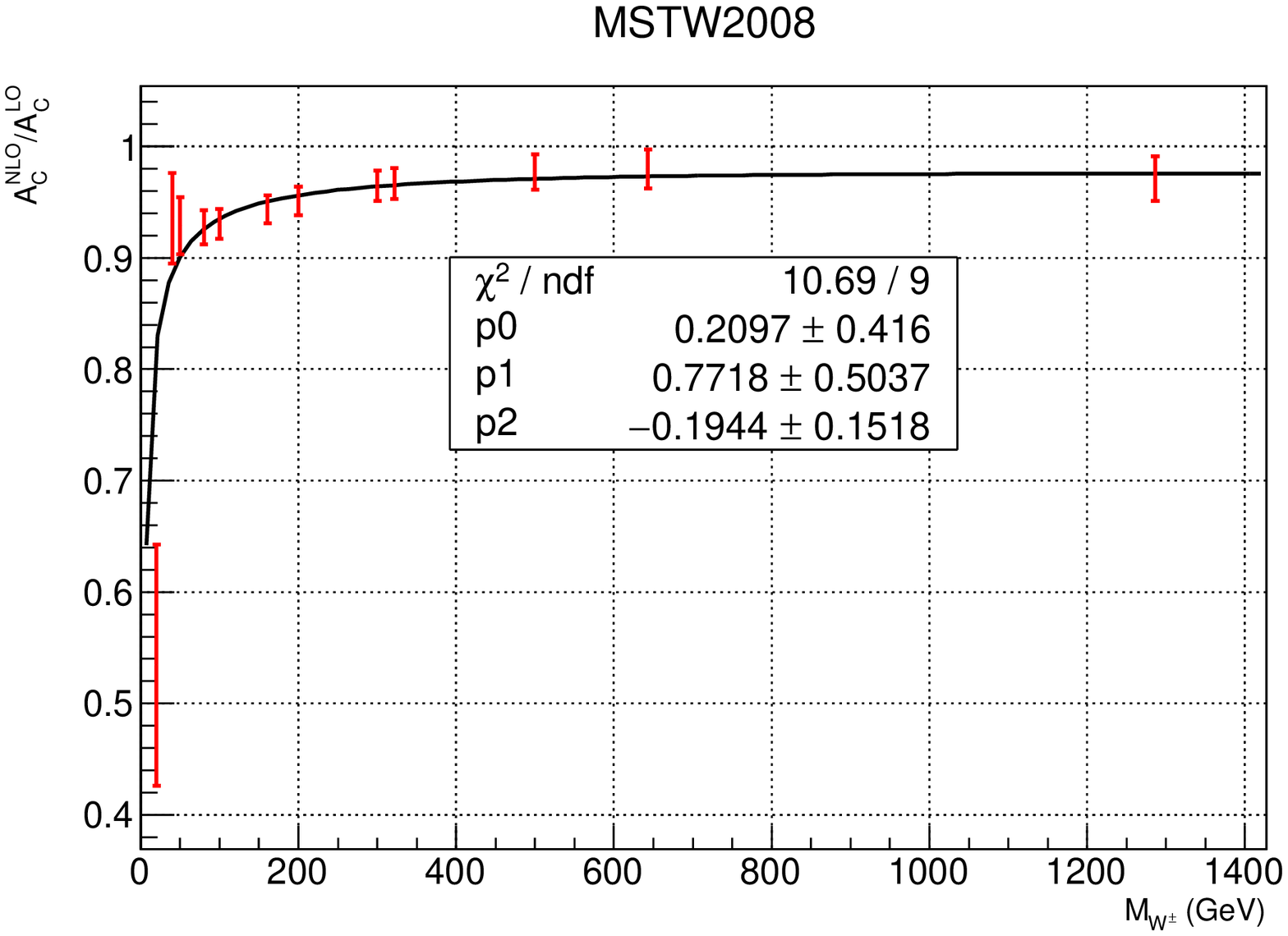}
\end{center}
\caption{\label{I-parton-LVL:Fig4} Comparison between the $A_{C}$ template curves.
The top LHS plot compares the LO PDFs: MRST2007lomod (blue, ref. curve), CTEQ6L1 (red), MSTW2008lo68cl (green).
The top RHS plot compares the NLO PDFs: MRST2004nlo (blue, ref. curve), CTEQ6.6 (red), MSTW2008nlo68cl (green).
The middle and the bottom rows display the $\frac{A_{C}^{NLO}}{A_{C}^{LO}}$ fitted by the same functional forms as the $A_{C}^{LO}$
template curves.}
\end{figure}

\vspace*{0.5mm}
\noindent
At this stage, it's interesting to compare the $A_{C}$ template curves produced with different PDFs using MCFM.  From figure \ref{I-parton-LVL:Fig4} 
we can see that the $A_{C}$ of the different PDF used at LO and at NLO are in agreement at the $\pm 2\sigma$ level, provided that we switch the reference to a PDF set containing uncertainty PDFs. This figure also displays the $\frac{A_{C}^{NLO}}{A_{C}^{LO}}$ ratios for the three families of PDFs used. These ratios are almost flat
with respect to $M_{W^{\pm}}$ over the largest part of our range of interest. However at the low mass ends they vary rapidly. As we illustrate in the Appendix \ref{AppendixA}, these integral charge asymmetry ratios can be fitted by the same functional forms as the $A_{C}^{LO}$ and $A_{C}^{NLO}$.

\subsection{Experimental Measurement of $A_{C}(W^{\pm}\to\ell^{\pm}\nu)$}
\label{sec:Part1-particle-LVL}
\par
The aim of this sub-section is to study the biases on $A_{C}$ due to two different sources: the event selection and the residual background remaining after the latter cuts are applied.

%%%%%%%%%%%%%%%%%%%%%%%%%%%%%%%%%%%%%%%%%%%%%%%%%%%%%%%%%%%%%%%%%%%%%%%%%

\vspace*{1.5mm}
\subsubsection{\label{sec:Part2:Herwig-Settings} Monte Carlo Generation}
\vspace*{0.5mm}

\par
To quantify these biases we generate Monte Carlo (MC) event samples using the following LO generator: Herwig++ v2.5.0 \cite{Gieseke:2011na}. 
We adopt a tune of the underlying event derived by the ATLAS collaboration
\cite{Aad:2011qe} and we use accordingly the MRST2007lomod \cite{Sherstnev:2007nd} PDF.

\par\noindent
Herwig++ mainly uses $2\to 2$ LO ME that we denote in the standard way: $1 + 2\to 3 +4$.
For all the non-resonant processes, 
the production is splitted into bins of $M$, where $M=M(3,4)$ is the
invariant mass of the two outgoing particles.
\par\noindent
For the single vector boson ("V+jets") production, where V stands for $W^{\pm}$ and $\gamma^{*}/Z$, we mix
in the same MC samples the contributions from the pure Drell-Yan process V+0Lp ME and the V+1Lp ME.
For all the SM processes a common cut of $M>10 \rm\ GeV$ is applied. 

%%%%%%%%%%%%%%%%%%%%%%%%%%%%%%%%%%%%%%%%%%%%%%%%%%%%%%%%%%%%%%%%%%%%%%%%%

\par\noindent
All the samples are normalized using the Herwig++ cross section multiplied by a K-factor that includes at least the NLO QCD corrections. We'll denote NLO
(respectively NNLO) K-factor the ratio: $\frac{\sigma_{NLO}}{\sigma_{LO}}$ (respectively $\frac{\sigma_{NNLO}}{\sigma_{LO}}$). We choose not the apply such higher order corrections to the normalization of the following non-resonant inclusive processes:

\begin{itemize}
\item light flavour QCD (denoted QCD LF): $2\to 2$ MEs involving $u/d/s/g$ partons
\item heavy flavour QCD  (denoted QCD HF): $c+\bar{c}$ and $b+\bar{b}$
\item prompt photon productions: $\gamma+jets$ and $\gamma+\gamma$
\end{itemize}

\noindent
Despite their large cross sections these non-resonant processes will turn out to have very low efficiencies and to represent a small fraction of the remaining background in the event selection used in the analyses we perform.

The NNLO K-factors for the $\gamma^{*}/Z(\to \ell^{\pm}\ell^{\mp})$ process are derived from PHOZR \cite{Hamberg:1990np} with  $\mu_{R}=\mu_{F}=M(\ell^{\pm}\ell^{\mp})$ and using the MSTW2008nnlo68cl PDF for $\sigma_{NNLO}$ and the MRST2007lomod one for $\sigma_{LO}$.

\noindent
The top pairs and single top \cite{Campbell:2004ch}\cite{Campbell:2009ss} NLO K-factors are obtained by running MCFM v5.8
using the MSTW2008nlo68cl and the MSTW2008lo68cl PDFs for the numerator and the denominator respectively, with the QCD scales set as follows: $\mu_{R}=\mu_{F}=\hat{s}$.

%%%%%%%%%%%%%%%%%%%%%%%%%%%%%%%%%%%%%%%%%%%%%%%%%%%%%%%%%%%%%%%%%%%%%%%%%

\vspace*{1.5mm}
\subsubsection{\label{sec:Part2:FastSim} Fast Simulation of the Detector Response}
\vspace*{0.5mm}
\par
We use the following setup of Delphes v1.9 \cite{Ovyn:2009tx} to get a fast simulation of the ATLAS detector response as well as a crude emulation of its trigger.
The generated MC samples are written in the HepMC v2.04.02 format \cite{Dobbs:2001ck} and passed through Delphes.

\par\noindent
For the object reconstruction we also use Delphes defaults, with the exception of utilizing the "anti-kT" jet finder \cite{Cacciari:2008gp} with a cone radius of $\Delta R=\sqrt{(\Delta\eta)^2+(\Delta\phi)^2}=0.4$.

%%%%%%%%%%%%%%%%%%%%%%%%%%%%%%%%%%%%%%%%%%%%%%%%%%%%%%%%%%%%%%%%%%%%%%%%%

\vspace*{1.5mm}
\subsubsection{\label{sec:Part2:W-Analysis} Analyses  of the $W^{\pm}\to\ell^{\pm}\nu$ Process}
\vspace*{0.5mm}

\par
We consider only the electron and the muon channels. For these analyses we set  the integrated luminosity to $\int {\cal L} dt =1\ fb^{-1}$.
\par
Instead of trying to derive unreliable systematic uncertainties for these analyses using Delphes, we choose to use realistic values as quoted in actual LHC data analysis publications. We choose the analyses with the largest data samples so as to reduce as much as possible the statistical uncertainties in their measurements
but also to benefit from the largest statistics for the data samples utilized to derive their systematic uncertainties. This choice leads us to quote systematic uncertainties from analyses performed by the CMS collaboration. Namely we use: 
 
 \begin{equation}
 \delta_{Syst} A_{C}(W^{\pm}\to e^{\pm}\nu_{e})=1.0\%
 \label{AC:Syst:Electron}
 \end{equation}
 
 \begin{equation}
 \delta_{Syst} A_{C}(W^{\pm}\to\mu^{\pm}\nu_{\mu})=0.4\%
 \label{AC:Syst:Muon}
\end{equation}

\noindent
The values quoted in equations  \ref{AC:Syst:Electron} and \ref{AC:Syst:Muon} come from references  \cite{Chatrchyan:2012xt} and \cite{Chatrchyan:2013mza}, respectively.

\noindent 
And to get an estimate of the uncertainty on a ratio of number of expected events we use the systematics related to the measurement of the following cross sections ratio
\begin{equation}
\sigma(pp\to W^{\pm}\to\ell^{\pm}\nu_{\ell})/\sigma(pp\to\gamma^{*}/Z\to\ell^{\pm}\ell^{\mp})
\label{Alpha:Syst}
\end{equation}
which amounts to $1.0\%$ \cite{CMS:2011aa}.

%%%%%%%%%%%%%%%%%%%%%%%%%%%%%%%%%%%%%%%%%%%%%%%%%%%%%%%%%%%%%%%%%%%%%%%%%

\vspace*{1.5mm}
\noindent
2.2.4.\ a.\ The\ Electron\ Channel
\vspace*{0.5mm}

\noindent 
2.2.4.\ a.1.\ Event\ Selection\ in\ the\ Electron\ Channel
\par
The following cuts are applied:
\begin{itemize}
\item $p_{T}(e^{\pm}) > 25 \rm\ GeV$
\item $|\eta(e^{\pm})|<1.37\rm\ or\ 1.53<|\eta(e^{\pm})|<2.4$
\item Tracker Isolation: reject events with additional tracks of $p_{T}>2$ GeV within a cone of $\Delta R=0.5$ around the direction of the $e^{\pm}$ track
\item Calorimeter Isolation: the ratio of, the scalar sum of  $E_{T} $ deposits in the calorimeter within a cone of $\Delta R=0.5$ around the direction of the
$e^{\pm}$, to the $p_{T}(e^{\pm})$, must be less than 1.2 
\item $\rlap{\kern0.25em/}E_{T} > 25 \rm\ GeV$
\end{itemize}

\begin{itemize}
\item $M_{T} = \sqrt{2p_{T}(\ell^{\pm})\rlap{\kern0.25em/}E_{T}[1-cos\Delta\phi(\ell^{\pm},\rlap{\kern0.25em/}E_{T})]} > 40 \rm\ GeV$   
\item Reject events with an additional leading isolated muon: $\mu^{\pm}_{1}$ 
\item Reject events with an additional trailing isolated electron: $e^{\pm}_{2}$ 
\item Reject events with an additional second track ($Track_{2}$) such that:
$$
         \begin{cases}
	 Q(e^{\pm}_{1}) = -Q(Track_{2})\\
	 3 < p_{T}(Track_{2}) < 10 \rm\ GeV\\
	M[e_{1}^{\pm},Track_{2}]> 50 \rm\ GeV
         \end{cases}
$$
\end{itemize}

\begin{figure}[h]
\begin{center}
\includegraphics[scale=0.35]{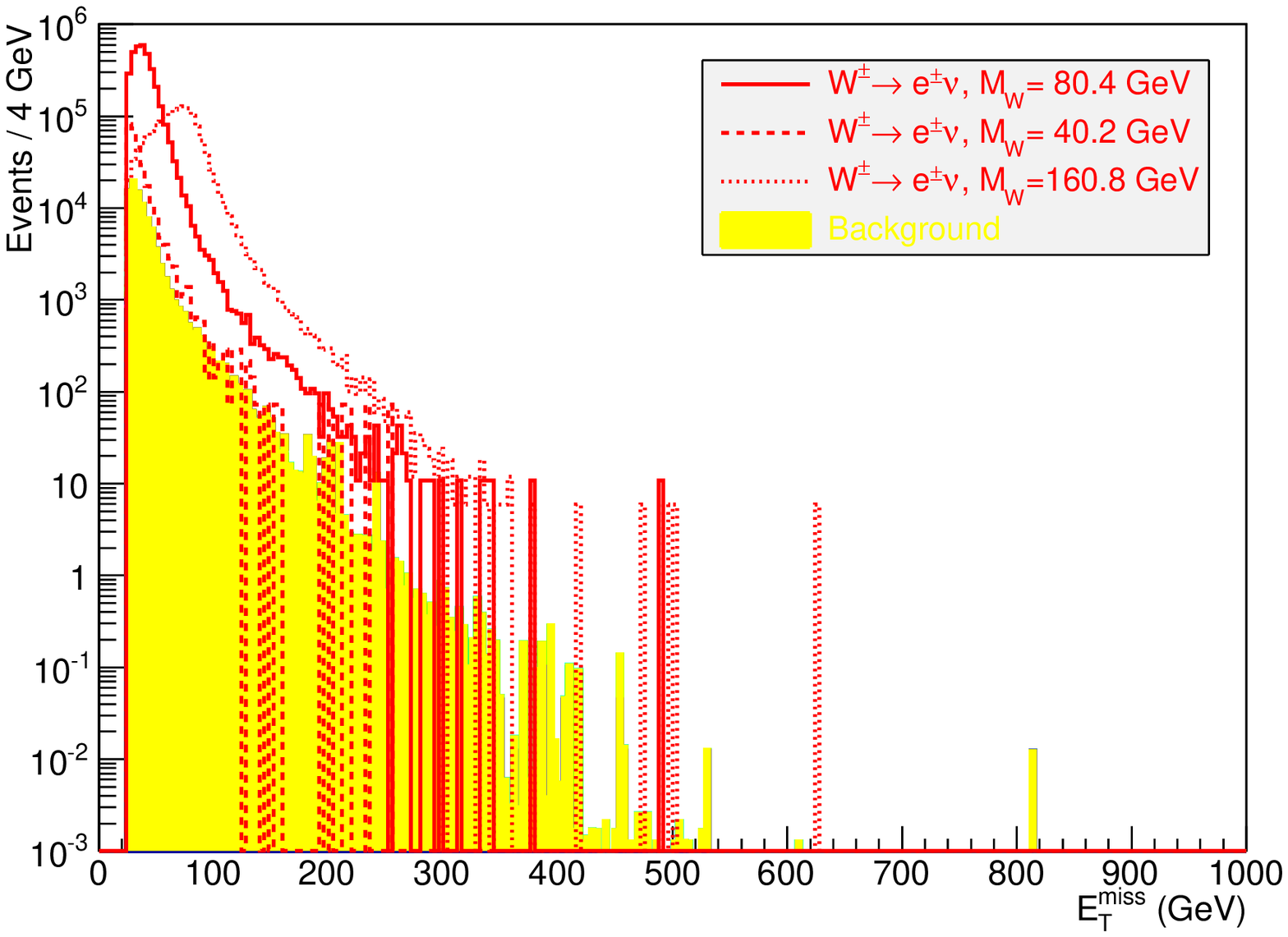}
\includegraphics[scale=0.35]{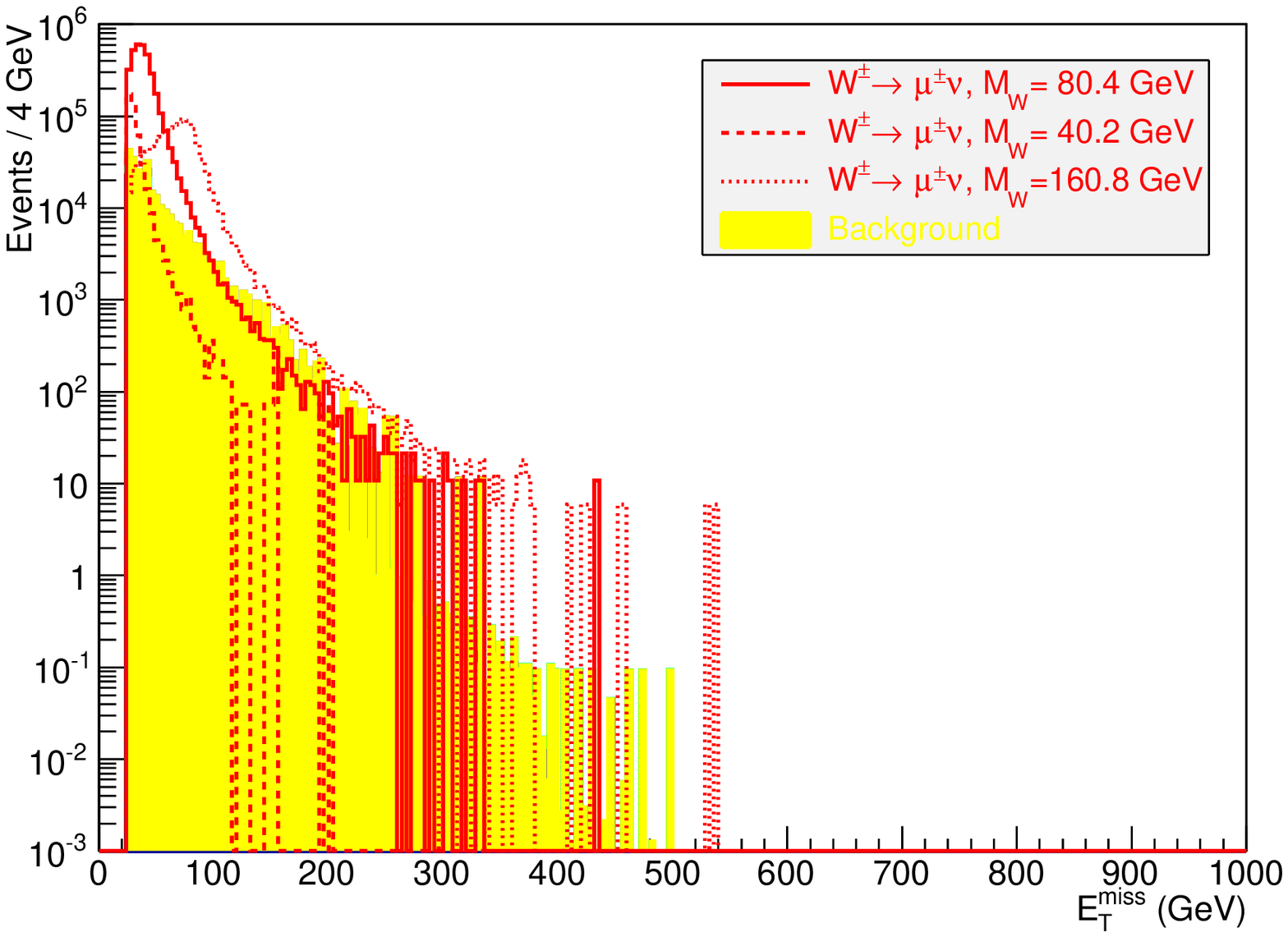}
\end{center}
\caption{\label{sec:Part2:w_lnu_mET_fig} $\rlap{\kern0.25em/}E_{T}$  distribution after the event selection is applied for the $W^{\pm}\to e^{\pm}\nu_{e}$ (LHS) and for the $W^{\pm}\to\mu^{\pm}\nu_{\mu}$ (RHS) analysis.}
\end{figure}

\noindent
The corresponding selection efficiencies and event yields (expressed in thousanths of events) are reported in table \ref{sec:Part2:w_enu_SEL_tab}. Figure \ref{sec:Part2:w_lnu_mET_fig} displays the $\rlap{\kern0.25em/}E_{T}$ distribution after the event selection in the electron channel (LHS) and in the muon channel (RHS).

\begin{table}[h]
\small\begin{center}
\begin{tabular}{|c|c|c|c|}
\hline\hline 
$Process$			& $\epsilon$		& $N_{exp}$          &	$A_{C}\pm\delta A_{C}^{Stat}$	\\ 
				& ($\%$)             	& (k evts)	     & 	($\%$)		\\
\hline\hline
Signal: $W^{\pm}\to e^{\pm}\nu_{e}$& 	        &		&			\\
$M(W^{\pm}) = 40.2 \rm\ GeV$          & $0.81\pm 0.01$       & 290.367     	    & $9.66\pm 1.57$  \\
$M(W^{\pm}) = 60.3 \rm\ GeV$          & $13.69\pm 0.05$      & 	2561.508   & $11.22\pm 0.38$  \\
$M(W^{\pm}) = \underline{80.4} \rm\ GeV$& $29.59\pm 0.04$    & 3343.195	    & $16.70\pm 0.18$  \\
$M(W^{\pm}) = 100.5 \rm\ GeV$         & $39.19\pm 0.07$      & 	2926.093    & $20.77\pm 0.22$  \\
$M(W^{\pm}) = 120.6 \rm\ GeV$         & $44.84\pm 0.07$      & 2357.557	   & $23.19\pm 0.21$  \\
$M(W^{\pm}) = 140.7 \rm\ GeV$         & $48.66\pm 0.07$      & 	1899.820   & $25.29\pm 0.20$  \\
$M(W^{\pm}) = 160.8 \rm\ GeV$         & $51.28\pm 0.07$      & 	1527.360  & $26.87\pm 0.19$  \\
$M(W^{\pm}) = 201.0 \rm\ GeV$         & $54.54\pm 0.07$      & 	1.032    	& $29.06\pm 0.18$  \\
\hline\hline
Background                       &	-               & $91.614\pm 1.706$	      & $10.07\pm 0.15$	\\
\hline
$W^{\pm}\to\mu^{\pm}\nu_{\mu}/\tau^{\pm}\nu_{\tau}/q\bar{q\prime}$ & $0.211\pm 0.003$ & $71.350$  & $12.92\pm 1.25$ \\
\hline
$t\bar t$                        &   $5.76\pm 0.02$     & 	6.600	      & $1.00\pm 0.37$  \\
$t+b,\ t+q(+b)$                  &   $3.59\pm 0.01$     & 	1.926	     & $28.97\pm 0.35$  \\
\hline 
$W+W,\ W+\gamma^{*}/Z,\ \gamma^{*}/Z+\gamma^{*}/Z$ 	& $2.94\pm 0.01$     & 2.331  & $10.65\pm 0.35$ \\
\hline
$\gamma+\gamma,\ \gamma+jets,\ \gamma+W^{\pm},\ \gamma+Z$ & $0.201\pm 0.001$ & 0.759  &  $17.25\pm 0.53$ \\
\hline
$\gamma^{*}/Z$                  &   $0.535\pm 0.001$    &   5.746	      & $4.43\pm 0.23$	\\
\hline
QCD HF                          & $(0.44\pm 0.17)\times 10^{-4}$ & 1.347      &  $14.29\pm 37.41$\\
QCD LF                          & $(0.87\pm 0.33)\times 10^{-4}$ &  1.555     &  $71.43\pm 26.45$ \\
\hline\hline
\end{tabular}       
\end{center}
\caption{\label{sec:Part2:w_enu_SEL_tab} Selection efficiencies, event yields and integral charge asymmetries for the $W^{\pm}\to e^{\pm}\nu_{e}$ analysis.}
\end{table}

\noindent
The non-resonant background processes represent just $\sim 4\%$ of the total background after the event selection, this justifies the approximation of not to include the NLO QCD corrections to their normalizations. 

\vspace{1.0cm}
\noindent 
2.2.4.\ a.2.\ Common\ Procedure\ for\ the Background\ Subtraction\ and the Propagation of the Experimental\ Uncertainty
\par 
If we were to apply such an analysis on real collider data, we would get in the end the measured integral charge asymmetry $A_{C}^{Meas}$ of the data sample passing the selection cuts. And obviously we wouldn't know which event come from which sub-process. Since the MC enables to separate the different contributing sub-processes, it's possible to extract the integral charge asymmetry of the signal (S), knowing that of the total background (B).
\noindent
If we denote $\alpha^{Exp} = \frac{N_{B}^{Exp}}{N_{S}^{Exp}}$ the ratio of the expected number of background events to the expected number of signal events, we can express $A_{C}^{Exp}(S+B)$, the integral charge asymmetry of all remaining events either from signal or from background, with respect to that quantity
for signal only events $A_{C}^{Exp}(S)$, and for background only events $A_{C}^{Exp}(B)$. This writes: 
\begin{equation}
A_{C}^{Exp}(S+B) = \frac{A_{C}^{Exp}(S)+\alpha^{Exp}\cdot A_{C}^{Exp}(B)}{1+\alpha^{Exp}}
\end{equation} 
where the upper script "Exp" stands for "Expected".

\noindent
This formula can easily be inverted to extract $A_{C}^{Exp}(S)$ in what we'll refer to as the "background subtraction equation": 
\begin{equation}
A_{C}^{Exp}(S) = (1+\alpha^{Exp})\cdot A_{C}^{Exp}(S+B)-\alpha^{Exp}\cdot A_{C}^{Exp}(B)
\label{Bkgd_Subtract}
\end{equation}

\noindent
Note that these expressions involve only ratios hence their experimental systematic uncertainty remains relatively small.
\noindent
The uncertainty on $A_{C}^{Exp}(S)$ is calculated by taking account the correlation between the uncertainties of $\alpha^{Exp}$, $A_{C}^{Exp}(B)$, and $A_{C}^{Exp}(S+B)$.
\begin{equation}
\begin{split}
\lbrack\delta A_{C}(S)\rbrack^{2} = \lbrack A_{C}(S+B)-A_{C}(B)\rbrack^{2}\cdot \lbrack\delta\alpha\rbrack^{2} + 
                                                                 (1+\alpha)^{2}\cdot \lbrack\delta A_{C}(S+B)\rbrack^{2} + \alpha^{2}\cdot \lbrack\delta A_{C}(B)\rbrack^{2} \\
								 + 2\cdot\lbrack A_{C}(S+B)-A_{C}(B)\rbrack\cdot (1+\alpha)\cdot COV[\alpha,A_{C}(S+B)] \\
								 - 2\cdot\lbrack A_{C}(S+B)-A_{C}(B)\rbrack\cdot \alpha\cdot COV[\alpha,A_{C}(B)]\\
								 - 2\cdot\alpha\cdot (1+\alpha)\cdot COV[A_{C}(B),A_{C}(S+B)]
\end{split}
\label{Uncert_Bkgd_Subtract}
\end{equation}

\noindent
In order to propagate the experimental uncertainties from equations \ref{AC:Syst:Electron}, \ref{AC:Syst:Muon}, and \ref{Alpha:Syst} to $\delta A_{C}(S)$, we perform
pseudo-experiments running 10,000,000 trials for each. In these trials all quantities involved in the background subtraction equation \ref{Bkgd_Subtract} is allowed to
fluctuate according to a gaussian smearing that has its central value as a mean and its total uncertainty as an RMS. In each of these pseudo-experiments, the signal S and the backrgound B float separately. For each of the events categories (S or B) separately, the numbers of positively and negatively charged events also fluctuate but in full anti-correlation. This procedure enables to estimate numerically the values of the variances and covariances appearing in equation \ref{Uncert_Bkgd_Subtract}.

\par
In a realistic analysis context, $A_{C}^{Exp}(S)$ can be obtained from a full simulation of the signal, $A_{C}^{Exp}(B)$ and $\alpha^{Exp}$ can also be obtained this way or through data-driven techniques. The experimental systematic uncertainties can be propagated as usually done to each of these quantities. And one can extract $A_{C}^{Obs}(S)$ from a data sample using the following form of equation \ref{Bkgd_Subtract}:
\begin{equation}
A_{C}^{Obs}(S) = (1+\alpha^{Meas})\cdot A_{C}(Data)-\alpha^{Meas}\cdot A_{C}^{Meas}(B)
\end{equation}
\noindent
provided  a good estimate of the number of remaining signal and background events after the event selection as well as the integral charge asymmetries of the signal and of the background are established. The upper script "Obs" stands for observed.

\vspace{1.0cm}
\noindent 
2.2.4.\ a.3.\ The\ Measured\ $A_{C}$\ in\ the\ Electron\ Channel

\par
For the nominal W mass, we calculate  $A_{C}^{Meas}(S)$ using the inputs from the analysis in the electron channel only with their statistical uncertainties:
\begin{itemize}
\item $A_{C}^{Exp}(S)	= (16.70\pm 0.18)\%$
\item $A_{C}^{Exp}(B)	= (10.07\pm 0.15)\%$
\item $A_{C}^{Exp}(S+B) = (16.52\pm 0.11)\%$
\item $\alpha^{Exp} = (2.74\pm 0.05)\times 10^{-2}$
\end{itemize}

\noindent
After the background subtraction and the propagation of the experimental systematic uncertainties, we get:
\begin{equation}
A_{C}^{Meas}(S) = (16.70\pm 0.76)\%
\label{Meas_AC_W_enu_80}
\end{equation} 

\vspace{1.0cm}
\noindent 
2.2.4.\ a.4.\ The\ $A_{C}$\ Template\ Curve\ in\ the\ Electron\ Channel
\par
\begin{figure}[h]
\begin{center}
\includegraphics[scale=0.35]{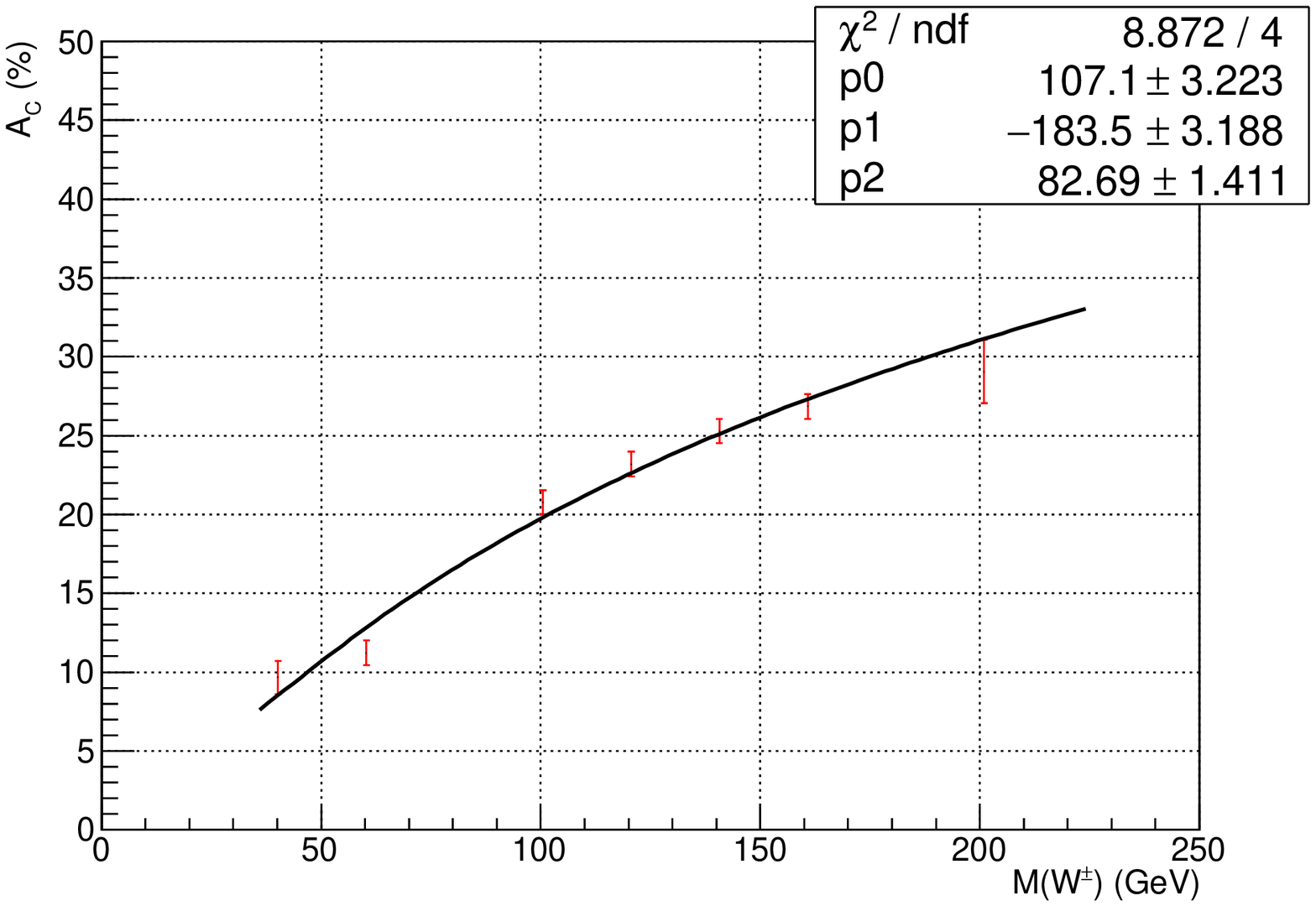}
\includegraphics[scale=0.35]{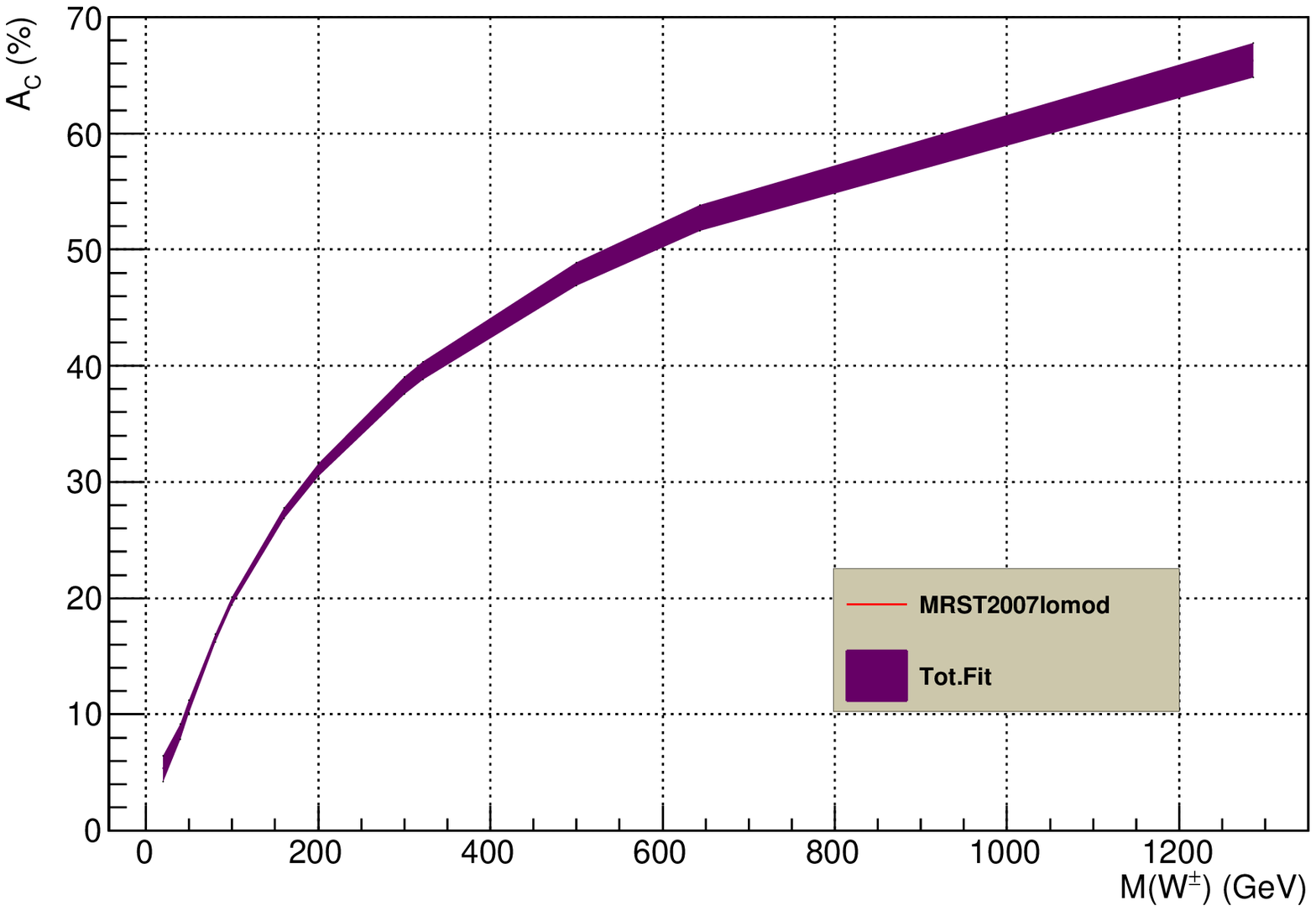}\\
\includegraphics[scale=0.35]{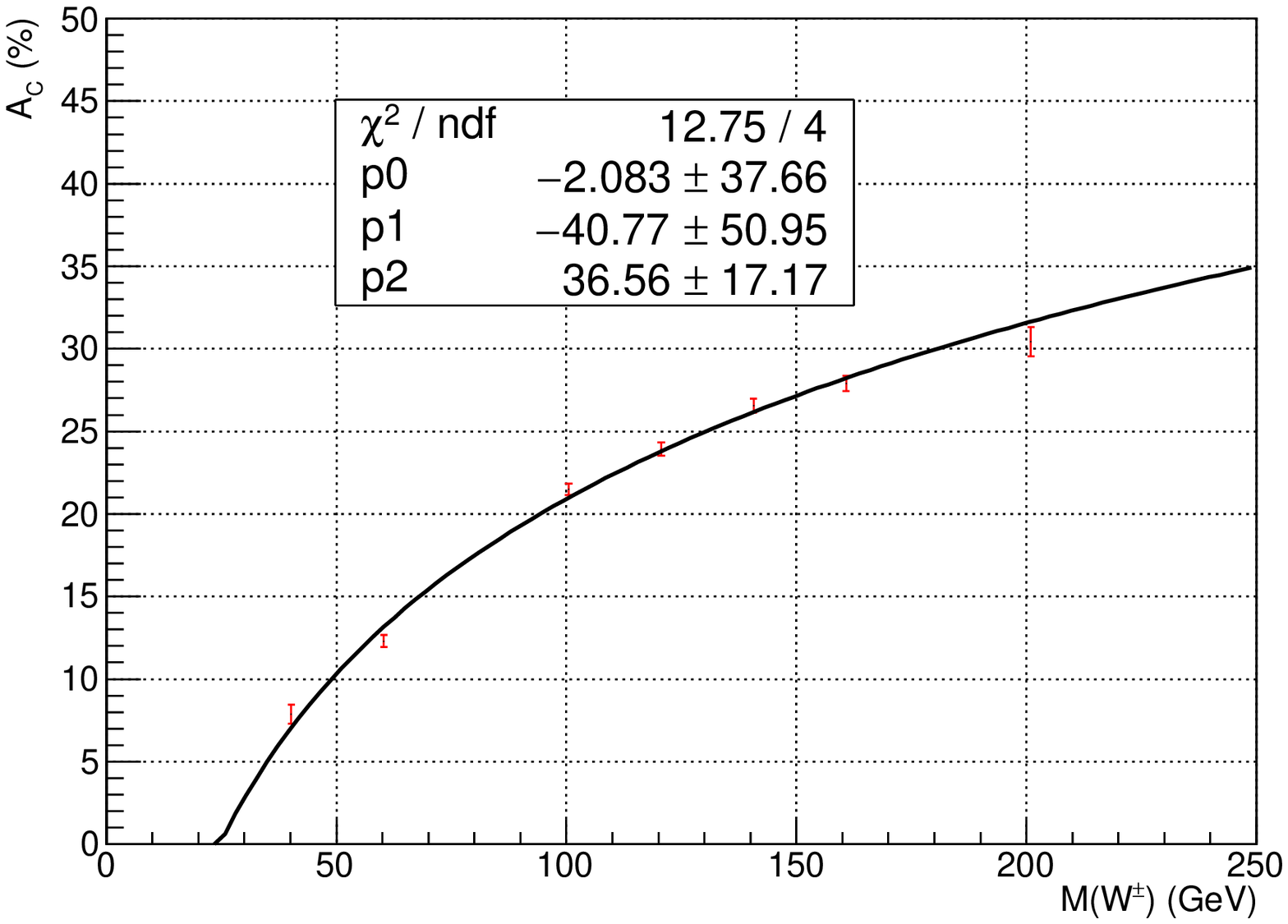}
\includegraphics[scale=0.35]{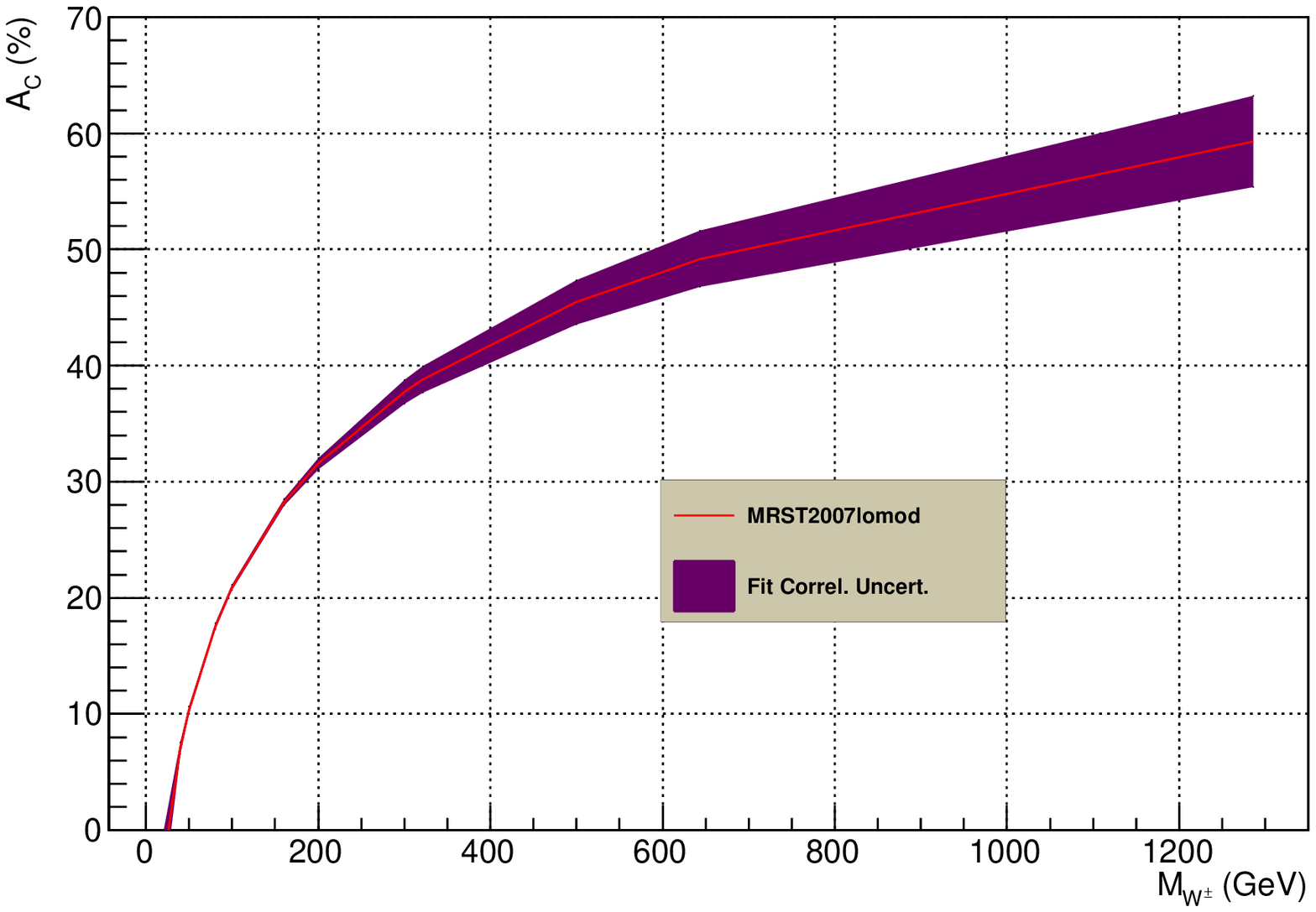}
\end{center}
\caption{\label{sec:Part2:GCATC_AC_W_lnu} The  $A_{C}^{Meas}$ template curves for the electron channel (top) and the muon channel (bottom). The fits to the
$A^{Meas}_{C}(S)$ are presented on the LHS. These fits with uncertainty bands accounting for the correlation between the uncertainties of the fit parameters are shown on the RHS.}
\end{figure}

In order to establish the experimental $A_{C}$ template curve, we apply a "multitag and probe method". We consider all the $W^{\pm}\to e^{\pm}\nu_{e}$ MC samples with a non-nominal W mass as the multitag and the one with the nominal W mass as the probe. We apply equation \ref{Bkgd_Subtract} to each of the multitag samples and plot their $A_{C}^{Meas}(S)$ as a function of $M_{W^{\pm}}$. A second degree polynomial of logarithms of logarithms is well suited to fit the template curve as shown in the LHS of figure \ref{sec:Part2:GCATC_AC_W_lnu}, for the electron channel. The fit to this template curve can expressed by equation \ref{sec:Part2:GCATC_AC_W_enu_Fit}.
Note that we do not include the probe sample in the template curve since we want to estimate the accuracy of its indirect mass measurement.

\begin{equation}
A_{C}^{Meas}(W^{\pm}\to e^{\pm}+\nu_{e}) = -107.1-183.5\times Log(Log(M_{W^{\pm}}))+82.69\times Log(Log(M_{W^{\pm}}))^{2}
\label{sec:Part2:GCATC_AC_W_enu_Fit}
\end{equation}

\begin{table}[h]
\small\begin{center}
\begin{tabular}{|c|c|c|c|c|c|}
\hline\hline 
$Process$  &  $\alpha^{Exp}\pm\delta \alpha^{Stat}$   &	$Z_{N}$ & $A_{C}^{Meas.}$ & $\delta A_{C}^{Meas.}$ & $\delta A_{C}^{Meas.Fit}$\\ 
				&  	     & ($\sigma$) &	($\%$)	&	($\%$)	&	($\%$)\\
\hline\hline
Signal: $W^{\pm}\to e^{\pm}\nu_{e}$            & 	                                                  &		        &                 &              &			\\
$M(W^{\pm}) = 40.2 \rm\ GeV$                      & $(31.55\pm 0.77)\times 10^{-2}$  &  37.25         &  $9.66$     &     1.05    & 0.60 \\
$M(W^{\pm}) = 60.3 \rm\ GeV$                      & $(  3.58\pm 0.07)\times 10^{-2}$  &  $>>5.00$  &  $11.22$   &       0.78 & 0.52 \\ 
$M(W^{\pm}) = \underline{80.4} \rm\ GeV$ & $(  2.74\pm 0.05)\times 10^{-2}$  &  $>>5.00$  &  $16.70$    &      0.76 & 0.35 \\ 
$M(W^{\pm}) = 100.5 \rm\ GeV$                    & $(  3.13\pm 0.06)\times 10^{-2}$  &  $>>5.00$   &  $20.77$   &       0.77  & 0.33 \\ 
$M(W^{\pm}) = 120.6 \rm\ GeV$                    & $(  3.89\pm 0.07)\times 10^{-2}$  &  $>>5.00$   & $23.19$    &      0.78   & 0.35 \\ 
$M(W^{\pm}) = 140.7 \rm\ GeV$                    & $(  4.82\pm 0.09)\times 10^{-2}$  &  $>>5.00$   &  $25.29$    &      0.78 & 0.39 \\ 
$M(W^{\pm}) = 160.8 \rm\ GeV$                    & $(  6.00\pm 0.11)\times 10^{-2}$  &  $>>5.00$   &  $26.86$    &      0.79 & 0.42 \\ 
$M(W^{\pm}) = 201.0 \rm\ GeV$                    & $(88.77\pm 1.66)\times 10^{0}$   &   0.19	       &  $ 29.07$     &     2.03 & 0.48 \\ 
\hline\hline
\end{tabular}       
\end{center}
\caption{\label{sec:Part2:Meas_AC_w_enu_tab} Noise to signal ratio, signal statistical significance, and expected and measured integral charge asymmetries for the signal after the event selection in the electron channel.}
\end{table}

\noindent
The values of the noise to signal ratio ($\alpha^{Exp}$), the signal statistical significance ($Z_{N}$, defined in the next paragraph), the expected ($A_{C}^{Exp}$), and the measured ($A_{C}^{Meas}$) integral charge asymmetries for the signal after the event selection in the electron channel are reported in table \ref{sec:Part2:Meas_AC_w_enu_tab}.
\par\noindent
The signal significances reported are calculated using a conversion of the confidence level of the signal plus background hypothesis $CL_{S+B}$ into an equivalent number of one-sided gaussian standard deviations $Z_{N}$ as proposed in \cite{2006sppp.conf..112C} and implemented in RooStats \cite{RooStats:Zn}. For these calculations the systematic uncertainty of the background was set to $5\%$, which completely covers the total
uncertainty for the measurement of the inclusive cross section $\sigma(p+p\to W^{\pm}\to\ell^{\pm}\nu)$ as reported in \cite{CMS:2011aa}.
\par\noindent
We recalculate the uncertainty on $A_{C}^{Meas}(S)$ accounting for the correlation between the parameters when fitting the $A_{C}^{Meas}(S)$ template curve by applying equation \ref{Uncert_Bkgd_Subtract}. This results in a slightly reduced uncertainty as shown in equation \ref{MeasFit_AC_W_enu_80}.
\begin{equation}
A_{C}^{Meas.Fit}(S) = (16.70\pm 0.35)\%
\label{MeasFit_AC_W_enu_80}
\end{equation} 

%%%%%%%%%%%%%%%%%%%%%%%%%%%%%%%%%%%%%%%%%%%%%%%%%%%%%%%%%%%%%%%%%%%%%%%%%

\vspace*{1.5mm}
\par\noindent
2.2.4.\ b.\ The\ Muon\ Channel
\vspace*{0.5mm}

\vspace{0.5 cm}
\par\noindent
2.2.4.\ b.1.\ Event\ Selection\ in\ the\ Muon\ Channel

\par
The following cuts are applied:
\begin{itemize}
\item $p_{T}(\mu) > 20 \rm\ GeV$
\item $|\eta(\mu)|<2.4$
\item Tracker Isolation: reject events with additional tracks of $p_{T}>2$ GeV within a cone of $\Delta R=0.5$ around the direction of the $\mu^{\pm}$ track
\item Calorimeter Isolation: the ratio of, the scalar sum of  $E_{T} $ deposits in the calorimeter within a cone of $\Delta R=0.5$ around the direction of the $\mu^{\pm}$, to the $p_{T}(\mu^{\pm})$ must be less than 0.25 
\item $\rlap{\kern0.25em/}E_{T} > 25 \rm\ GeV$
\end{itemize}

\begin{itemize}
\item $M_{T} > 40 \rm\ GeV$   
\item Reject events with an additional trailing isolated muon: $\mu^{\pm}_{2}$ 
\item Reject events with an additional leading isolated electron: $e^{\pm}_{1}$ 
\item Reject events with an additional second track ($Track_{2}$) such that :
$$
         \begin{cases}
          Q(\mu^{\pm}_{1}) = -Q(Track_{2}) \\
         3 < p_{T}(Track_{2}) < 10 \rm\ GeV \\
         M[\mu_{1}^{\pm},Track_{2}]> 50 \rm\ GeV
         \end{cases}
$$
\end{itemize}

\noindent
The corresponding selection efficiencies and event yields are reported in table \ref{sec:Part2:w_munu_SEL_tab}.
The RHS of figure \ref{sec:Part2:w_lnu_mET_fig} displays the $\rlap{\kern0.25em/}E_{T}$ distribution after the event selection.
The non-resonant background processes represent $\sim 3\%$ of the total background after the event selection.  

\begin{table}[h]
\small\begin{center}
\begin{tabular}{|c|c|c|c|}
\hline\hline 
$Process$		& $\epsilon$		& $N_{exp}$     &   $A_{C}(S)\pm\delta A_{C}^{Stat}(S)$	\\ 
				& ($\%$)             	& (k evts)	     &	($\%$)		\\
\hline\hline
Signal: $W^{\pm}\to\mu^{\pm}\nu_{\mu}$ & 	        &		  &			\\
$M(W^{\pm}) = 40.2 \rm\ GeV$          & $1.22\pm 0.02$       & 439.192	  & $7.86\pm 1.28$	\\
$M(W^{\pm}) = 60.3 \rm\ GeV$          & $12.27\pm 0.05$      & 2295.224	   & $12.30\pm 0.40$	\\
$M(W^{\pm}) = \underline{80.4} \rm\ GeV$& $29.32\pm 0.04$    & 3313.642	  & $17.42\pm 0.18$  \\
$M(W^{\pm}) = 100.5 \rm\ GeV$         & $54.03\pm 0.07$	& 4034.779	 & $21.48\pm 0.19$  \\
$M(W^{\pm}) = 120.6 \rm\ GeV$         & $31.30\pm 0.07$	& 1645.675	  & $23.93\pm 0.25$  \\
$M(W^{\pm}) = 140.7 \rm\ GeV$         & $33.71\pm 0.07$	& 1316.121	  & $26.56\pm 0.23$  \\
$M(W^{\pm}) = 160.8 \rm\ GeV$         & $35.37\pm 0.07$	& 1053.514	 & $27.90\pm 0.23$  \\   
$M(W^{\pm}) = 201.0 \rm\ GeV$         & $82.84\pm 0.05$	&   1.568  	   & $30.44\pm 0.15$	\\
\hline\hline
Background                       &	-               & $277.787\pm 21.555$	    & $7.36\pm 0.15$	\\
\hline
$W^{\pm}\to e^{\pm}\nu_{e}/\tau^{\pm}\nu_{\tau}/q\bar{q\prime}$ &    $0.291\pm 0.003$    &  177.500     &   $8.70\pm 1.07$    \\
\hline
$t\bar t$                        &   $4.27\pm 0.02$     &   4.895	       &    $-0.14\pm 0.43$    \\
$t+b,\ t+q(+b)$            &   $0.485\pm 0.005$     &  0.264              & $27.14\pm 0.96$	\\
\hline 
$W+W,\ W+\gamma^{*}/Z,\ \gamma^{*}/Z+\gamma^{*}/Z$ &   $3.25\pm 0.01$   & 2.478    &   $11.39\pm 0.33$	\\
\hline
$\gamma+\gamma,\ \gamma+jets,\ \gamma+W^{\pm},\ \gamma+Z$ &  $0.135\pm 0.001$  &  0.497   &  $17.48\pm 0.65$   \\
\hline
$\gamma^{*}/Z$            &   $0.727\pm 0.001$  &  43.382  & $5.79\pm 0.20$	\\
\hline
QCD HF                          & $(2.13\pm 0.37)\times 10^{-4}$ &  17.983    &	$-17.65\pm16.88$ \\
QCD LF                          & $(1.38\pm 0.41)\times 10^{-4}$ &  30.788    & 	$9.09\pm 30.03$ \\
\hline\hline
\end{tabular}       
\end{center}
\caption{\label{sec:Part2:w_munu_SEL_tab} Event selection efficiencies, event yields and integral charge asymmetries for the 
$W^{\pm}\to\mu^{\pm}\nu_{\mu}$ analysis.}
\end{table}

\vspace{0.5 cm}
\par\noindent
2.2.4.\ b.2.\ The\ Measured\ $A_{C}$\ in\ the\ Muon\ Channel

The $A_{C}^{Meas}(S)$  treatment described in paragraph 2.2.4. a.2. is applied to the probe sample in the muon channel, starting from the following inputs:

\begin{itemize}
\item $A_{C}^{Exp}(S)	= (17.42\pm 0.18)\%$
\item $A_{C}^{Exp}(B)	= ( 7.36\pm 0.15)\%$
\item $A_{C}^{Exp}(S+B) = (16.64\pm 0.12)\%$
\item $\alpha^{Exp} = (8.38\pm 0.65)\times 10^{-2}$
\end{itemize}

\noindent
For the nominal W mass, this leads to a measured integral charge asymmetry of:
\begin{equation}
A_{C}^{Meas}(S) = (17.42\pm 0.34)\%
\end{equation}
\noindent
where the uncertainty is also dominated by the value in equation \ref{AC:Syst:Muon}.

\vspace{0.5 cm}
\par\noindent
2.2.4.\ b.3.\ The\ Template\ Curve\ in\ the\ Muon\ Channel

After applying the $A_{C}^{Meas}(S)$  treatment to the tag samples in the muon channel, we get the $A_{C}^{Meas}(S)$
template curve shown in the RHS of figure \ref{sec:Part2:GCATC_AC_W_lnu}. The fit to this template curve is reported in equation
\ref{sec:Part2:GCATC_AC_W_munu_Fit}.

\begin{equation}
A_{C}^{Meas}(W^{\pm}\to\mu^{\pm}\nu_{\mu}) = -2.08-40.77\times Log(Log(M_{W^{\pm}}))+36.56\times Log(Log(M_{W^{\pm}}))^{2}
\label{sec:Part2:GCATC_AC_W_munu_Fit}
\end{equation}

\noindent
The values of the noise to signal ratio ($\alpha^{Exp}$), the signal statistical significance ($Z_{N}$), and the expected ($A_{C}^{Exp}$) and the measured ($A_{C}^{Meas}$) integral charge asymmetries for the signal after the event selection in the muon channel are reported in table \ref{sec:Part2:Meas_AC_w_munu_tab}.

\begin{table}[h]
\small\begin{center}
\begin{tabular}{|c|c|c|c|c|c|}
\hline\hline 
$Process$  &  $\alpha^{Exp}\pm\delta \alpha^{Stat}$   &	$Z_{N}$ & $A_{C}^{Meas.}$ & $\delta A_{C}^{Meas.}$ & $\delta A_{C}^{Meas.Fit}$\\ 
				&  	     & ($\sigma$) &	($\%$)	&	($\%$) &	($\%$)	\\
\hline\hline
Signal: $W^{\pm}\to\mu^{\pm}\nu_{\mu}$   & 	                                                      &		           &            &                    &			         \\
$M(W^{\pm}) = 40.2 \rm\ GeV$                       & $(63.25\pm  4.97)\times 10^{-2}$  &  11.19         & 7.86  &  0.59     & 0.45 \\
$M(W^{\pm}) = 60.3 \rm\ GeV$                       & $(12.10\pm  0.94)\times 10^{-2}$  & 2295.22      &  12.30 & 0.37  & 0.27 \\
$M(W^{\pm}) = \underline{80.4} \rm\ GeV$ & $(8.38\pm  0.65)\times 10^{-2}$    & 3313.64      &  17.42 & 0.34  & 0.27 \\
$M(W^{\pm}) = 100.5 \rm\ GeV$                    & $(6.88\pm  0.53)\times 10^{-2}$     &  4034.78     &  21.48 & 0.35   & 0.22 \\
$M(W^{\pm}) = 120.6 \rm\ GeV$                    & $(16.88\pm  1.31)\times 10^{-2}$   &  1645.68     & 23.93 &0.40   & 0.19 \\ 
$M(W^{\pm}) = 140.7 \rm\ GeV$                    & $(21.11\pm  1.64)\times 10^{-2}$   &   1316.12    &  26.56& 0.42  & 0.22 \\
$M(W^{\pm}) = 160.8 \rm\ GeV$                    & $(26.37\pm  2.05)\times 10^{-2}$   &    1053.51   & 27.90 & 0.45   & 0.27 \\ 
$M(W^{\pm}) = 201.0 \rm\ GeV$                    & $(17.72\pm  1.37)\times 10^{1}$    &    1.57 	&  30.44 & 0.87   & 0.40 \\
\hline\hline
\end{tabular}       
\end{center}
\caption{\label{sec:Part2:Meas_AC_w_munu_tab} Noise to signal ratio, signal statistical significance, and expected and measured integral charge asymmetries for the signal after the event selection in the muon channel.}
\end{table}

\par\noindent
Again, accounting for the correlation between the parameters when fitting the $A_{C}^{Meas}(S)$ template curve enables to reduce the uncertainty as shown in equation \ref{MeasFit_AC_W_munu_80}.
\begin{equation}
A_{C}^{Meas.Fit}(S) = (17.42\pm 0.27)\%
\label{MeasFit_AC_W_munu_80}
\end{equation}

\newpage
\subsection{Indirect Determination of $M_{W^{\pm}}$}
\label{sec:Part1-mass-constraint}
\vspace*{1.5mm}
\subsubsection{\label{sec:Part3:W-enu-munu-Individual} Results in the Individual Channels}
\vspace*{0.5mm}

The $A_{C}^{Meas}(S)\pm\delta A_{C}^{Meas.Fit}(S)$ in the electron and in the muon channels translate into indirect 
$M^{Meas.Fit}_{W^{\pm}}\pm\delta M_{W^{\pm}}$ measurements using the experimental $A_{C}$ template curves from the RHS of figure \ref{sec:Part2:GCATC_AC_W_lnu}
in each of these channels:

\begin{equation}
A_{C}^{Meas.Fit}(S) = (16.70\pm 0.35)\% \Rightarrow M^{Meas.Fit}(W^{\pm}\to e^{\pm}\nu_{e})=81.07^{+2.06}_{-2.01}\rm\ GeV,
\end{equation}

\begin{equation} 
A_{C}^{Meas.Fit}(S) = (17.42\pm 0.27)\% \Rightarrow M^{Meas.Fit}(W^{\pm}\to\mu^{\pm}\nu_{\mu})=79.67^{+3.56}_{-1.39}\rm\ GeV.
\end{equation}

%%%%%%%%%%%%%%%%%%%%%%%%%%%%%%%%%%%%%%%%%%%%%%%%%%%%%%%%%%%%%%%%%%%%%%%%%

\vspace*{1.5mm}
\subsubsection{\label{sec:Part3:W-enu-munu-Combination} Combination of the Electron and the Muon Channels}
\vspace*{0.5mm}

We combine the electron and muon channels using a weighted mean for the measured $W^{\pm}$ mass, the weight is the inverse of the uncertainty on 
the measured mass. In order to account for the asymmetric uncertainties, we slightly modify the expressions for the weighted mean and the weighted RMS of a quantity $x$ as follows:

\begin{equation}
 <x> = \frac{\sum_{i=1}^{N}\frac{x_{i}}{\delta^{2}_{i}}}{\sum_{i=1}^{N}\frac{1}{\delta^{2}_{i}}} 
\rm\ \ \ \to\rm\ \ \  
   <x> = \frac{\sum_{i=1}^{N}[\frac{x_{i}}{(\delta^{Up}_{i})^{2}}+\frac{x_{i}}{(\delta^{Down}_{i})^{2}}]}
   {\sum_{i=1}^{N}[\frac{1}{(\delta^{Up}_{i})^{2}}+\frac{1}{(\delta^{Down}_{i})^{2}}]}
\end{equation}

\begin{equation}
\delta^{2}(<x>)=\frac{1}{\sum_{i=1}^{N}\frac{x_{i}}{\delta^{2}_{i}}}
\rm\ \ \ \to\rm\ \ \
\delta^{2}(<x>)=\frac{1}{\sum_{i=1}^{N}[\frac{x_{i}}{(\delta^{Up}_{i})^{2}}+\frac{x_{i}}{(\delta^{Down}_{i})^{2}}]}  
\end{equation}

\noindent
where $x_{i}$, $\delta^{Up}_{i}$ and $\delta^{Down}_{i}$ are  respectively the central value, the upward uncertainty and the downward uncertainty of the mass derived in the channel $i$.   

\noindent
The result of the combination is:
\begin{equation}
M^{Comb.Meas.}(W^{\pm})=80.30\pm 0.96\rm\ GeV\rm\ [Expt.\ Comb.].
\end{equation}

\subsection{Final Result for MRST2007lomod}
\label{sec:Part1-MRST2007}
\noindent
The next step is to estimate the theoretical uncertainty corresponding to the measured mass and to combine it with the experimental uncertainty. We simply use the central value of the measured $W^{\pm}$ mass and we read-off the theoretical template curve the intervals, defined by the intercepts with upper and lower fit curves.
\begin{equation}
M_{Theory}(W^{\pm})=80.30^{+0.19}_{-0.21}\rm\ GeV\rm\ [MRST2007lomod]
\end{equation}

\noindent
Finally we just sum in quadrature the theoretical and experimental upward and downward 
uncertainties: 

\begin{equation}
\delta_{Tot.} M(W^{\pm})=80.30
\begin{cases}
+\sqrt{(0.96)^{2}+(0.19)^{2}}= & +0.98\\
-\sqrt{(0.96)^{2}+(0.21)^{2}} = & -0.98
\end{cases}
  \rm\ GeV
\end{equation}

\noindent
Therefore the final result for the MRST2007lomod PDF reads:
\begin{equation}
M_{W^{\pm}}=80.30^{+0.98}_{-0.98}\rm\ GeV\ [Total\ MRST2007lomod].
\end{equation}

\noindent
This constitutes an indirect $M_{W^{\pm}}$ mesurement with a relative accuracy of $1.2\%$, where the experimental 
uncertainty largely dominates over the (underestimated) theoretical uncertainty.

\subsection{Final Results for the Other Parton Density Functions}
\label{sec:Part1-PDF-Info}
%%%%%%%%%%%%%%%%%%%%%%%%%%%%%%%%%%%%%%%%%%%%%%%%%%%%%%%%%%%%%%%%%%%%%%%%%

Since Delphes v1.9 does not store the set of variables $(x_{1},x_{2},flav_{1},flav_{2},Q^{2})$ necessary to access the PDF information from the generator, we slightly modify it so as to retrieve the \\ "HepMC::PdfInfo" object from the HepMC event record and to store it within the Delphes GEN branch as described in \cite{PDF-Info-Fix}.

\par\noindent
Based upon these variables we can apply PDF re-weightings so as to make experimental $A_{C}$ predictions for the CTEQ6L1 and the MSTW2008lo68cl PDFs. The new event weight is calculated in the standard way:

\begin{equation}
\rm PDFweight(New\ PDF)=
\frac{f^{New\ PDF}_{Flav_{1}}(x_{1},Q^{2})}{f^{Old\ PDF}_{Flav_{1}}(x_{1},Q^{2})}
\times
\frac{f^{New\ PDF}_{Flav_{2}}(x_{2},Q^{2})}{f^{Old\ PDF}_{Flav_{2}}(x_{2},Q^{2})}
\end{equation}
where the "Old PDF" is the default one, MRST2007lomod, and the "New PDF" is either CTEQ6L1 or MSTW2008lo68cl.

\noindent
We re-run the electron and muon channel analyses and just change the weights of all the selected events. This results in signal
event yields, and $A_{C}^{Exp}(S)$,  $A_{C}^{Exp}(B)$ as reported in tables \ref{PDF-RW:S:CTEQ6L1} and \ref{PDF-RW:B:CTEQ6L1}
for the CTEQ6L1 PDF and in tables \ref{PDF-RW:S:MSTW2008lo68cl} and \ref{PDF-RW:B:MSTW2008lo68cl} for the MSTW2008lo68cl one.

\begin{table}[h]
\begin{center}
\begin{tabular}{|c|c|c|}
\hline\hline 
$\rm M_{W^{\pm}}$		&	$N_{Exp}(S)$	&	$A_{C}^{Exp}(S)$	\\
(GeV)						&	(k Evts)		&	($\%$)		\\
\hline\hline
$40.2\ ^{e^{\pm}}_{ \mu^{\pm}}$ & 
$^{288.688\pm 5.866}_{ 947.643\pm 11.535 }$ &
$^{ 11.26\pm 2.06  }_{ 7.86\pm 1.28 }$ \\
\hline	
$60.3\ ^{e^{\pm}}_{ \mu^{\pm}}$ &
$^{ 2491.955\pm 10.746}_{ 5285.294\pm 16.847 }$ &
$^{10.65\pm 0.49  }_{ 12.30\pm 0.40  }$ \\
\hline	
$80.4\ ^{e^{\pm}}_{\mu^{\pm}}$ &
$^{ 3766.569\pm 8.423}_{ 5551.710\pm 6.752 }$ &
$^{ 15.78\pm 0.29 }_{  17.42\pm 0.18 }$ \\
\hline	
$100.5\ ^{e^{\pm}}_{\mu^{\pm}}$ &
$^{ 4106.984\pm 5.009 }_{ 4188.292\pm 4.997 }$ &
$^{20.64\pm 0.19}_{ 21.48\pm 0.19  }$ \\
\hline	
$120.6\ ^{e^{\pm}}_{\mu^{\pm}}$ &
$^{  2739.825\pm 4.796}_{ 3777.497\pm 4.730 }$ &
$^{23.54\pm 0.26}_{  23.93\pm 0.25 }$ \\
\hline	
$140.7\ ^{e^{\pm}}_{\mu^{\pm}}$ &
$^{ 2284.590\pm 3.512 }_{ 3020.544\pm 3.268 }$ &
$^{25.52\pm 0.25}_{26.56\pm 0.23}$ \\
\hline	
$160.8\ ^{e^{\pm}}_{\mu^{\pm}}$ &
$^{ 1584.146\pm 2.512 }_{ 2461.819\pm 2.255 }$ &
$^{27.07\pm 0.24}_{ 27.90\pm 0.23 }$ \\
\hline	
$201.0\ ^{e^{\pm}}_{\mu^{\pm}}$ &
$^{ 1.259\pm 0.002 }_{  1.628\pm 0.001 }$ &
$^{29.57\pm 0.23}_{ 30.64\pm 0.15 }$ \\
\hline\hline
\end{tabular}       
\end{center}
\caption{\label{PDF-RW:S:CTEQ6L1} Number of expected signal events and expected signal $A_{C}$ as a function of $M(W^{\pm})$ for the electron and muon analyses
reweighted to the CTEQ6L1 PDF predictions.}
\end{table}

\begin{table}[h]
\begin{center}
\begin{tabular}{|c|c|c|}
\hline\hline 
$W^{\pm}$ Decay Channel	&	$N_{Exp}(B)$	      &	$A_{C}^{Exp}(B)$    \\
					         &	(k Evts)		      &	($\%$)		    \\
\hline
$e^{\pm}$		               &   $352.660\pm 7.996$     & $9.74\pm 0.23$  \\
$\mu^{\pm}$	                    &	$707.617\pm 29.944$   &	$7.45\pm 0.15$	\\
\hline\hline
\end{tabular}       
\end{center}
\caption{\label{PDF-RW:B:CTEQ6L1} Number of expected background events and expected background $A_{C}$ for the electron (upper line)
and the muon (lower line) analyses reweighted to the CTEQ6L1 PDF predictions.}
\end{table}

\begin{figure}[h]
\begin{center}
\includegraphics[scale=0.35]{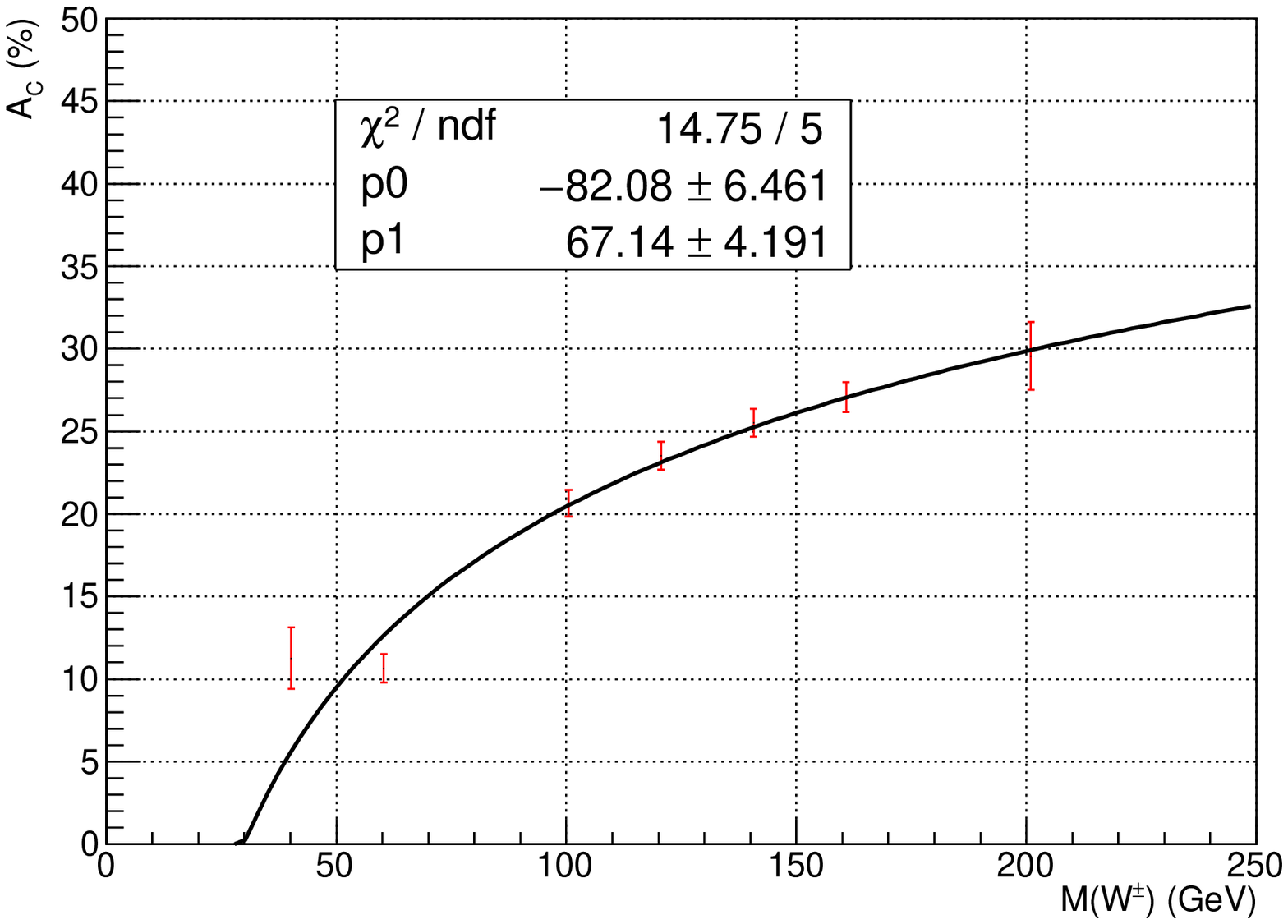}
\includegraphics[scale=0.35]{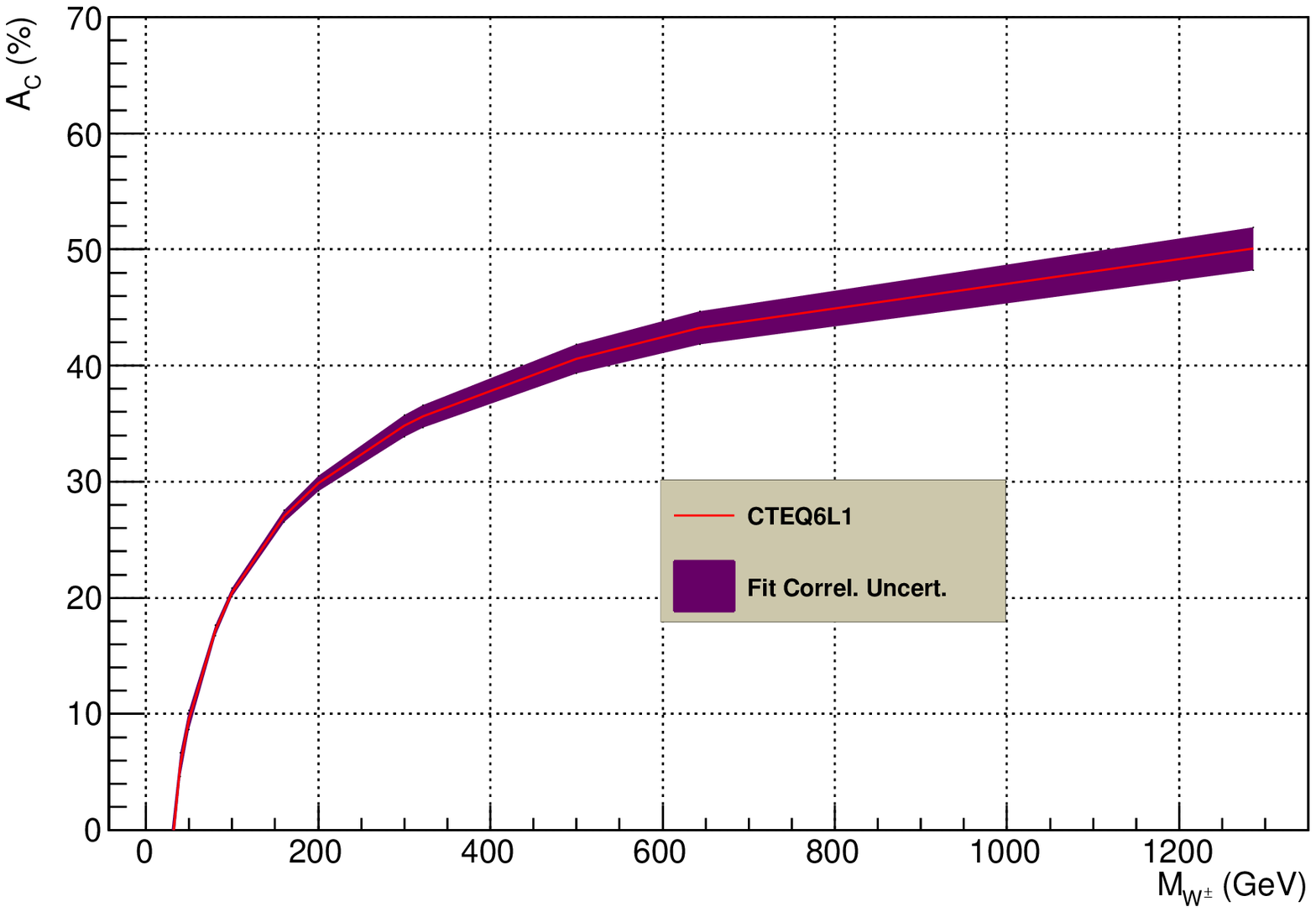}\\
\includegraphics[scale=0.35]{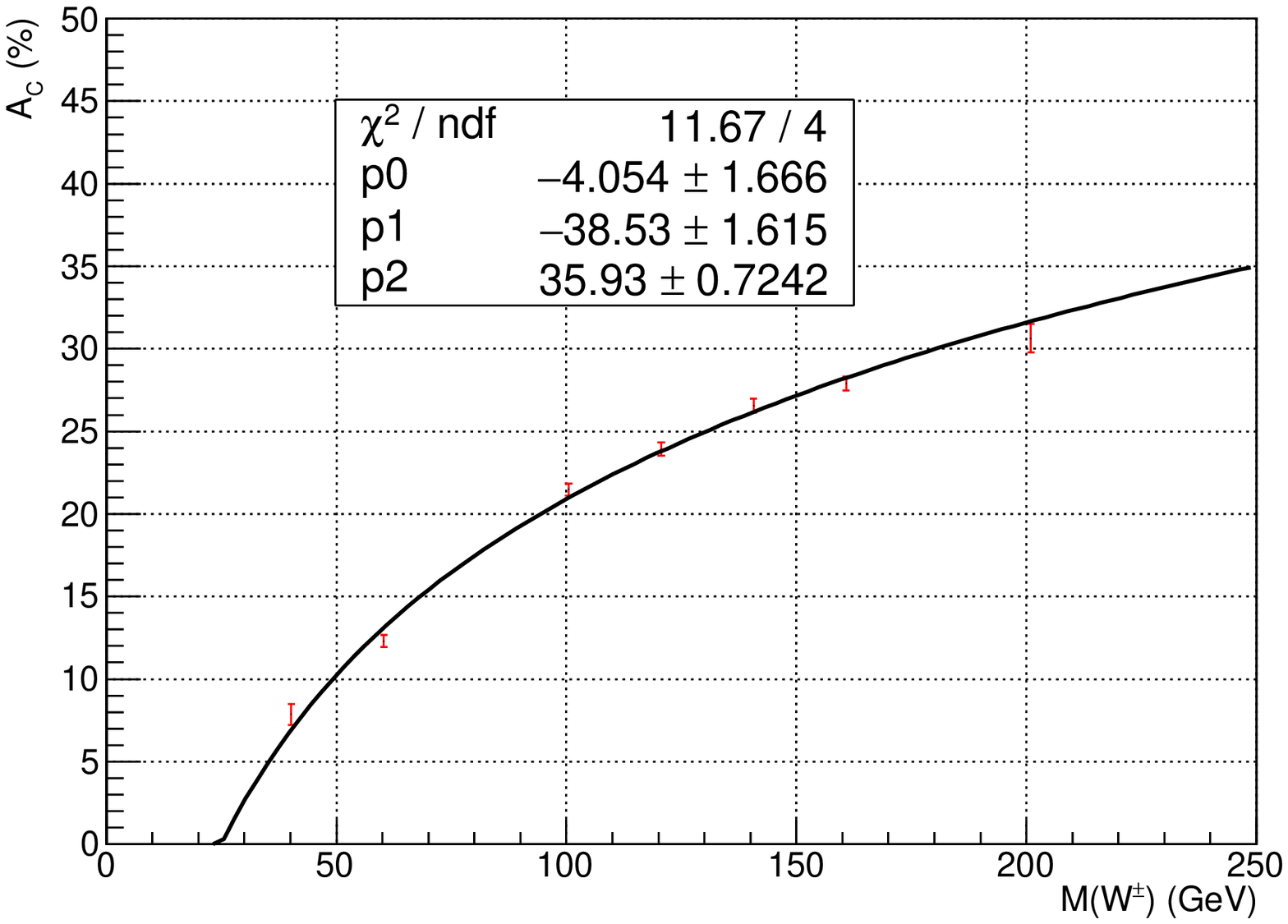}
\includegraphics[scale=0.35]{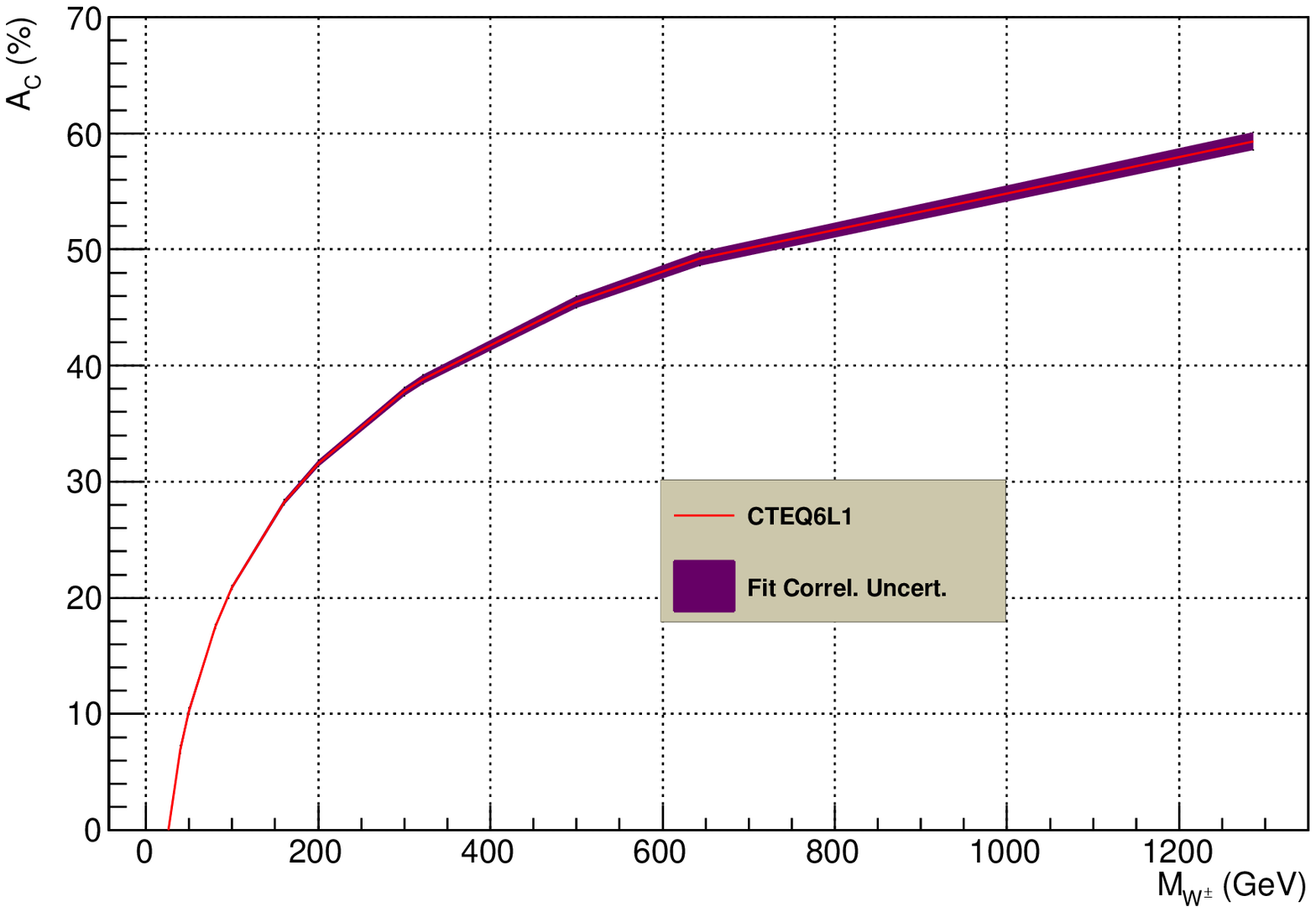}
\end{center}
\caption{\label{sec:Part2:PDF-RW-GCATC-e} The CTEQ6L1 $A_{C}$ template curves for the $W^{\pm}\to e^{\pm}\nu_{e}$ (top) and the $W^{\pm}\to\mu^{\pm}\nu_{\mu}$ (bottom) analyses. The fits to the
$A^{Exp}_{C}(S)$ are presented on the LHS. These fits with uncertainty bands accounting for the correlation between the uncertainties of the fit parameters are shown on the RHS.}
\end{figure}

\begin{figure}[h]
\begin{center}
\includegraphics[scale=0.35]{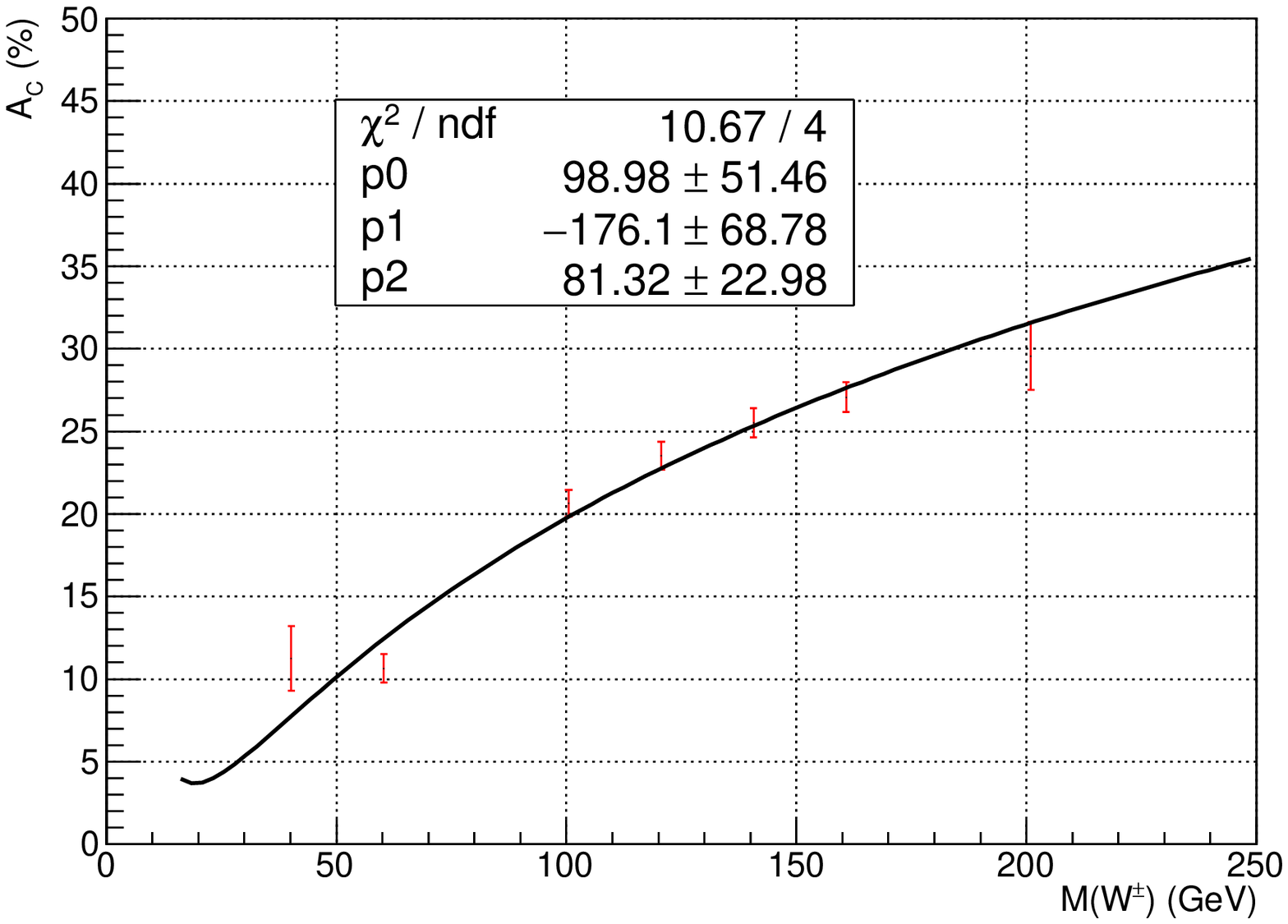}
\includegraphics[scale=0.35]{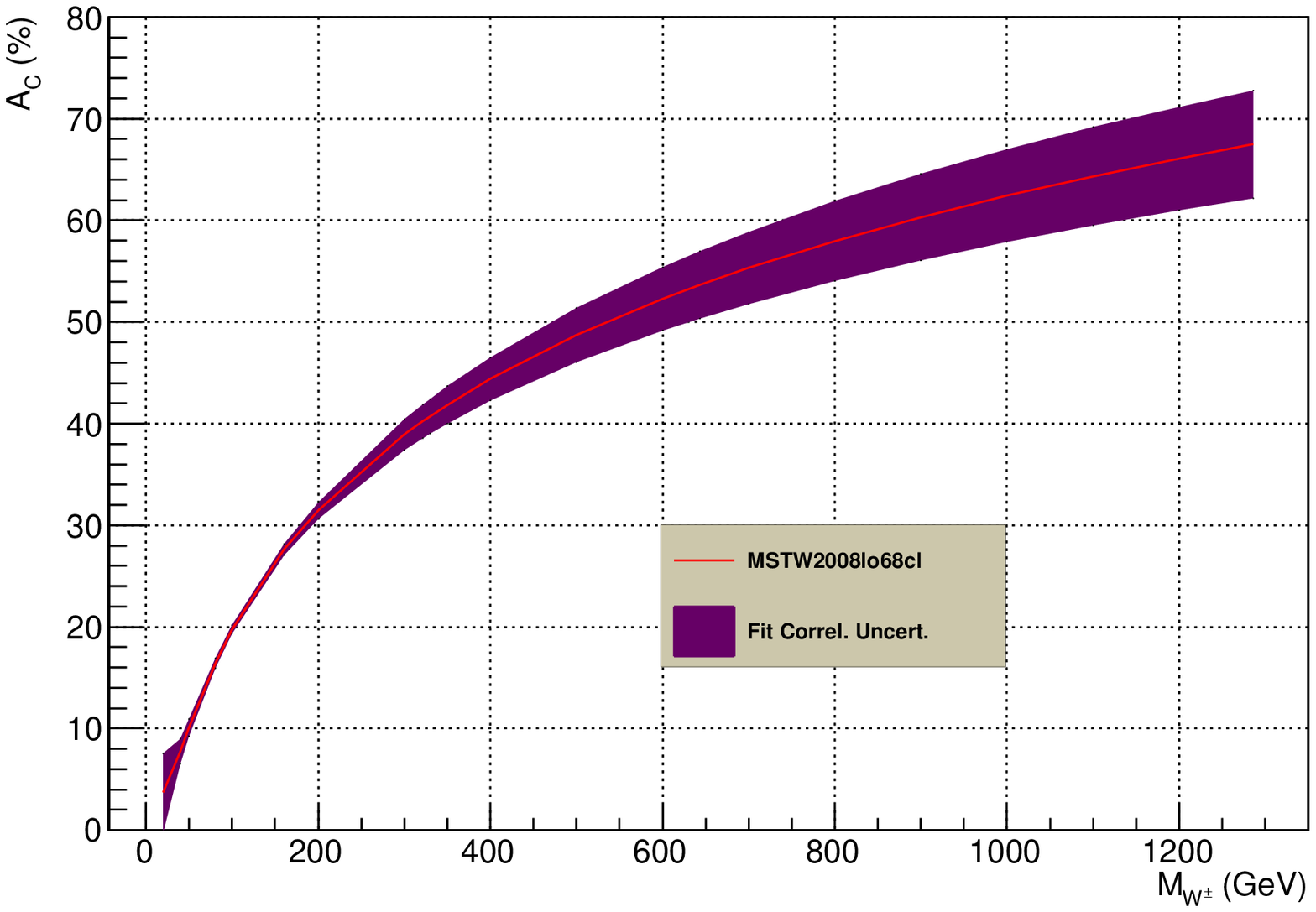}\\
\includegraphics[scale=0.35]{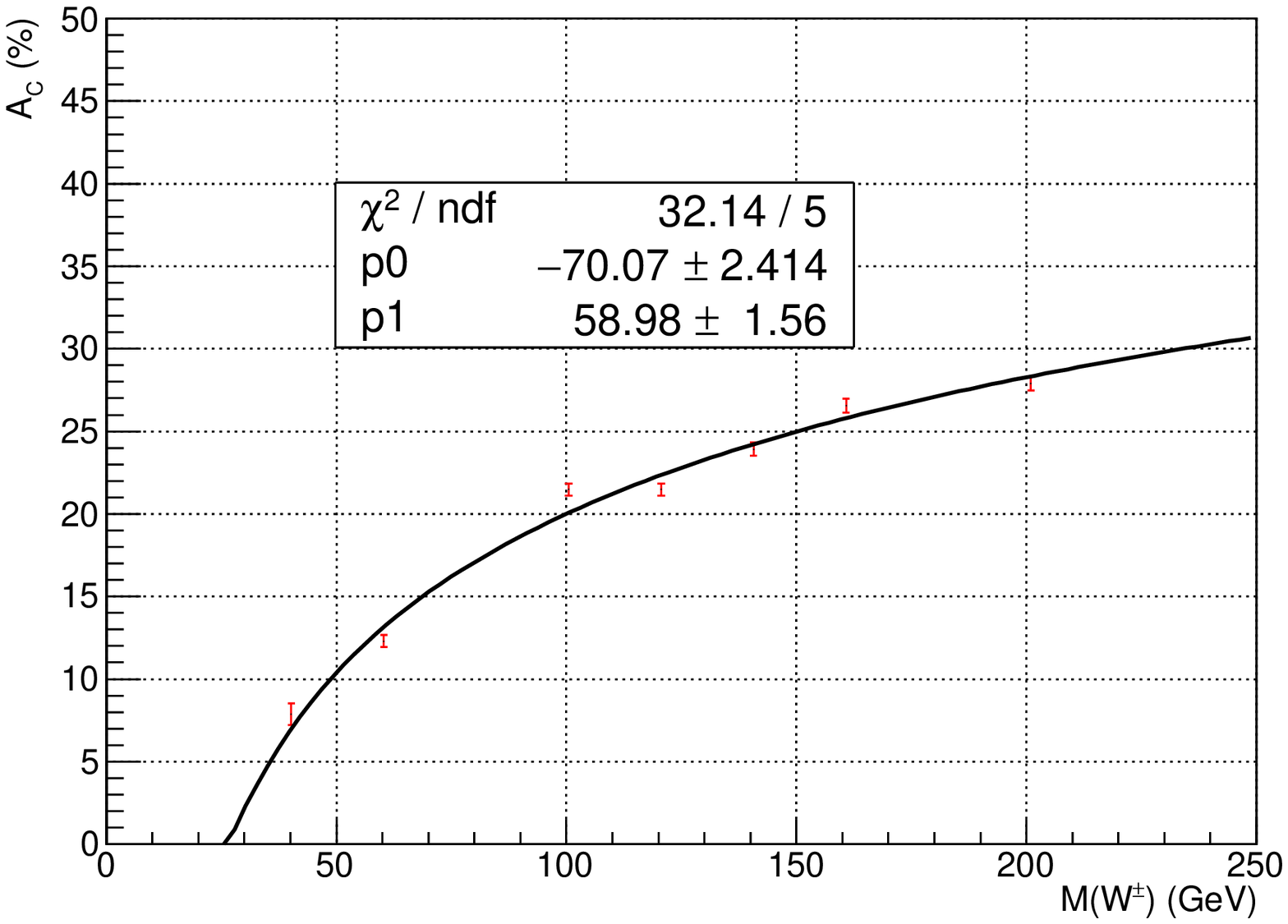}
\includegraphics[scale=0.35]{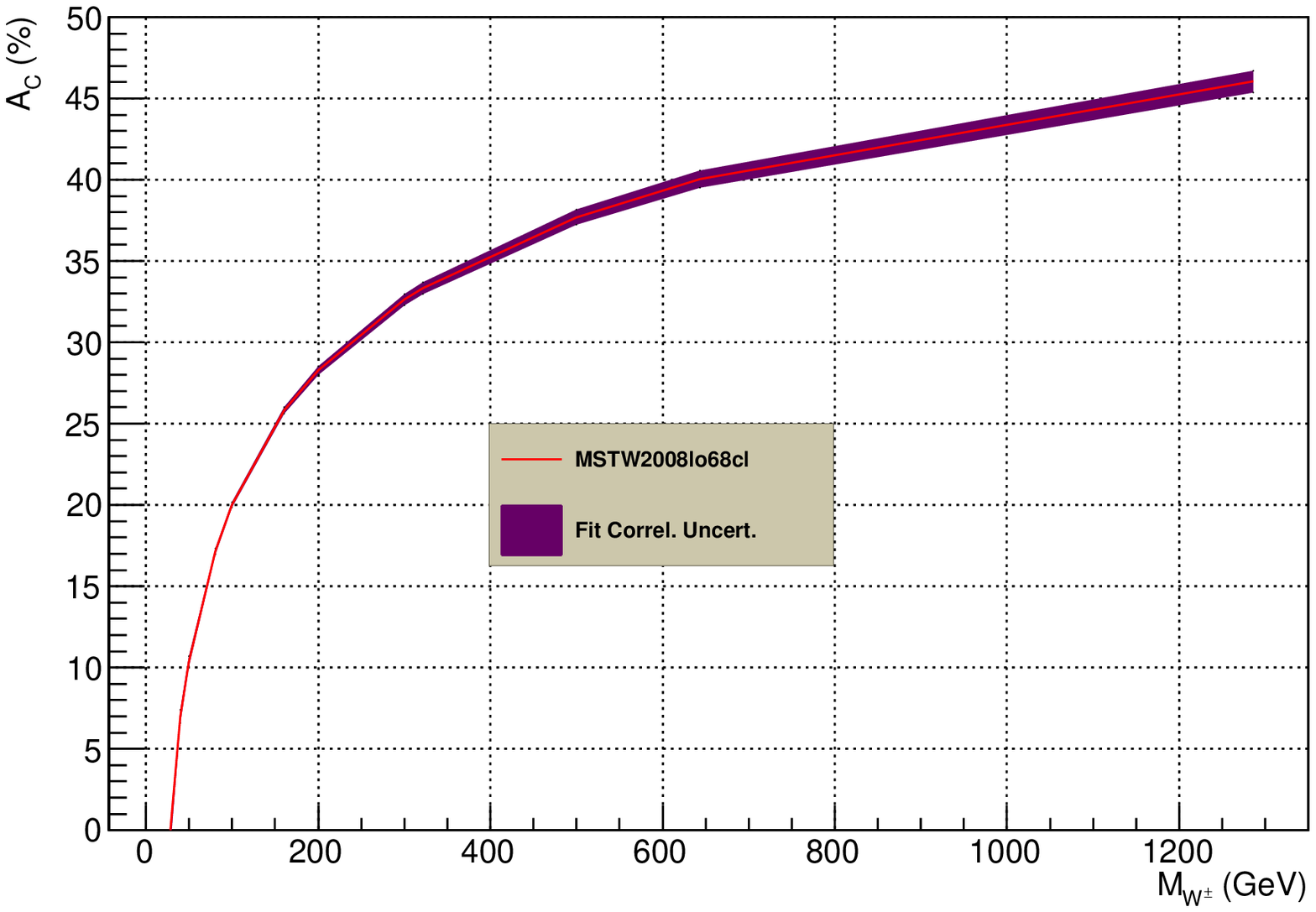}
\end{center}
\caption{\label{sec:Part2:PDF-RW-GCATC-mu} The MSTW2008lo68cl $A_{C}$ template curves for the $W^{\pm}\to e^{\pm}\nu_{e}$ (top) and  the $W^{\pm}\to\mu^{\pm}\nu_{\mu}$ (bottom) analyses. The fits to the
$A^{Exp}_{C}(S)$ are presented on the LHS. These fits with uncertainty bands accounting for the correlation between the uncertainties of the fit parameters are shown on the RHS.}
\end{figure}

\noindent
Then we produce the experimental $A_{C}$ template curves for CTEQ6L1 and MSTW2008lo68cl and both analysis channels as displayed in figures
\ref{sec:Part2:PDF-RW-GCATC-e} and  \ref{sec:Part2:PDF-RW-GCATC-mu}.

\noindent
For the CTEQ6L1 PDF, we find:
\begin{equation}
A_{C}^{Meas.Fit}(S) = (15.78\pm 0.50)\% \Rightarrow M^{Meas}(W^{\pm}\to e^{\pm}\nu_{e})=73.39^{+2.40}_{-2.30}\rm\ GeV,
\end{equation}
\begin{equation}
A_{C}^{Meas.Fit}(S) = (17.42\pm 0.18)\% \Rightarrow M^{Meas}(W^{\pm}\to\mu^{\pm}\nu_{\mu})=79.82^{+0.94}_{-0.92}\rm\ GeV
\end{equation}
which leads to the following combined value:
\begin{equation}
M^{Comb.Meas.}(W^{\pm}\to\ell^{\pm}\nu_{\ell})=(78.95\pm 0.61)\rm\ GeV\ [Expt.\ CTEQ6L1]
\end{equation}

\noindent
To this measured central value of the mass correspond the following theoretical uncertainties:
\begin{equation}
M(W^{\pm})=78.95^{+0.11}_{-0.13}\rm\ GeV\rm\ [Theory\ CTEQ6L1],
\end{equation}

\noindent
Therefore the final result for the CTEQ6L1 PDF reads:
\begin{equation}
M(W^{\pm})=78.95^{+0.62}_{-0.62}\rm\ GeV\rm\ [Total\ CTEQ6L1]
\end{equation}

\noindent
and it's dominant uncertainty is also experimental, since its theoretical uncertainty is underestimated. This represents an indirect measurement of the $W^{\pm}$ mass with a relative accuracy of $0.8\%$. 

\begin{table}[h]
\begin{center}
\begin{tabular}{|c|c|c|}
\hline\hline 
$\rm M_{W^{\pm}}$		&	$N_{Exp}(S)$	&	$A_{C}^{Exp}(S)$	\\
(GeV)						&	(k Evts)		&	($\%$)		\\
\hline\hline
$40.2\ ^{e^{\pm}}_{\mu^{\pm}}$ &
$^{  280.257\pm 5.781   }_{  913.868\pm 11.334 }$ &
$^{ 11.26\pm 2.06 }_{  7.86\pm 1.28 }$ \\
\hline	
$60.3\ ^{e^{\pm}}_{\mu^{\pm}}$ &
$^{ 2469.515\pm 10.705 }_{ 5219.408\pm 16.783 }$ &
$^{ 10.65\pm 0.49 }_{  12.30\pm 0.40 }$ \\
\hline	
$80.4\ ^{e^{\pm}}_{\mu^{\pm}}$ &
$^{  3663.615\pm 8.363 }_{ 5711.468\pm 6.753 }$ &
$^{  15.78\pm 0.29 }_{ 17.42\pm 0.18  }$ \\
\hline	
$100.5\ ^{e^{\pm}}_{\mu^{\pm}}$ &
$^{ 4053.288\pm 5.016 }_{ 4165.175\pm 5.000 }$ &
$^{ 20.64\pm 0.19 }_{ 21.48\pm 0.19  }$ \\
\hline	
$120.6\ ^{e^{\pm}}_{\mu^{\pm}}$ &
$^{ 2665.994\pm 4.800 }_{ 3811.380\pm 4.697 }$ &
$^{ 23.54\pm 0.26 }_{  23.93\pm 0.25 }$ \\
\hline	
$140.7\ ^{e^{\pm}}_{\mu^{\pm}}$ &
$^{ 2221.101\pm 3.530 }_{ 3033.091\pm 3.252 }$ &
$^{ 25.52\pm 0.25 }_{  26.56\pm 0.23 }$ \\
\hline	
$160.8\ ^{e^{\pm}}_{\mu^{\pm}}$ &
$^{ 1539.501\pm 2.516 }_{ 2446.996\pm 2.280 }$ &
$^{ 27.07\pm 0.24 }_{ 27.90\pm 0.23  }$ \\
\hline	
$201.0\ ^{e^{\pm}}_{\mu^{\pm}}$ &
$^{ 1.230\pm 0.002 }_{ 1.645\pm 0.001 }$ &
$^{ 29.57\pm 0.23 }_{  30.64\pm 0.15 }$ \\
\hline\hline
\end{tabular}       
\end{center}
\caption{\label{PDF-RW:S:MSTW2008lo68cl} Number of expected signal events and expected signal $A_{C}$ as a function of $M(W^{\pm})$ for the electron and muon analyses reweighted to the MSTW2008lo68cl PDF predictions.}
\end{table}

\noindent
For the MSTW2008lo68cl PDF:

\begin{table}[h]
\begin{center}
\begin{tabular}{|c|c|c|}
\hline\hline 
$W^{\pm}$ Decay Channel &	$N_{Exp}(B)$	&	$A_{C}^{Exp}(B)$  \\
						 &	(k Evts)		&	($\%$)		  \\
\hline
$e^{\pm}$		               &   $371.956\pm 8.081$   &  $9.74\pm 0.23$	 \\
$\mu^{\pm}$	                   &	$721.196\pm 29.968$   &  $7.45\pm 0.15$  \\
\hline\hline
\end{tabular}       
\end{center}
\caption{\label{PDF-RW:B:MSTW2008lo68cl} Number of expected background events and expected background $A_{C}$ for the electron (upper line) and muon (lower line) analyses reweighted to the  MSTW2008lo68cl PDF predictions.}
\end{table}

\begin{equation}
A_{C}^{Meas.Fit}(S) = (15.78\pm 0.52)\% \Rightarrow M^{Meas}(W^{\pm}\to e^{\pm}\nu_{e})=76.91^{+2.80}_{-2.74}\rm\ GeV,
\end{equation}
\begin{equation}
A_{C}^{Meas.Fit}(S) = (17.42\pm 0.18)\% \Rightarrow M^{Meas}(W^{\pm}\to\mu^{\pm}\nu_{\mu})=82.07^{+1.11}_{-1.10}\rm\ GeV
\end{equation}

\noindent
which leads to the following combined value:

\begin{equation}
M^{Comb.Meas.}(W^{\pm}\to\ell^{\pm}\nu_{\ell})=(81.36\pm 0.73)\rm\ GeV
\end{equation}

\noindent
The corresponding theoretical uncertainties are:

\begin{equation}
M(W^{\pm})=81.36^{+1.50}_{-1.32}\rm\ GeV\rm\ [Theory\ MSTW2008lo68cl],
\end{equation}

\noindent
Therefore the final result for the MSTW2008lo68cl PDF reads:
\begin{equation}
M(W^{\pm})=81.36^{+1.67}_{-1.51}\rm\ GeV\rm\ [Total\ MSTW2008lo68cl]
\end{equation}
\noindent
and it's dominant uncertainty comes from  $\delta^{Theory}_{PDF} A_{C}$ . In this case, this represents an indirect measurement of the $W^{\pm}$ mass with a relative accuracy of $2.1\%$.

\subsection{Summary of the $M_{W^{\pm}}$ Measurements and their Accuracy}
\label{sec:Part1-Summary-Info}
\noindent
We sum up the indirect mass measurements of $M_{W^{\pm}}$ extracted from the integral charge asymmetry 
of the $W^{\pm}\to\ell^{\pm}\nu$ inclusive process within table \ref{W-Summary:Tab}. Therein we also present
a few figures of merit of the accuracy of these measurements:

\begin{enumerate}
\item $\frac{\delta M^{Fit}_{W^{\pm}}}{M^{Fit}_{W^{\pm}}}$
\item $\frac{(M^{Fit}_{W^{\pm}}-M^{True}_{W^{\pm}})}{M^{True}_{W^{\pm}}}$
\item $\frac{(M^{Fit}_{W^{\pm}}-M^{True}_{W^{\pm}})}{\delta M^{Fit}_{W^{\pm}}}$
\end{enumerate}

\noindent
In this notation, $M^{Fit}_{W^{\pm}}$ and $\delta M^{Fit}_{W^{\pm}}$ represent the indirectly measured  $M_{W^{\pm}}$ and its uncertainty,
and $M^{True}_{W^{\pm}}$ stands for the nominal $W^{\pm}$ boson mass.
\par\noindent
The first figure of merit (1.) reflects  the intrinsic resolution power of the indirect mass measurement, irrespective of its possible biases, it's expressed in $\%$. The second and the third ones measure the accuracy with respect to the nominal $W^{\pm}$ boson mass: firstly as a relative uncertainty in $\%$ irrespective of the precision of the method (2.) and secondly as a compatibility between the nominal and the predicted masses given the precision of the method (3.), expressed in number of standard deviations ($\sigma$).

\begin{table}[h]
\begin{center}
\begin{tabular}{|c|c|c|c|}
\hline\hline
Figures of   Merit          & \multicolumn{3}{c|}{Considered LO PDFs} \\ 
of the Accuracy             & MRST2007lomod	   & CTEQ6L1    & MSTW2008lo68cl		   \\
\hline
1. $\frac{\delta M^{Fit}_{W^{\pm}}}{M^{Fit}_{W^{\pm}}} $    & $1.2\%$    &  $0.8\%$   &  $2.1\%$  \\
\hline
2. $\frac{(M^{Fit}_{W^{\pm}}-M^{True}_{W^{\pm}})}{M^{True}_{W^{\pm}}} $    & $-0.1\%$    &  $-1.8\%$   &  $+1.2\%$  \\
\hline
3. $\frac{(M^{Fit}_{W^{\pm}}-M^{True}_{W^{\pm}})}{\delta M^{Fit}_{W^{\pm}}} $    & $-0.1\sigma$    &  $-2.3\sigma$   &  $+0.6\sigma$  \\
\hline\hline
\end{tabular}       
\end{center}
\caption{\label{W-Summary:Tab} Summary of the indirect mass measurements of $M_{W^{\pm}}$ extracted from the integral charge asymmetry 
of the $W^{\pm}\to\ell^{\pm}\nu$ process. Different figures of merit of the accuracy of these measurements are presented.}
\end{table}

\noindent
The values of the figures of merit in table \ref{W-Summary:Tab} show that already at LO, this new method enables to get a good estimate of the $W^{\pm}$ boson mass.

%%%%%%%%%%%%%%%%%%%%%%%%%%%%%%%%%%%%%%%%%%%%%%%%%%%%%%%%%%%%%%%%%%%%%%%%%%%%%%%%%%%%%%%%%%%%%%% 
%%% PART II:  W1 + Z2 -> 3l + mET
%%%%%%%%%%%%%%%%%%%%%%%%%%%%%%%%%%%%%%%%%%%%%%%%%%%%%%%%%%%%%%%%%%%%%%%%%%%%%%%%%%%%%%%%%%%%%%%
\newpage
\vspace*{5mm}
\section{\label{sec:next} Inclusive Production of $\tilde\chi^{\pm}_{1}+\tilde\chi^{0}_{2}\to 3\ell^{\pm}+\rlap{\kern0.25em/}E_{T}$ }
\vspace*{2.5mm}

\subsection{ Theoretical Prediction of $A_{C}(\tilde\chi^{\pm}_{1}+\tilde\chi^{0}_{2})$}
\label{sec:next-parton-LVL}
\par
In this section we repeat the types of calculations done in section \ref{sec:Part1-parton-LVL} but now for a process
of interest in R-parity conserving SUSY searches, namely the $p+p\to\tilde\chi^{\pm}_{1}+\tilde\chi^{0}_{2}\to 3\ell^{\pm}+\rlap{\kern0.25em/}E_{T}$ inclusive production. 
\par\noindent
We use Resummino v1.0.0 \cite{Fuks:2013vua} to calculate the $p+p\to\tilde\chi^{\pm}_{1}+\tilde\chi^{0}_{2}$ cross sections at different levels of theoretical accuracy. At fixed order in QCD we run these calculations at the LO and the NLO. In addition, we also run them starting from the NLO MEs and including the "Next-to-Leading Log" (NLL) analytically resummed corrections. The latter, sometimes refered to as "NLO+NLL" will simply be denoted "NLL" in the following.
\par\noindent
We calculate these cross sections at $\sqrt{s}=8$ TeV using "Simplified Models" \cite{Alwall:2008ag} for the following masses: 
$$M_{\tilde\chi^{\pm}_{1}}=M_{\tilde\chi^{0}_{2}}=100,105,115,125,135,145,150,200,250,300,400,500,600,700\rm\ GeV$$ and
using the PDFs reported in table \ref{Tab:AC_PDF}. We set the QCD scales as $\mu_{R}=\mu_{F}=\mu_{0}=M_{\tilde\chi^{\pm}_{1}}+M_{\tilde\chi^{0}_{2}}$. 
Regarding the phase space sampling, a statistical precision of $0.1\%$ is requested for the numerical integration of the cross sections.

\begin{table}[h]
\begin{center}
\begin{tabular}{|c|c|}
\hline
LO				&	NLO $\&$ NLL	\\
\hline\hline
MRST2007lomod	&	MRST2004nlo	\\
%\hline
CTEQ6L1			&	CTEQ6.1	\\
%\hline
MSTW2008lo68cl	&	MSTW2008nlo68cl	\\
\hline
\end{tabular}
\end{center}
\caption{\label{Tab:AC_PDF} PDFs used for the calculations of $\sigma(\tilde\chi^{\pm}_{1}+\tilde\chi^{0}_{2})$ at the LO in QCD and
the NLO and the NLL.}
\end{table}

\noindent
The integral charge asymmetries as functions of $M_{\tilde\chi^{\pm}_{1}}+M_{\tilde\chi^{0}_{2}}$ for this process are presented in tables \ref{II-parton-LVL:Tab:AC_MRST}, \ref{II-parton-LVL:Tab:AC_CTEQ6},  and \ref{II-parton-LVL:Tab:AC_MSTW} for the MRST2007lomod/MRST2004nlo, the CTEQ6L1/CTEQ61, and the MSTW2008lo68cl/MSTW2008nlo68cl PDFs, respectively.

%%%%%%%%%%%%%%%%%%%%%%%%%%%%%%%%%%%%%%%%%%%%%%%%%%%%%%%%%%%%%%%%%%%%%%%%%%
\newpage
\vspace*{1.5mm}
\subsubsection{\label{II-parton-LVL:MRST} $A_{C}(\tilde\chi^{\pm}_{1}+\tilde\chi^{0}_{2})$ Template Curves for MRST}
\begin{figure}[htbp]
\begin{center}
\includegraphics[scale=0.35]{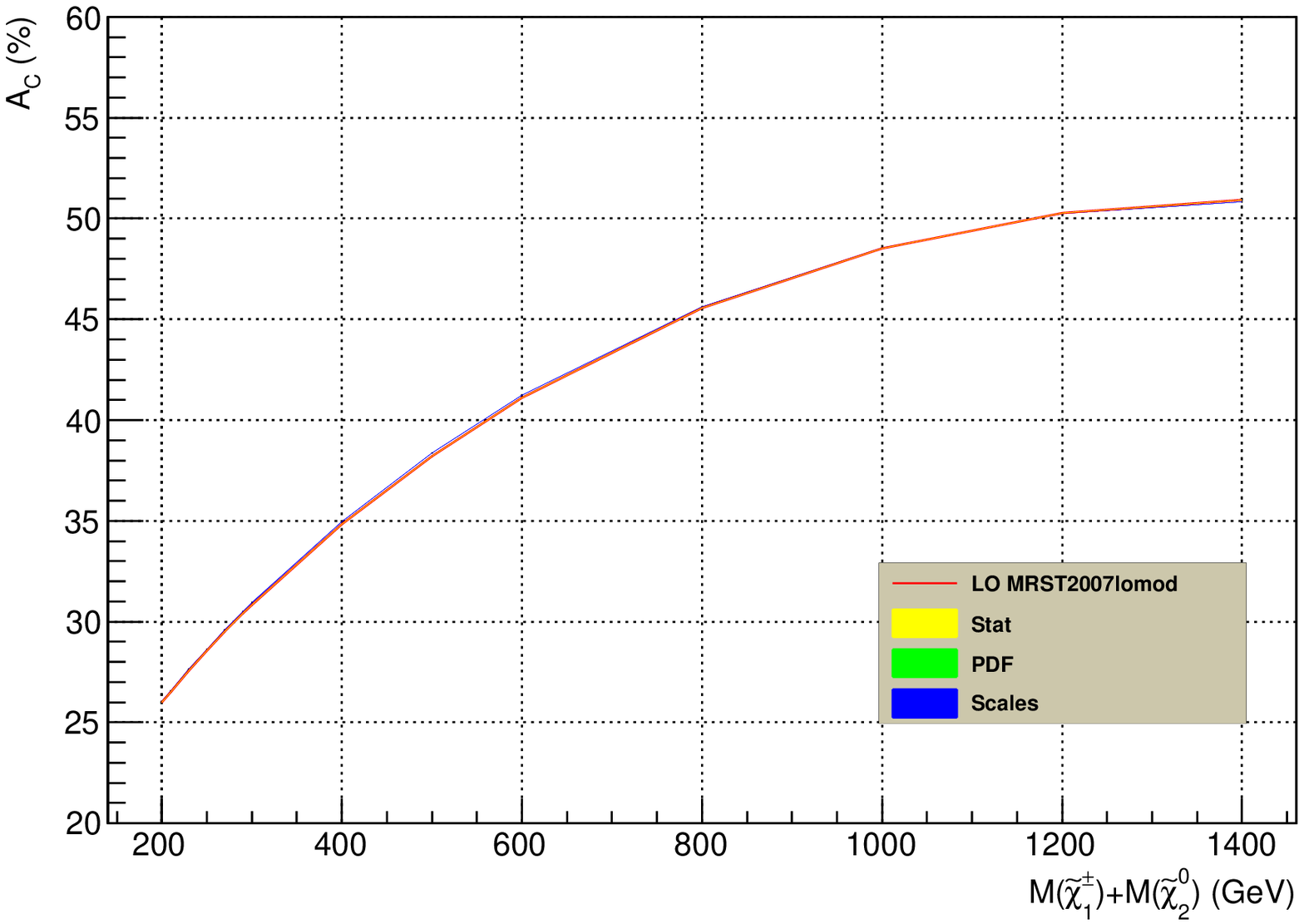}
\includegraphics[scale=0.35]{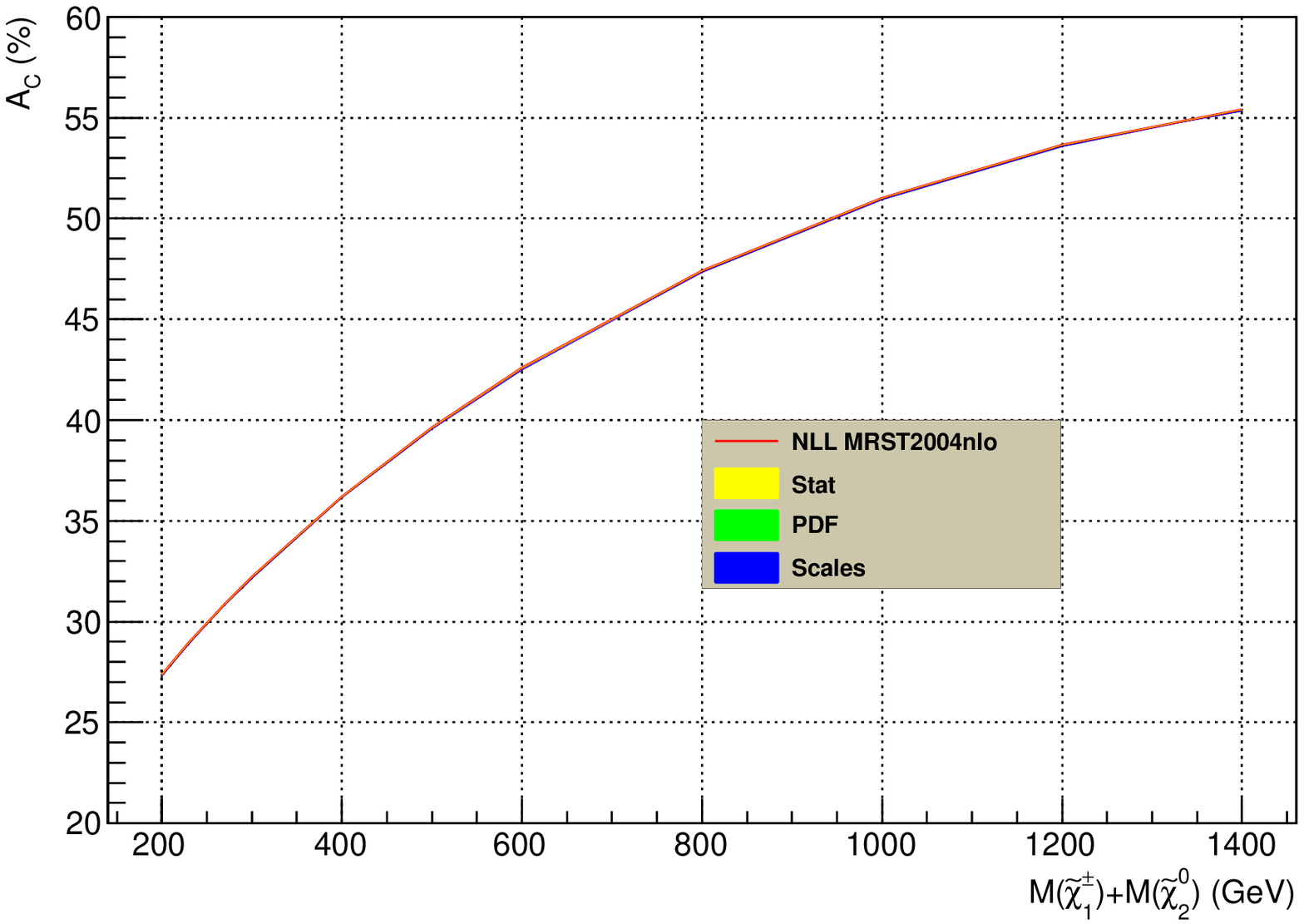}\\
\includegraphics[scale=0.33]{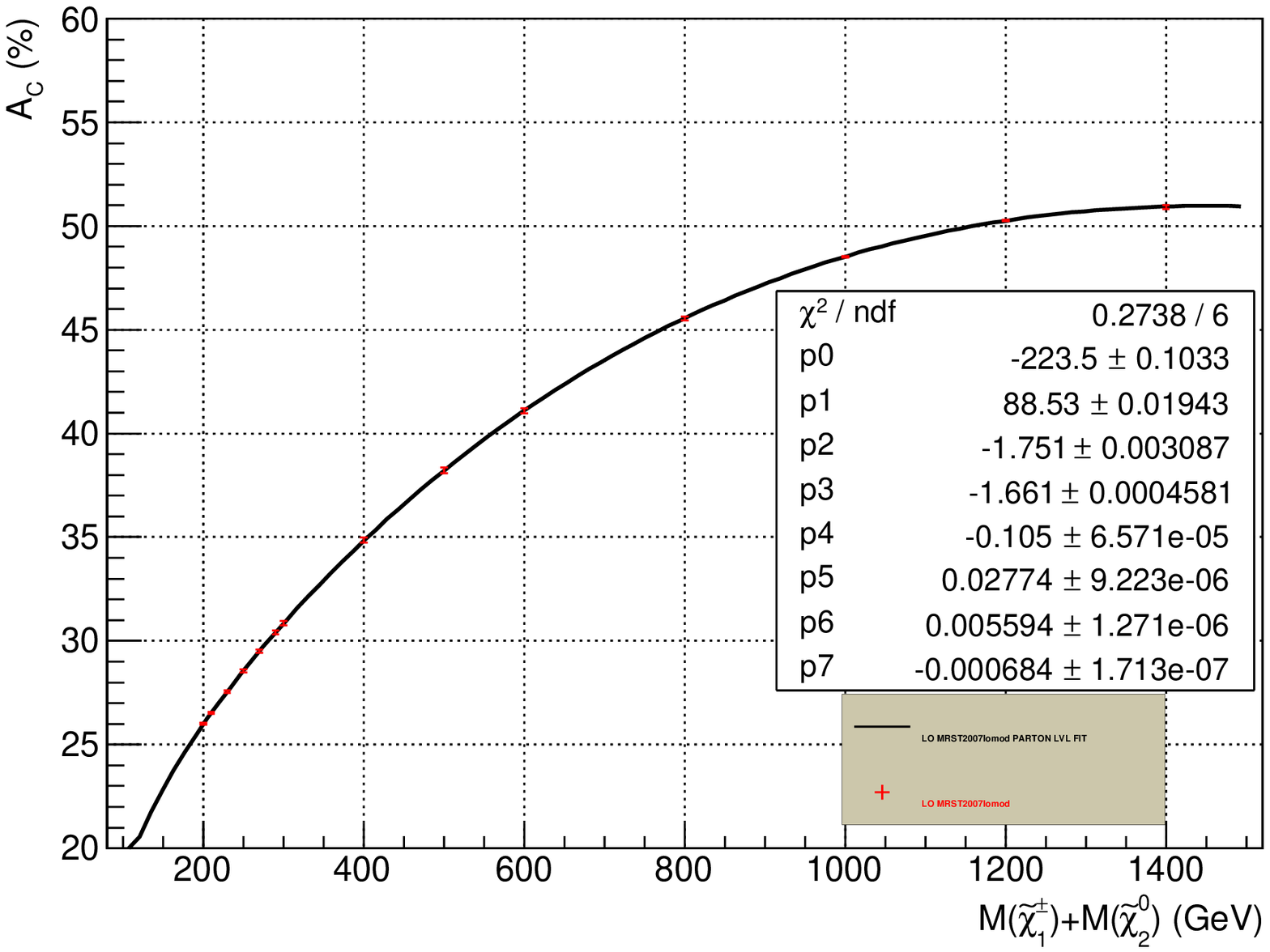}
\includegraphics[scale=0.33]{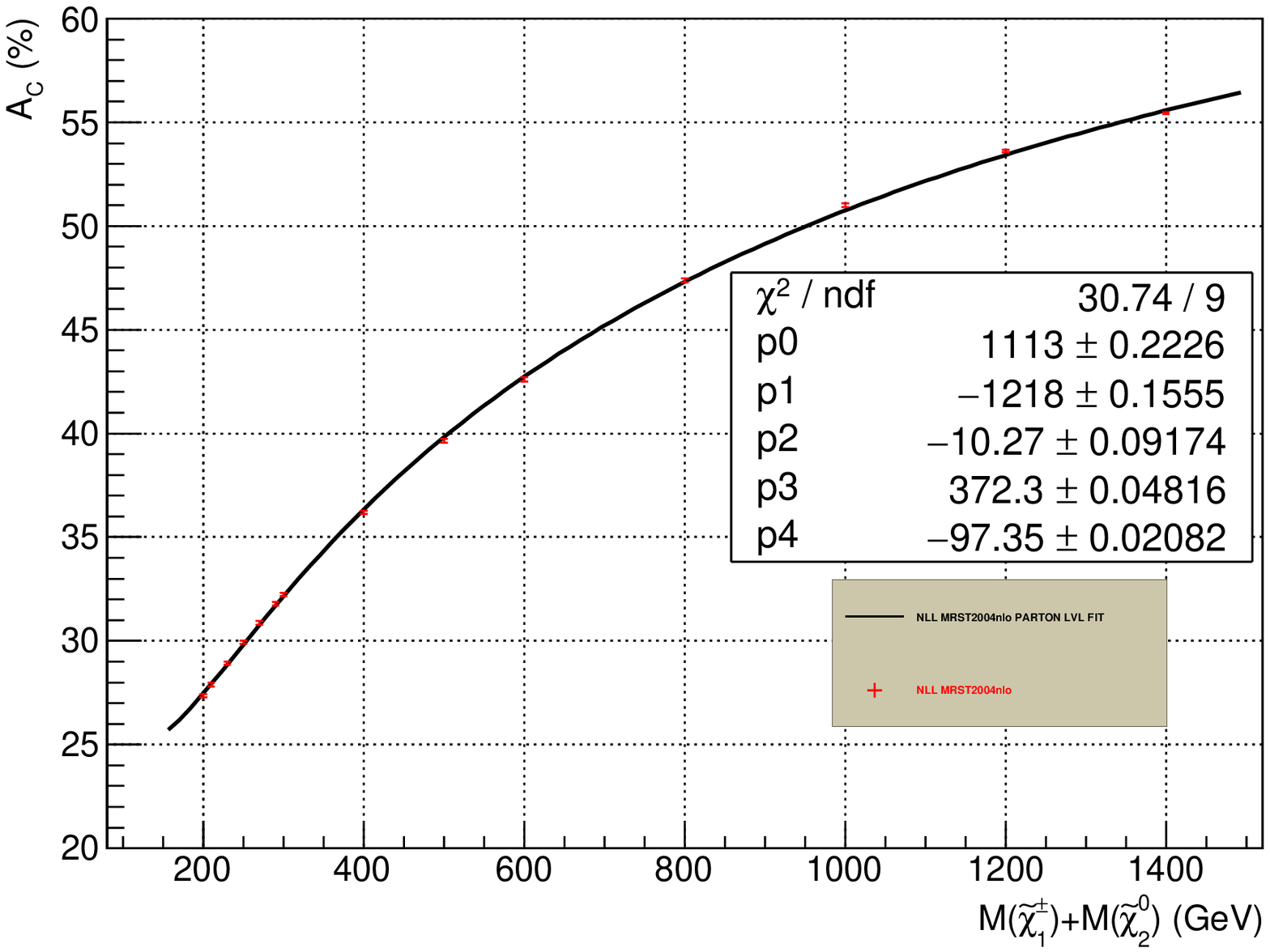}\\
\includegraphics[scale=0.33]{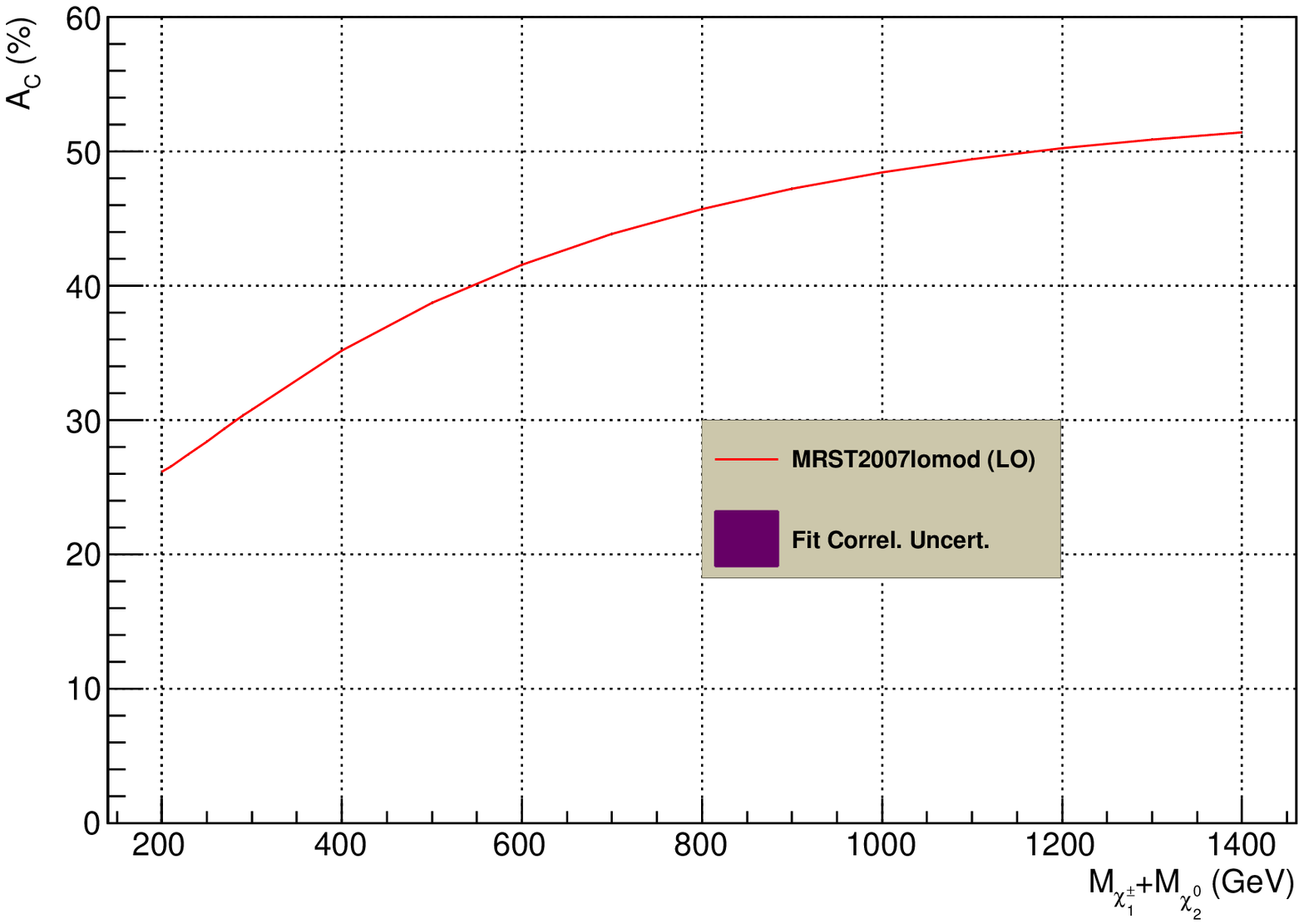}
\includegraphics[scale=0.33]{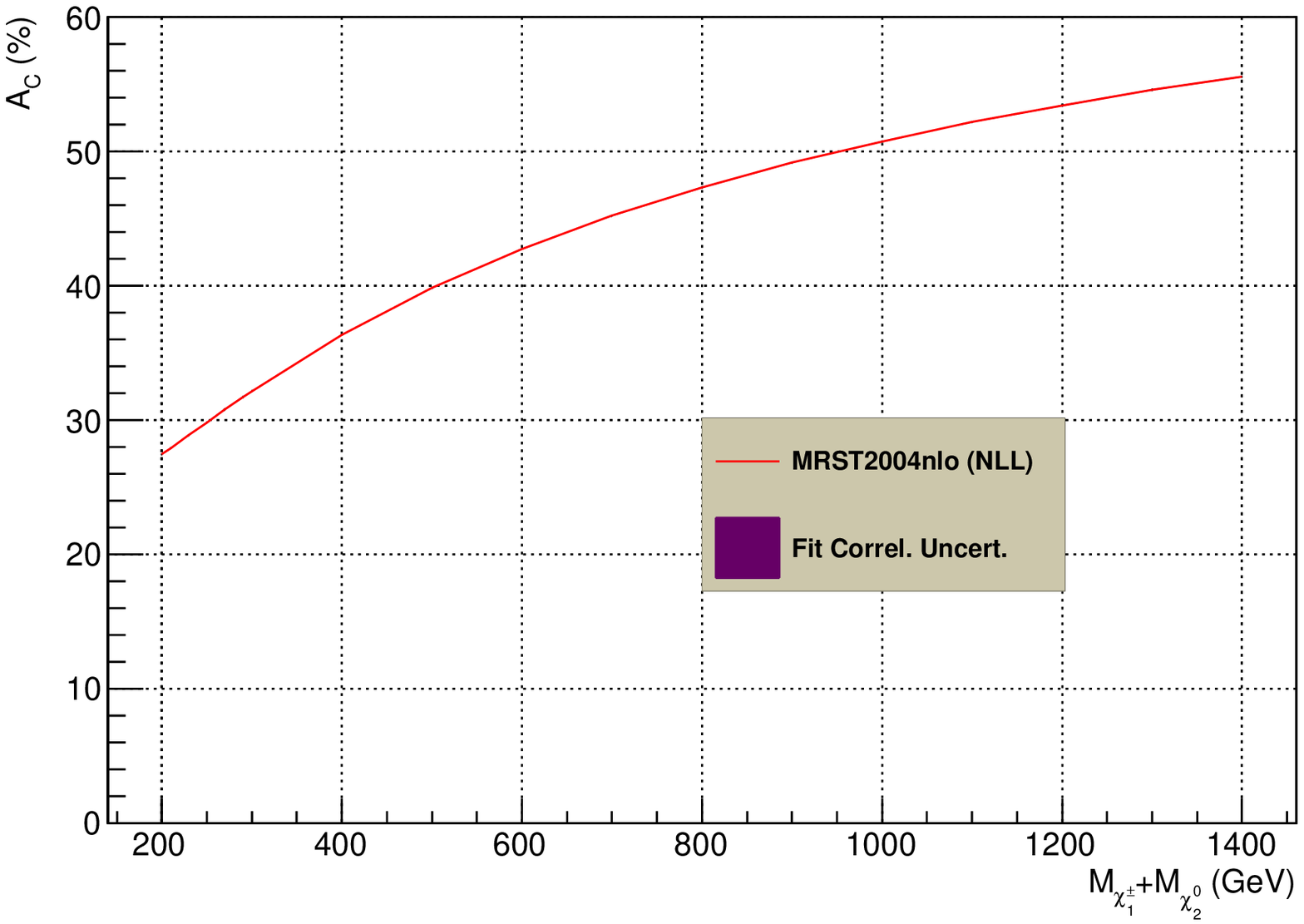}
\end{center}
\caption{\label{II-parton-LVL:Fig2} The theoretical MRST $A_{C}$ template curves. The raw curve with its uncertainty bands and the corresponding fitted curve wtih uncorrelated and with correlated uncertainties are displayed on the top, the middle and the bottom rows, respectively. The LHS concerns the LO calculations based upon the MRST2007lomod PDF and the RHS concerns the NLL calculations using the MRST2004nlo PDF.}
\end{figure}

\vspace*{0.5mm}
\noindent
The theoretical  MRST $A_{C}$ template curves are obtained by computing the $A_{C}$ based upon the cross sections of the signed processes used for table
\ref{II-parton-LVL:Tab:AC_MRST}. They are displayed in figure  \ref{II-parton-LVL:Fig2}.

\begin{table}[htbp]
\begin{center}
\begin{tabular}{|c|c|c|c|c|c|}
\hline
$\rm M_{\chi^{\pm}_{1}}+M_{\chi^{0}_{2}}$  & $A_{C}$ & $\delta (A_{C})_{Stat}$ & $\delta (A_{C})_{Scale}$ & $\delta (A_{C})_{PDF}$ & $\delta (A_{C})_{Total}$\\ 
                                         (GeV)                                &   ($\%$)  &        ($\%$)                        &   ($\%$)                              &               ($\%$)                  &                          ($\%$)      \\
\hline
200.             &      LO: 25.991   & $\pm  0.004$     & $ ^{-0.037}_{+0.056}$ 
             &  	0.000       & $ ^{+0.037}_{-0.056}$ \\
%200.      & NLO: 27.282   & $\pm  0.011$     & $ ^{+0.014}_{-0.027}$ 
%             & not quoted      & $ ^{+0.018}_{-0.029}$ \\
             &  NLL: 27.363  & $\pm  0.011$     & $ ^{+0.092}_{-0.074}$ 
             & not quoted     & $ ^{+0.093}_{-0.075} $ \\
\hline
210.              &     LO:  26.52   & $\pm  0.003$     & $ ^{-0.046}_{+0.063}$ 
                                & 0.000                  & $  ^{+0.046}_{-0.063}$ \\
%  210.     & NLO:   27.825& $\pm  0.009$      & $ ^{+0.015}_{-0.023}$ 
%                                & not quoted        & $ ^{+0.018}_{-0.024}$ \\
              &  NLL:  27.904& $\pm  0.009$      & $  ^{+0.100}_{-0.066}$ 
                                & not quoted        & $  ^{+0.101}_{-0.067}$ \\
\hline
 230.             &    LO:   27.562& $\pm  0.002$    & $  ^{-0.061}_{+0.074  }$ 
                                & 0.000                 & $ ^{+0.061 }_{-0.074}$ \\
%  230.    & NLO:    28.868& $\pm  0.006$    & $ ^{+0.012}_{-0.015}$ 
%                                & not quoted      & $ ^{+0.014}_{-0.016}$ \\
              &  NLL:  28.938& $\pm  0.006$    & $ ^{+0.098}_{ -0.056}$ 
                                & not quoted      & $ ^{+0.099}_{ -0.057}$ \\
\hline
250.              &  LO:   28.549  & $\pm  0.002$    & $ ^{-0.073}_{+0.085}$ 
                                & 0.000                 & $ ^{+0.074 }_{-0.085 }$ \\
%250.       & NLO: 29.865  & $\pm  0.004$    & $ ^{+0.006 }_{-0.014 }$ 
%                                & not quoted      & $ ^{+0.008}_{-0.014}$ \\
              & NLL: 29.934  & $\pm  0.004$    & $ ^{+0.084}_{-0.072}$ 
                                & not quoted      & $ ^{+0.084}_{-0.073}$ \\
\hline
270.              &  LO:   29.495                   & $\pm  0.001$               & $ ^{-0.084}_{+0.094}$ 
                          & 0.000                      & $ ^{+0.084}_{-0.094}$ \\
%   270.    &  NLO: 30.816                   & $\pm  0.003$               & $ ^{+0.001}_{-0.009}$ 
%                          & not quoted           & $ ^{+0.003}_{-0.010}$ \\
               & NLL:  30.877                  & $\pm  0.003$               & $ ^{+0.085}_{-0.088}$ 
                          & not quoted           & $ ^{+0.085}_{-0.088}$ \\
\hline
290.              &   LO:    30.403          & $\pm  0.001$               & $ ^{-0.094}_{+0.102}$ 
              & 0.000                        & $ ^{+0.094}_{-0.102}$ \\
%  290.     & NLO:    31.731       & $\pm  0.002$               & $ ^{-0.006}_{-0.009}$ 
%              & not quoted             & $ ^{+0.007}_{-0.009}$ \\
              &    NLL:   31.786      & $\pm  0.002$               & $ ^{+0.079 }_{-0.091 }$ 
              & not quoted             & $ ^{+0.079 }_{-0.091 }$ \\
\hline
 300.              & LO:    30.844            & $\pm  0.001$               & $ ^{-0.098}_{+0.106}$ 
              & 0.000                        & $ ^{+0.098}_{-0.106}$ \\
%  300.     &  NLO: 32.176         & $\pm  0.002$               & $ ^{-0.010}_{-0.011}$ 
%              & not quoted             & $ ^{+0.010 }_{-0.011 }$ \\
              &    NLL:  32.229      & $\pm  0.002$               & $ ^{+0.076}_{-0.093}$ 
              & not quoted             & $ ^{+0.076}_{-0.093}$ \\
\hline	      
400.              &  LO:    34.847           & $\pm  0.000$               & $ ^{-0.125}_{+0.126}$ 
              & 0.000                  & $ ^{+0.125 }_{-0.126 }$ \\
%  400.     &  NLO: 36.201            & $\pm  0.001$               & $ ^{-0.007 }_{-0.005 }$ 
%              & not quoted        & $ ^{+0.007}_{-0.005}$ \\
              &    NLL: 36.213          & $\pm  0.001$               & $ ^{+0.086}_{-0.069}$ 
              & not quoted        & $ ^{+0.086 }_{-0.069 }$ \\
\hline
500.              &  LO:     38.230          & $\pm   0.000$               & $ ^{-0.132 }_{+0.131 }$ 
              & 0.000                  & $ ^{+0.132}_{-0.131}$ \\
%  500.    & NLO: 39.651             & $\pm   0.000$               & $ ^{-0.022 }_{+0.010 }$ 
%              & not quoted        & $ ^{+0.022 }_{-0.010 }$ \\
              &     NLL:  39.648        & $\pm   0.000$               & $ ^{ +0.101}_{ -0.100}$ 
              & not quoted        & $ ^{+0.101}_{-0.100}$ \\
\hline
 600.               &   LO:   41.101           & $\pm  0.000$               & $ ^{-0.127}_{+0.124}$ 
              & 0.000                  & $ ^{+0.127}_{-0.124}$ \\
%  600.     &   NLO:  42.614          & $\pm  0.000$               & $ ^{-0.021}_{+0.009}$ 
%              & not quoted        & $ ^{+0.021}_{-0.009}$ \\
              &   NLL:   42.600         & $\pm  0.000$               & $ ^{+0.104}_{-0.129}$ 
             & not quoted        & $ ^{+0.104}_{-0.129}$ \\
\hline
 800.              & LO:      45.548          & $\pm  0.000$               & $ ^{-0.091}_{+0.086}$ 
              & 0.000                  & $ ^{+0.091}_{-0.086}$ \\
%   800.    &  NLO:  47.393           & $\pm  0.000$               & $ ^{-0.001}_{+0.016}$ 
%              & not quoted        & $ ^{+0.001 }_{-0.016 }$ \\
              &   NLL:  47.420          & $\pm  0.000$               & $ ^{+0.118 }_{-0.073 }$ 
              & not quoted        & $ ^{+0.118}_{-0.073}$ \\
\hline  
1000.              &  LO:   48.528            & $\pm  0.000$               & $ ^{-0.038}_{+0.033}$ 
              & 0.000                  & $ ^{ +0.038}_{-0.033 }$ \\
%  1000.   &  NLO:  50.958           & $\pm  0.000$               & $ ^{+0.005}_{+0.007}$ 
%              & not quoted        & $ ^{+0.005}_{-0.007}$ \\
              &   NLL:  51.035          & $\pm  0.000$               & $ ^{+0.116}_{-0.063}$ 
              & not quoted        & $ ^{ +0.116}_{ -0.063}$ \\
\hline
 1200.             &  LO:    50.264           & $\pm  0.000$               & $ ^{+0.024}_{-0.025}$ 
              & 0.000                  & $ ^{+0.024 }_{-0.025 }$ \\
 %1200.    &  NLO:  53.540           & $\pm  0.000$               & $ ^{+0.003}_{+0.033}$ 
%              & not quoted        & $ ^{+0.003}_{-0.033}$ \\
              &    NLL: 53.658          & $\pm  0.000$               & $ ^{ +0.101}_{+0.021 }$ 
             & not quoted        & $ ^{ +0.101}_{ -0.021}$ \\
\hline
 1400.              & LO:    50.924            & $\pm  0.000$               & $ ^{+0.088}_{-0.081}$ 
              & 0.000                  & $ ^{+0.088}_{-0.081}$ \\
%1400.     &  NLO:  55.357           & $\pm  0.000$               & $ ^{-0.022}_{-0.074}$ 
%              & not quoted        & $ ^{+0.022}_{-0.074}$ \\
              &   NLL:  55.404          & $\pm  0.000$               & $ ^{+0.008}_{-0.083}$ 
              & not quoted        & $^{+0.008 }_{-0.083}$ \\
\hline
\end{tabular}       
\end{center}
\caption{\label{II-parton-LVL:Tab:AC_MRST} The MRST $A_{C}(\tilde\chi^{\pm}_{1}\tilde\chi^{0}_{2})$ table with the breakdown of the different sources of theoretical uncertainty.}
\end{table}

\begin{table}[htbp]
\begin{center}
\begin{tabular}{|c|c|c|}
\hline
$\rm M_{\chi^{\pm}_{1}}+M_{\chi^{0}_{2}}\ (GeV) $   & $A_{C}^{Fit}$  &  $\delta A^{Fit}_{C}$  \\ 
                                                   (GeV)                                     &             ($\%$)    &          ($\%$)                    \\ 
\hline
200.             &      LO:  25.984  & $\pm  0.025$      \\
                    &  NLL:  27.435   & $\pm  0.031$      \\
\hline
210.            &      LO:  26.530   & $\pm  0.024$      \\
                   &  NLL:  27.927    & $\pm 0.030 $      \\
\hline
230.           &      LO:   27.571  & $\pm  0.024$      \\
                  &  NLL: 28.904     & $\pm  0.028$      \\
 \hline
250.          &      LO:  28.557   & $\pm  0.023$      \\
                 &  NLL: 29.866      & $\pm  0.027$      \\
\hline
270.          &      LO:  29.498   & $\pm  0.023$      \\
                &  NLL: 30.807      & $\pm  0.027$      \\
\hline
290.         &      LO:  30.400  & $\pm 0.022 $      \\
                &  NLL: 31.724      & $\pm  0.026$      \\
\hline
 300.       &      LO:  30.838  & $\pm  0.022$      \\
               &  NLL:  32.172   & $\pm  0.026$      \\
\hline	      
400.       &      LO:  34.824  & $\pm 0.021 $      \\
              &  NLL:  36.286   & $\pm  0.025$      \\
\hline
500.      &      LO:  38.215  & $\pm 0.020 $      \\
             &  NLL: 39.768    & $\pm  0.027$      \\
\hline
 600.      &      LO:  41.102  & $\pm  0.019$      \\
              &  NLL: 42.720    & $\pm  0.029$      \\
\hline
 800.     &      LO:  45.562  & $\pm  0.016$      \\
             &  NLL:  47.400   & $\pm  0.034$      \\
\hline  
1000.    &      LO:  48.532  & $\pm  0.015$      \\
             &  NLL:  50.881   & $\pm   0.041$      \\  
\hline
1200.    &      LO:  50.261  & $\pm  0.017$      \\
             &  NLL: 53.508    & $\pm  0.049$      \\ 
\hline
1400.    &      LO:   50.945 & $\pm  0.022$      \\
             &  NLL:  55.501   & $\pm  0.057$      \\
\hline
\end{tabular}       
\end{center}
\caption{\label{II-parton-LVL:Tab:AC_Fit_MRST} The MRST $A_{C}^{Fit}(\tilde\chi^{\pm}_{1}\tilde\chi^{0}_{2})$ table with its theoretical uncertainty accounting for the correlations between the parameters fitting the $A_{C}^{Raw}$ template curves.}
\end{table}

%%%%%%%%%%%%%%%%%%%%%%%%%%%%%%%%%%%%%%%%%%%%%%%%%%%%%%%%%%%%%%%%%%%%%%%%%%

\vspace*{1.5mm}
\subsubsection{\label{II-parton-LVL:CTEQ} $A_{C}(\tilde\chi^{\pm}_{1}+\tilde\chi^{0}_{2})$ Template Curves for CTEQ6}
\begin{figure}[htbp]
\begin{center}
\includegraphics[scale=0.35]{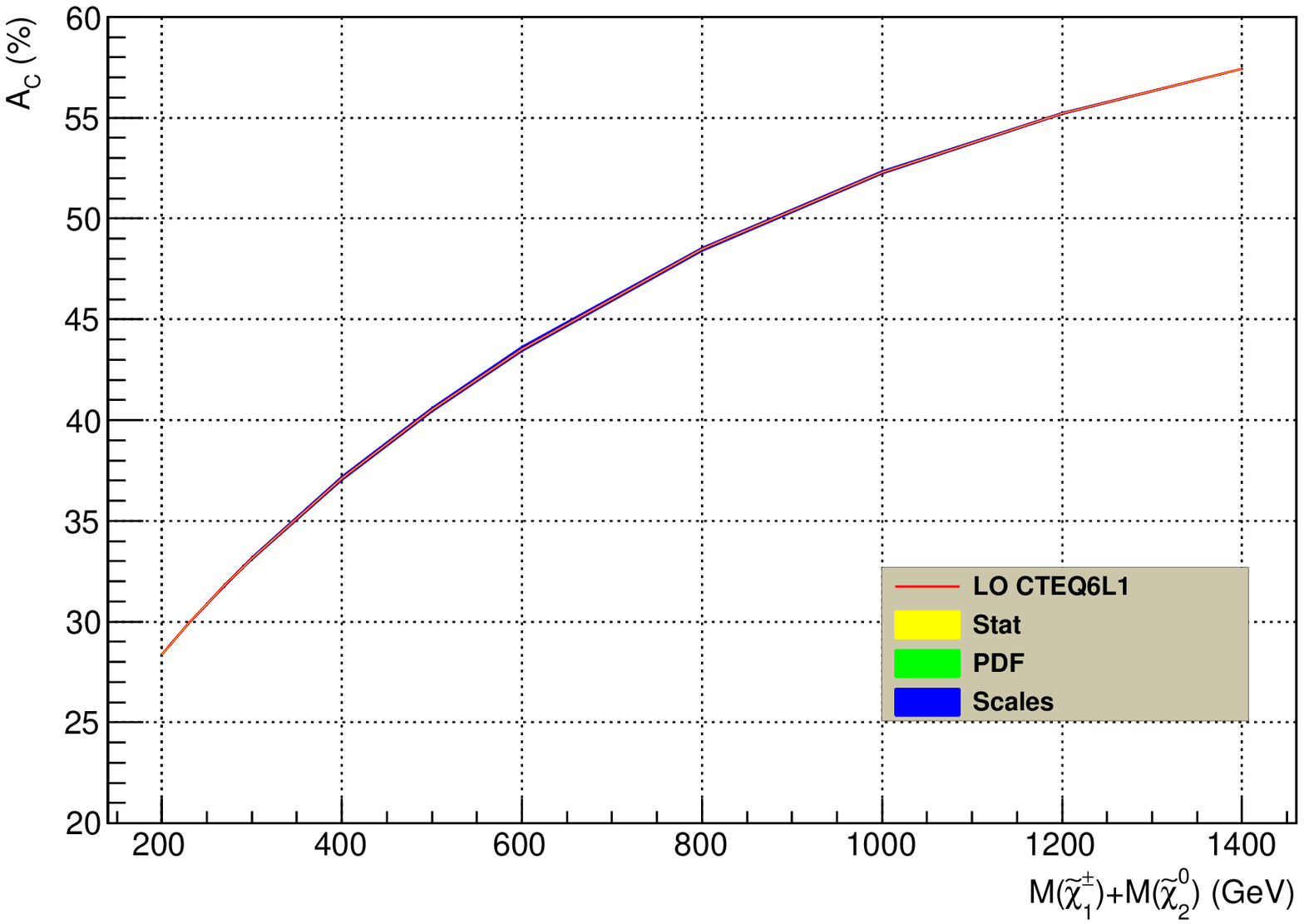}
\includegraphics[scale=0.35]{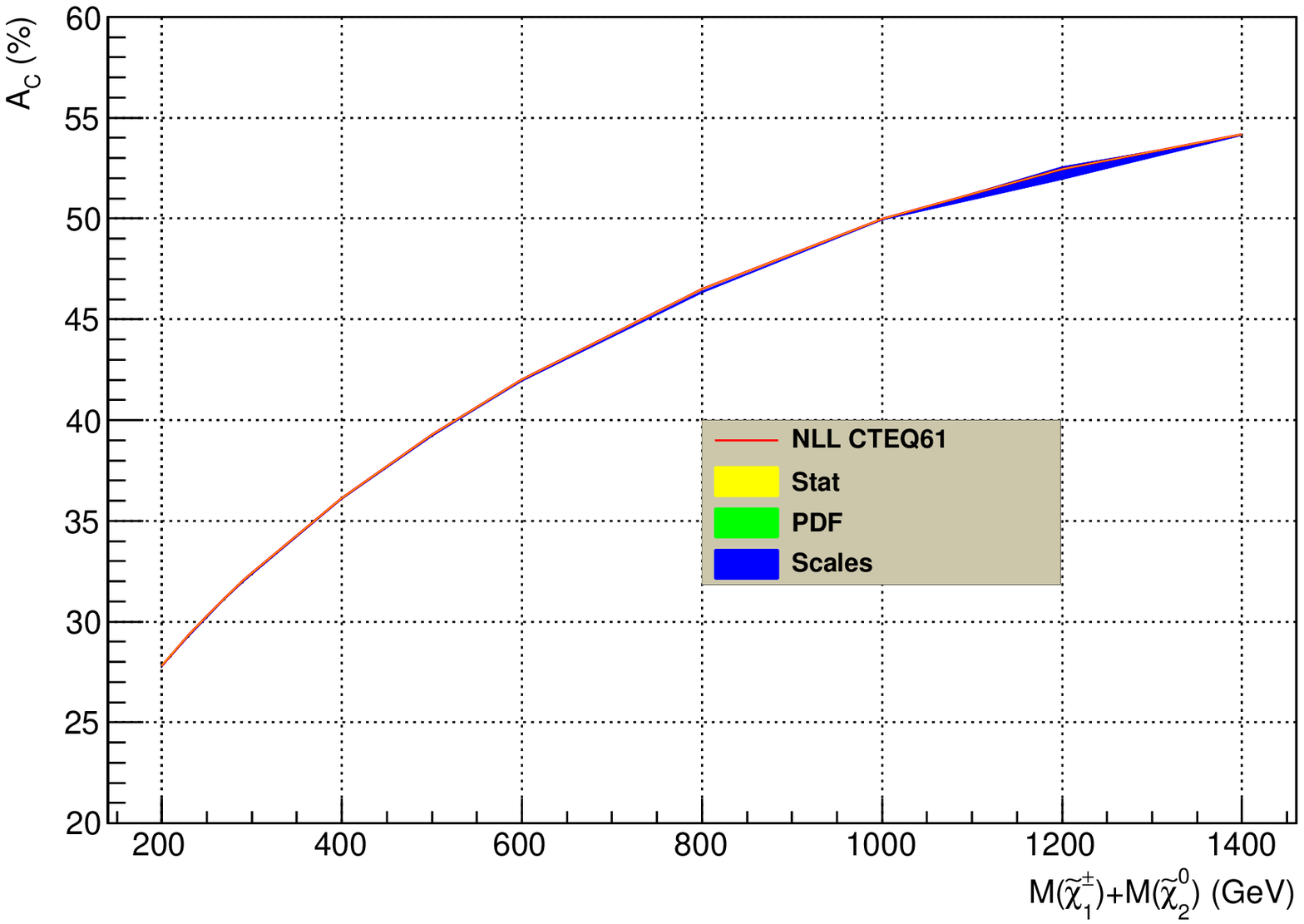}\\
\includegraphics[scale=0.33]{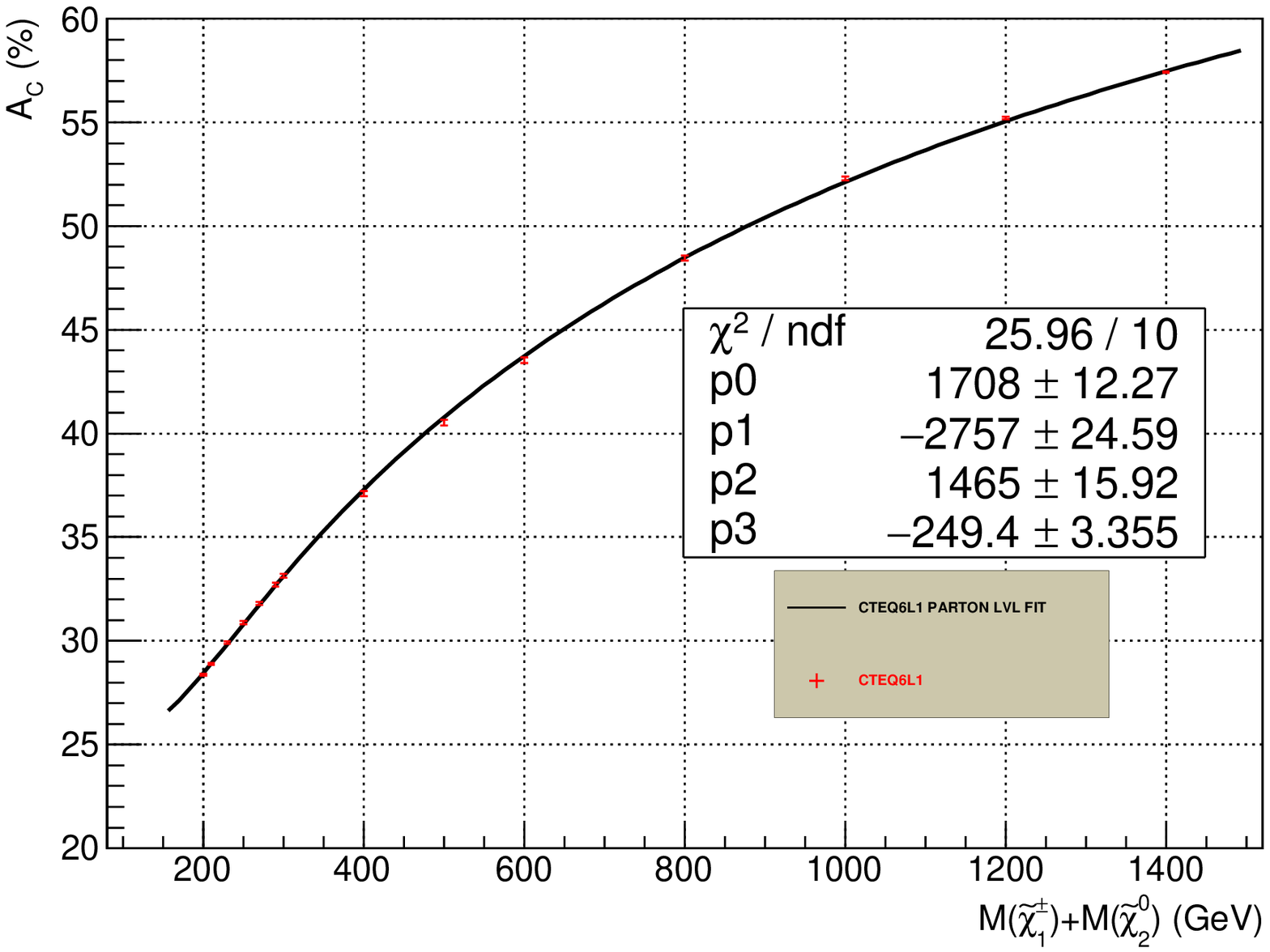}
\includegraphics[scale=0.33]{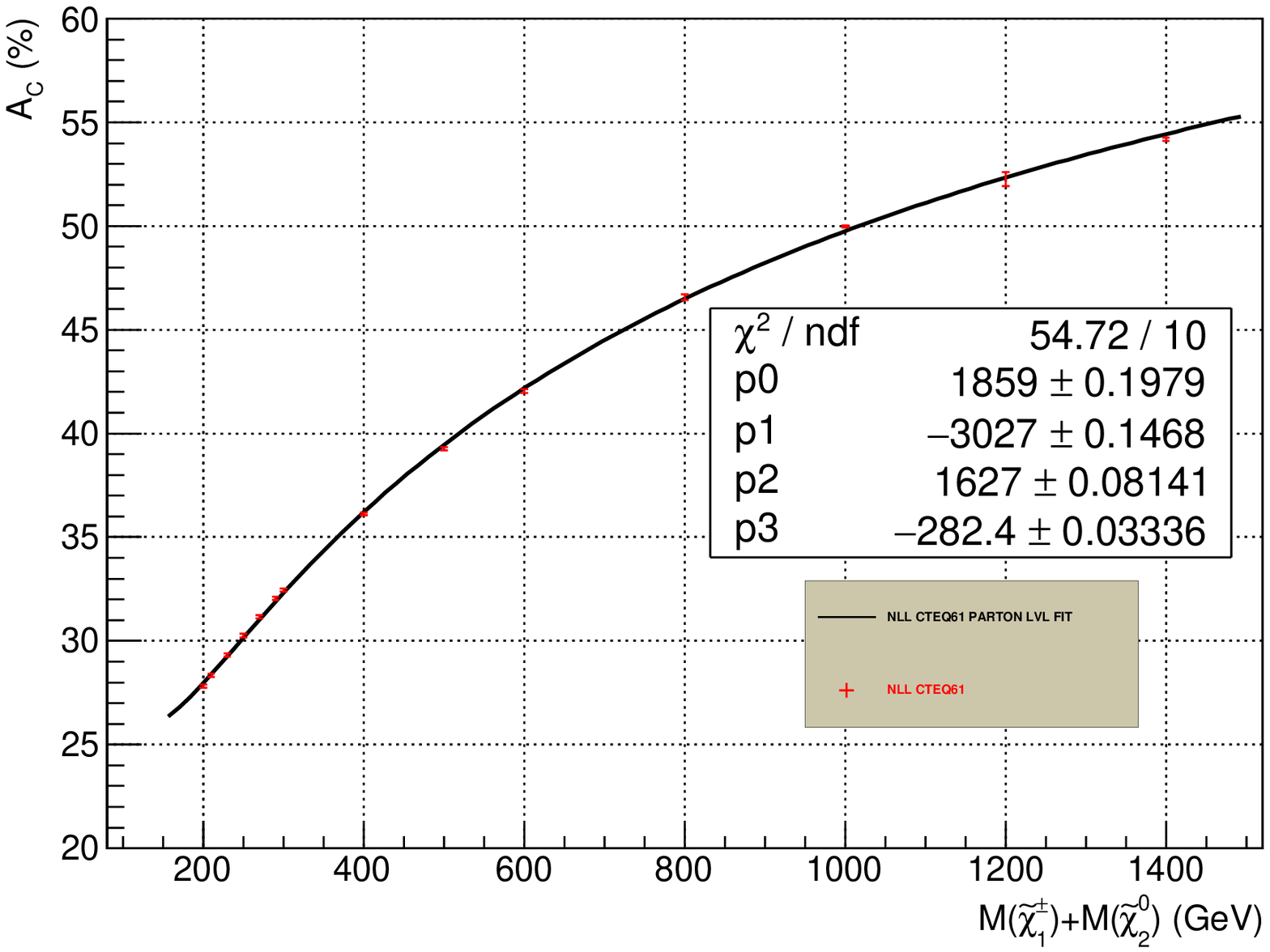}\\
\includegraphics[scale=0.33]{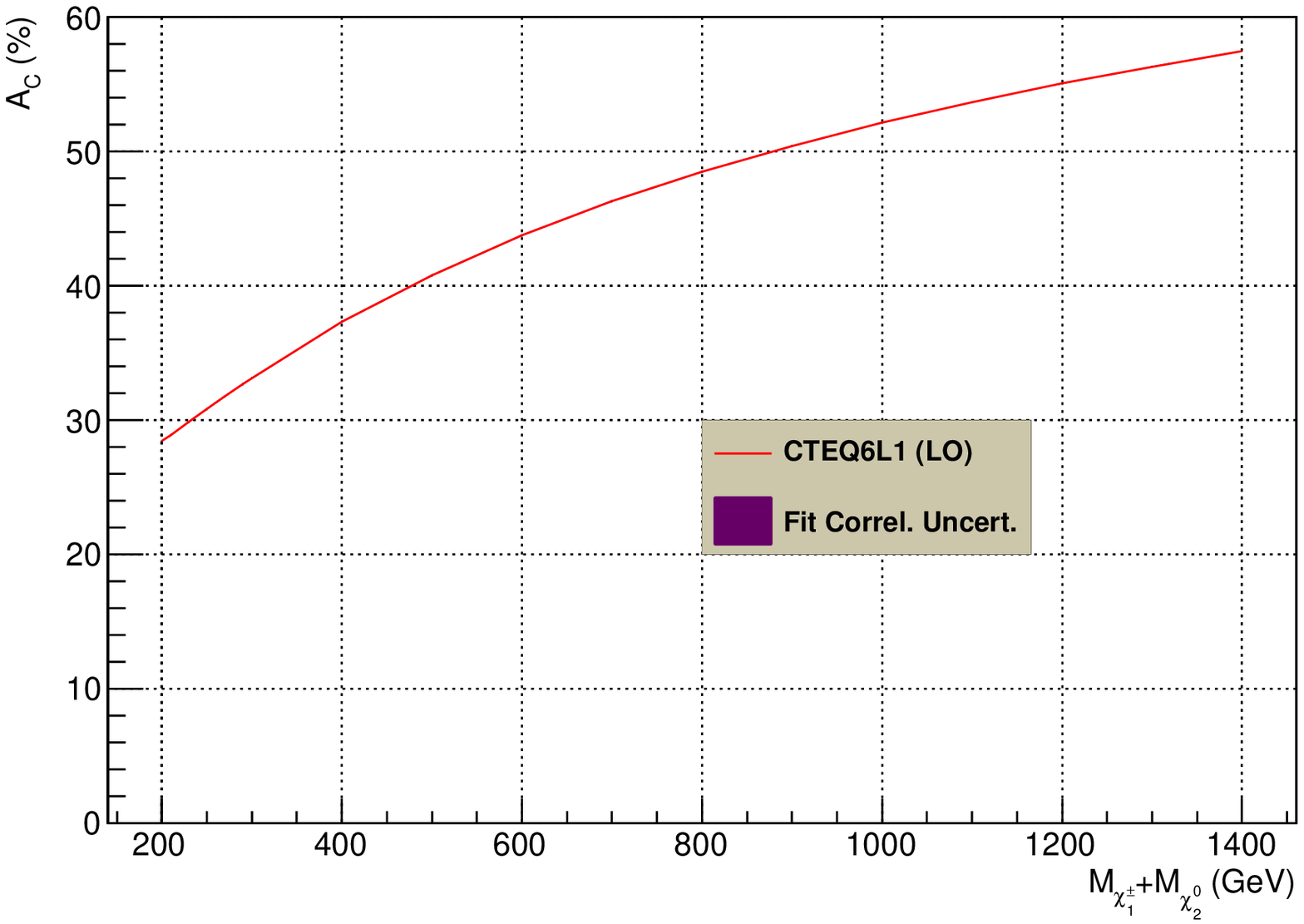}
\includegraphics[scale=0.33]{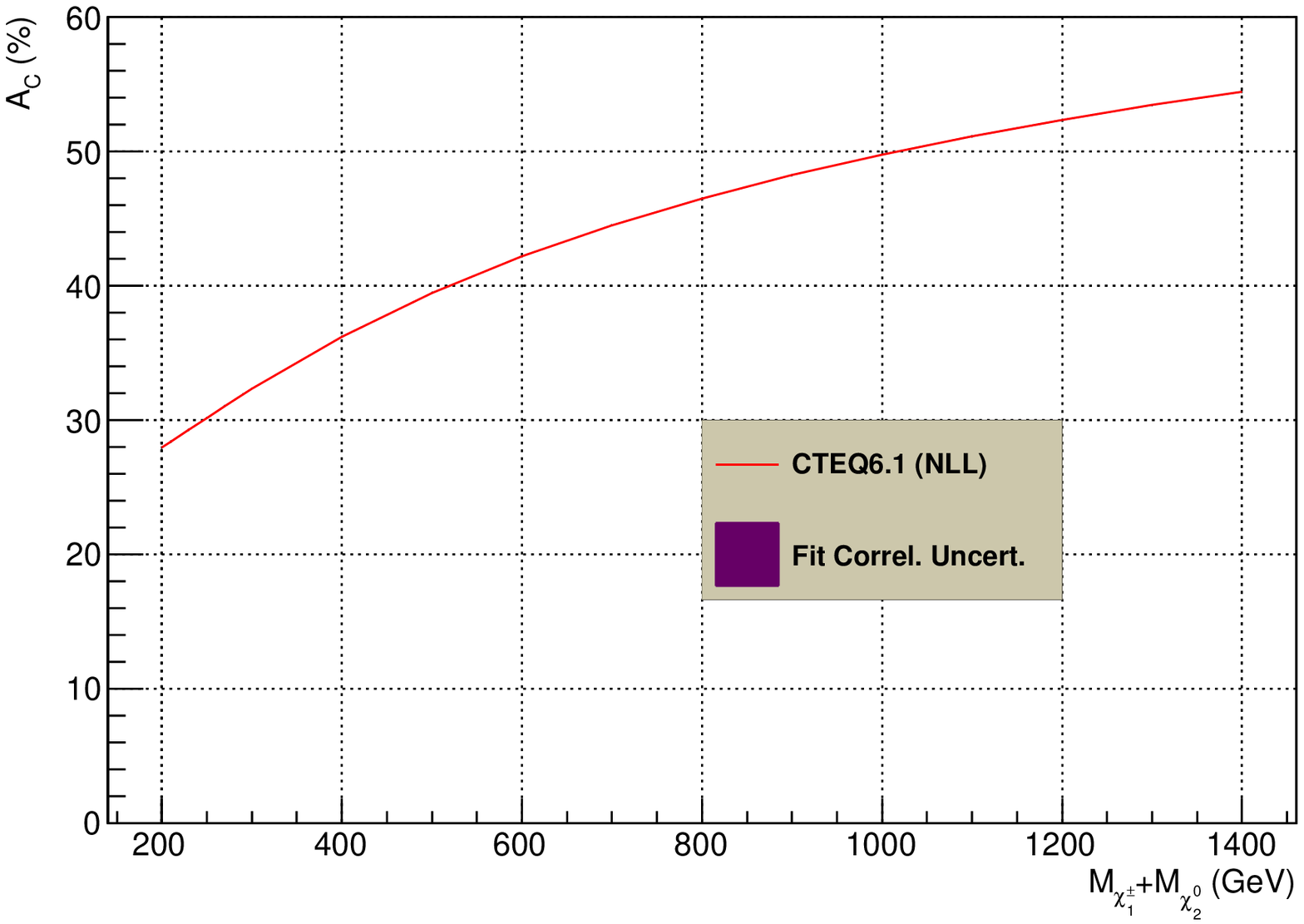}
\end{center}
\caption{\label{II-parton-LVL:Fig1} The theoretical CTEQ6 $A_{C}$ template curves. The raw curve with its uncertainty bands and the corresponding fitted curve wtih uncorrelated and with correlated uncertainties are displayed on the top, the middle and the bottom rows, respectively. The LHS concerns the LO calculations based upon the CTEQ6L1 PDF and the RHS concerns the NLL calculations using the CTEQ6.1 PDF.}
\end{figure}

\vspace*{0.5mm}
\noindent
The theoretical  CTEQ6 $A_{C}$ template curves are obtained by computing the $A_{C}$ based upon the cross sections of the signed processes used for table
\ref{II-parton-LVL:Tab:AC_CTEQ6}. They are displayed in figure  \ref{II-parton-LVL:Fig1}.

\newpage
\begin{table}[htbp]
\begin{center}
\begin{tabular}{|c|c|c|c|c|c|}
\hline
$\rm M_{\chi^{\pm}_{1}}+M_{\chi^{0}_{2}}$  & $A_{C}$ & $\delta (A_{C})_{Stat}$ & $\delta (A_{C})_{Scale}$ & $\delta (A_{C})_{PDF}$ & $\delta (A_{C})_{Total}$\\ 
                                         (GeV)                                &   ($\%$)  &        ($\%$)                        &   ($\%$)                              &               ($\%$)                  &                          ($\%$)      \\
\hline
200.              &   LO:  28.367    & $\pm  0.003$               & $ ^{-0.030}_{+0.045}$ 
              &   0.000              & $ ^{+0.030}_{-0.045}$ \\
%  200.     &   NLO: 27.735 & $\pm  0.010$               & $ ^{+0.024}_{-0.035}$ 
%              &   not quoted    & $ ^{+0.026}_{-0.037}$ \\
              &   NLL:  27.822 & $\pm  0.010$               & $ ^{+0.076}_{-0.074}$ 
              &   not quoted    & $ ^{+0.077}_{-0.075}$ \\	      
\hline
210.              &    LO:  28.896               & $\pm  0.003$               & $ ^{-0.038}_{+0.051}$ 
              &   0.000              & $ ^{+0.038}_{-0.051}$ \\
%210.       &   NLO: 28.259             & $\pm  0.008$               & $ ^{+0.024}_{-0.033}$ 
%              &   not quoted    & $ ^{+0.025}_{-0.035}$ \\
              &   NLL: 28.345             & $\pm  0.008$               & $ ^{+0.084}_{-0.069}$ 
              &   not quoted    & $ ^{+0.084}_{-0.069}$ \\	      
\hline
230.              &   LO:    29.911 & $\pm  0.002$               & $ ^{-0.053}_{+0.064}$ 
              &   0.000              & $ ^{+0.053}_{-0.064}$ \\
%230.      &   NLO: 29.255 & $\pm  0.006$               & $ ^{+0.026}_{-0.027}$ 
%              &   not quoted    & $ ^{+0.026}_{-0.028}$ \\
              &   NLL:  29.333 & $\pm  0.006$               & $ ^{+0.102}_{-0.054}$ 
              &   not quoted    & $ ^{+0.102}_{-0.054}$ \\	      
\hline
250.              &    LO:   30.880  & $\pm  0.001$               & $ ^{-0.066}_{+0.074}$ 
              &   0.000              & $ ^{+0.066}_{+0.074}$ \\
%250.      &   NLO:  30.199 & $\pm  0.004$               & $ ^{+0.023}_{-0.020}$ 
%              &   not quoted    & $ ^{+0.023}_{-0.020}$ \\
              &   NLL:  30.273& $\pm  0.004$               & $ ^{+0.093}_{-0.064}$ 
              &   not quoted    & $ ^{+0.093}_{-0.064}$ \\	      
\hline
270.               &   LO:  31.808               & $\pm  0.001$               & $ ^{-0.077}_{+0.084}$ 
              &   0.000              & $ ^{+0.077}_{-0.084}$ \\
%270.       &   NLO:  31.101           & $\pm  0.003$               & $ ^{+0.011}_{-0.019}$ 
%              &   not quoted    & $ ^{+0.011}_{-0.020}$ \\
              &   NLL:  31.169           & $\pm  0.003$               & $ ^{+0.078}_{-0.070}$ 
              &   not quoted    & $ ^{+0.078}_{-0.070}$ \\	      
\hline
290.              &    LO:   32.701             & $\pm  0.001$               & $ ^{-0.087}_{+0.092}$ 
              &   0.000              & $ ^{+0.087}_{-0.092}$ \\
%290.      &   NLO:  31.960           & $\pm  0.002$               & $ ^{+0.005}_{-0.017}$ 
%              &   not quoted    & $ ^{+0.005}_{-0.017}$ \\
              &   NLL: 32.026            & $\pm  0.002$               & $ ^{+0.065}_{-0.090}$ 
              &   not quoted    & $ ^{+0.065}_{-0.090}$ \\	      
\hline
300.               &   LO:   33.135              & $\pm  0.001$               & $ ^{-0.091}_{+0.096}$ 
              &   0.000              & $ ^{+0.091}_{-0.096}$ \\
%300.       &   NLO:  32.374           & $\pm  0.002$               & $ ^{+0.003}_{-0.012}$ 
%              &   not quoted    & $ ^{+0.003}_{-0.012}$ \\
              &   NLL:  32.434           & $\pm  0.002$               & $ ^{+0.065}_{-0.089}$ 
              &   not quoted    & $ ^{+0.065}_{-0.089}$ \\	      
\hline
 400.              &   LO:   37.104              & $\pm  0.000$               & $ ^{-0.121}_{+0.121}$ 
              &   0.000              & $ ^{+0.121}_{-0.121}$ \\
%400.       &   NLO: 36.117            & $\pm  0.001$               & $ ^{+0.001}_{-0.004}$ 
%              &   not quoted    & $ ^{+0.001}_{-0.005}$ \\
              &   NLL:  36.136           & $\pm  0.001$               & $ ^{+0.080}_{-0.055}$ 
              &   not quoted    & $ ^{+0.080}_{-0.055}$ \\	      
\hline
500.               &    LO:  40.531              & $\pm  0.000$               & $ ^{-0.134}_{+0.131}$ 
              &   0.000              & $ ^{+0.134}_{-0.131}$ \\
%500.      &   NLO:  39.293           & $\pm  0.000$               & $ ^{-0.016}_{+0.003}$ 
%              &   not quoted    & $ ^{+0.016}_{-0.003}$ \\
              &   NLL:  39.285           & $\pm 0.000 $               & $ ^{+0.088}_{-0.057}$ 
              &   not quoted    & $ ^{+0.088}_{-0.057}$ \\	      
\hline
 600.             &   LO:   43.527              & $\pm  0.000$               & $ ^{-0.137}_{+0.132}$ 
              &   0.000              & $ ^{+0.137}_{-0.132}$ \\
%600.       &   NLO:  42.027           & $\pm  0.000$               & $ ^{-0.017}_{+0.004}$ 
%              &   not quoted    & $ ^{+0.017}_{-0.004}$ \\
              &   NLL:  42.023           & $\pm  0.000$               & $ ^{+0.056}_{-0.119}$ 
              &   not quoted    & $ ^{+0.056}_{-0.119}$ \\	      
\hline
800.              &  LO:    48.473              & $\pm  0.000$               & $ ^{-0.121}_{+0.116}$ 
              &   0.000              & $ ^{+0.121}_{-0.116}$ \\
%800.       &   NLO:   46.464          & $\pm  0.000$               & $ ^{+0.004}_{+0.010}$ 
%              &   not quoted    & $ ^{+0.004}_{-0.010}$ \\
              &   NLL:  46.514           & $\pm 0.000 $               & $ ^{+0.094}_{-0.194}$ 
              &   not quoted    & $ ^{+0.094}_{-0.194}$ \\	      
\hline  
1000.              &  LO:   52.293               & $\pm  0.000$               & $ ^{-0.094}_{+0.090}$ 
              &   0.000              & $ ^{+0.094}_{-0.090}$ \\
%1000.     &   NLO:  49.824           & $\pm  0.000$               & $ ^{+0.007}_{+0.001}$ 
%              &   not quoted    & $ ^{+0.007}_{-0.001}$ \\
              &   NLL: 49.985            & $\pm  0.000$               & $ ^{+0.054}_{-0.053}$ 
              &   not quoted    & $ ^{+0.054}_{-0.053}$ \\	      
\hline
1200.              &   LO:   55.219              & $\pm  0.000$               & $ ^{-0.063}_{+0.061}$ 
              &   0.000              & $ ^{+0.063}_{-0.061}$ \\
%1200.     &   NLO: 52.255            & $\pm  0.000$               & $ ^{+0.010}_{+0.020}$ 
%              &   not quoted    & $ ^{+0.010}_{-0.020}$ \\
              &   NLL:  52.447           & $\pm  0.000$               & $ ^{+0.528}_{+0.147}$ 
              &   not quoted    & $ ^{+0.528}_{-0.147}$ \\	      
\hline
1400.              &   LO:   57.428              & $\pm  0.000$               & $ ^{-0.034}_{+0.033}$ 
              &   0.000              & $ ^{+0.034}_{-0.033}$ \\
%1400.     &   NLO: 53.906            & $\pm  0.000$               & $ ^{-0.006}_{-0.047}$ 
%              &   not quoted    & $ ^{+0.006}_{-0.047}$ \\
              &   NLL: 54.190            & $\pm  0.000$               & $ ^{+0.069}_{-0.081}$ 
              &   not quoted    & $ ^{+0.069}_{-0.081}$ \\	      
\hline
\end{tabular}       
\end{center}
\caption{\label{II-parton-LVL:Tab:AC_CTEQ6} The CTEQ6 $A_{C}(\tilde\chi^{\pm}_{1}\tilde\chi^{0}_{2})$ table with the breakdown of the different sources of theoretical uncertainty.}
\end{table}

\begin{table}[htbp]
\begin{center}
\begin{tabular}{|c|c|c|}
\hline
$\rm M_{\chi^{\pm}_{1}}+M_{\chi^{0}_{2}}\ (GeV) $   & $A_{C}^{Fit}$  &  $\delta A^{Fit}_{C}$  \\ 
                                                   (GeV)                                     &             ($\%$)    &          ($\%$)                    \\ 
\hline
200.      &      LO:  28.407  & $\pm  0.034$      \\
             &  NLL: 27.811    & $\pm  0.027$      \\
\hline
210.      &      LO:  28.900  & $\pm  0.027$      \\
             &  NLL:  28.340   & $\pm  0.026$      \\
\hline
 230.     &      LO:   29.876 & $\pm 0.023 $      \\
             &  NLL: 29.342    & $\pm 0.024 $      \\
 \hline
250.      &      LO:  30.832  & $\pm  0.027$      \\
             &  NLL: 30.282    & $\pm  0.023$      \\
\hline
270.      &      LO:  31.766  & $\pm  0.032$      \\
             &  NLL:  31.172   & $\pm  0.022$      \\
\hline
290.      &      LO:  32.674  & $\pm 0.037$      \\
             &  NLL:  32.018   & $\pm  0.022$      \\
\hline
 300.     &      LO:  33.119  & $\pm  0.038$      \\
             &  NLL:  32.428   & $\pm  0.022$      \\
\hline	      
400.       &      LO:  37.203  & $\pm  0.046$      \\
             &  NLL:  36.126    & $\pm  0.023$      \\
\hline
500.      &      LO:  40.687  & $\pm  0.048$      \\
             &  NLL:   39.287  & $\pm  0.026$      \\
\hline
 600.     &      LO:  43.675  & $\pm 0.052$      \\
             &  NLL:  42.041   & $\pm 0.027$      \\
\hline
 800.     &      LO:  48.507  & $\pm  0.058$      \\
             &  NLL:  46.558   & $\pm  0.030$      \\
\hline  
1000.    &      LO:  52.220  & $\pm 0.052$      \\
             &  NLL:  49.977   & $\pm  0.033$      \\  
\hline
 1200.   &      LO:  55.133  & $\pm 0.034$      \\
             &  NLL:  52.477   & $\pm 0.041$      \\ 
\hline
 1400.   &      LO:  57.447  & $\pm 0.032$      \\
             &  NLL:  54.189   & $\pm 0.052$      \\
\hline
\end{tabular}       
\end{center}
\caption{\label{II-parton-LVL:Tab:AC_Fit_CTEQ} The CTEQ $A_{C}^{Fit}(\tilde\chi^{\pm}_{1}\tilde\chi^{0}_{2})$ table with its theoretical uncertainty accounting for the correlations between the parameters fitting the $A_{C}^{Raw}$ template curves.}
\end{table}

%%%%%%%%%%%%%%%%%%%%%%%%%%%%%%%%%%%%%%%%%%%%%%%%%%%%%%%%%%%%%%%%%
\clearpage\newpage
\vspace*{1.5mm}
\subsubsection{\label{II-parton-LVL:MSTW} $A_{C}(\tilde\chi^{\pm}_{1}+\tilde\chi^{0}_{2})$ Template Curves for MSTW2008}
\begin{figure}[ht]
\begin{center}
\includegraphics[scale=0.35]{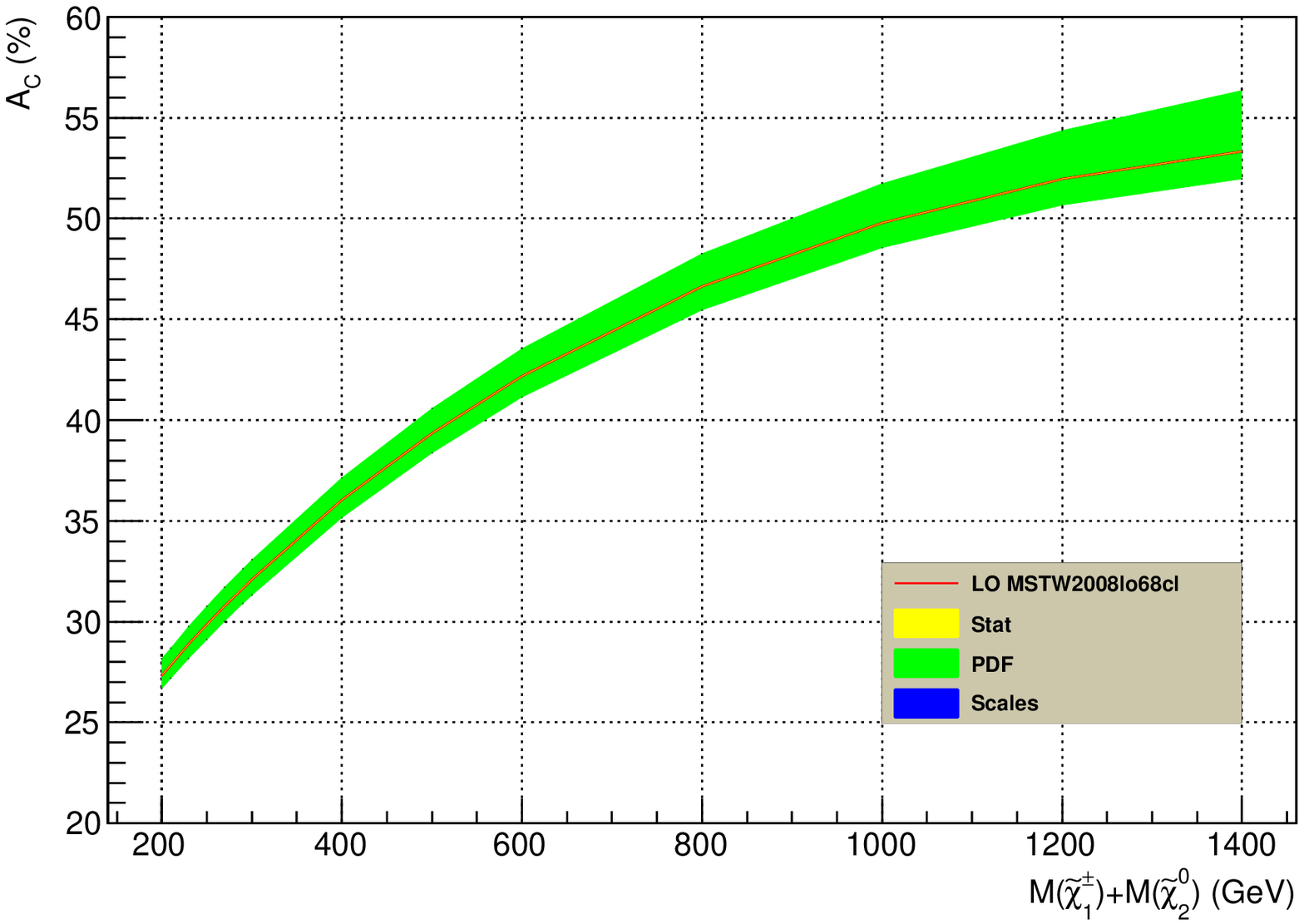}
\includegraphics[scale=0.35]{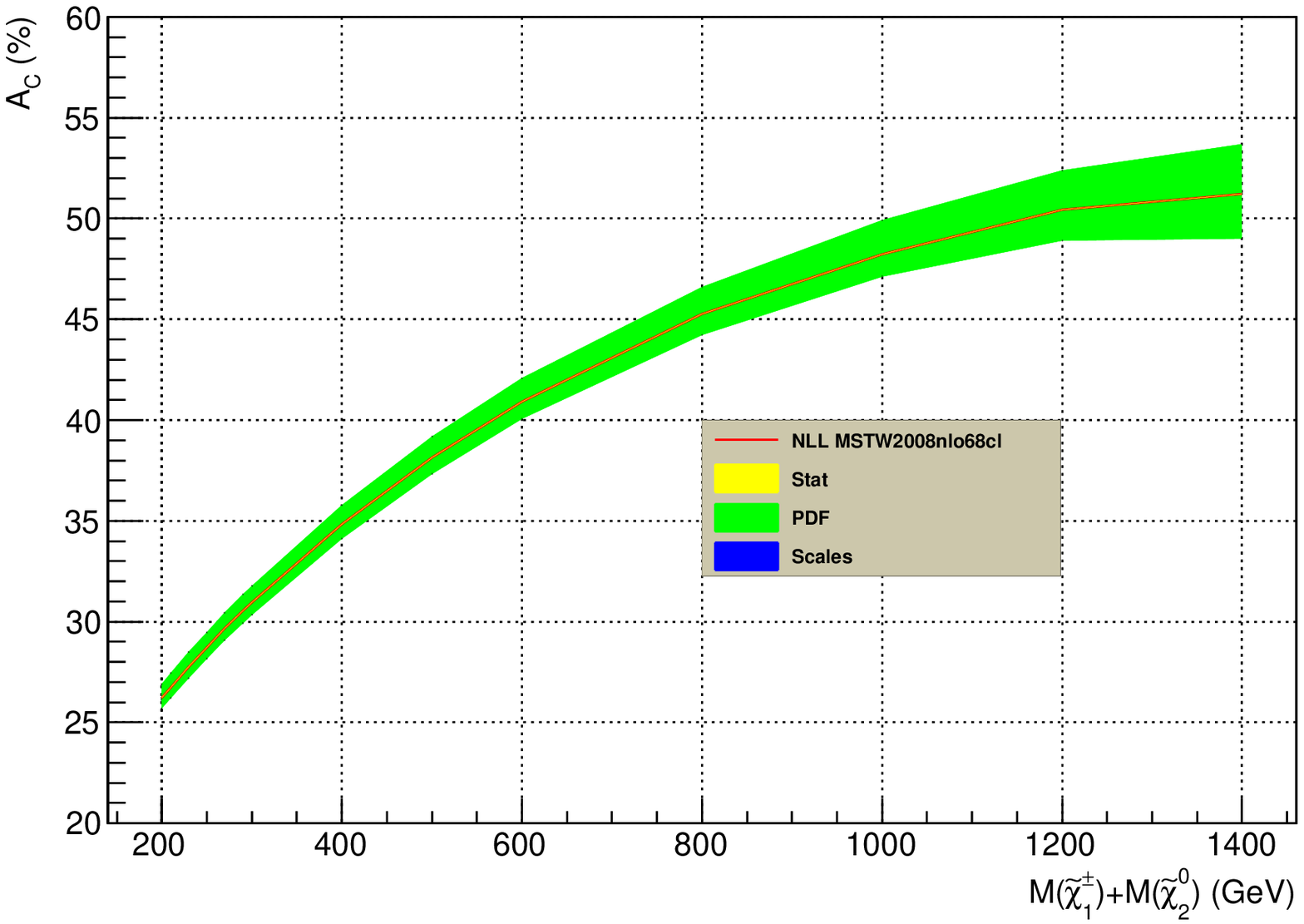}\\
\includegraphics[scale=0.33]{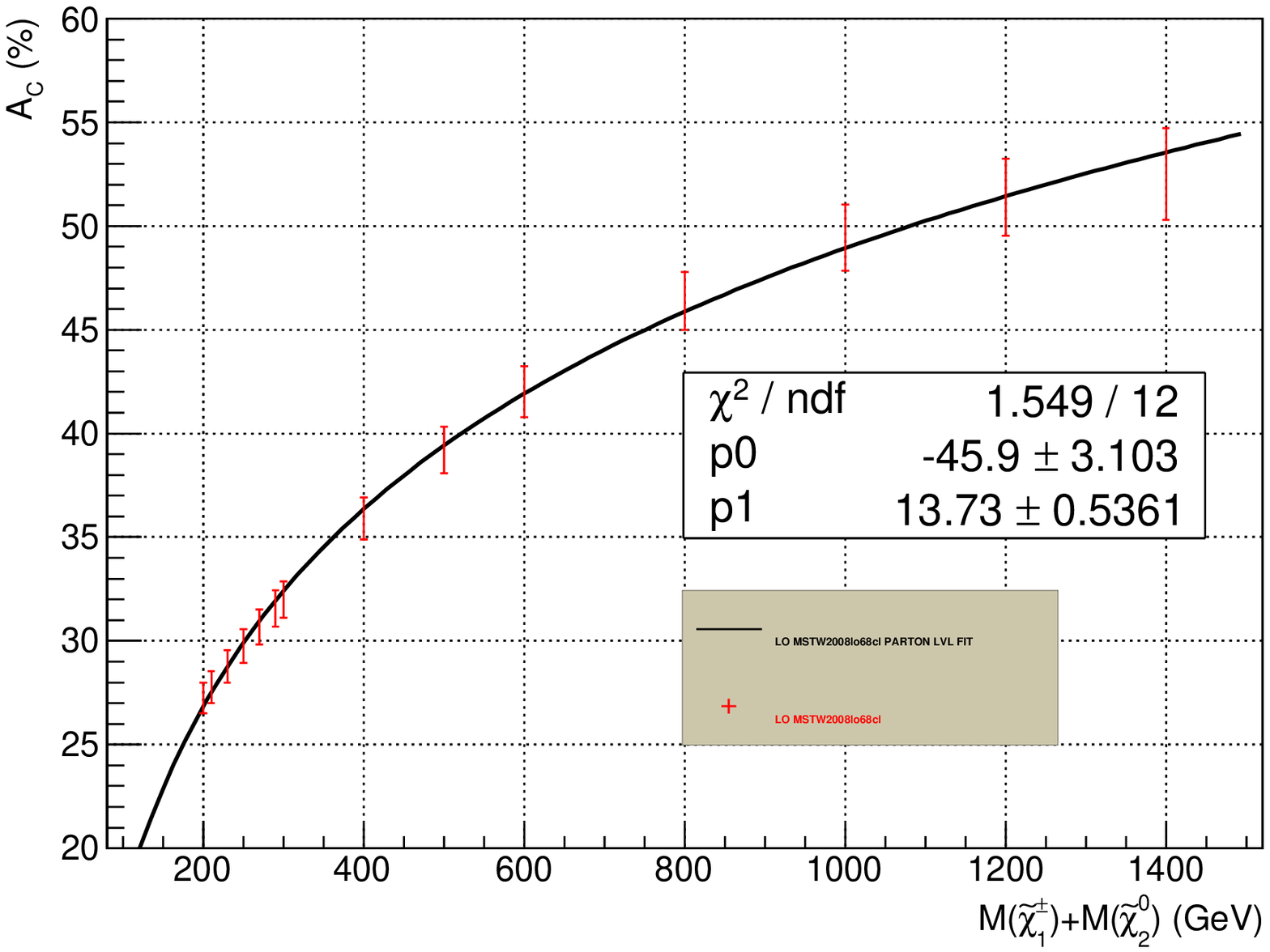}
\includegraphics[scale=0.33]{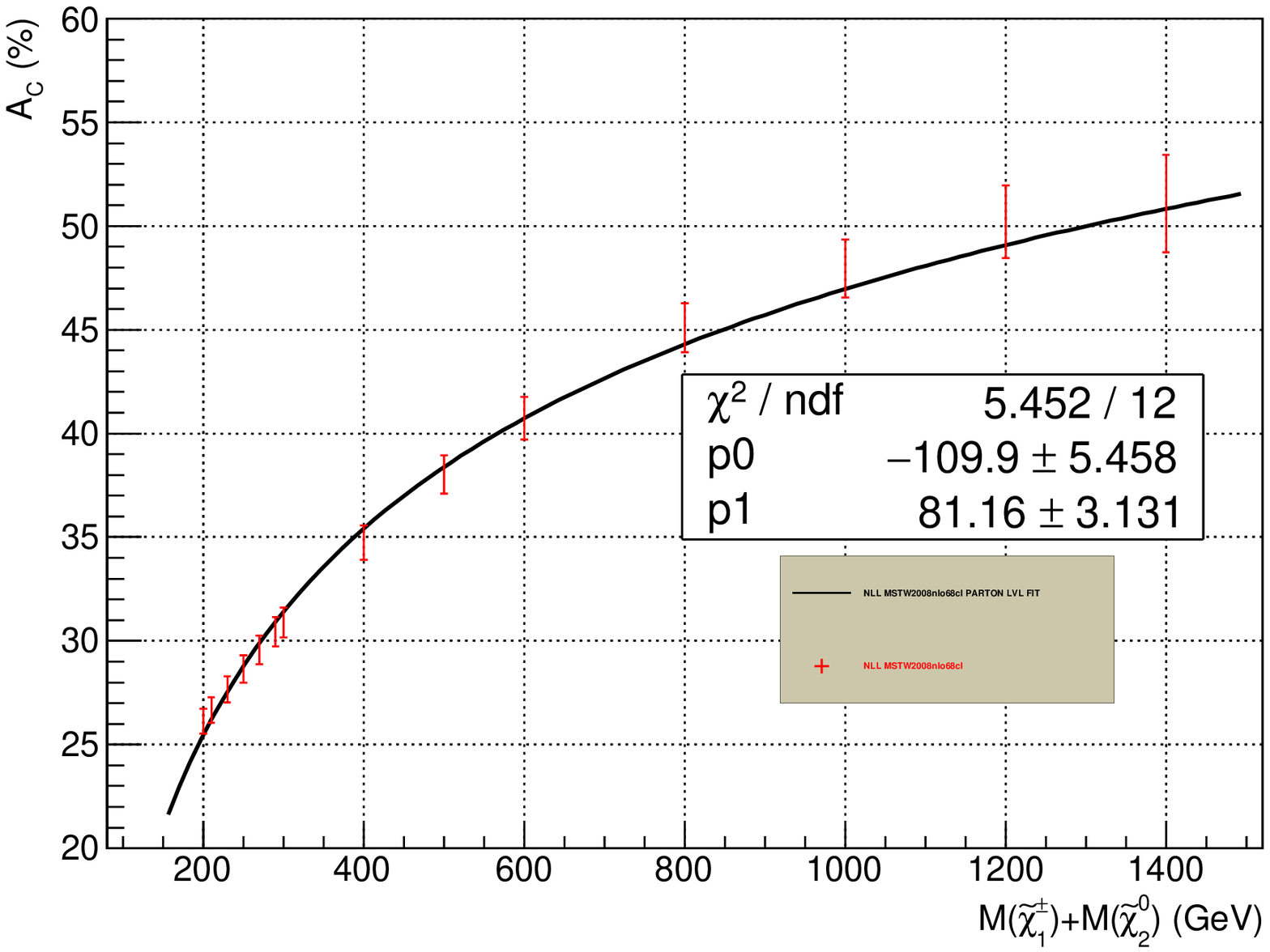}\\
\includegraphics[scale=0.33]{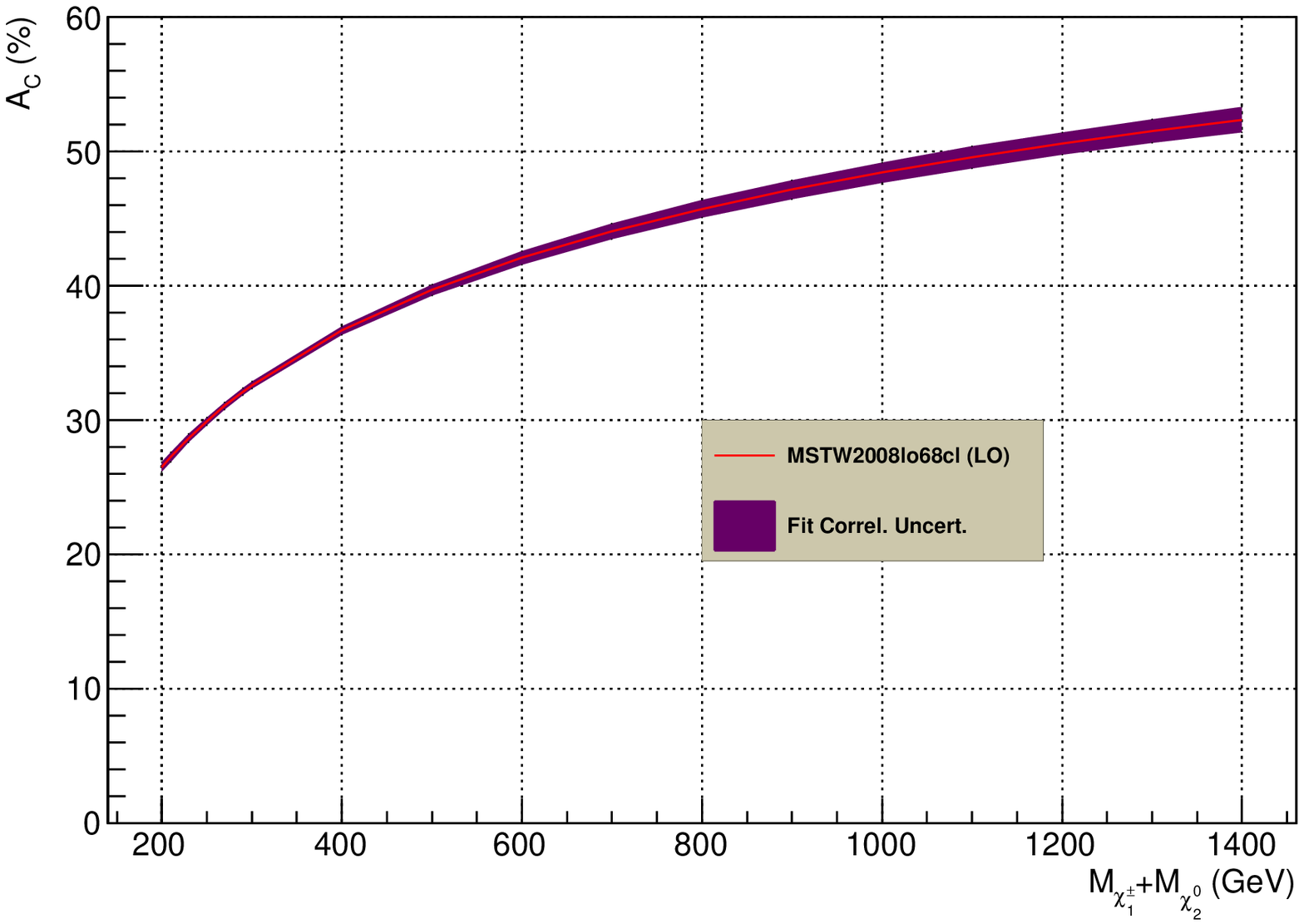}
\includegraphics[scale=0.33]{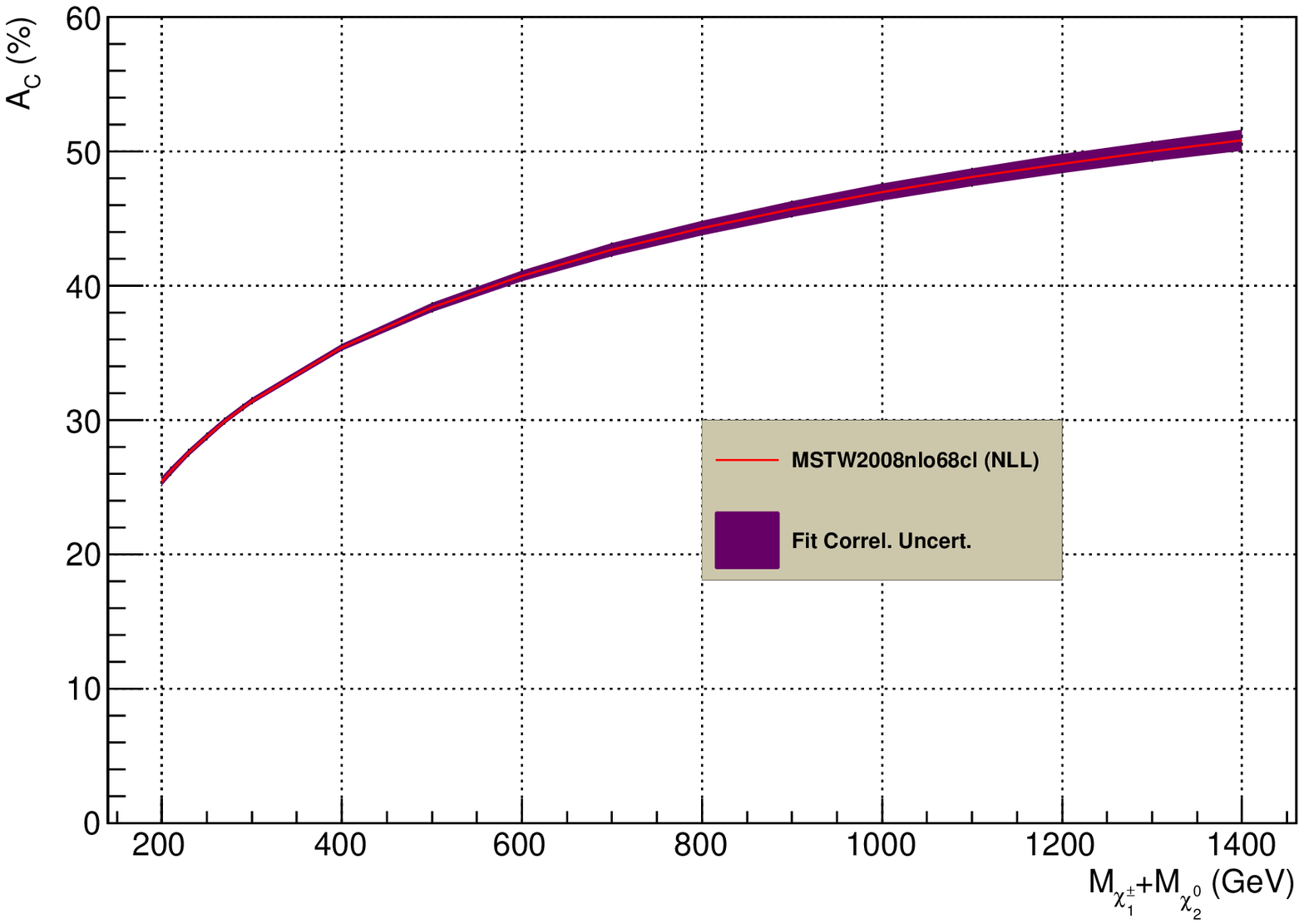}
\end{center}
\caption{\label{II-parton-LVL:Fig3} The theoretical MSTW2008 $A_{C}$ template curves. The raw curve with its uncertainty bands and the corresponding fitted curve wtih uncorrelated and with correlated uncertainties are displayed on the top, the middle and the bottom rows, respectively. The LHS concerns the LO calculations based upon the MSTW2008lo68cl PDF and the RHS concerns the NLL calculations using the MSTW2008nlo68cl PDF.}
\end{figure}

\vspace*{0.5mm}
\noindent
The theoretical  MSTW2008lo68cl $A_{C}$ template curves are obtained by computing the $A_{C}$ based upon the cross sections of the signed processes used for table
\ref{II-parton-LVL:Tab:AC_MSTW}. They are displayed in figure  \ref{II-parton-LVL:Fig3}.

\newpage
\begin{table}[htbp]
\begin{center}
\begin{tabular}{|c|c|c|c|c|c|}
\hline
$\rm M_{\chi^{\pm}_{1}}+M_{\chi^{0}_{2}}$  & $A_{C}$ & $\delta (A_{C})_{Stat}$ & $\delta (A_{C})_{Scale}$ & $\delta (A_{C})_{PDF}$ & $\delta (A_{C})_{Total}$\\ 
                                         (GeV)                                &   ($\%$)  &        ($\%$)                        &   ($\%$)                              &               ($\%$)                  &                          ($\%$)      \\
\hline
200.       &     LO:  27.330              & $\pm  0.003$               & $ ^{-0.034}_{+0.049}$ 
              &     $ ^{+0.827}_{-0.649}$             & $ ^{+0.828}_{-0.651}$ \\
%200.       &   NLO:  26.140           & $\pm  0.011$               & $ ^{+0.012}_{-0.016}$ 
%              &      $ ^{+0.679}_{-0.518}$ &  $ ^{+0.679}_{-0.518}$ \\
              &   NLL:  26.215           & $\pm  0.011$               & $ ^{+0.091}_{-0.067}$ 
              &   $ ^{+0.682}_{-0.518}$    & $ ^{+0.688}_{-0.522}$ \\	      
\hline
210.       &   LO:   27.857               & $\pm  0.003$               & $ ^{-0.042}_{+0.056}$ 
              &   $ ^{+0.845}_{-0.663}$              & $ ^{+0.846}_{-0.665}$ \\
%210.       &   NLO:  26.672           & $\pm  0.009$               & $ ^{+0.009}_{-0.014}$ 
%              &   $ ^{+0.692}_{-0.529}$    & $ ^{+0.692}_{-0.529}$ \\
              &   NLL:  26.744           & $\pm  0.009$               & $ ^{+0.080}_{-0.056}$ 
              &   $ ^{+0.694}_{-0.530}$    & $ ^{+0.698}_{-0.533}$ \\	      
\hline
230.              &   LO:   28.872               & $\pm  0.002$               & $ ^{-0.056}_{-0.068}$ 
              &   $ ^{+0.878}_{-0.690}$              & $ ^{+0.880}_{-0.693}$ \\
%230.       &   NLO:  27.695           & $\pm  0.006$               & $ ^{+0.001}_{-0.003}$ 
%              &   $ ^{+0.719}_{-0.550}$    & $ ^{+0.719}_{-0.550}$ \\
              &   NLL: 27.757            & $\pm  0.006$               & $ ^{+0.085}_{-0.040}$ 
              &   $ ^{+0.722}_{-0.549}$    & $ ^{+0.727}_{-0.550}$ \\	      
\hline
250.              &    LO:   29.842              & $\pm  0.001$               & $ ^{-0.069}_{+0.078}$ 
              &   $ ^{+0.911}_{-0.716}$              & $ ^{+0.913}_{-0.720}$ \\
%250.       &   NLO: 28.671            & $\pm  0.004$               & $ ^{-0.011}_{+0.007}$ 
%              &   $ ^{+0.745}_{-0.570}$    & $ ^{+0.745}_{-0.570}$ \\
              &   NLL:  28.730           & $\pm 0.004$               & $ ^{+0.073}_{-0.053}$ 
              &   $ ^{+0.747}_{-0.573}$    & $ ^{+0.751}_{-0.575}$ \\	      
\hline
 270.              &    LO:   30.770              & $\pm  0.001$               & $ ^{-0.080}_{+0.087}$ 
              &   $ ^{+0.942}_{-0.742}$              & $ ^{+0.945}_{-0.747}$ \\
%270.       &   NLO: 29.602            & $\pm  0.003$               & $ ^{-0.019}_{+0.006}$ 
%              &   $ ^{+0.773}_{-0.589}$    & $ ^{+0.773}_{-0.589}$ \\
              &   NLL:   29.658          & $\pm  0.003$               & $ ^{+0.063}_{-0.069}$ 
              &   $ ^{+0.773}_{-0.595}$    & $ ^{+0.775}_{-0.599}$ \\	      
\hline
290.              &   LO:    31.662              & $\pm  0.001$               & $ ^{-0.088}_{+0.094}$ 
              &   $ ^{+0.972}_{-0.766}$              & $ ^{+0.976}_{-0.772}$ \\
%290.       &   NLO:  30.492           & $\pm  0.002$               & $ ^{-0.020}_{+0.007}$ 
%              &   $ ^{+0.800}_{-0.608}$    & $ ^{+0.800}_{-0.608}$ \\
              &   NLL:  30.540           & $\pm  0.002$               & $ ^{+0.058}_{-0.080}$ 
              &   $ ^{+0.802}_{-0.608}$    & $ ^{+0.804}_{-0.613}$ \\	      
\hline
300.              &   LO:   32.096               & $\pm  0.001$               & $ ^{-0.092}_{+0.097}$ 
              &   $ ^{+0.987}_{-0.778}$              & $ ^{+0.991}_{-0.784}$ \\
%300.               &   NLO:  30.923           & $\pm  0.002$               & $ ^{-0.016}_{+0.007}$ 
%              &   $ ^{+0.813}_{-0.616}$    & $ ^{+0.813}_{-0.616}$ \\
              &   NLL:  30.969           & $\pm  0.002$               & $ ^{+0.068}_{-0.089}$ 
              &   $ ^{+0.802}_{-0.625}$    & $ ^{+0.805}_{-0.632}$ \\	      
\hline	      
400.               &    LO:   36.028              & $\pm  0.000$               & $ ^{-0.117}_{+0.117}$ 
              &   $ ^{+1.123}_{-0.885}$              & $ ^{+1.129}_{-0.893}$ \\
%400.       &   NLO: 34.843            & $\pm 0.001$               & $ ^{-0.015}_{+0.020}$ 
%              &   $ ^{+0.927}_{-0.708}$    & $ ^{+0.927}_{-0.708}$ \\
              &   NLL:  34.846           & $\pm  0.001$               & $ ^{+0.105}_{-0.043}$ 
              &   $ ^{+0.929}_{-0.713}$    & $ ^{+0.935}_{-0.714}$ \\	      
\hline
500.              &   LO:    39.351                & $\pm  0.000$               & $ ^{-0.123}_{+0.122}$ 
              &   $ ^{+1.250}_{-0.971}$              & $ ^{+1.256}_{-0.979}$ \\
%500.       &   NLO: 38.161            & $\pm  0.000$               & $ ^{-0.030}_{+0.017}$ 
%              &   $ ^{+1.043}_{-0.781}$    & $ ^{+1.044}_{-0.781}$ \\
              &   NLL: 38.145            & $\pm  0.000$               & $ ^{+0.097}_{-0.093}$ 
              &   $ ^{+1.042}_{-0.803}$    & $ ^{+1.047}_{-0.808}$ \\	      
\hline
600.              &     LO:   42.179             & $\pm  0.000$               & $ ^{-0.118}_{+0.116}$ 
              &   $ ^{+1.372}_{-1.043}$              & $ ^{+1.377}_{-1.050}$ \\
%600.       &   NLO: 40.948             & $\pm  0.000$               & $ ^{-0.026}_{+0.021}$ 
%              &   $ ^{+1.158}_{-0.845}$    & $ ^{+1.158}_{-0.846}$ \\
              &   NLL:  40.906           & $\pm  0.000$               & $ ^{+0.121}_{-0.103}$ 
              &   $ ^{+1.171}_{-0.841}$    & $ ^{+1.177}_{-0.847}$ \\	      
\hline
800.              &    LO:   46.628              & $\pm  0.000$               & $ ^{-0.088}_{+0.085}$ 
              &   $ ^{+1.627}_{-1.161}$              & $ ^{+1.629}_{-1.164}$ \\
%800.       &   NLO:  45.283           & $\pm  0.000$               & $ ^{-0.008}_{+0.023}$ 
%              &   $ ^{+1.382}_{-0.995}$    & $ ^{+1.382}_{-0.995}$ \\
              &   NLL:  45.265           & $\pm  0.000$               & $ ^{+0.101}_{-0.080}$ 
              &   $ ^{+1.352}_{-1.027}$    & $ ^{+1.356}_{-1.030}$ \\	      
\hline  
1000.              &    LO:  49.793               & $\pm  0.000$               & $ ^{-0.051}_{+0.046}$ 
              &   $ ^{+1.953}_{-1.242}$              & $ ^{+1.953}_{-1.243}$ \\
%1000.     &   NLO: 48.238            & $\pm  0.000$               & $ ^{-0.016}_{+0.005}$ 
%              &   $ ^{+1.663}_{-1.177}$    & $ ^{+1.663}_{-1.177}$ \\
              &   NLL: 48.243            & $\pm  0.000$               & $ ^{+0.112}_{-0.019}$ 
              &   $ ^{+1.674}_{-1.124}$    & $ ^{+1.678}_{-1.125}$ \\	      
\hline
1200.              &  LO:    51.956               & $\pm  0.000$               & $ ^{-0.014}_{+0.013}$ 
              &   $ ^{+2.407}_{-1.301}$              & $ ^{+2.408}_{-1.301}$ \\
%1200.     &   NLO: 50.045            & $\pm  0.000$               & $ ^{-0.028}_{-0.029}$ 
%              &   $ ^{+2.042}_{-1.471}$    & $ ^{+2.042}_{-1.472}$ \\
              &   NLL: 50.430            & $\pm  0.000$               & $ ^{+0.031}_{-0.000}$ 
              &   $ ^{+1.966}_{-1.534}$    & $ ^{+1.966}_{-1.534}$ \\	      
\hline
 1400.             &  LO:      53.328              & $\pm  0.000$               & $ ^{+0.018}_{-0.013}$ 
              &   $ ^{+3.019}_{-1.375}$              & $ ^{+3.019}_{-1.375}$ \\
%1400.     &   NLO:  50.760           & $\pm  0.000$               & $ ^{+0.043}_{-0.060}$ 
%              &   $ ^{+2.569}_{-1.950}$    & $ ^{+2.569}_{-1.951}$ \\
              &   NLL:   51.216          & $\pm  0.000$               & $ ^{-0.082}_{+0.060}$ 
              &   $ ^{+2.470}_{-2.216}$    & $ ^{+2.472}_{-2.217}$ \\	      
\hline
\end{tabular}       
\end{center}
\caption{\label{II-parton-LVL:Tab:AC_MSTW} The MSTW2008 $A_{C}(\tilde\chi^{\pm}_{1}\tilde\chi^{0}_{2})$ table with the breakdown of the different sources of theoretical uncertainty.}
\end{table}

\begin{table}[htbp]
\begin{center}
\begin{tabular}{|c|c|c|}
\hline
$\rm M_{\chi^{\pm}_{1}}+M_{\chi^{0}_{2}}\ (GeV) $   & $A_{C}^{Fit}$  &  $\delta A^{Fit}_{C}$  \\ 
                                                   (GeV)                                     &             ($\%$)    &          ($\%$)                    \\ 
\hline
200.      &      LO:  26.841746  &  $\pm  0.358$      \\
             &  NLL: 25.767          &  $\pm  0.304$      \\
\hline
210.      &      LO:  27.512        &  $\pm 0.341$      \\
             &  NLL:  26.426         &  $\pm  0.286$      \\
\hline
230.      &      LO:   28.761       & $\pm  0.310$      \\
             &  NLL:  27.656         & $\pm  0.257$      \\
 \hline
250.      &      LO:   29.905       & $\pm  0.287$      \\
             &  NLL:    28.783       & $\pm  0.235$      \\
\hline
270.      &      LO:   30.962      & $\pm  0.271$      \\
             &  NLL:  29.824        &  $\pm 0.220$      \\
\hline
290.      &      LO:  31.943       & $\pm 0.261$      \\
             &  NLL: 30.790         & $\pm  0.212$      \\
\hline
 300.     &      LO:     32.409    & $\pm 0.258$      \\
             &  NLL:       31.248   & $\pm  0.211$      \\
\hline	      
400.      &      LO:     36.358    & $\pm  0.282$      \\
             &  NLL:      35.138    & $\pm  0.251$      \\
\hline
500.      &      LO:     39.422    & $\pm  0.350$      \\
             &  NLL:     38.1545   &  $\pm  0.328$      \\
\hline
 600.     &      LO:     41.925    & $\pm 0.423$      \\
             &  NLL:       40.619   & $\pm 0.405$      \\
\hline
 800.     &      LO:     45.875    & $\pm  0.554$      \\
             &  NLL:     44.509     & $\pm  0.537$      \\
\hline  
1000.    &      LO:    48.939     & $\pm  0.663$      \\
             &  NLL:     47.526     & $\pm  0.644$      \\  
\hline
 1200.   &      LO:    51.442     & $\pm  0.754$      \\
             &  NLL:     49.991     & $\pm  0.733$      \\ 
\hline
 1400.   &      LO:     53.559    & $\pm  0.832$      \\
             &  NLL:      52.075    & $\pm  0.810$      \\
\hline
\end{tabular}       
\end{center}
\caption{\label{II-parton-LVL:Tab:AC_Fit_MSTW} The MSTW $A_{C}^{Fit}(\tilde\chi^{\pm}_{1}\tilde\chi^{0}_{2})$ table with its theoretical uncertainty accounting for the correlations between the parameters fitting the $A_{C}^{Raw}$ template curve.}
\end{table}

%%%%%%%%%%%%%%%%%%%%%%%%%%%%%%%%%%%%%%%%%%%%%%%%%%%%%%%%%%%%%%%%%

\vspace*{1.5mm}
\subsubsection{\label{II-parton-LVL:Comp} Comparing the different $A_{C}$ Template Curves}
\vspace*{0.5mm}

\begin{figure}[h]
\begin{center}
\includegraphics[scale=0.35]{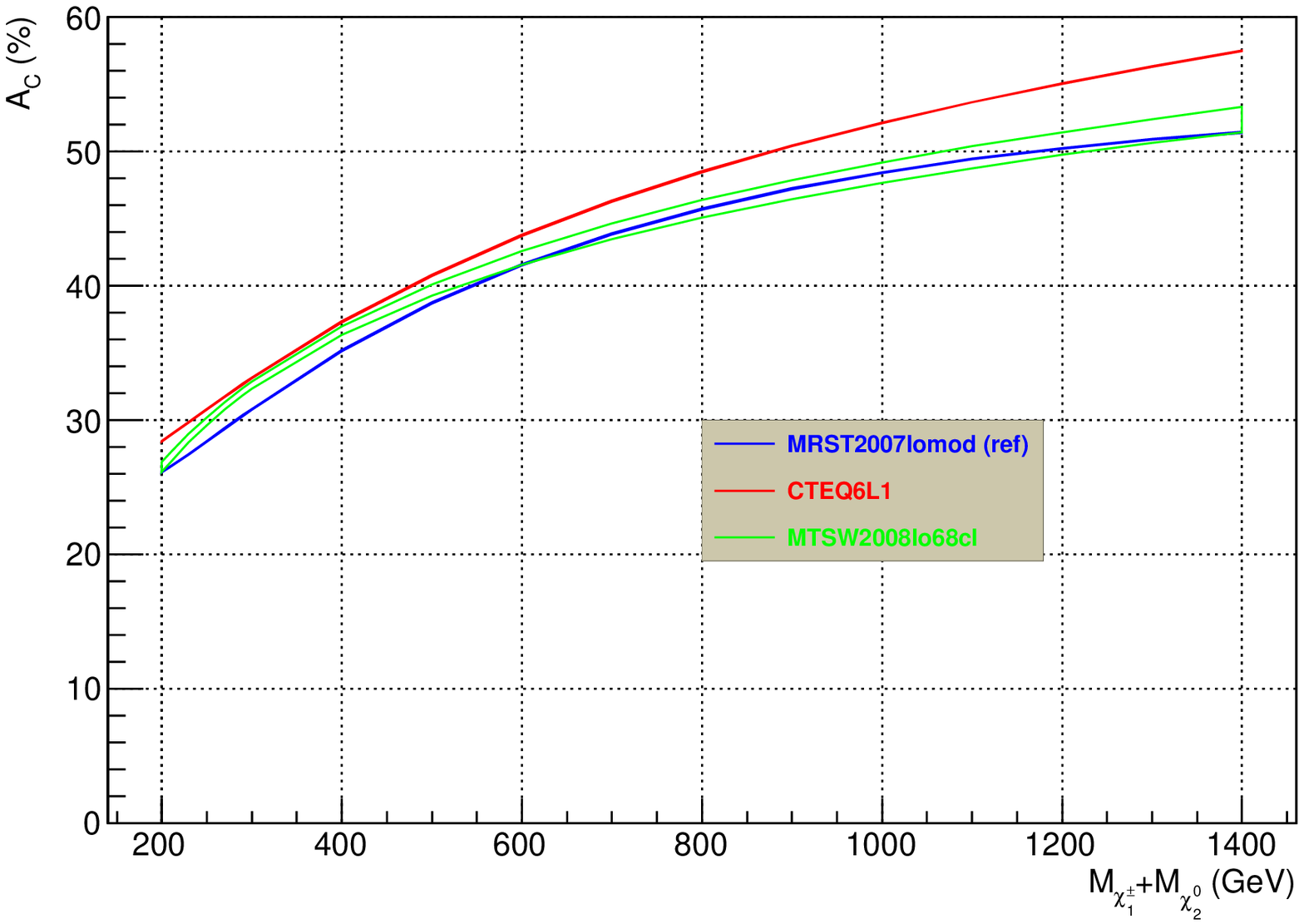}
\includegraphics[scale=0.35]{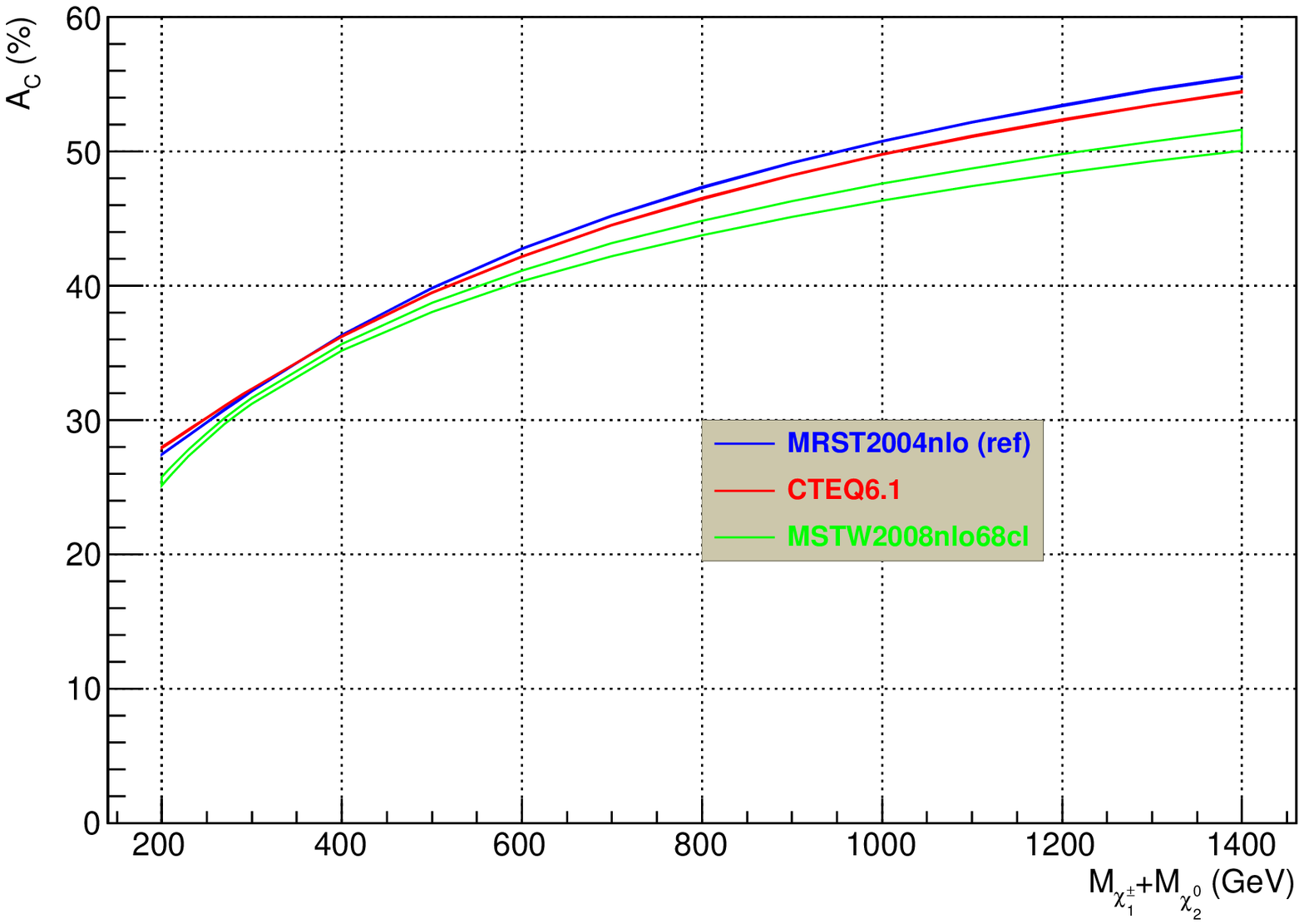}\\
\includegraphics[scale=0.35]{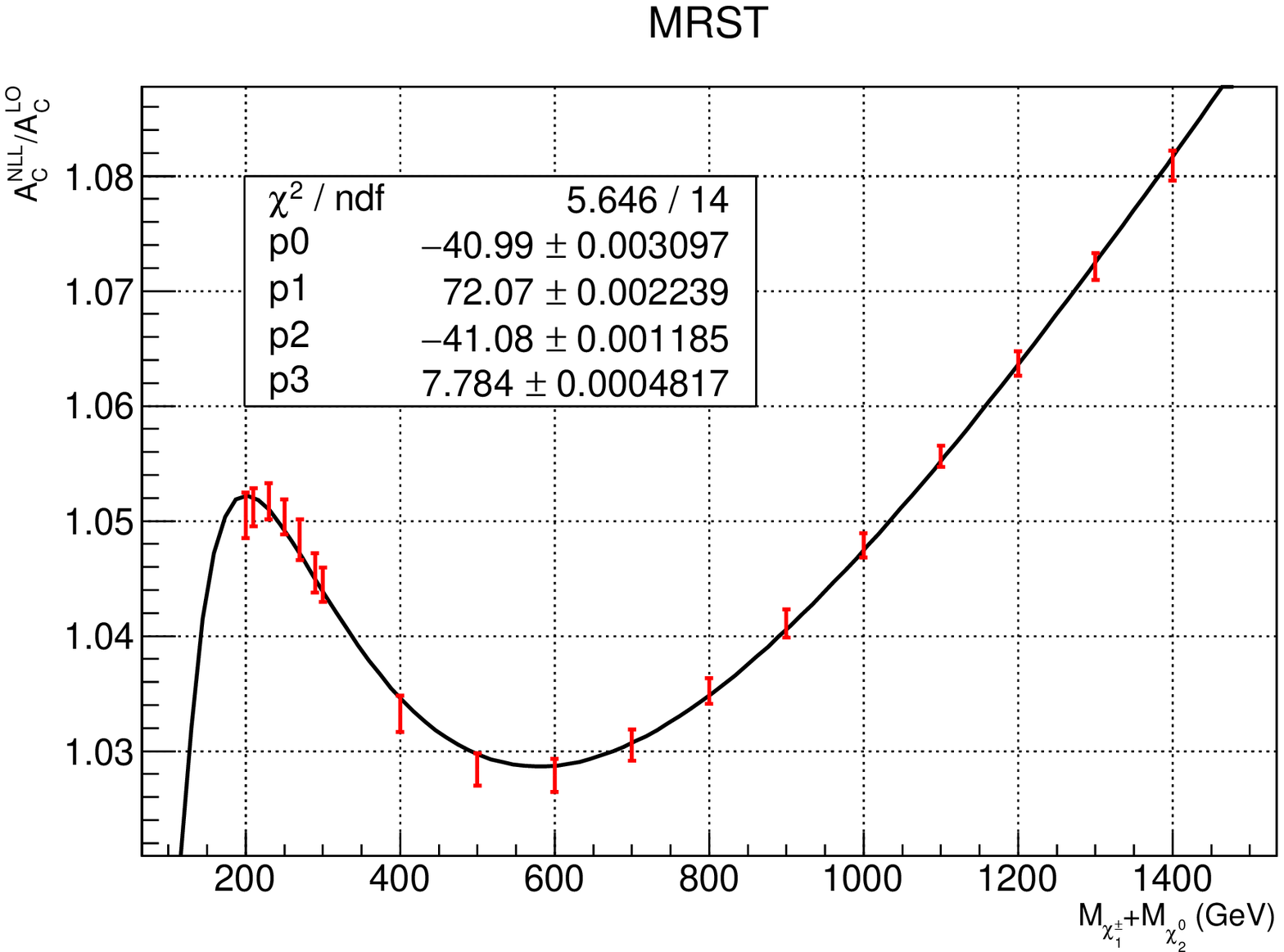}
\includegraphics[scale=0.35]{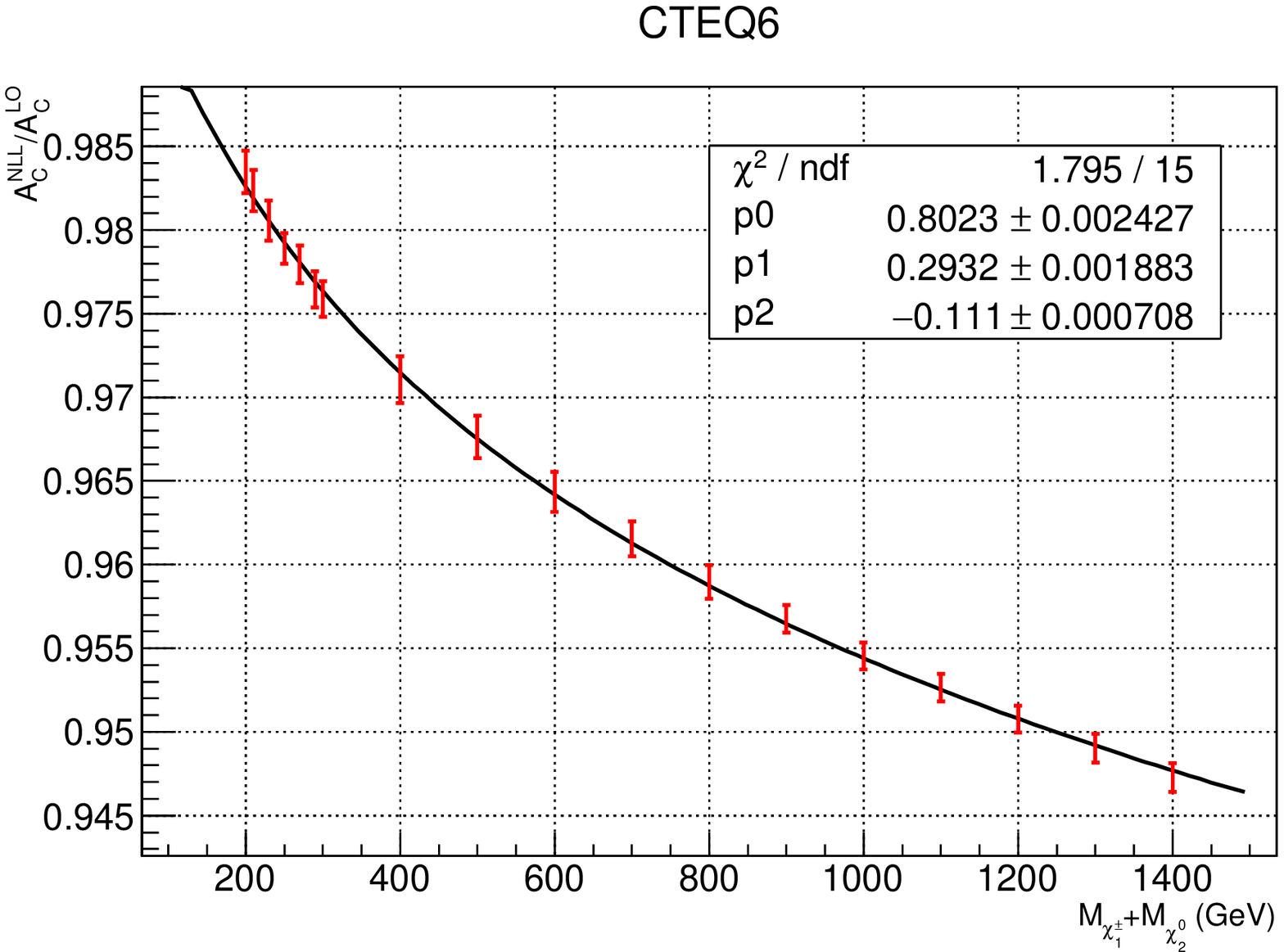}\\
\includegraphics[scale=0.35]{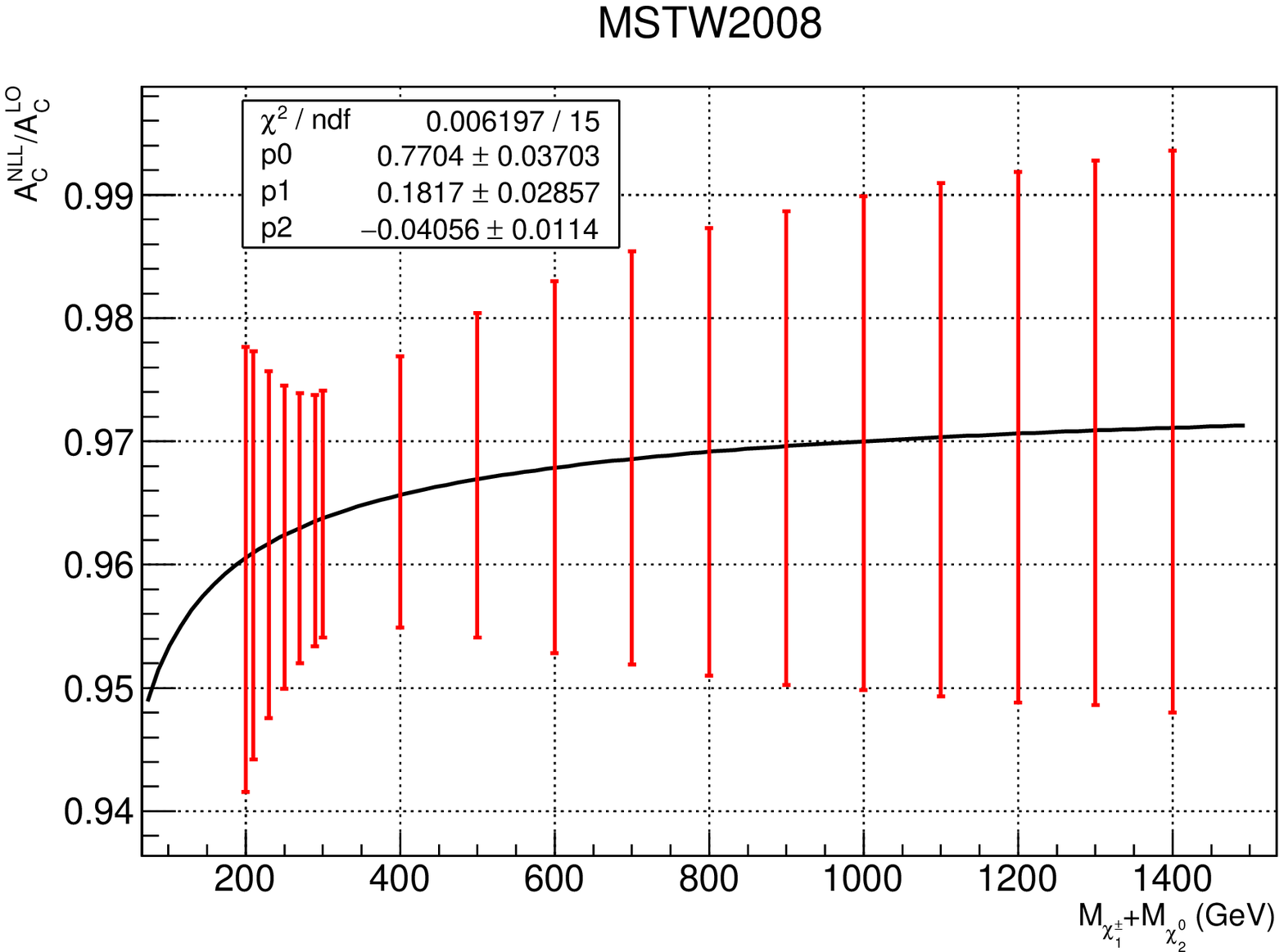}
\end{center}
\caption{\label{AC_w1z2_Compare} Comparison between the $A_{C}$ template curves.
The top LHS plot compares the LO PDFs: MRST2007lomod (blue, ref. curve), CTEQ6L1 (red), MSTW2008lo68cl (green).
The top RHS plot compares the NLO PDFs: MRST2004nlo (blue, ref. curve), CTEQ6.1 (red), MSTW2008nlo68cl (green).
The middle and the bottom rows display the $\frac{A_{C}^{NLL}}{A_{C}^{LO}}$ fitted by the same functional forms as the $A_{C}^{LO}$
template curves.}
\end{figure}

\noindent
Here again we compare the $A_{C}$ template curves produced with different PDFs using Resummino this time.  From figure \ref{AC_w1z2_Compare} 
we can see that the $A_{C}$ of the different PDF used at LO and at NLO are in agreement only at the $\pm 3\sigma$ level. This figure also displays the $\frac{A_{C}^{NLL}}{A_{C}^{LO}}$ ratios for the three families of PDFs used.

\newpage
\subsection{ Experimental Measurement of $A_{C}(\tilde\chi^{\pm}_{1}+\tilde\chi^{0}_{2}\to 3\ell^{\pm}+\rlap{\kern0.25em/}E_{T})$}
\label{sec:next-particle-LVL}
The aim of this sub-section is to repeat, in the context of the considered SUSY signal, a study similar
to that of section \ref{sec:Part1-particle-LVL}.
\par
\begin{figure}[h]
\begin{center}
\includegraphics[scale=0.35]{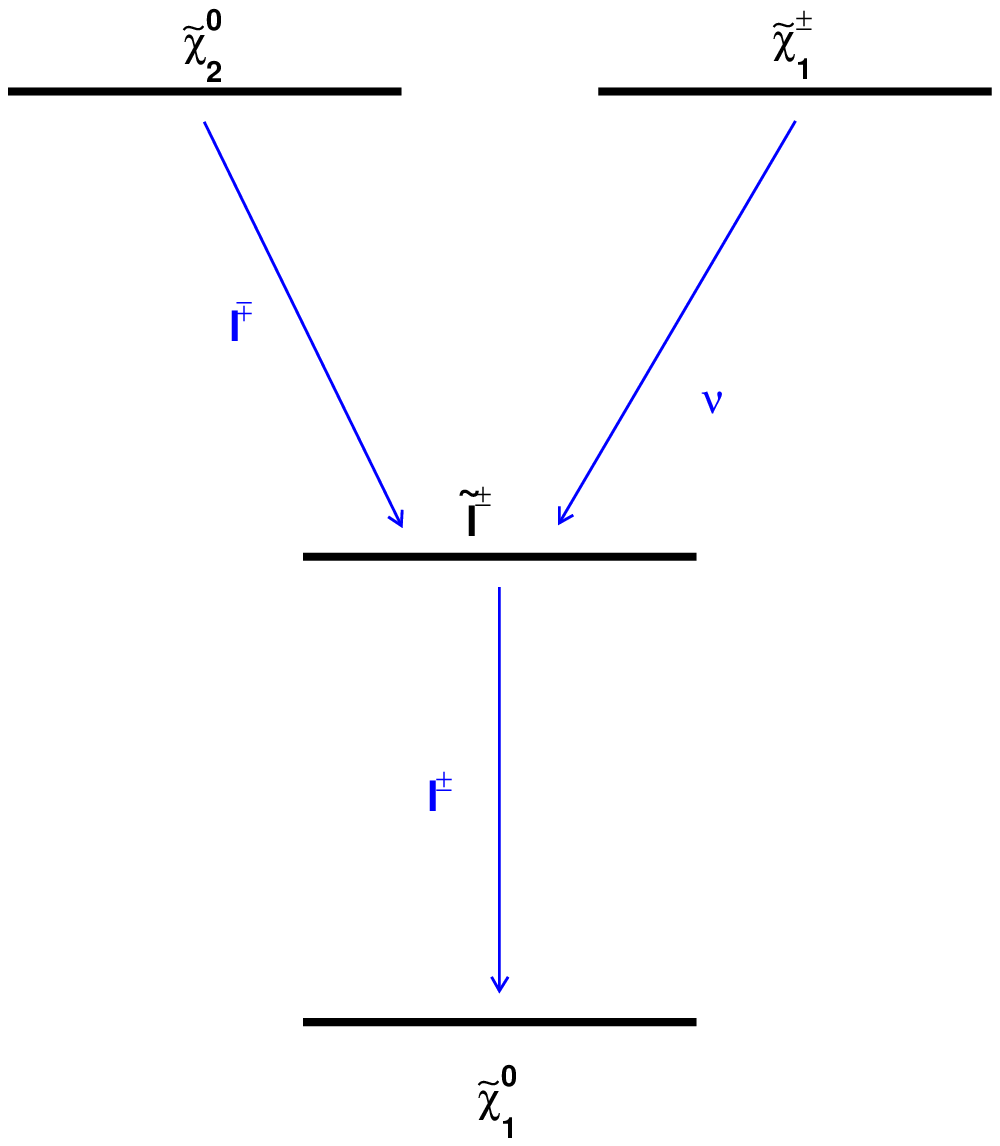}
\includegraphics[scale=0.35]{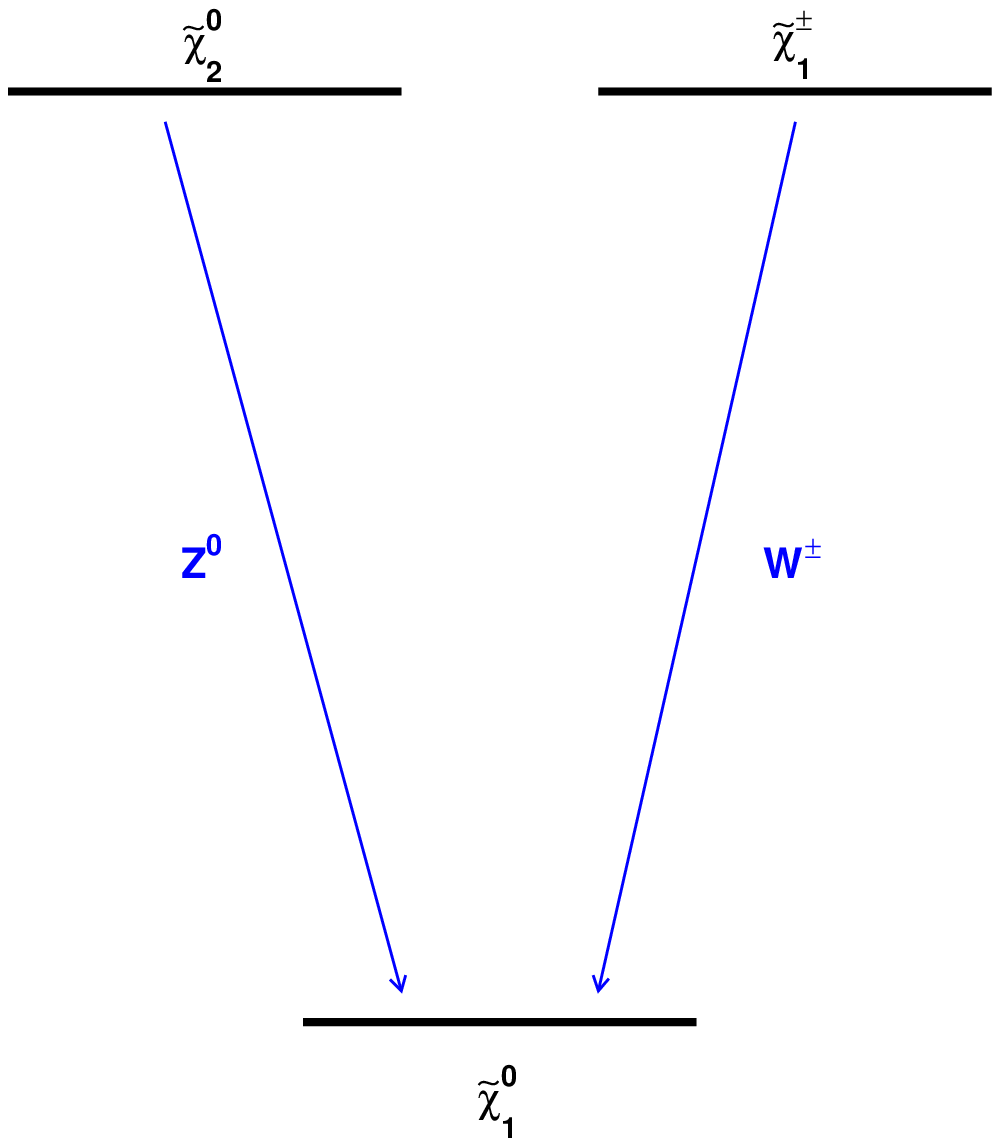}
\end{center}
\caption{\label{sec:Part2:w1z2_Simpl_Models} The sketch of the Simplified Models used to generate the signal samples:
the S1 signal (LHS) has a $\tilde\ell^{\pm}$ NLSP whereas for the S2 signal (RHS) the mass degenerate 
$\tilde\chi^{\pm}_{1}$ and $\tilde\chi^{0}_{2}$ are the NLSPs. Both signals share the $\tilde\chi^{0}_{1}$ as the LSP.}
\end{figure}

\noindent
We use Simplified Models to generate our signal in the two configurations shown in figure \ref{sec:Part2:w1z2_Simpl_Models}. 
\par\noindent
The first signal configuration, denoted S1, supposes that the lightest part of the SUSY mass spectrum is made
of $\tilde\chi^{\pm}_{1}$, $\tilde\chi^{0}_{2}$, $\tilde\ell^{\pm}$ (i.e. $\tilde e^{\pm}$ or $\tilde\mu^{\pm}$), and $\tilde\chi^{0}_{1}$, in order of decreasing mass. In addition, the following decays (and their charge conjugate) are all supposed to have a braching ratio of $100\%$: $\tilde\chi^{\pm}_{1}\to\tilde\ell^{\pm}(\to\ell^{\pm}\tilde\chi^{0}_{1})+\nu$, $\tilde\chi^{0}_{2}\to\tilde\ell^{\pm}(\to\ell^{\pm}\tilde\chi^{0}_{1})+\ell^{\mp}$. In practice, within the MSSM, very large braching ratios for these decays are guaranteed by the envisaged mass hierarchy.
 \par\noindent
The second signal configuration, denoted S2, supposes that the lightest part of the SUSY mass spectrum is made
of $\tilde\chi^{\pm}_{1}$, $\tilde\chi^{0}_{2}$, and $\tilde\chi^{0}_{1}$, in order of decreasing mass. The charged sleptons
are supposed to be much heavier. In addition, the following SUSY decays are all supposed to have a braching ratio of $100\%$: $\tilde\chi^{\pm}_{1}\to W^{\pm}(\to\ell^{\pm}\nu)+\tilde\chi^{0}_{1}$, $\tilde\chi^{0}_{2}\to Z^{0}(\to\ell^{\pm}\ell^{\mp})+\tilde\chi^{0}_{1}$. In practice, within the MSSM,  these braching ratios not only
depend on the envisaged mass hierarchy, but also on the fields composition of the $\tilde\chi^{0}_{2}$, the $\tilde\chi^{\pm}_{1}$, and the
$\tilde\chi^{0}_{1}$. Regarding the SM leptonic decays of the $W^{\pm}$ and the $Z^{0}$ gauge bosons, we used 
their actual SM branching ratios. This will have the obvious consequence of a much smaller event yield for the S2 signals
compared to the S1 signals of same mass.
\par\noindent
The hypotheses common to configurations S1 and S2 are that the lightest SUSY particle (LSP) is the $\tilde\chi^{0}_{1}$, and that the
$\tilde\chi^{0}_{2}$ and the $\tilde\chi^{\pm}_{1}$ are mass degenerate. 

\vspace*{1.5mm}
\subsubsection{\label{sec:Part4:Herwig-Settings} Monte Carlo Generation}
\vspace*{0.5mm}

We generate a new set of MC samples. We report here only the MC parameters that are different from those used in sub-section \ref{sec:Part2:Herwig-Settings}. We use the following LO generator: Herwig++ v2.5.2 for the SUSY signal and for most of the background processes.
\par
The other background processes:  $W^{+}+W^{-}+W^{\pm}$, $W^{+}+W^{-}+\gamma^{*}/Z$, $W^{\pm}+\gamma^{*}/Z+\gamma^{*}/Z$
$\gamma^{*}/Z+\gamma^{*}/Z+\gamma^{*}/Z$, $W^{\pm}+1c+0Lp$,  $W^{\pm}+1c+1Lp$,
$W^{\pm}+c\bar{c}+0Lp$, $W^{\pm}+b\bar{b}+0Lp$, $W^{\pm}+t\bar{t}+0Lp$ are generated using Alpgen v2.14 at the parton level.
Those samples are passed on to Pythia v8.170 for the parton showering,
the fragmentation of the colored particles, the modelling of the underlying event and the decay of the unstable particles. 
\par\noindent
For the $W^{\pm}+HF$ process, and the VVV processes in Alpgen the only decay mode generated is 
$\gamma^{*}/Z(\to f\bar f)$ where $f=\ell^{\pm}, \tau^{\pm}, \nu, q$ and $75<M(f\bar{f})<125$ GeV, whereas for the
$W^{\pm}(\to e^{\pm}\nu_{e}/\mu^{\pm}\nu_{\mu}/\tau^{\pm}\nu_{\tau})$ process no mass cuts are applied.

\par\noindent
For the $W+HF$ processes, the renormalization scale is set to $$\mu_{R}=\mu_{F}=\sqrt{M^{2}(W)+\sum^{N^{FS}_{p}}_{i=1} M^{2}_{T}(i)}$$ 
\noindent
where the i index runs over the number of FS partons $N^{FS}_{p}$, and where $M^{2}_{T}=M^{2}+p^{2}_{T}$. 

%%%%%%%%%%%%%%%%%%%%%%%%%%%%%%%%%%%%%%%%%%%%%%%%%%%%%%%%%%%%%%%%%%%%%%%%%

\par\noindent
In particular for the signal samples, we test distinct mass hypotheses in different configurations.
\par\noindent
For the S1 signal, we vary $M_{\tilde\chi^{0}_{2}}$ in the range [100,700] GeV by steps of
100 GeV, and we set $M_{\tilde\chi^{0}_{1}}=M_{\tilde\chi^{0}_{2}}/2$ and $M_{\tilde\ell^{\pm}}=[M_{\tilde\chi^{0}_{2}}+M_{\tilde\chi^{\pm}_{1}}]/2$.
\par\noindent
For the S2 signal, we produce a single "S2a" sample, i.e. with $M_{\tilde\chi^{0}_{2}}-M_{\tilde\chi^{0}_{1}} < M_{Z}$, for which
we set $M_{\tilde\chi^{0}_{2}}=100$ GeV, $M_{\tilde\chi^{0}_{1}}=50$ GeV. This enables to explore the case where the
$\tilde\chi^{\pm}_{1}$  and the $\tilde\chi^{0}_{2}$ decay through a $W^{\pm}$ and through a $Z$ that are both off-shell. 
For the other S2 samples, denoted "S2b" and described in the following paragraph, both the $W^{\pm}$ and the $Z$ bosons are on-shell. 
In addition, we vary $M_{\tilde\chi^{0}_{2}}$ in the range [200,700] GeV by steps of 100 GeV, setting $M_{\tilde\chi^{0}_{1}}=M_{\tilde\chi^{0}_{2}}/2$. We also vary $M_{\tilde\chi^{0}_{2}}$ in the range [105,145] GeV by steps of 10 GeV with a fixed value of $M_{\tilde\chi^{0}_{1}}=13.8$ GeV. And finally, we added two samples: $[M_{\tilde\chi^{0}_{2}},M_{\tilde\chi^{0}_{1}}]=$ [150,50] GeV and [250,125] GeV.

%%%%%%%%%%%%%%%%%%%%%%%%%%%%%%%%%%%%%%%%%%%%%%%%%%%%%%%%%%%%%%%%%%%%%%%%%

\vspace*{1.5mm}
\subsubsection{\label{sec:Part4:W-Analysis} Analysis of the $\tilde\chi^{\pm}_{1}\tilde\chi^{0}_{2}\to 3\ell^{\pm}+\rlap{\kern0.25em/}E_{T}$ Process}
\vspace*{0.5mm}

We considered only the electron and the muon channels. For these analyses we set  the integrated luminosity to $\int {\cal L} dt =20\ \ fb^{-1}$.

%%%%%%%%%%%%%%%%%%%%%%%%%%%%%%%%%%%%%%%%%%%%%%%%%%%%%%

\vspace*{0.5mm}
\par\noindent 
1). Event Selection in the Trilepton Channel
\par
A first set of requirements related to the leptons are applied for the event selection as mentioned hereafter:
\begin{enumerate}
\item $N(\ell^{\pm})\geq 3$
%\item $|\sum^{3}_{i=1}Q(\ell^{\pm}_{i})|= 1$
\item Electron candidates:
  \begin{enumerate} 
     \item $|\eta(e^{\pm})|<1.37\ or\ 1.53<|\eta(e^{\pm})|<2.47$
     \item $p_{T}(e^{\pm}) > 10$ GeV
     %\item $ETRatio > 1.5$ (see \ref{Calo-Isol})
     %\item IsolFlag=1 (see \ref{Tracker-Isol})
  \end{enumerate}
\item Muon candidates:
  \begin{enumerate} 
     \item $|\eta(\mu^{\pm})|<2.4$
     \item $p_{T}(\mu^{\pm}) > 10$ GeV
     %\item $ETRatio > 0.5$
     %\item IsolFlag=1
  \end{enumerate}
\item $p_{T}(\ell^{\pm}_{1}) > 20$ GeV
\item $p_{T}(\ell^{\pm}_{2}) > 10$ GeV
\item $p_{T}(\ell^{\pm}_{3}) > 10$ GeV
\item Tracker Isolation: reject events with additional tracks of $p_{T}>2$ GeV within a cone of $\Delta R=0.5$ around the direction of the $\ell^{\pm}$ track
\item Calorimeter Isolation: ratio of the scalar sum of  $E_{T} $ deposits in the calorimeter within a cone of $\Delta R=0.5$ around the direction of the $\ell^{\pm}$, to the $p_{T}(\ell^{\pm})$ must be less than 1.2 for $e^{\pm}$ and less than 0.25 for $\mu^{\pm}$
\item $\rlap{\kern0.25em/}E_{T} > 35 \rm\ GeV$
\item $M_{T2} > 75$ GeV
\end{enumerate}

\noindent
The latter cut is applied on the so-called "stransverse mass":  $M_{T2}$. We used a boost-corrected calculation of this 
variable as described in \cite{Polesello:2009rn} and implemented in MCTLib \cite{Tovey:MCTLib}.

\begin{figure}[h]
\begin{center}
\includegraphics[scale=0.7]{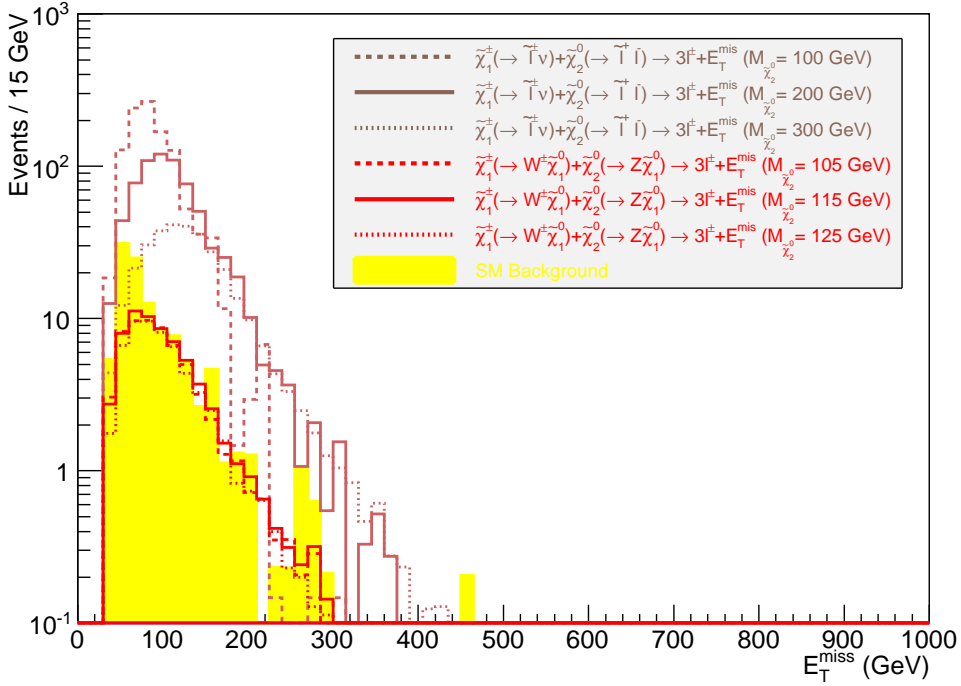}
\end{center}
\caption{\label{sec:Part4:mET_SEL} Distribution of the $\rlap{\kern0.25em/}E_{T}$ after the event selection. The background, the S1, and the S2 signals are the filled yellow, the hollow brown, and the hollow red histograms, respectively.}
\end{figure}

%%%%%%%%%%%%%%%%%%%%%%%%%%%%%%%%%%%%%%%%%%%%%%%%%%%%%%

\begin{table}[h]
\begin{center}
\begin{tabular}{|c|c|c|c|c|}
\hline\hline 
$Process$			& $\epsilon$		& $N_{exp}$    &  $Z_{N}$    &	$A_{C}^{Exp}\pm\delta A_{C}^{Stat}$	\\ 
				       & ($\%$)             	 & (Evts)	    &                      &	($\%$)		                                    \\
\hline\hline
\underline{S1 Signal} & 	          &		&     &		   \\
$[M_{\tilde\chi^{0}_{2}},M_{\tilde\ell^{\pm}},M_{\tilde\chi^{0}_{1}}]$  \rm\ GeV	         &		  &		&    &	       \\ 
$[100,75,50]$	    & $0.45\pm 0.01$	      &  1097.43       &   31.70   &   $(7.70\pm 0.27)$ \\
$[200,150,100]$   & $4.39\pm 0.02$	    &  702.98	      &   23.86  &   $(16.06\pm 0.20)$ \\
$[300,225,150]$   & $11.41\pm 0.03$	    &  319.48	      &    13.79   & $(21.30\pm 0.17)$ \\
$[400,300,200]$   & $16.15\pm 0.04$	    &  113.02 &    6.04   & $(24.40\pm 0.18)$ \\
$[500,375,250]$   & $18.98\pm 0.04$	    &  37.96  &    2.25   & $(27.21\pm 0.16)$ \\
$[600,450,300]$   & $21.01\pm 0.04$	    &  12.60  &    0.74   & $(27.20\pm 0.14)$ \\
$[700,525,350]$   & $22.66\pm 0.04$	    &  4.53   &    0.23   & $(29.06\pm 0.15)$ \\
\hline
\underline{S2 Signal} & 	          &		&     &		   \\
$[M_{\tilde\chi^{0}_{2}},M_{\tilde\chi^{0}_{1}}]$  \rm\ GeV       & 	           &		&     &		\\   
$[100,50]$     & $9.33\pm 0.18$	      &  0.14	     &	 -0.06   & $(7.62\pm 0.38)$ \\
$[105,13.8]$  & $2.10\pm 0.01$	     &  61.75	   &   3.55	& $(7.84\pm 0.23)$ \\
$[115,13.8]$  & $3.17\pm 0.02$	     &  65.46	   &	3.74   & $(7.73\pm 0.21)$ \\
$[125,13.8]$  & $3.85\pm 0.02$	     &  57.49	   &	3.32   & $(9.34\pm 0.21)$ \\
$[135,13.8]$  & $4.95\pm 0.02$	     &  54.84	   &   3.18    & $(10.43\pm 0.17)$ \\
$[145,13.8]$  & $5.85\pm 0.02$	     &  49.05	   &   2.87   & $(11.50\pm 0.19)$ \\
$[150,50]$    & $3.90\pm 0.02$	     &  28.65	    &	1.71  & $(12.06\pm 0.19)$ \\
$[200,100]$   & $4.59\pm 0.02$	     & 10.70	   &   0.62	& $(16.66\pm 0.20)$ \\
$[250,125]$   & $8.53\pm 0.03$	     &  7.79	   &   0.44	& $(18.28\pm 0.18)$ \\
$[300,150]$   & $12.42\pm 0.03$	     &  5.06	   &   0.26	& $(20.98\pm 0.18)$ \\
$[400,200]$   & $17.67\pm 0.04$	     &  1.80	   &   0.05	& $(24.11\pm 0.17)$ \\
$[500,250]$   & $20.09\pm 0.04$	     &  0.58	    &  -0.03	& $(27.51\pm 0.16)$ \\
$[600,300]$   & $21.70\pm 0.04$	     &  0.19	   &  -0.06	 & $(27.25\pm 0.18)$ \\
$[700,350]$      & $22.17\pm 0.04$        &  0.06    &	   -0.07   & $(27.91\pm 0.17)$ \\  
\hline\hline
Background              &	-              & 	109.51        & - & $(28.04\pm 0.20)$	\\
\hline
$W^{\pm}(\to e^{\pm}\nu_{e}/\mu^{\pm}\nu_{\mu}/\tau^{\pm}\nu_{\tau}/q\bar{q^{'}})+LF$ & $0.00\pm 0.00$ &0.00 & - & - \\
$W^{\pm}(\to e^{\pm}\nu_{e}/\mu^{\pm}\nu_{\mu}/\tau^{\pm}\nu_{\tau})+HF$ & $0.082\pm 0.004$ & 0.96  &-& $(36.93\pm 1.76)$ \\
\hline
$t\bar t$                              &   $0.00\pm 0.00$     & 	0.00	     &-& - \\
$t+b,\ t+q(+b)$                  &   $0.00\pm 0.00$     & 	0.00	       & - & - \\
\hline 
$W+W,\ W+\gamma^{*}/Z,\ \gamma^{*}/Z+\gamma^{*}/Z$ 	& $0.283\pm 0.002$     &  106.78 &-& $(26.95\pm 0.25)$ \\
$W^{+}+W^{-}+W^{\pm},\ W^{+}+W^{-}+\gamma^{*}/Z,$ 	& $0.576\pm 0.004$     &   1.77 &-& $(29.84\pm 0.34)$ \\
$W^{\pm}+\gamma^{*}/Z+\gamma^{*}/Z,\ \gamma^{*}/Z+\gamma^{*}/Z+\gamma^{*}/Z$ &      & & -   &  \\
\hline
$\gamma+\gamma,\ \gamma+jets,\ \gamma+W^{\pm},\ \gamma+Z$ & $0.00\pm 0.00$ &  0.00  & - & - \\
\hline
$\gamma^{*}/Z+LF$                  &   $0.00\pm 0.00$    &   0.00	      & - & - \\
$\gamma^{*}/Z+HF$                  &   $0.00\pm 0.00$    &   0.00	      & - & - \\
\hline
QCD HF                          & $0.00\pm 0.00$ &        0.00 & - & - \\
QCD LF                          & $0.00\pm 0.00$ &        0.00 & - & - \\
\hline\hline
\end{tabular}       
\end{center}
\caption{\label{trilepton_SEL_tab} Event selection efficiencies, event yields, signal significances and charge asymmetries for the $p+p\to\tilde\chi^{\pm}_{1}\tilde\chi^{0}_{2}\to 3\ell^{\pm}+\rlap{\kern0.25em/}E_{T}$ analysis.}
\end{table}

\begin{table}[h]
\begin{center}
\begin{tabular}{|c|c|c|c|c|c|}
\hline\hline 
$Process$  &  $\alpha^{Exp}\pm\delta \alpha^{Stat}$   &	$Z_{N}$ & $A_{C}^{Meas.}$ & $\delta A_{C}^{Tot.}$ & $\delta A_{C}^{Meas.Fit}$\\ 
				&  	     & ($\sigma$) &	($\%$)	&	($\%$)&	($\%$)	\\
\hline\hline
\underline{S1 Signal} & 	          &		&     &		&   \\
$[M_{\tilde\chi^{0}_{2}},M_{\tilde\ell^{\pm}},M_{\tilde\chi^{0}_{1}}]$  \rm\ GeV	         &		  &	&	&    &	       \\ 
$[100,75,50]$	    & $(9.98\pm 0.26)\times 10^{-2}$              &   31.70   &     7.70 & 0.83    &  0.74   \\
$[200,150,100]$   & $(15.58\pm 0.36)\times 10^{-2}$	       &   23.86   &   16.06 &0.85 &  0.44  \\
$[300,225,150]$   & $(34.28\pm 0.79)\times 10^{-2}$	       &    13.79  &   21.30 &0.96 &  0.48   \\
$[400,300,200]$   & $(96.89\pm 2.22)\times 10^{-2}$	       &    6.04    &   24.40 & 1.29 &  0.58   \\
$[500,375,250]$   & $(288.49\pm 6.61)\times 10^{-2}$	       &    2.25   &   27.21 & 1.75 &   0.69  \\
$[600,450,300]$   & $(869.13\pm 19.89)\times 10^{-2}$	      &    0.74   &   27.20 &1.97 &  0.77   \\
$[700,525,350]$   & $(241.74\pm 5.55)\times 10^{-1}$	       &    0.23   &   29.06 & 2.02 &  0.85   \\
\hline
\underline{S2 Signal}      & 	           &	&	     &		&   \\
$[M_{\tilde\chi^{0}_{2}},M_{\tilde\chi^{0}_{1}}]$  \rm\ GeV  & &	     &                  &		&                \\   
$[100,50]$     & $(78.22\pm 6989.64)\times 10^{1}$	           &  -0.06 &   7.62           &	 0.88   &   0.59               \\   
$[105,13.8]$  & $(177.34\pm 4.21)\times 10^{-2}$	      	   &   3.55  &   7.85           &	 1.58   &    0.56              \\  
$[115,13.8]$  & $(167.29\pm 3.91)\times 10^{-2}$	      	   &   3.74  &    7.73          &	 1.55   &     0.52               \\    
$[125,13.8]$  & $(190.49\pm 4.44)\times 10^{-2}$	      	   &   3.32  &    9.34           &	1.60   &       0.49             \\    
$[135,13.8]$  & $(199.69\pm 4.61)\times 10^{-2}$	      	   &   3.18  &    10.43         &	 1.62 &      0.46             \\     
$[145,13.8]$  & $(223.26\pm 5.16)\times 10^{-2}$	      	   &   2.87  &    11.50         &	 1.67 &      0.45             \\    
$[150,50]$     & $(382.23\pm 8.90)\times 10^{-2}$	            &	1.71  &   12.06          &	 1.85 &        0.44             \\   
$[200,100]$   & $(102.35\pm 2.34)\times 10^{-1}$	     	   &   0.62  &    16.66         &	2.00 &       0.46              \\   
$[250,125]$   & $(140.58\pm 3.23)\times 10^{-1}$	      	   &   0.44  &   18.28          &	 2.01 &     0.52               \\   
$[300,150]$   & $(216.42\pm 4.96)\times 10^{-1}$	      	   &   0.26  &    20.98         &	2.02 &      0.60               \\   
$[400,200]$   & $(608.39\pm 13.89)\times 10^{-1}$	      	  &   0.05  &    24.11         &	2.03 &     0.74               \\   
$[500,250]$   & $(18.88\pm 0.43)\times 10^{-5}$	      	           &  -0.03  &   27.51           &	2.03  &      0.86              \\   
$[600,300]$   & $(57.64\pm 1.32)\times 10^{-5}$	      	           &  -0.06  &   27.25           &	 2.03  &      0.96             \\   
$[700,350]$   & $(182.52\pm 4.17)\times 10^{-5}$                 &  -0.07  &    27.91           &	 2.03 &      1.04               \\   
\hline\hline
\end{tabular}       
\end{center}
\caption{\label{sec:Part4:trilepton_AC} Noise to signal ratio, signal statistical significance, and expected and measured integral charge asymmetries for the S1 and S2 signal samples for the $p+p\to\tilde\chi^{\pm}_{1}\tilde\chi^{0}_{2}\to 3\ell^{\pm}+\rlap{\kern0.25em/}E_{T}$ analysis.}
\end{table}

\par\noindent
The event selection efficiencies, event yields, signal significances and the expected integral charge asymmetries are reported in table \ref{trilepton_SEL_tab}. Figure \ref{sec:Part4:mET_SEL} displays the $\rlap{\kern0.25em/}E_{T}$ distribution after the event selection.
\par\noindent
We note that the S1 signal significance exceeds $5\sigma$ for $M_{\tilde\chi^{0}_{2}}=M_{\tilde\chi^{\pm}_{1}}$ in the [100,400] GeV interval, whereas the S2 signal significance reaches only the $3\sigma$ for 100 $<M_{\tilde\chi^{0}_{2}}=M_{\tilde\chi^{\pm}_{1}}<$ 150 GeV.
\par
\begin{figure}[h]
\begin{center}
\includegraphics[scale=0.225]{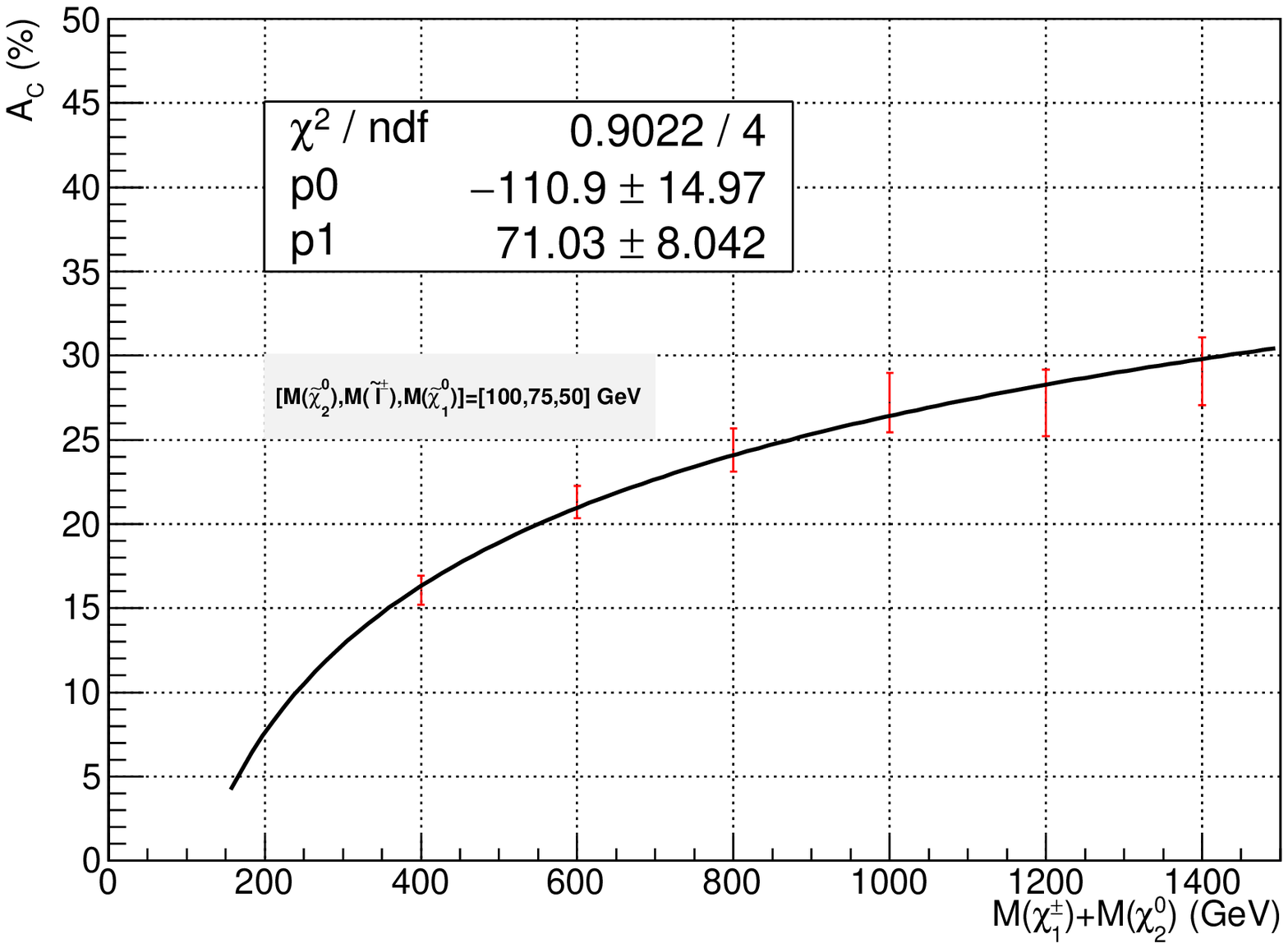}
\includegraphics[scale=0.225]{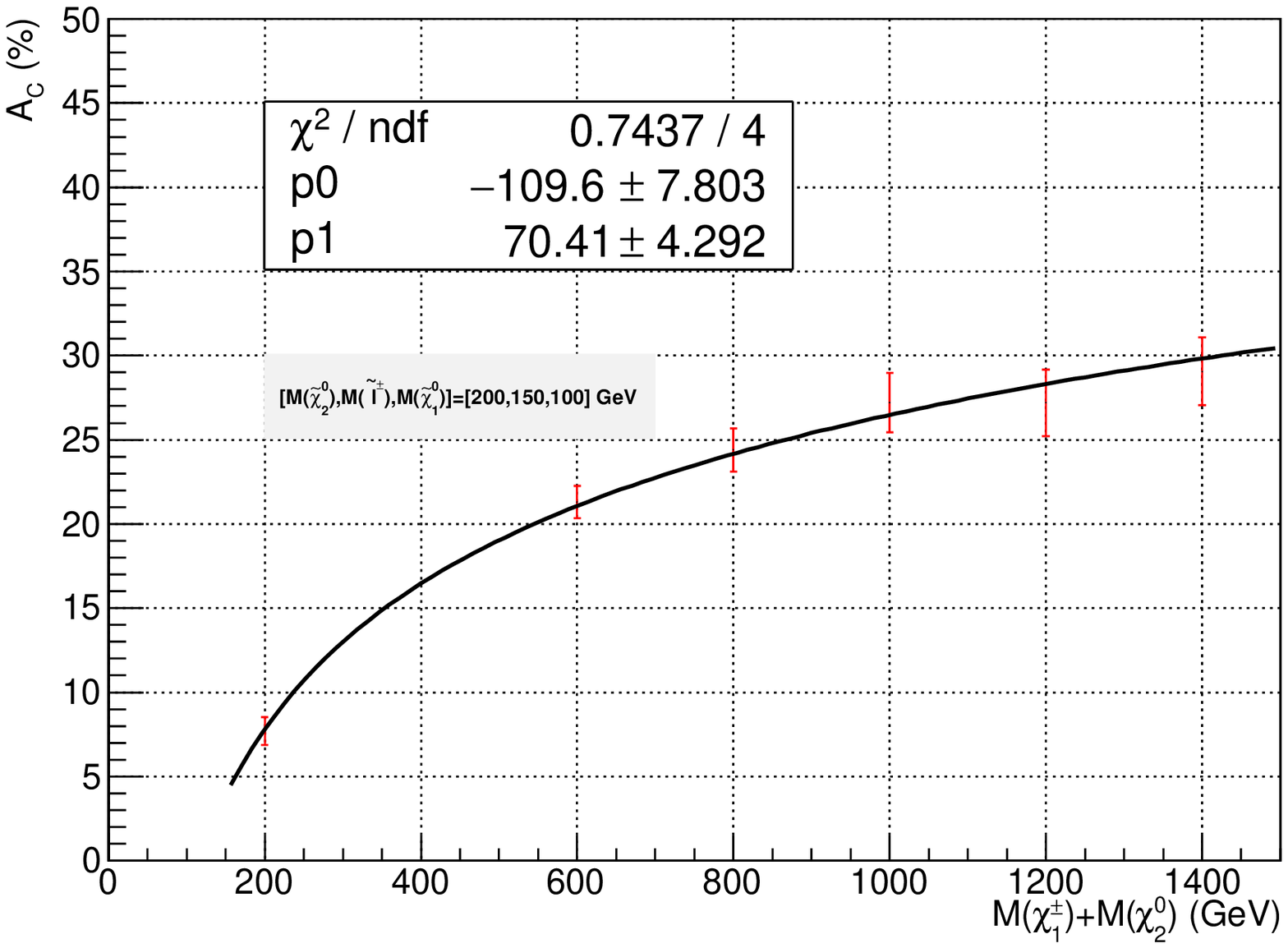}
\includegraphics[scale=0.225]{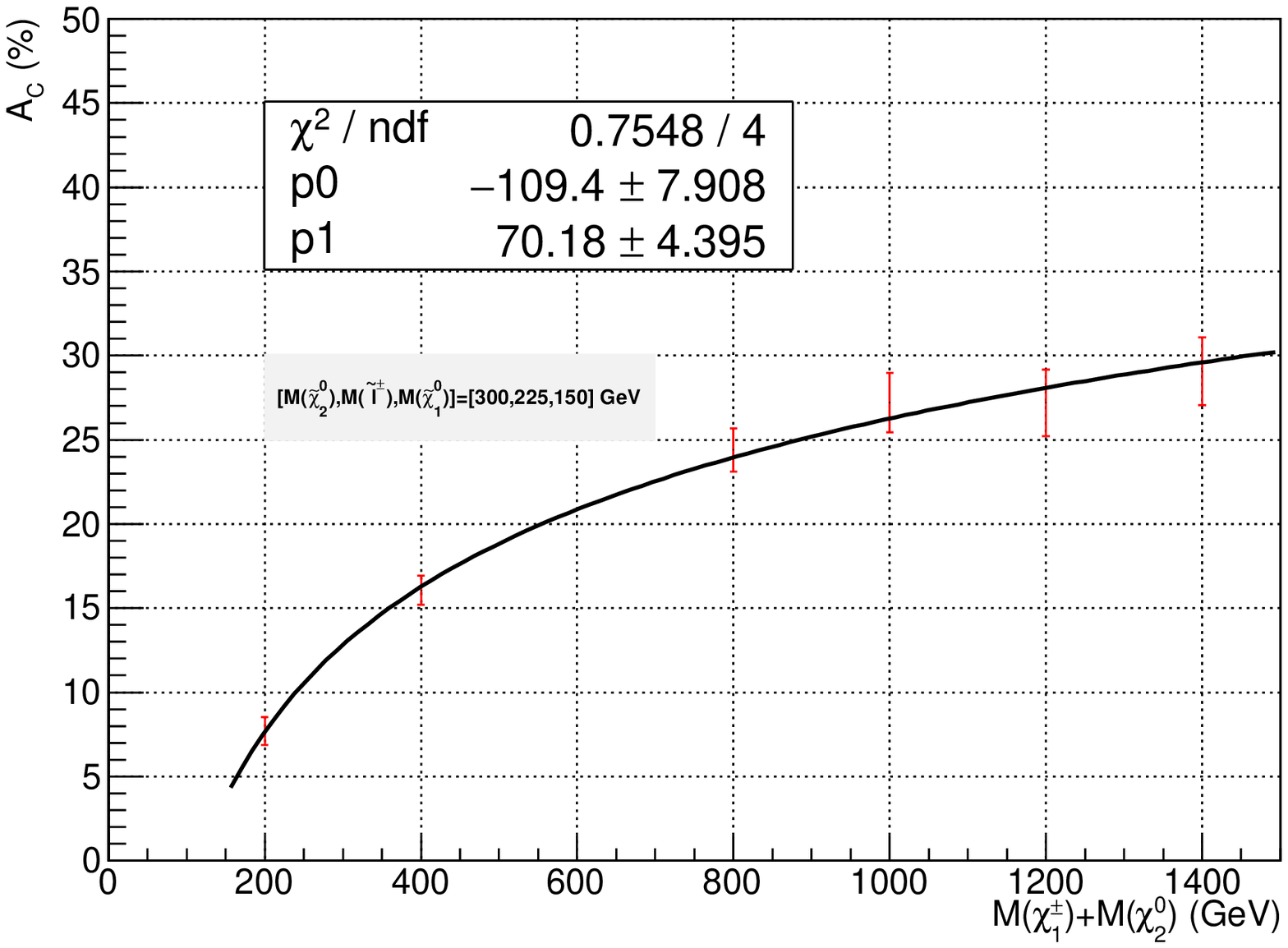} \\
\includegraphics[scale=0.225]{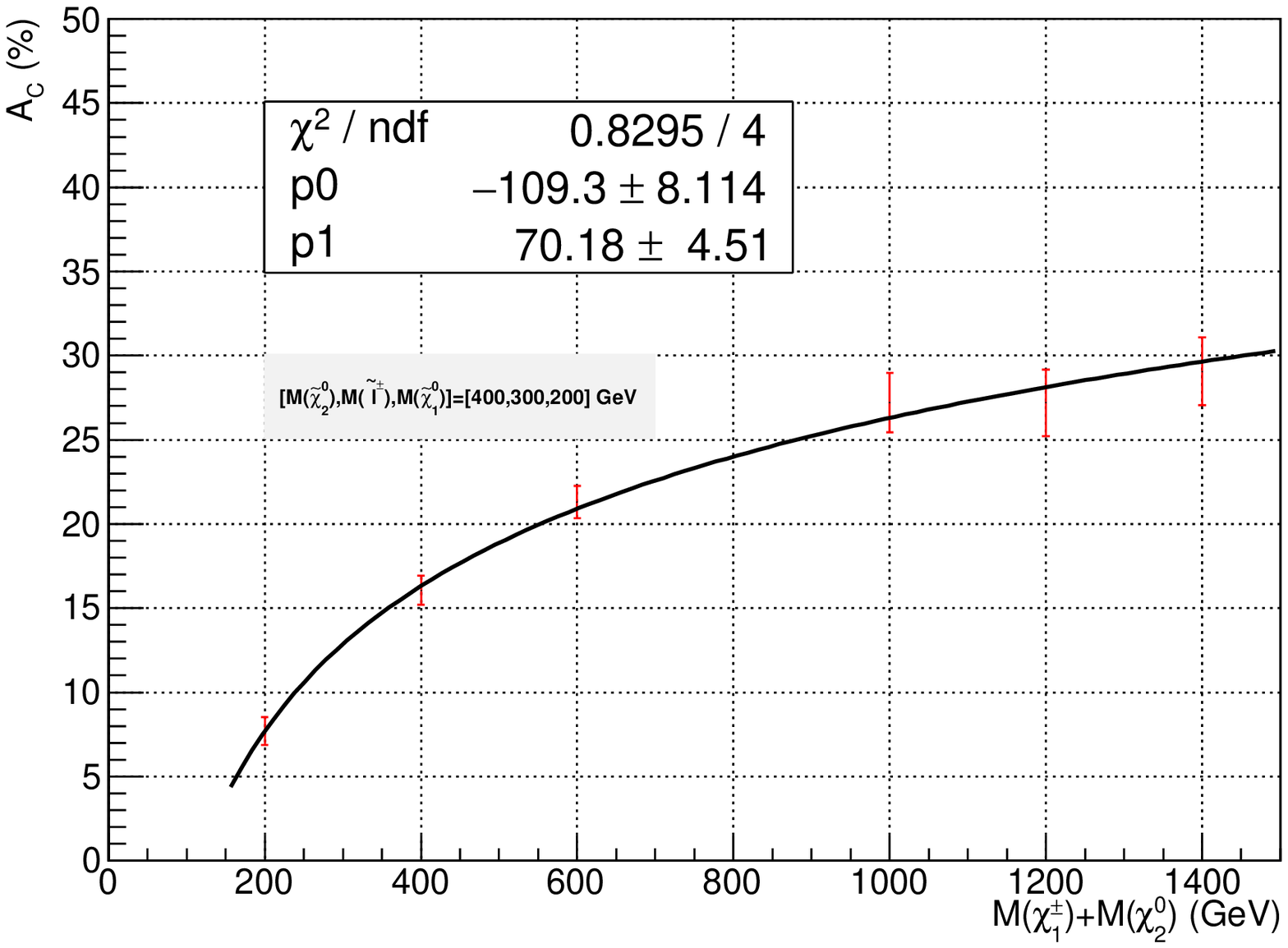}
\includegraphics[scale=0.225]{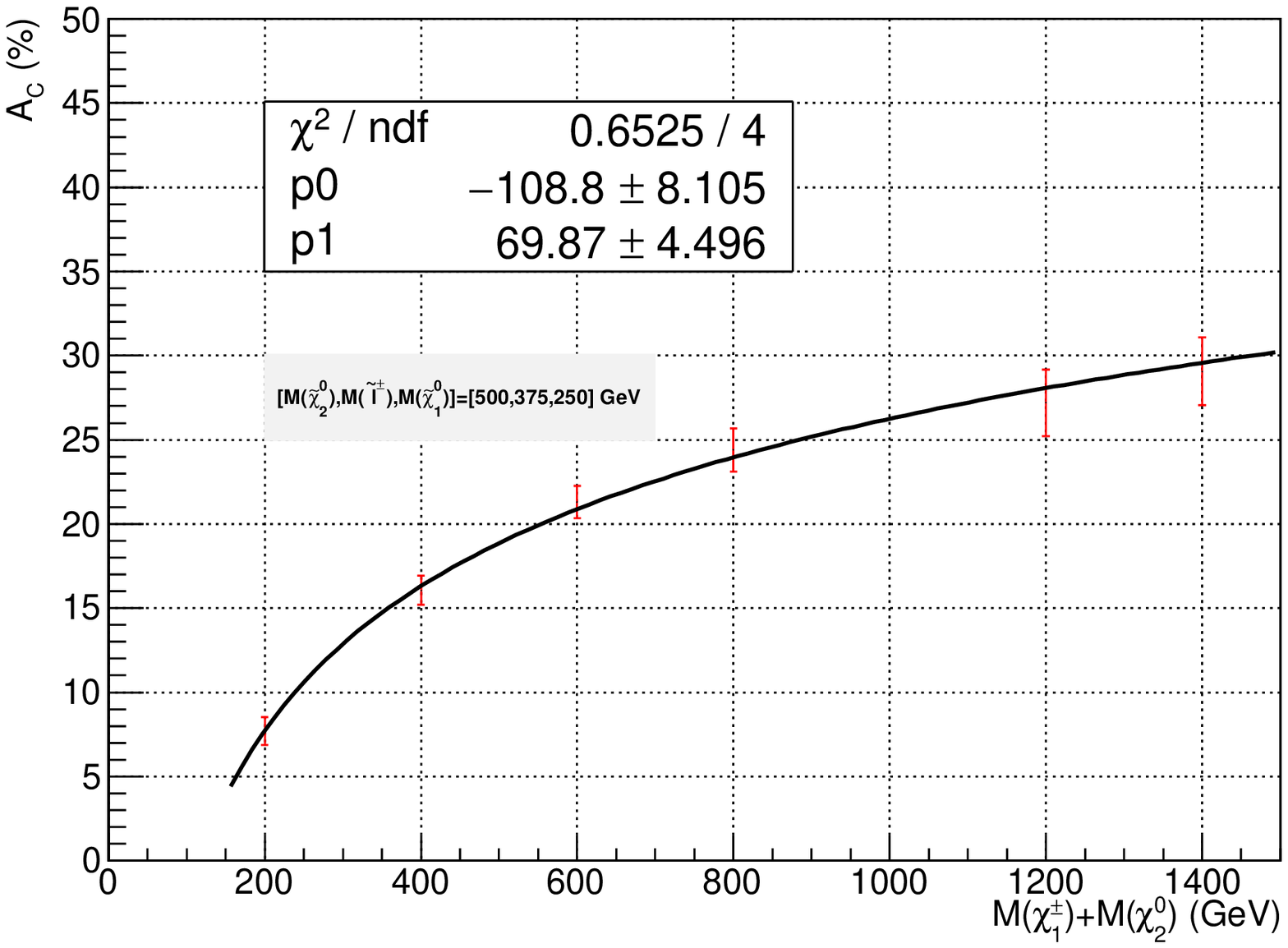}
\includegraphics[scale=0.225]{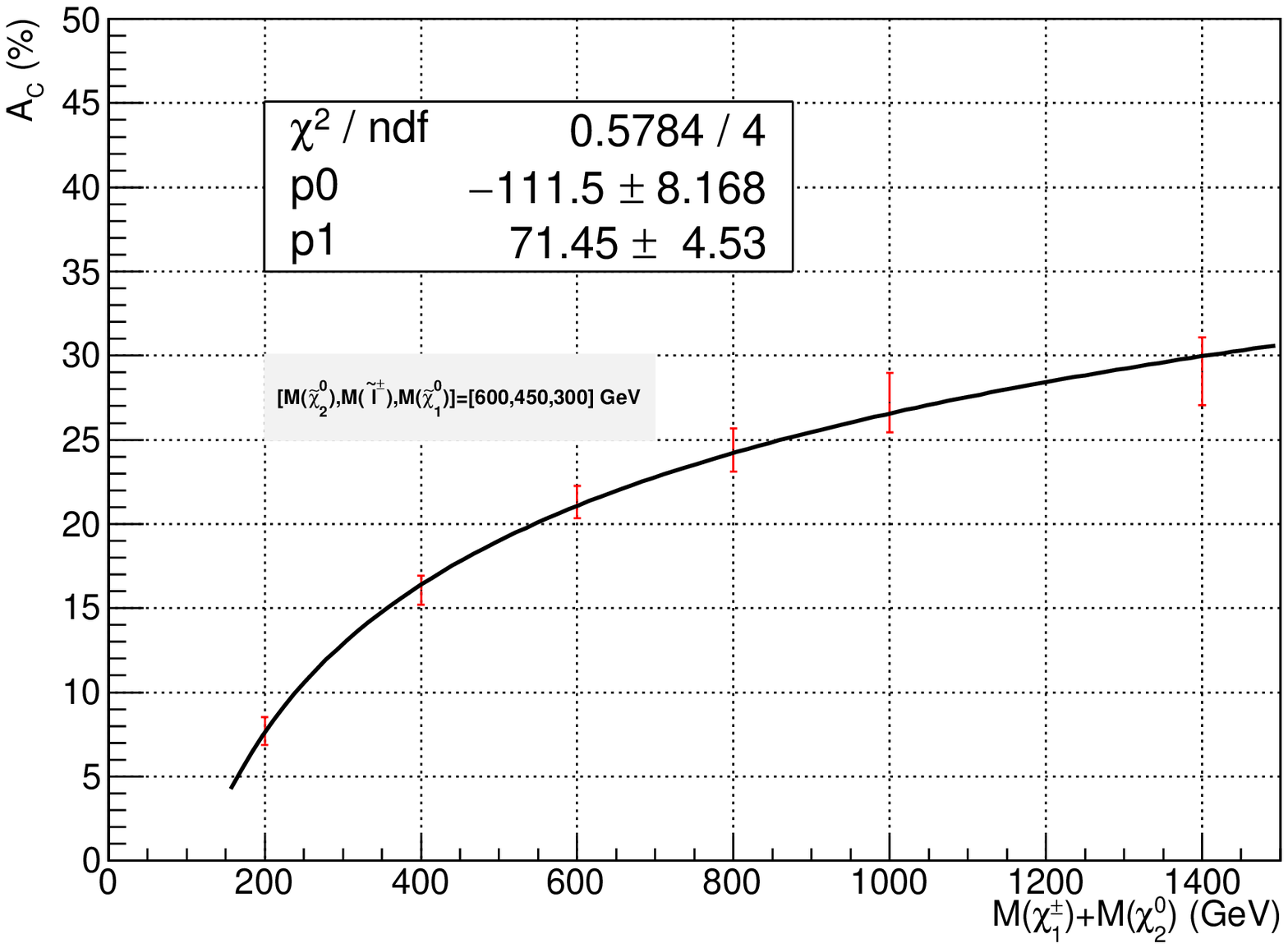} \\
\includegraphics[scale=0.225]{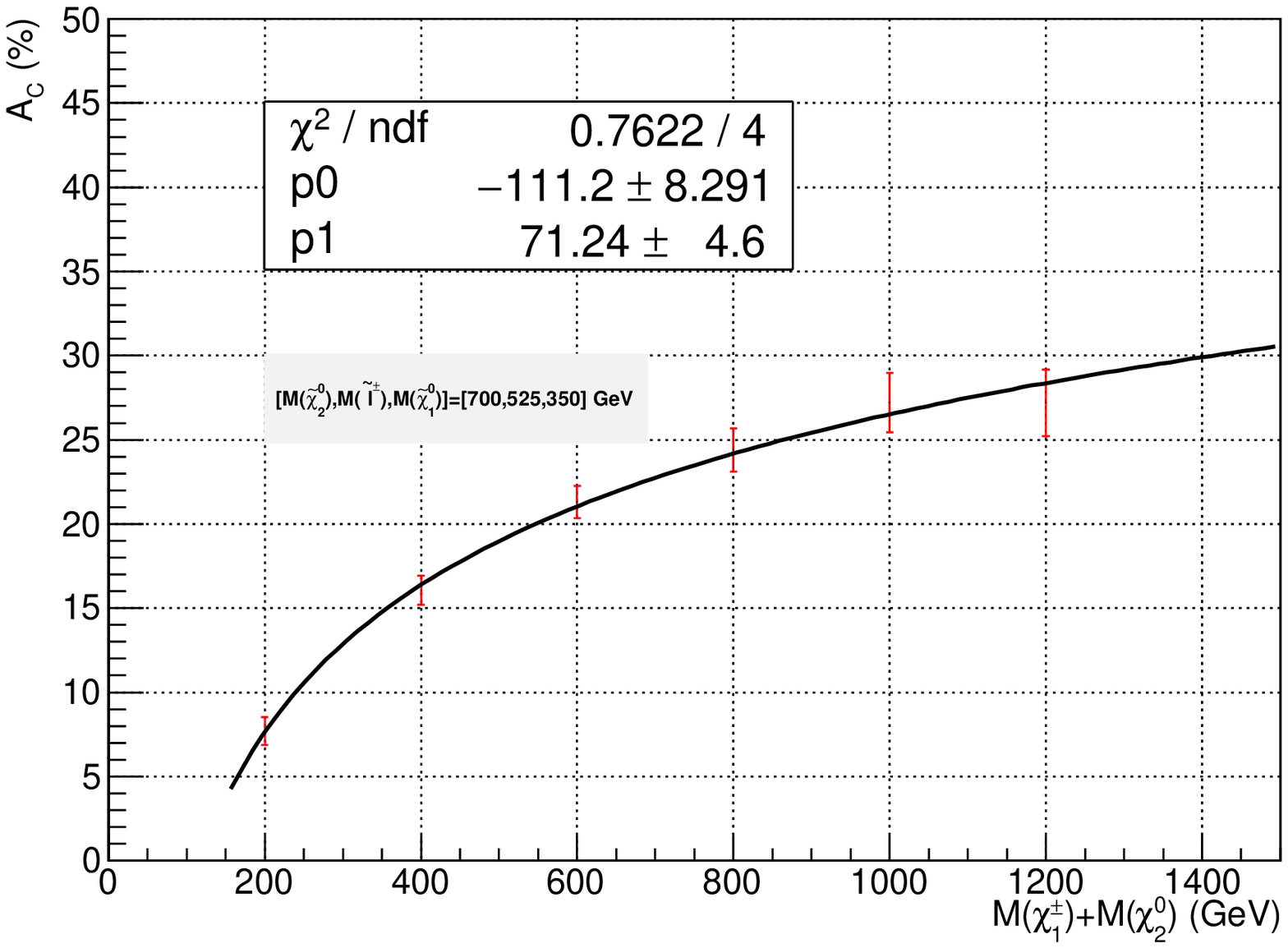}
\end{center}
\caption{\label{sec:Part4:GCATC_S1_Reco} Experimental $A_{C}$ template curves for the S1 signal samples, as they are listed, in table \ref{trilepton_SEL_tab} from the top to the bottom rows. Here, they appear ordered by increasing $\tilde\chi^{0}_{2}$ mass, from the top to the bottom row and from left to right.}
\end{figure}

\begin{figure}[h]
\begin{center}
\includegraphics[scale=0.225]{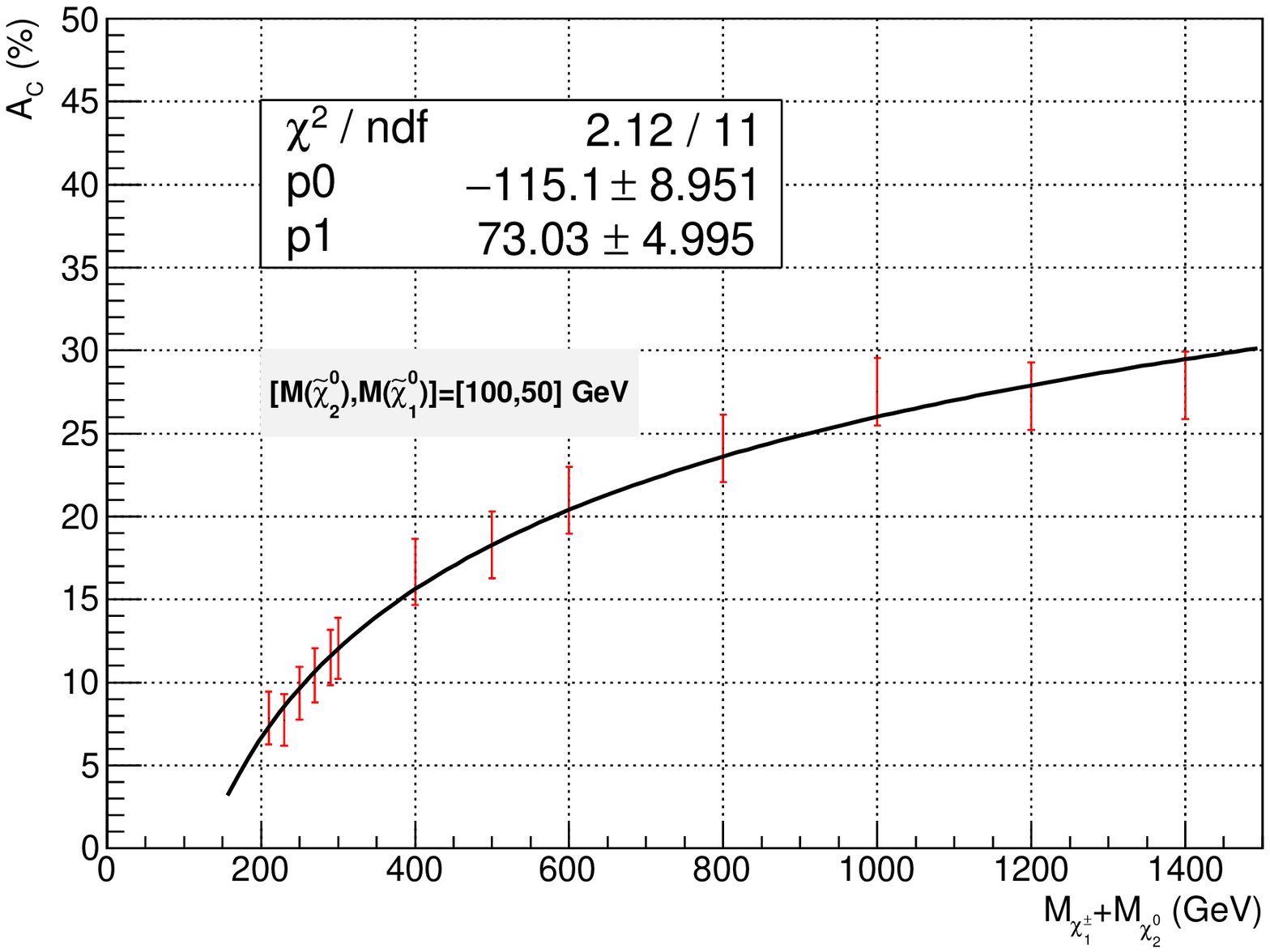}
\includegraphics[scale=0.225]{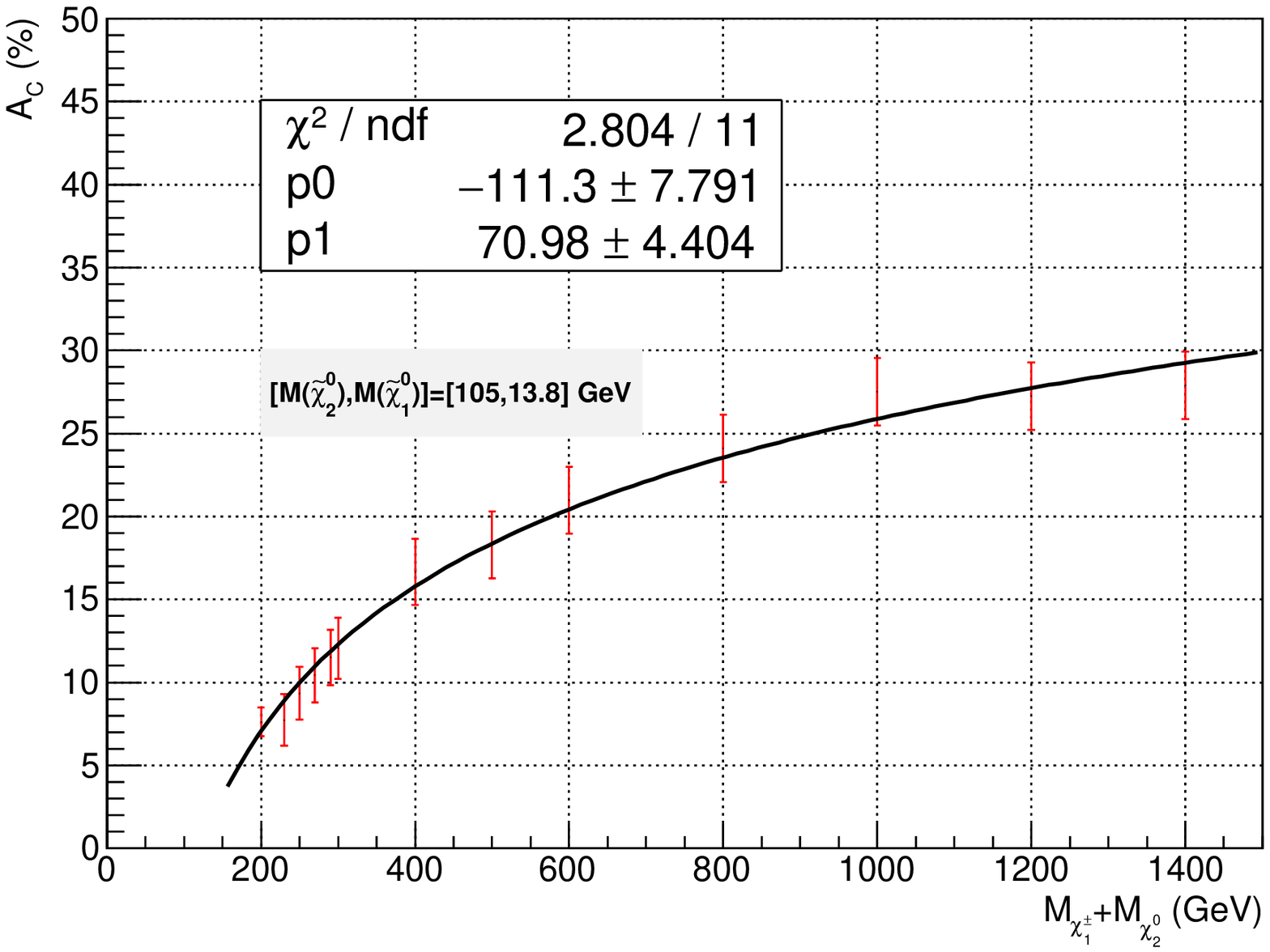}
\includegraphics[scale=0.225]{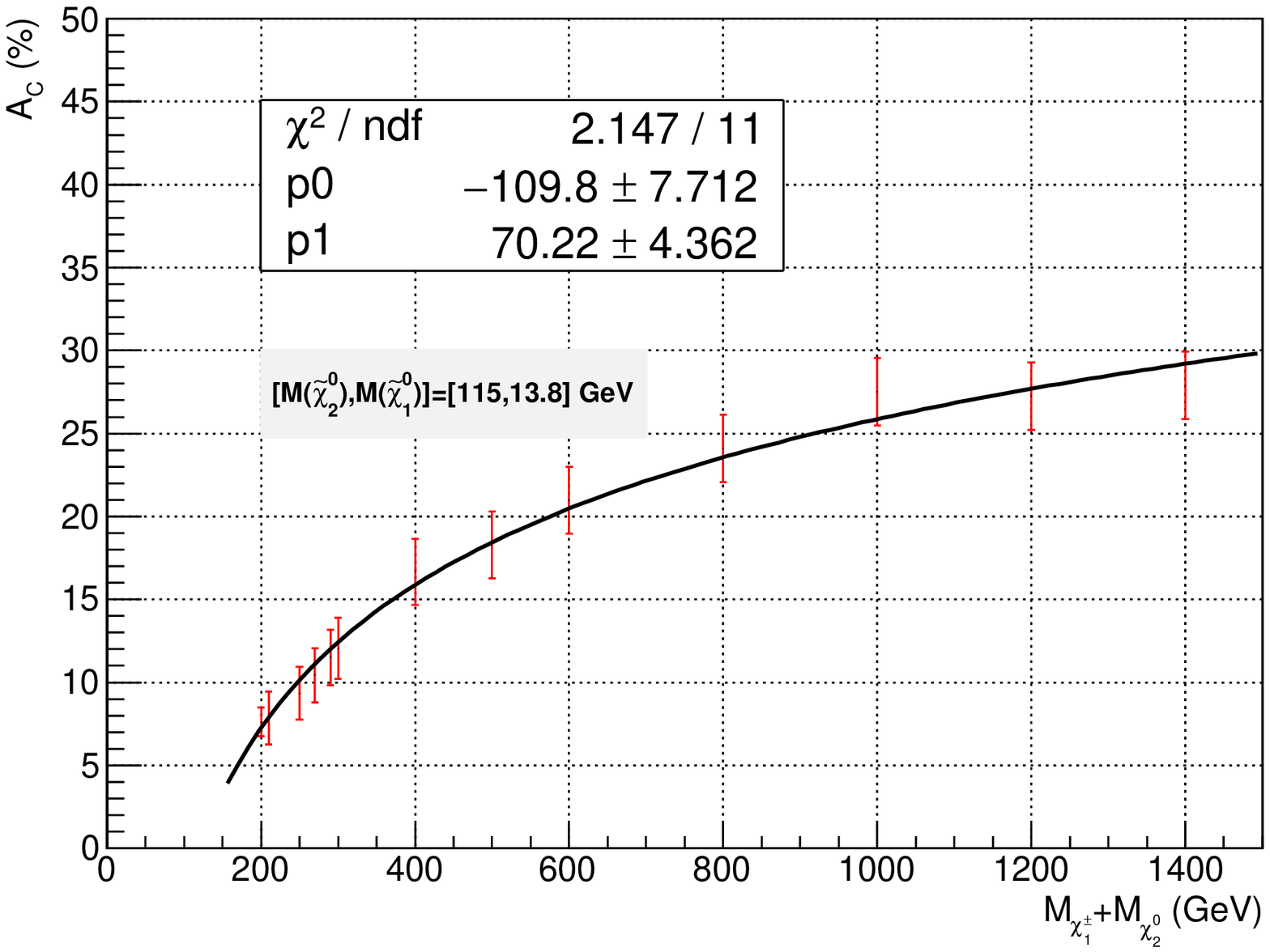} \\
\includegraphics[scale=0.225]{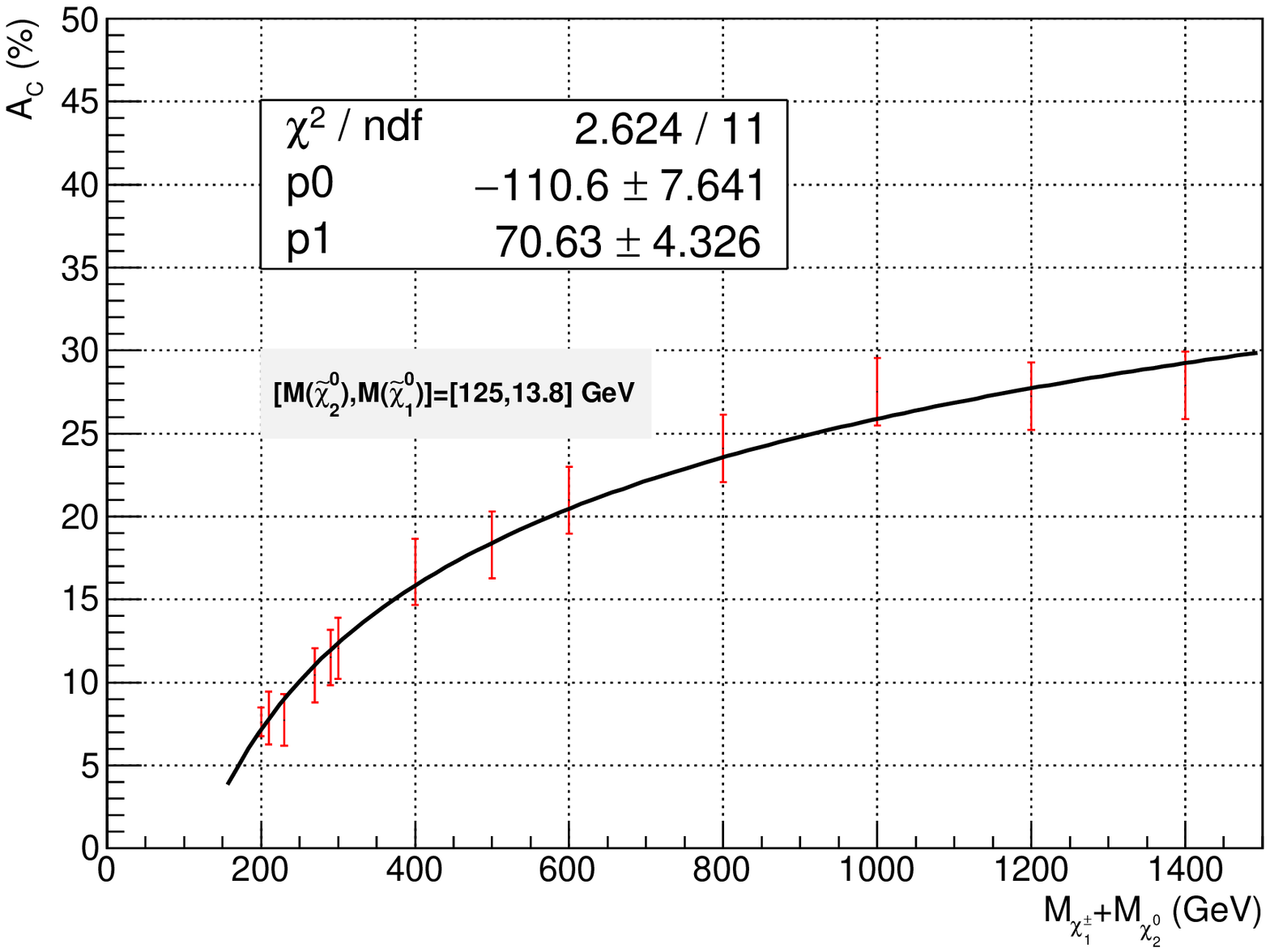}
\includegraphics[scale=0.225]{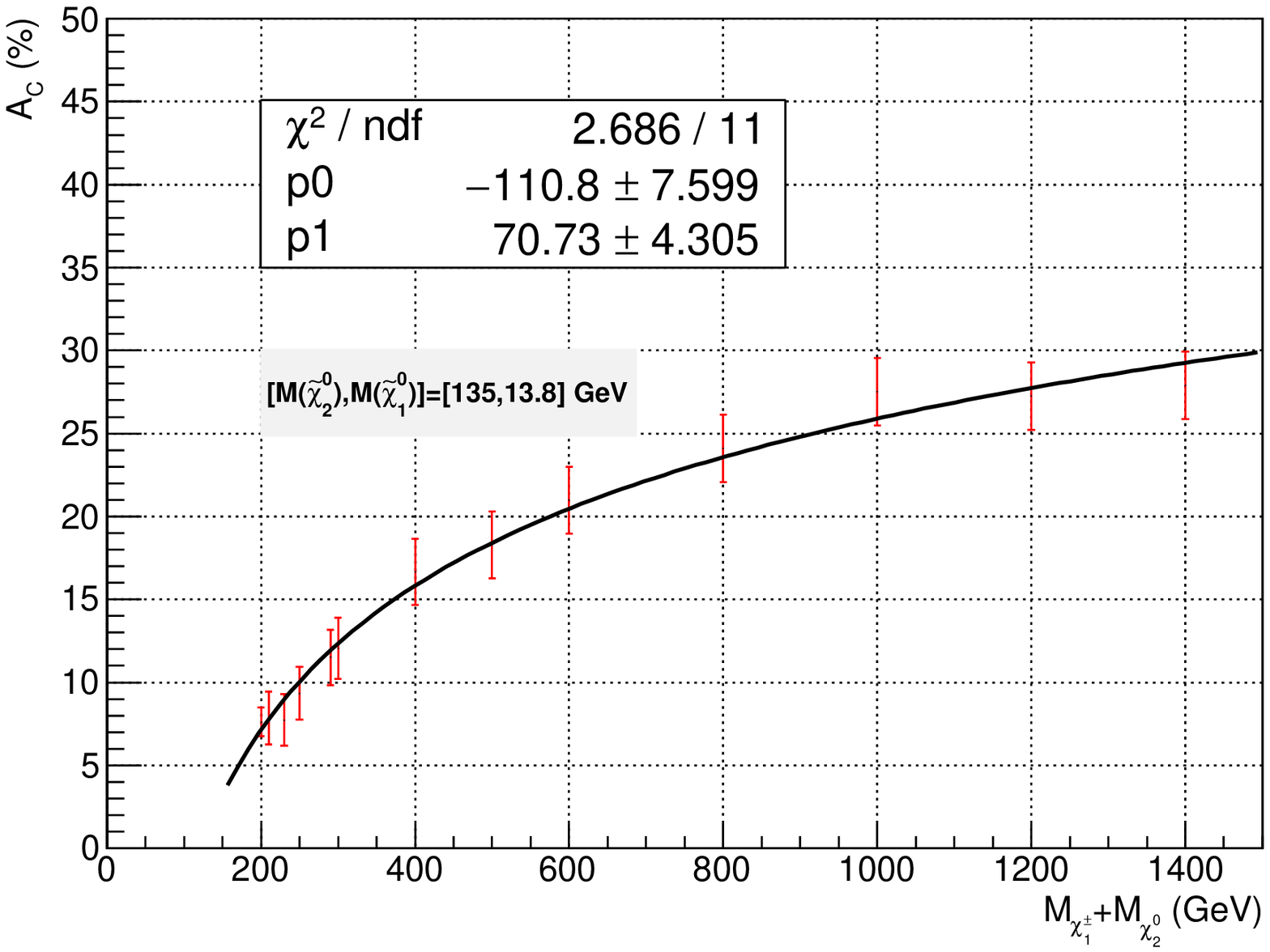}
\includegraphics[scale=0.225]{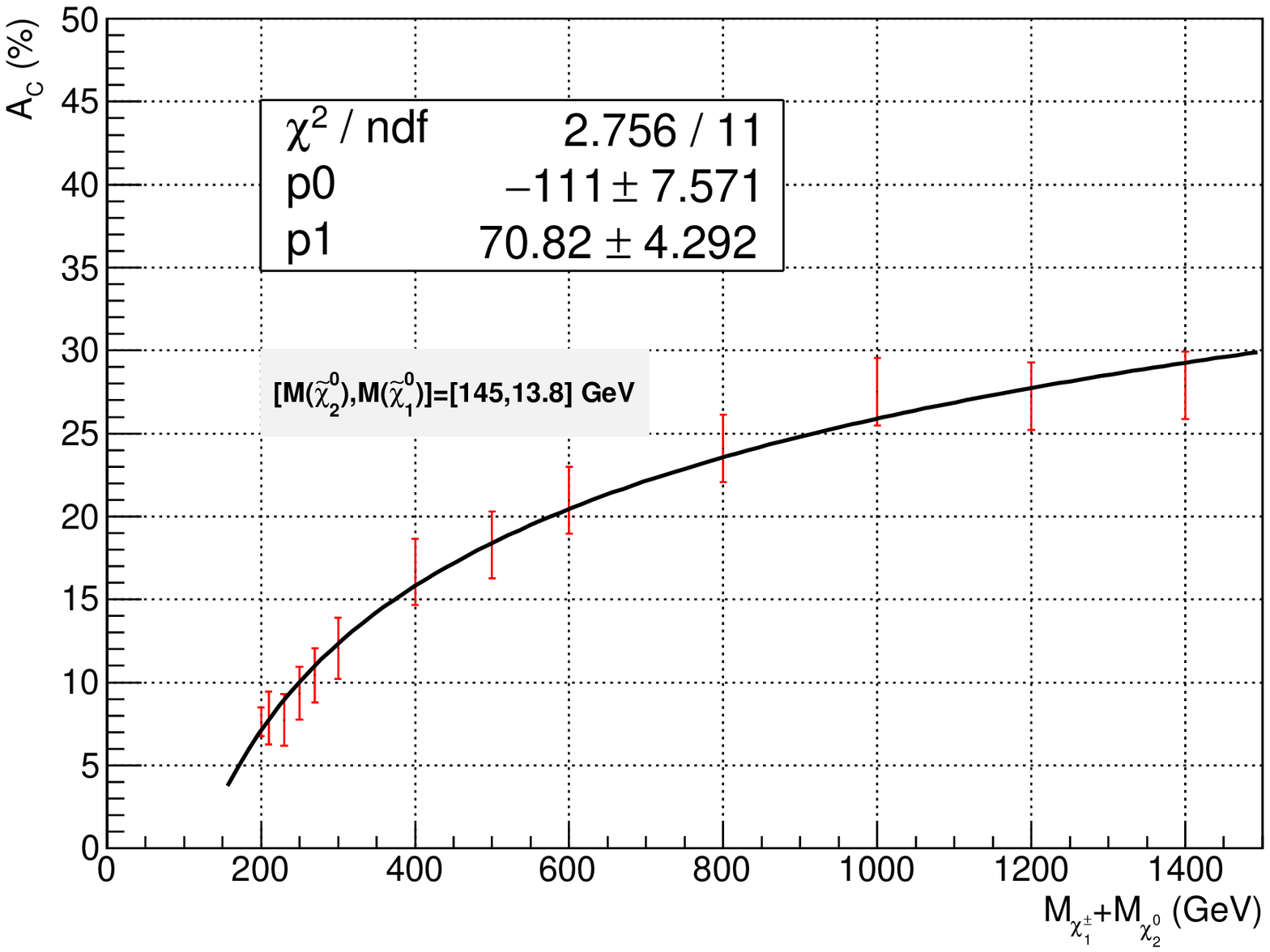} \\
\includegraphics[scale=0.225]{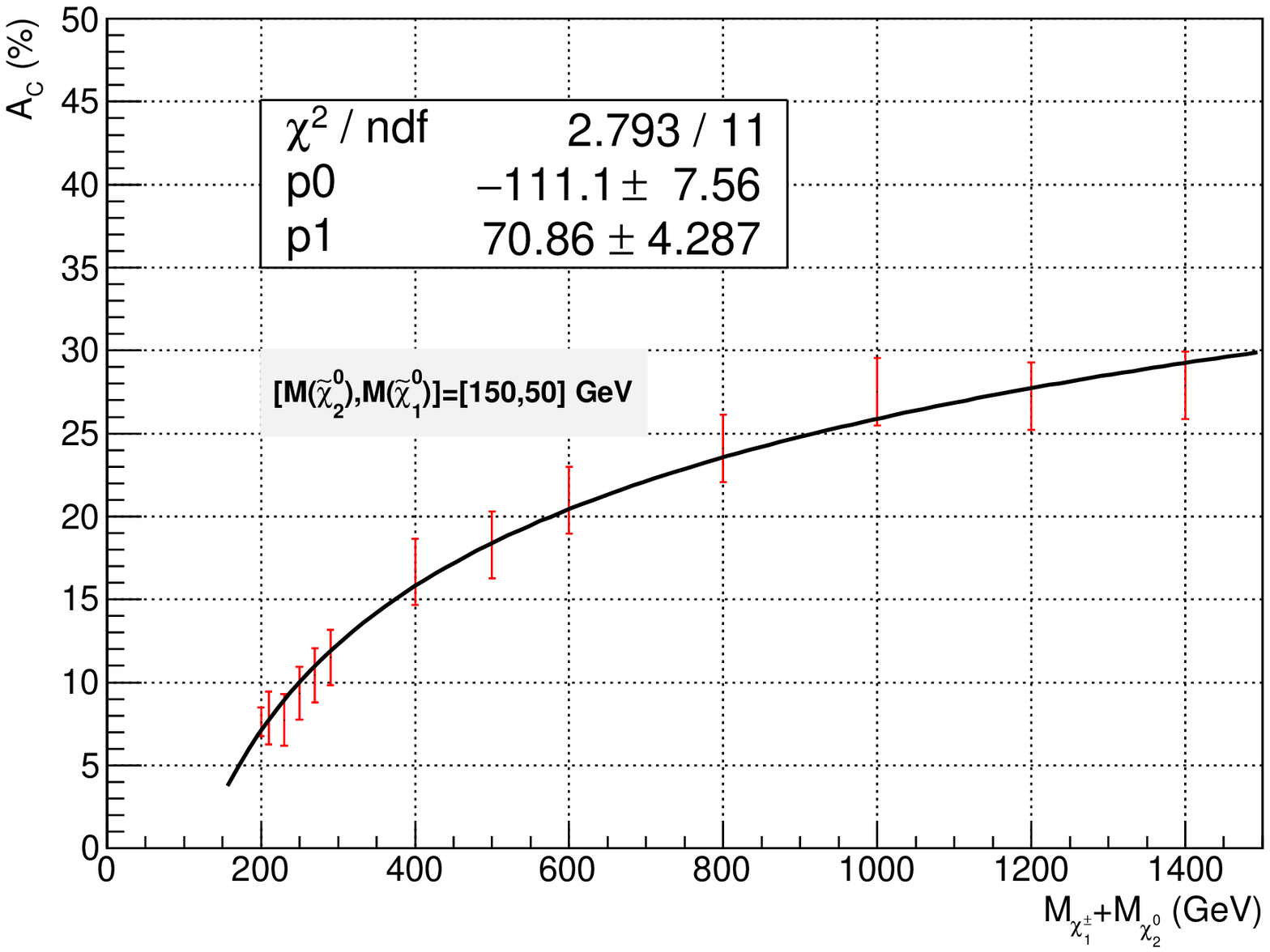}
\includegraphics[scale=0.225]{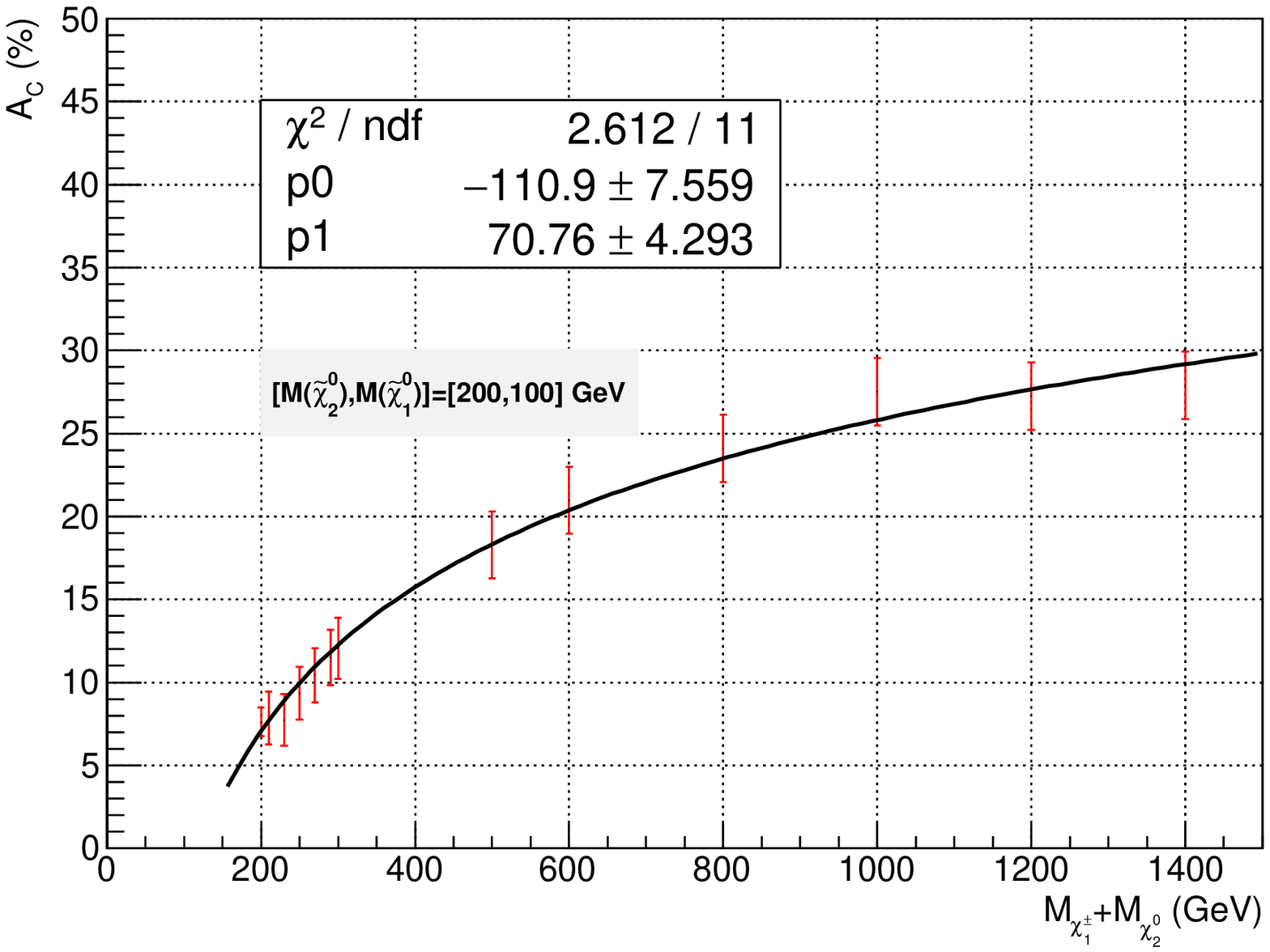}
\includegraphics[scale=0.225]{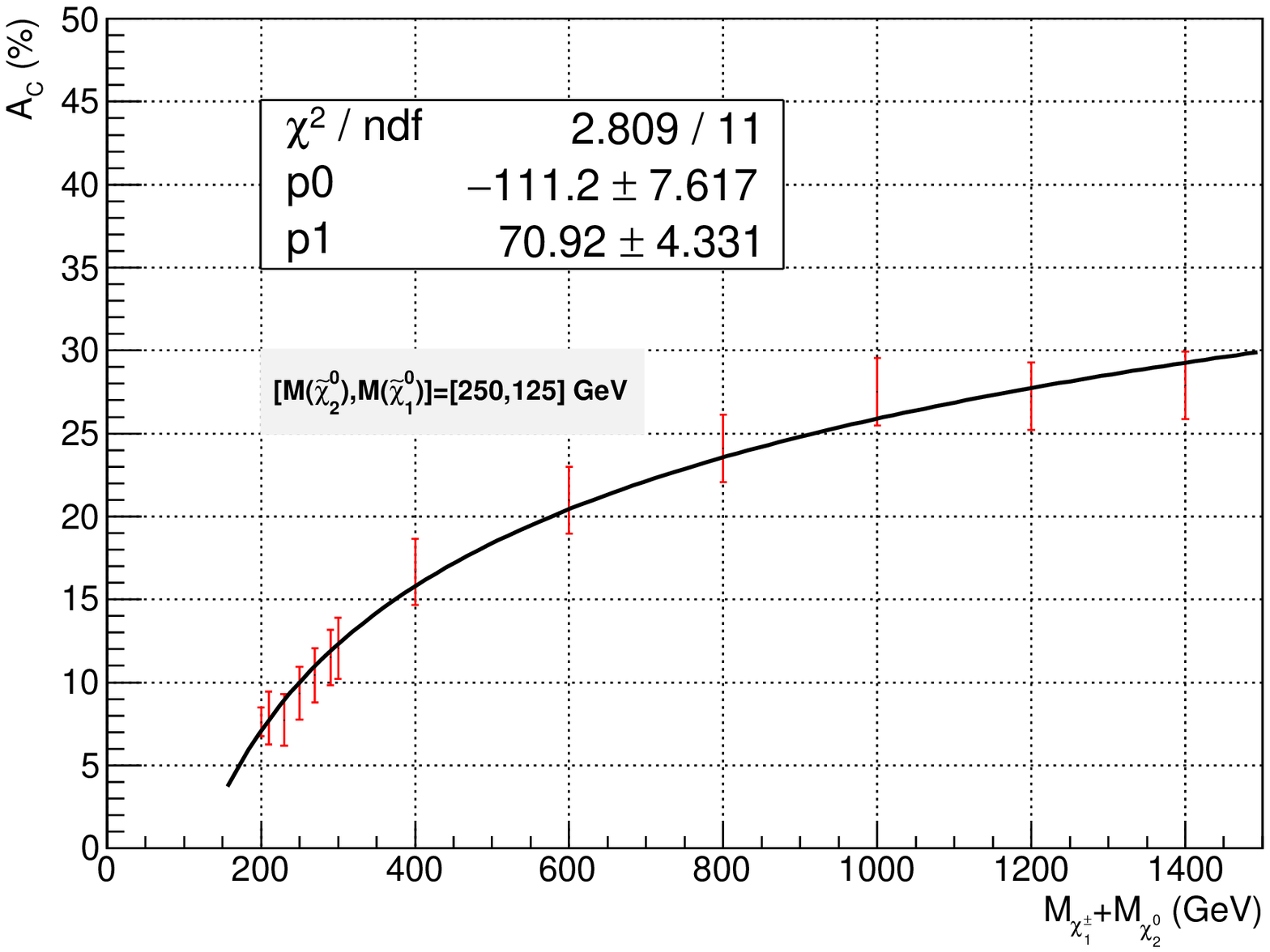} \\
\includegraphics[scale=0.225]{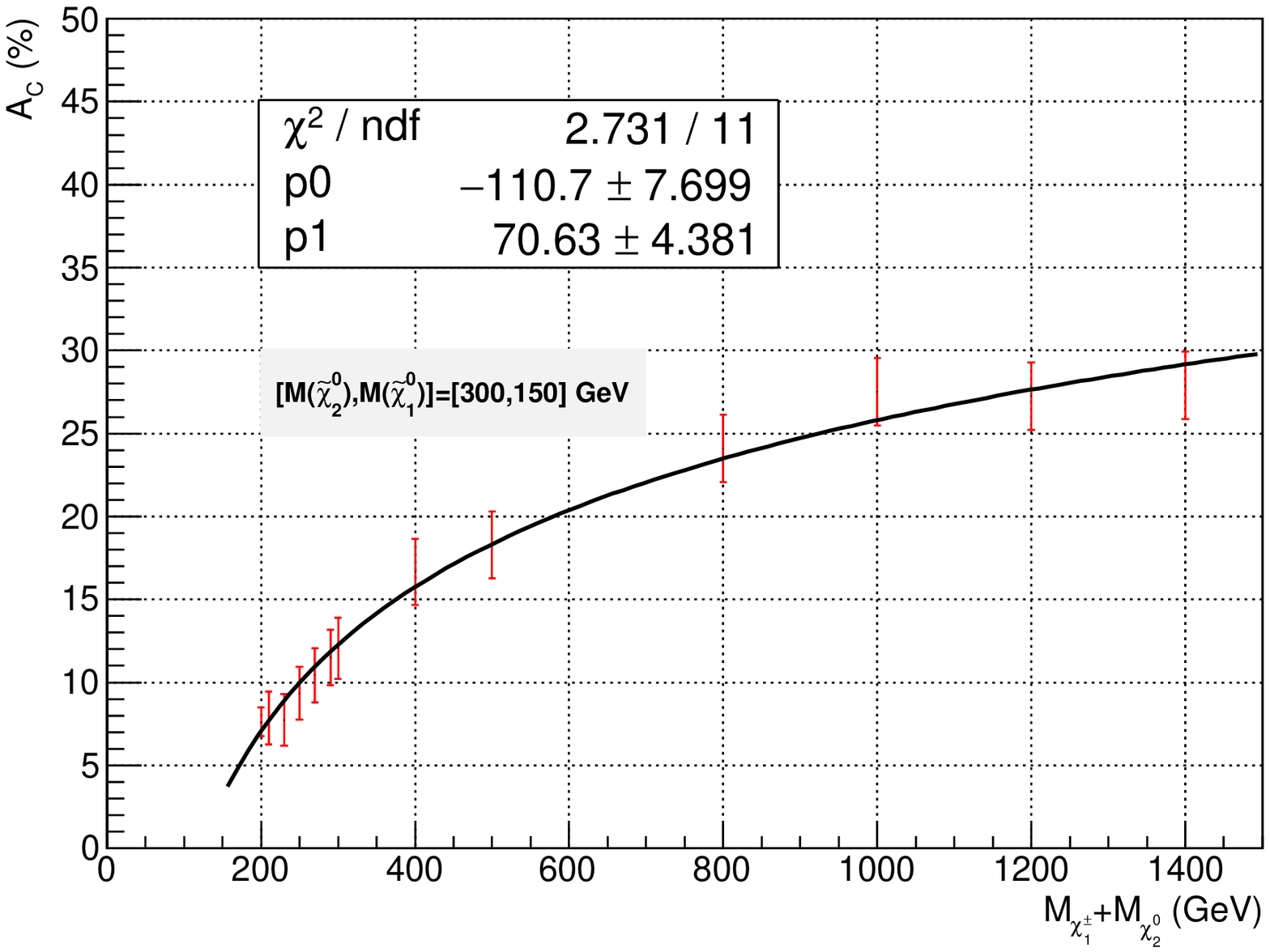}
\includegraphics[scale=0.225]{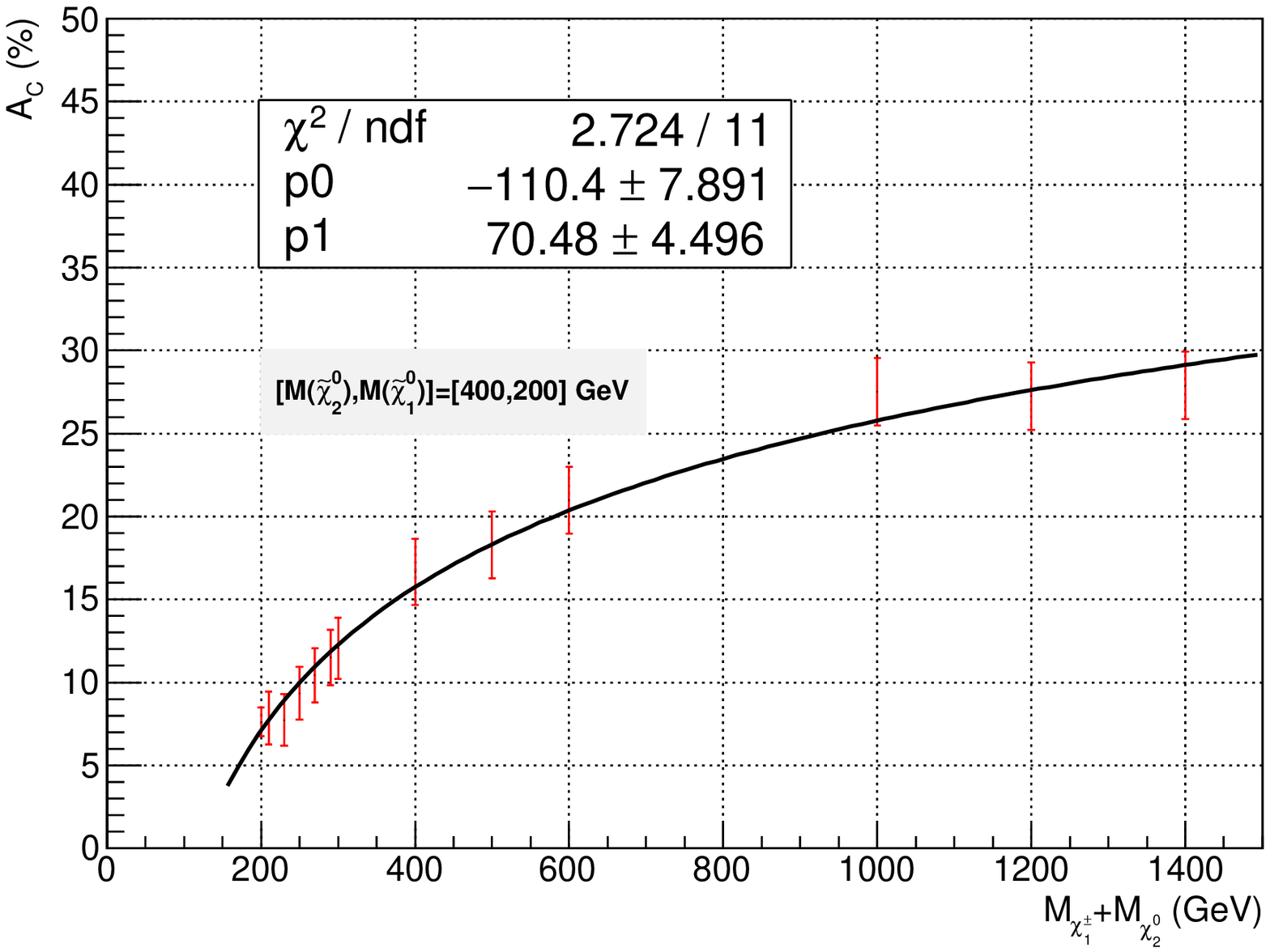}
\includegraphics[scale=0.225]{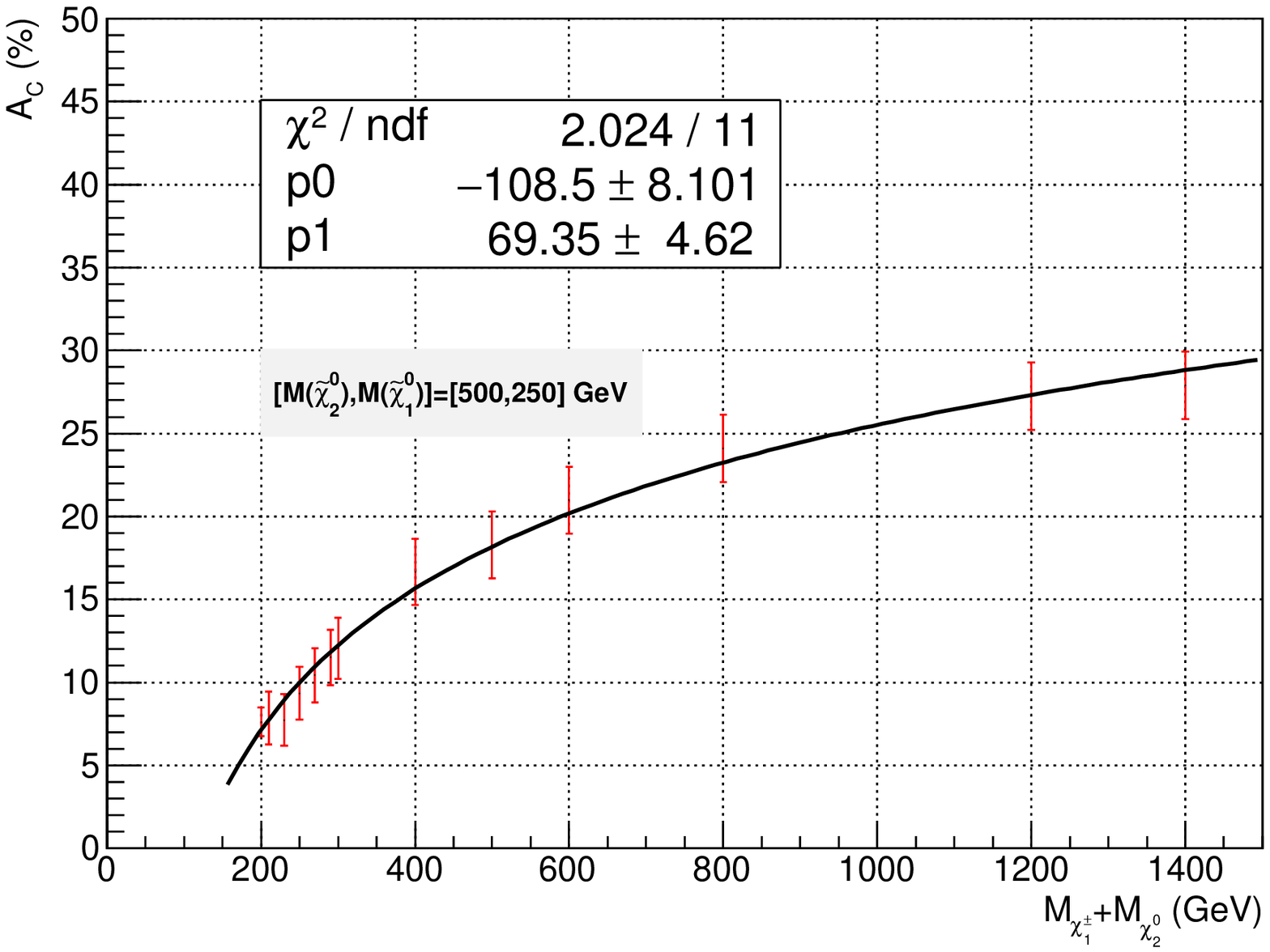} \\
\includegraphics[scale=0.225]{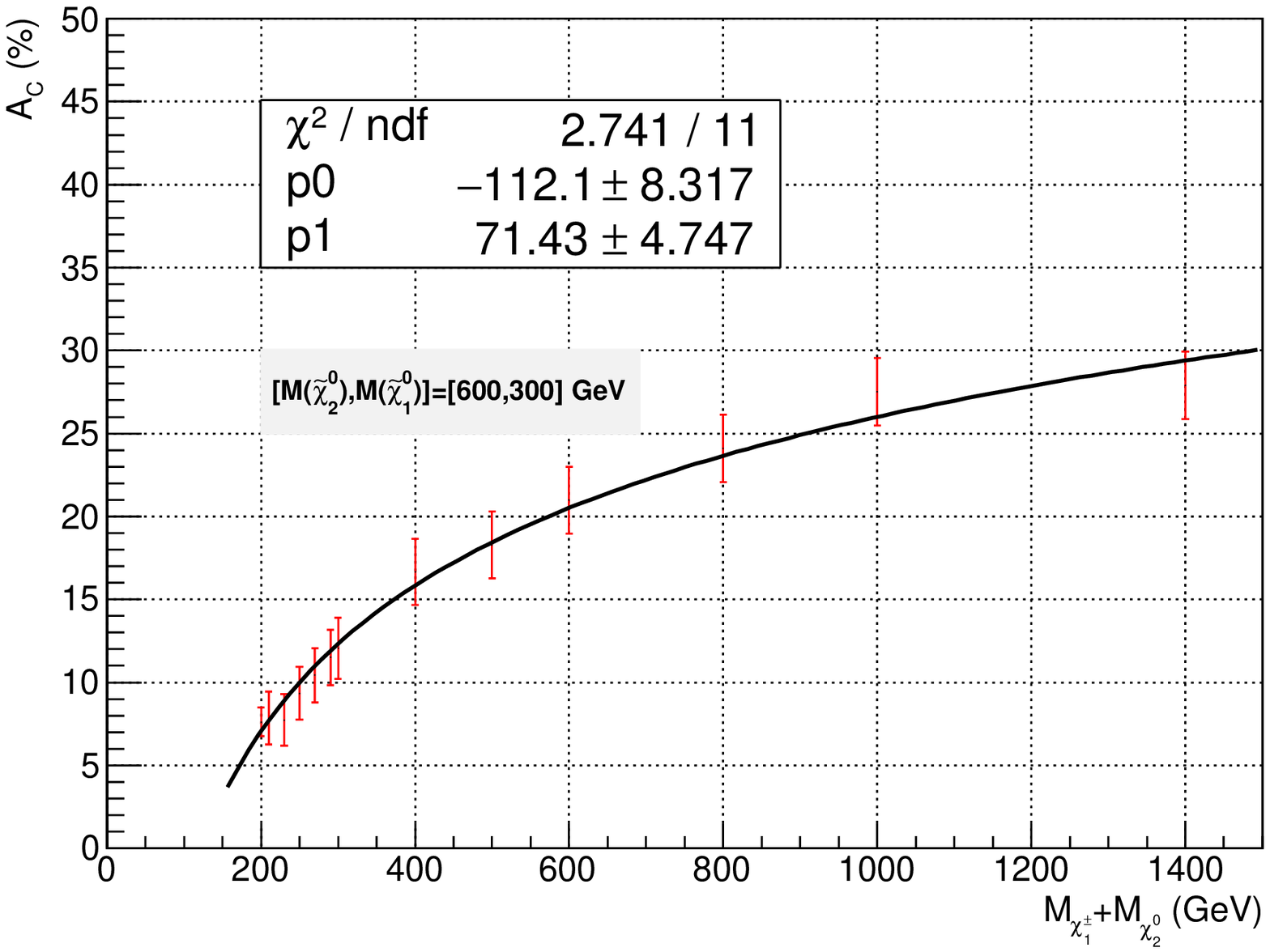}
\includegraphics[scale=0.225]{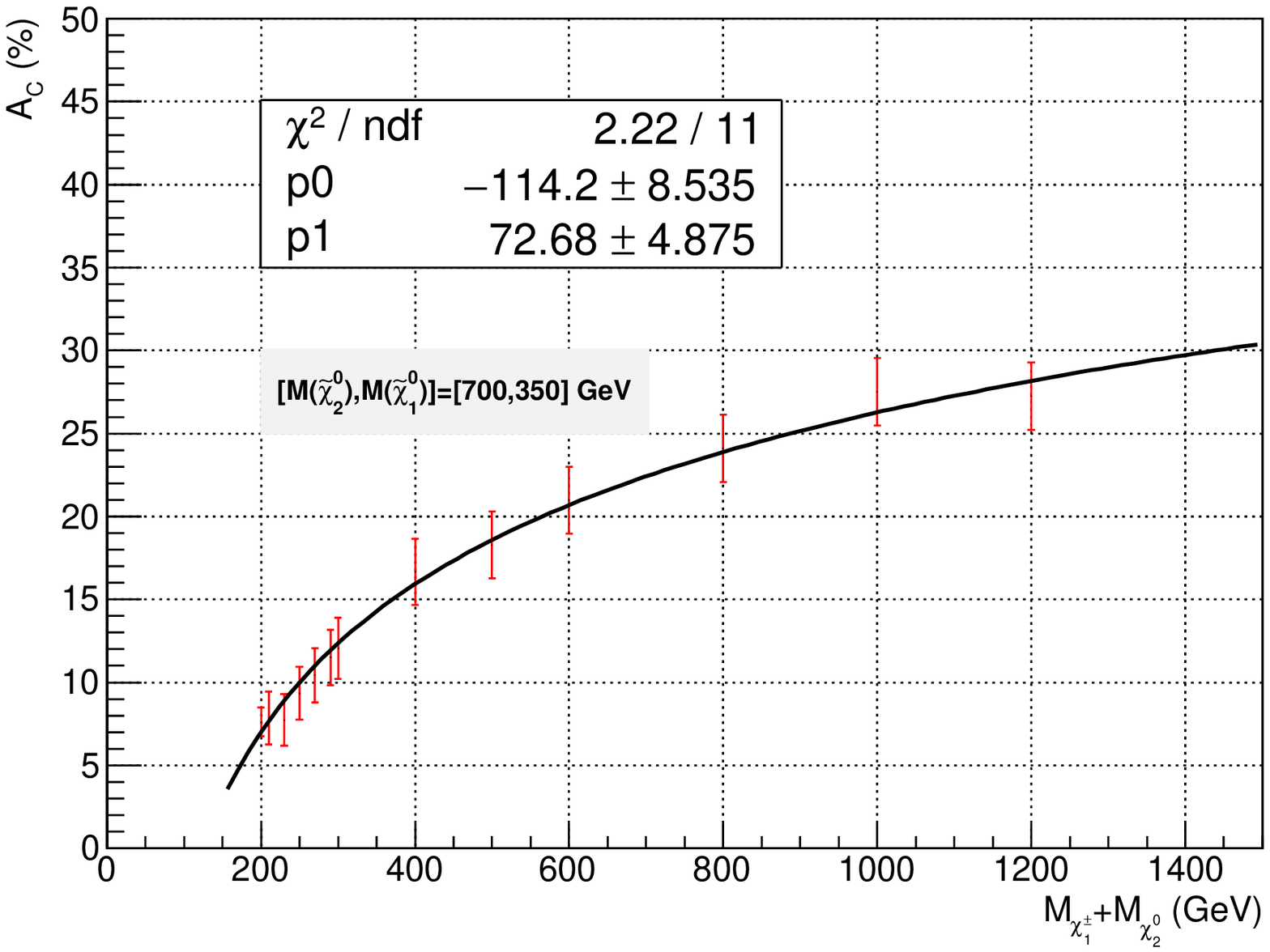}
\end{center}
\caption{\label{sec:Part4:GCATC_S2_Reco} Experimental $A_{C}$ template curves for the S2 signal samples, as they are listed, in table \ref{trilepton_SEL_tab} from the top to the bottom rows. Here, they appear ordered by increasing $\tilde\chi^{0}_{2}$ mass, from the top to the bottom row and from left to right.}
\end{figure}

\begin{figure}[h]
\begin{center}
\includegraphics[scale=0.35]{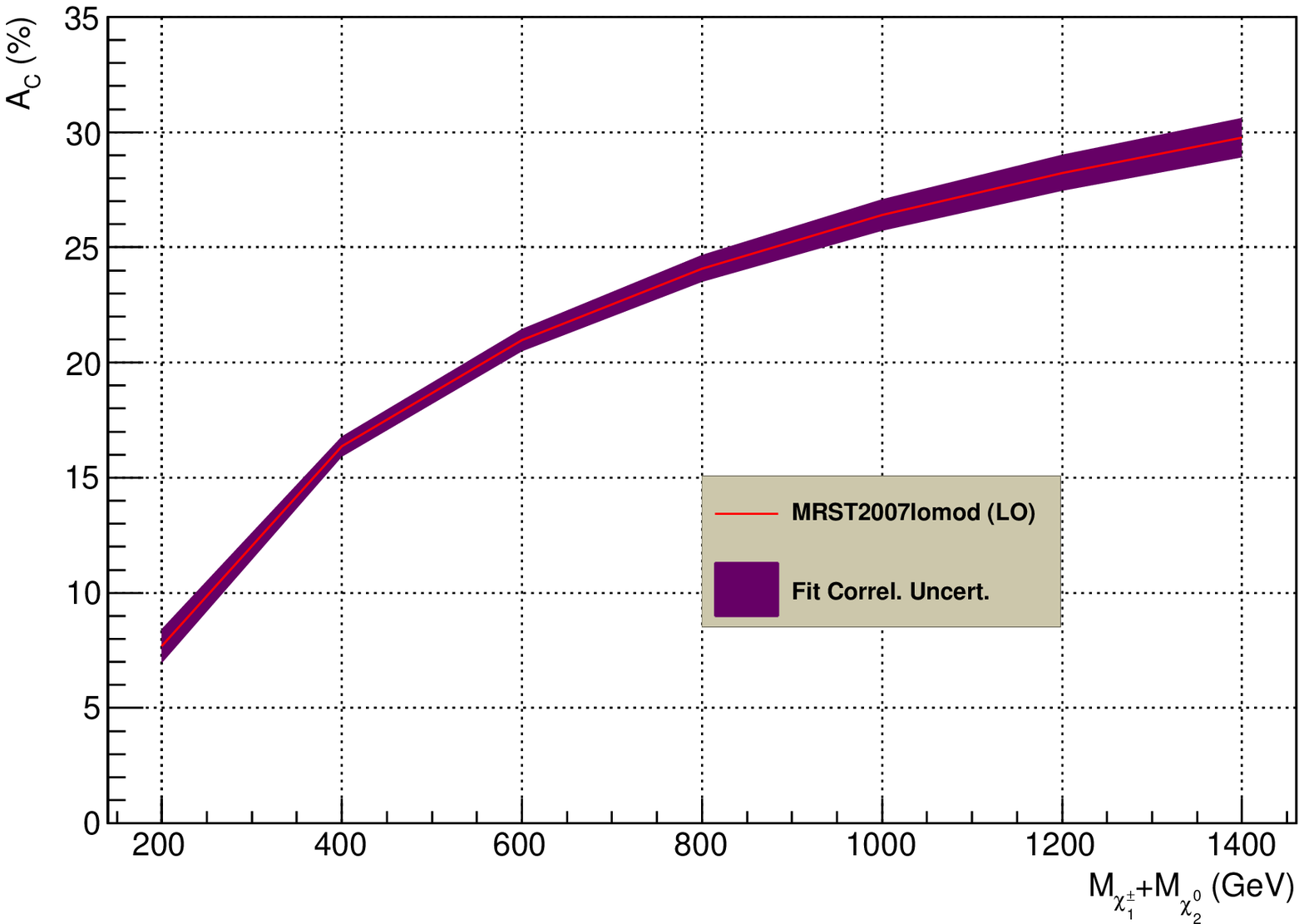}
\includegraphics[scale=0.35]{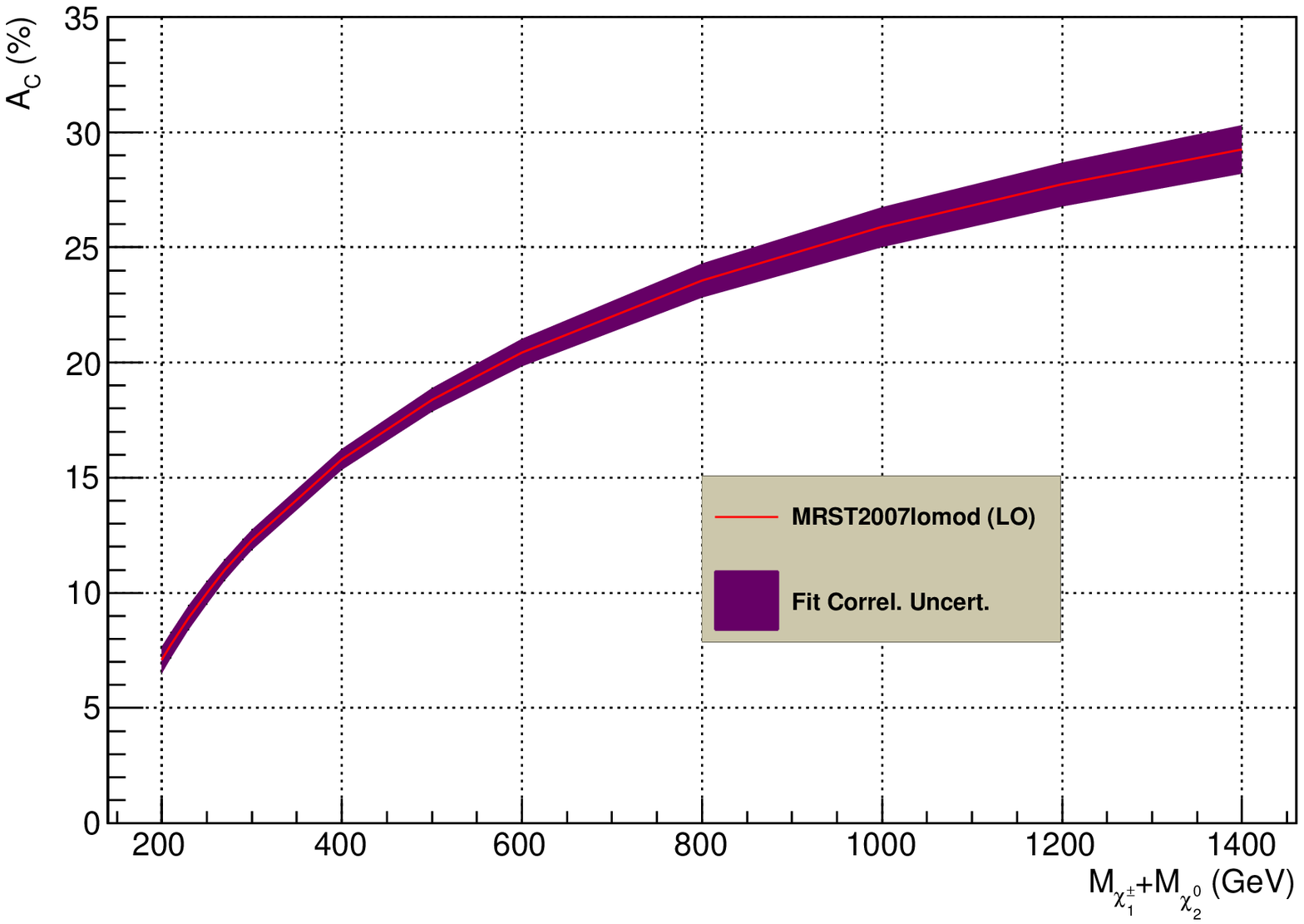}
\end{center}
\caption{\label{sec:Part4:GCATC_S1S2_Reco_Fit} Fitted $A_{C}$ template curves for the S1 (LHS) and the S2 (RHS) signal samples. The uncertainty accounts
for the correlations between the parameters used to fit the $A_{C}^{Meas}$ template curves.}
\end{figure}

\noindent
In this simple version of the analysis, we keep the same event selection for both teh S1 and the S2 signals. Therefore these signals samples share the same residual background
as well as the same bias from the event selection. In these conditions, we could use a common $A_{C}$ template curve for both of them. However, because we choose many overlapping masses between these two signal samples, we split them into two seperate sets of experimental $A_{C}$ template curves. The S1 $A_{C}$ template curve, that include the propagation of the realistic experimental uncertainties into each term of equation \ref{Bkgd_Subtract}, are displayed in figure \ref{sec:Part4:GCATC_S1_Reco}, the S2 ones are displayed in figure \ref{sec:Part4:GCATC_S2_Reco}. And the final signal template curves for which the uncertainties account
for the correlations between the parameters used to fit the $A_{C}^{Meas}$ template curves are shown in figure \ref{sec:Part4:GCATC_S1S2_Reco_Fit}, on the LHS for S1 and on the RHS for S2.

\newpage
\subsection{Indirect Determination of $M_{\tilde\chi^{\pm}_{1}}+M_{\tilde\chi^{0}_{2}}$}
\label{sec:next-mass-constraint}
\vspace*{1.5mm}
\subsubsection{Experimental Result for the S1 Signal}
\vspace*{0.5mm}

\noindent
Using the S1 signal experimental $A_{C}$ template curves of figure \ref{sec:Part4:GCATC_S1_Reco}, we can get the central values and the uncertainties of the indirectly measured $M_{\tilde\chi^{\pm}_{1}}+M_{\tilde\chi^{0}_{2}}$ for each input mass as reported in table \ref{sec:Part4:Final_Results_S1}. 

\begin{table}[h]
\begin{center}
\begin{tabular}{|c|c|c|}
\hline\hline 
$M_{\tilde\chi^{\pm}_{1}}+M_{\tilde\chi^{0}_{2}}$	 & $A_{C}^{Meas.}\pm\delta A_{C}^{Meas.Fit}$ & $M_{\tilde\chi^{\pm}_{1}}+M_{\tilde\chi^{0}_{2}}$	\\ 
Input Mass (GeV)                                                          & ($\%$)             	                                                  &          Measured Mass (GeV) \\
\hline\hline
200.	 &  $7.70\pm 0.74$    &  $200.37^{+11.51}_{-10.78}$ \\
\hline
400.     &  $16.06\pm 0.44$  &  $390.18^{+14.83}_{-14.21}$ \\
\hline
600.     &  $21.30\pm 0.48$  &  $617.94^{+27.70}_{-26.34}$ \\
\hline
800.    &  $24.40\pm 0.58$  &   $824.61^{+46.98}_{-44.09}$ \\
\hline
1000.  &  $27.21\pm 0.69$  &   $1083.15^{+76.95}_{-71.18}$ \\
\hline
1200.  & $27.20\pm 0.77$  &    $1082.08^{+86.18}_{-78.99}$ \\
\hline
1400.  & $29.06\pm 0.85$   &   $1304.01^{+118.38}_{-107.31}$ \\
\hline\hline
\end{tabular}       
\end{center}
\caption{\label{sec:Part4:Final_Results_S1} Measured $A_{C}(S)$ of the S1 signal samples with their full experimental uncertainty. Indirect mass measurement and their full experimental uncertainty as a function of the signal sample.}
\end{table}

\begin{figure}[hbp]
\begin{center}
\includegraphics[scale=0.5]{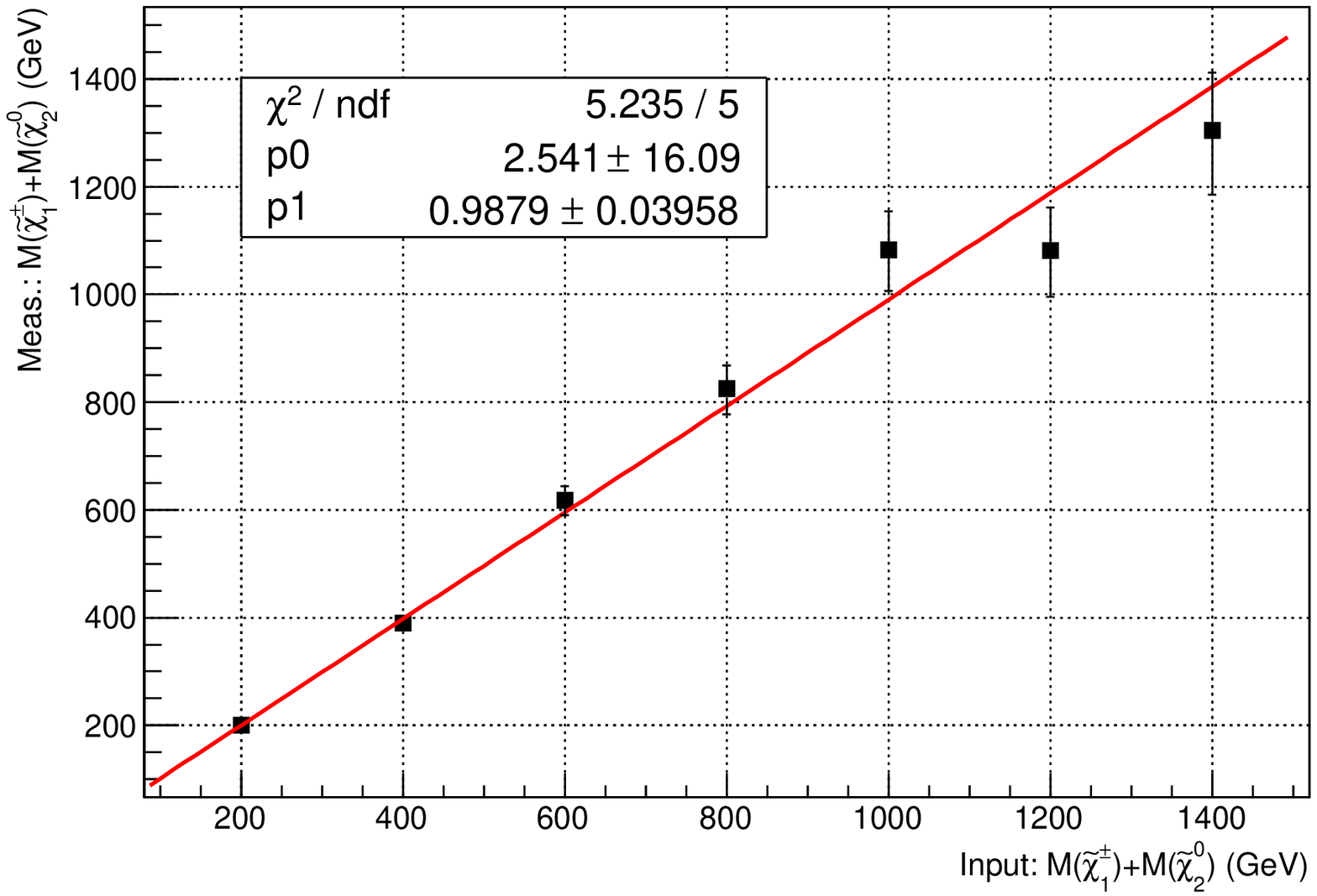}
\includegraphics[scale=0.5]{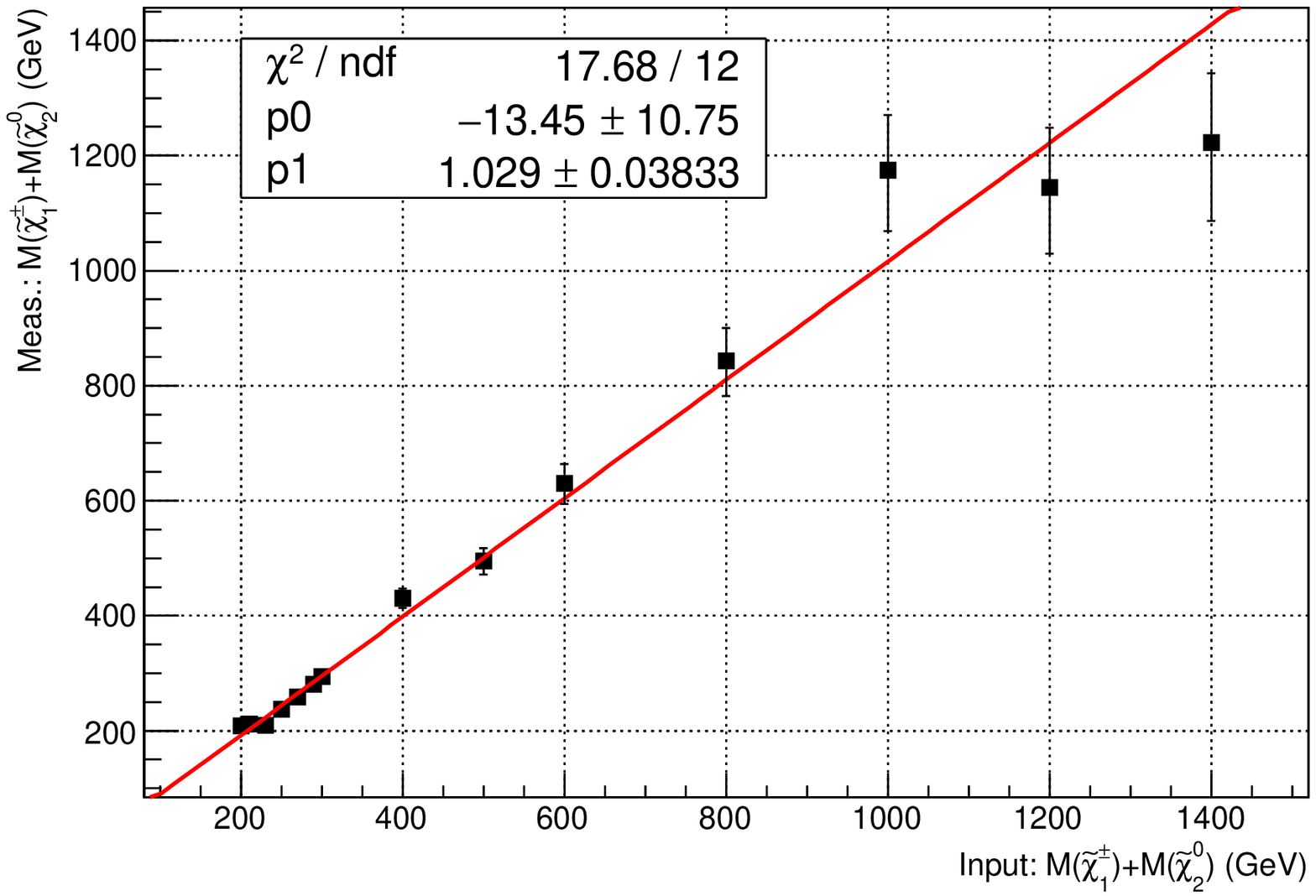}
\caption{\label{sec:Part4:Closure_Expt} Closure test of the indirect measurement of $M_{\tilde\chi^{\pm}_{1}}+M_{\tilde\chi^{0}_{2}}$ for the S1 (top) and S2 (bottom) signal samples with only experimental uncertainties.}
\end{center}
\end{figure}

\noindent
This enables us to perform a closure test of our method on the S1 signal sample as displayed at the top of figure \ref{sec:Part4:Closure_Expt}, where we can fit of the input versus the measured  $M_{\tilde\chi^{\pm}_{1}}+M_{\tilde\chi^{0}_{2}}$ by a linear function.

\noindent
This fit indicates, given the uncertainties, that the indirect measurement is:
\begin{equation}
\begin{cases} 
\rm linear:\ the\ slope\ of\ the\ fit\ function\ is\ compatible\ with\ 1\\
\rm unbiased:\ the\ y-intercept\ of\ the\ fit\ function\ is\ compatible\ with\ 0
\end{cases} 
\end{equation}

\noindent
Further elementary checks, forcing the parameters of the fit functions, tend to confirm these  indications, as presented in table \ref{sec:Part4:Forced_Fit_S1}.

\begin{table}[h]
\begin{center}
\begin{tabular}{|c|c|c|c|}
\hline\hline 
Forced Parameter   &  Fit                       & Fit                           &  Fit     \\ 
                                  & $\chi^{2}/Ndof$  & Y-Intercept            &  Slope \\
\hline\hline
Slope                         &   5.328/6             & $-1.67\pm 8.26$ & $1.0\pm 0.0$ \\
Y-Intercept              &   5.260/6             & $0.0\pm 0.0$          & $0.9933\pm 0.0203$ \\
\hline\hline
\end{tabular}       
\end{center}
\caption{\label{sec:Part4:Forced_Fit_S1}
Closure tests with a forced fit parameter for the S1 signal samples.}
\end{table}

%%%%%%%%%%%%%%%%%%%%%%%%%%%%%%%%%%%%%%%%%%%%%%%%%%%%

\vspace*{1.5mm}
\subsubsection{Experimental Result for the S2 Signal}
\vspace*{0.5mm}

\noindent
As in the previous sub-section, using the S2 signal $A_{C}$ template curves \ref{sec:Part4:GCATC_S2_Reco}, we can get the results reported in table \ref{sec:Part4:Final_Results_S2}. The closure test on the S2 signal samples is displayed at the bottom of figure \ref{sec:Part4:Closure_Expt}.

\vfill\eject\clearpage

\begin{table}[h]
\begin{center}
\begin{tabular}{|c|c|c|}
\hline\hline 
$M_{\tilde\chi^{\pm}_{1}}+M_{\tilde\chi^{0}_{2}}$	 & $A_{C}^{Meas.}\pm\delta A_{C}^{Meas.Fit}$ & $M_{\tilde\chi^{\pm}_{1}}+M_{\tilde\chi^{0}_{2}}$	\\ 
Input Mass (GeV)                                                          & ($\%$)             	                                                  &          Measured Mass (GeV) \\
\hline\hline
200.    &    $  7.62\pm  0.59$     &   $  208.34^{+9.51}_{-9.01}$ \\\hline
210.    &    $  7.85\pm  0.56$     &   $  211.99^{+9.20}_{-8.75}$ \\\hline
230.    &    $  7.73\pm  0.52$     &   $  210.08^{+8.43}_{-8.05}$ \\\hline
250.    &    $  9.34\pm  0.49$     &   $  237.72^{+9.01}_{-8.97}$ \\\hline
270.    &    $10.43\pm  0.46$     &   $  258.55^{+9.52}_{-9.13}$ \\\hline
290.    &    $11.50\pm  0.45$     &   $  281.34^{+10.29}_{-9.86}$ \\\hline
300.    &    $12.06\pm  0.44$     &   $  294.21^{+10.60}_{-10.17}$ \\\hline
400.    &    $16.66\pm  0.46$     &   $  430.69^{+17.35}_{-16.57}$ \\\hline
500.    &    $18.28\pm  0.52$     &   $  495.51^{+23.17}_{-21.97}$ \\\hline
600.    &    $20.98\pm 0.60$      &   $  630.50^{+35.51}_{-33.34}$ \\\hline
800.    &    $24.11\pm 0.74$      &   $  843.48^{+61.79}_{-57.00}$ \\\hline
1000.  &    $27.51\pm 0.86$      &   $1174.45^{+105.82}_{-95.96}$ \\\hline
1200.  &    $27.25\pm 0.96$      &   $1144.45^{+115.34}_{-103.44}$ \\\hline
1400.  &    $27.91\pm 1.04$      &   $1222.38^{+135.40}_{-120.22}$ \\\hline
\hline\hline
\end{tabular}       
\end{center}
\caption{\label{sec:Part4:Final_Results_S2} Measured $A_{C}(S)$ of the S2 signal samples with their full experimental uncertainty. Indirect mass measurement and their full experimental uncertainty as a function of the signal sample.}
\end{table}

\noindent
Here again the fit indicates, within the uncertainties, that the indirect mass measurement is linear and unbiased.
\noindent
The checks, forcing the parameters of the fit functions, tend to confirm these  indications,
as presented in table \ref{sec:Part4:Forced_Fit_S2}.

\begin{table}[h]
\begin{center}
\begin{tabular}{|c|c|c|c|}
\hline\hline 
Forced Parameter   &  Fit                       & Fit                            &  Fit     \\ 
                                  & $\chi^{2}/Ndof$  & Y-Intercept            &  Slope \\
\hline\hline
Slope                         &    18.27/13            & $-5.601\pm 3.349$ & $1.0\pm 0.0$ \\
Y-Intercept              &   19.25/13             & $0.0\pm 0.0$           & $0.9838\pm 0.0120$ \\
\hline\hline
\end{tabular}       
\end{center}
\caption{\label{sec:Part4:Forced_Fit_S2} Closure tests with a forced fit parameter for the S2 signal samples.}
\end{table}

\newpage
\subsection{Final Result for MRST2007lomod}
\label{sec:Part2-MRST2007}
\vspace*{1.5mm}
\subsubsection{Final Result for the S1 Signal}
\vspace*{0.5mm}

\begin{table}[h]
\begin{center}
\begin{tabular}{|c|c|c|c|}
\hline\hline 
Meas. $M_{\tilde\chi^{\pm}_{1}}+M_{\tilde\chi^{0}_{2}}$ & Expt. Uncert.     &   Theor. Uncert.  & Total Uncert.  \\
                                                     (GeV)                                         &      (GeV)             &        (GeV)           &       (GeV)         \\
\hline\hline
200.37 & $^{+11.51}_{-10.78}$ & $^{+0.90}_{-0.90}$ & $^{+11.55}_{-10.82}$\\
\hline
390.18 & $^{+14.83}_{-14.21}$ & $^{+1.07}_{-1.12}$ & $^{+14.87}_{-14.25}$\\
\hline
617.94 & $^{+27.70}_{-26.34}$ & $^{+2.15}_{-2.24}$ & $^{+27.78}_{-26.44}$\\
\hline
824.61 & $^{+46.98}_{-44.09}$ & $^{+2.69}_{-2.70}$ & $^{+47.06}_{-44.17}$\\
\hline
1083.15 & $^{+76.95}_{-71.18}$ & $^{+2.13}_{-2.24}$ & $^{+76.98}_{-71.22}$\\
\hline
1082.08 & $^{+86.18}_{-78.99}$ & $^{+2.16}_{-2.24}$ & $^{+86.21}_{-79.02}$\\
\hline
1304.01 & $^{+118.38}_{-107.31}$ & $^{+5.76}_{-5.38}$ & $^{+118.52}_{-107.44}$\\
\hline\hline
\end{tabular}       
\end{center}
\caption{\label{sec:Part4:Final_Results_S1_Full} Final results for the S1 samples with experimental and theoretical uncertainties.}
\end{table}

\begin{figure}[h]
\begin{center}
\includegraphics[scale=0.75]{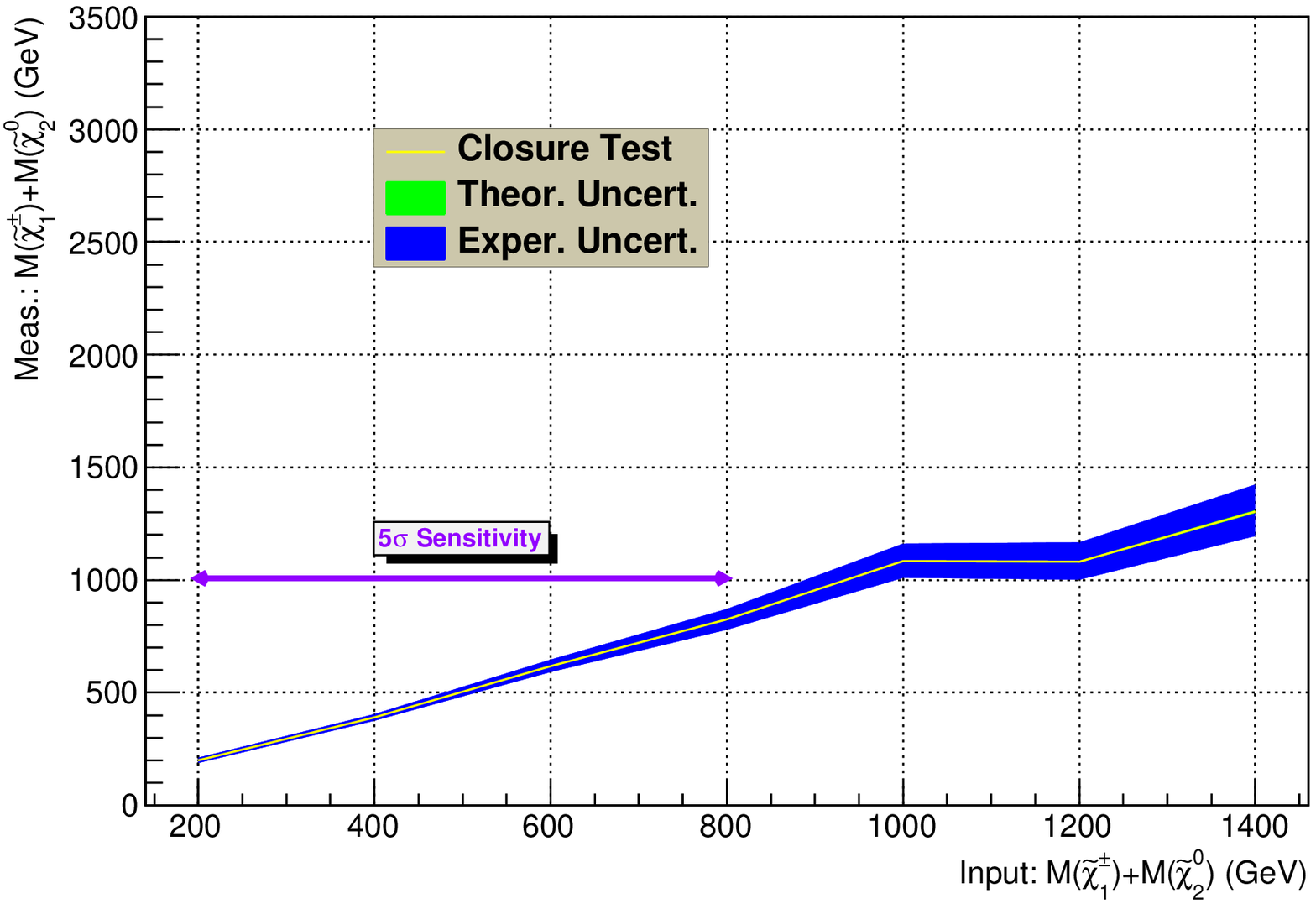}
\end{center}
\caption{\label{sec:Part4:Closure_S1} Closure test of the indirect measurement of $M_{\tilde\chi^{\pm}_{1}}+M_{\tilde\chi^{0}_{2}}$ for the S1 signal samples with both theoretical and experimental uncertainties. The sub-range with a signal sensitivity of $5\sigma$ is highlighted.}
\end{figure}

\noindent
For the S1 sub-samples with a signal significance in excess of $5\sigma$, the indirect measurements of $M_{\tilde\chi^{\pm}_{1}}+M_{\tilde\chi^{0}_{2}}$
are performed with an overall accuracy better than $6\%$ for input masses $M_{\tilde\chi^{0}_{2}}=M_{\tilde\chi^{\pm}_{1}}$ in the [100,300] GeV interval, and better than $10\%$ for $M_{\tilde\chi^{0}_{2}}=M_{\tilde\chi^{\pm}_{1}}\geq 400$ GeV. This is reported in table \ref{sec:Part4:Final_Results_S1_Full}
and displayed in figure \ref{sec:Part4:Closure_S1}.

%%%%%%%%%%%%%%%%%%%%%%%%%%%%%%%%%%%%%%%%%%%%%%%%%%%%

\vspace*{1.5mm}
\subsubsection{Final Result for the S2 Signal}
\vspace*{0.5mm}

\begin{table}[h]
\begin{center}
\begin{tabular}{|c|c|c|c|}
\hline\hline 
Meas. $M_{\tilde\chi^{\pm}_{1}}+M_{\tilde\chi^{0}_{2}}$ & Expt. Uncert.     &   Theor. Uncert.  & Total Uncert.  \\
                                                     (GeV)                                         &      (GeV)             &        (GeV)           &       (GeV)         \\
\hline\hline
208.34 & $^{+9.51}_{-9.01}$ & $^{+0.70}_{-0.76}$ & $^{+9.54}_{-9.04}$\\
\hline
211.99 & $^{+9.20}_{-8.75}$ & $^{+0.66}_{-0.69}$ & $^{+9.22}_{-8.78}$\\
\hline
210.08 & $^{+8.43}_{-8.05}$ & $^{+0.55}_{-0.76}$ & $^{+8.45}_{-8.09}$\\
\hline
237.72 & $^{+9.01}_{-8.97}$ & $^{+0.61}_{-0.64}$ & $^{+9.03}_{-8.99}$\\
\hline
258.55 & $^{+9.52}_{-9.13}$ & $^{+0.65}_{-0.76}$ & $^{+9.54}_{-9.16}$\\
\hline
281.34 & $^{+10.29}_{-9.86}$ & $^{+0.77}_{-0.86}$ & $^{+10.32}_{-9.90}$\\
\hline
294.21 & $^{+10.60}_{-10.17}$ & $^{+0.86}_{-0.87}$ & $^{+10.63}_{-10.21}$\\
\hline
430.69 & $^{+17.35}_{-16.57}$ & $^{+1.34}_{-1.44}$ & $^{+17.40}_{-16.63}$\\
\hline
495.51 & $^{+23.17}_{-21.97}$ & $^{+1.37}_{-1.46}$ & $^{+23.21}_{-22.02}$\\
\hline
630.50 & $^{+35.51}_{-33.34}$ & $^{+2.12}_{-2.24}$ & $^{+35.57}_{-33.42}$\\
\hline
843.48 & $^{+61.79}_{-57.00}$ & $^{+2.57}_{-2.74}$ & $^{+61.84}_{-57.07}$\\
\hline
1174.45 & $^{+105.82}_{-95.96}$ & $^{+2.44}_{-2.47}$ & $^{+105.85}_{-95.99}$\\
\hline
1144.45 & $^{+115.34}_{-103.44}$ & $^{+2.40}_{-2.53}$ & $^{+115.36}_{-103.47}$\\
\hline
1222.38 & $^{+135.40}_{-120.22}$ & $^{+3.38}_{-3.34}$ & $^{+135.44}_{-120.27}$\\
\hline\hline
\end{tabular}       
\end{center}
\caption{\label{sec:Part4:Final_Results_S2_Full} Final results for the S2 samples with experimental and theoretical uncertainties.}
\end{table}

\begin{figure}[h]
\begin{center}
\includegraphics[scale=0.75]{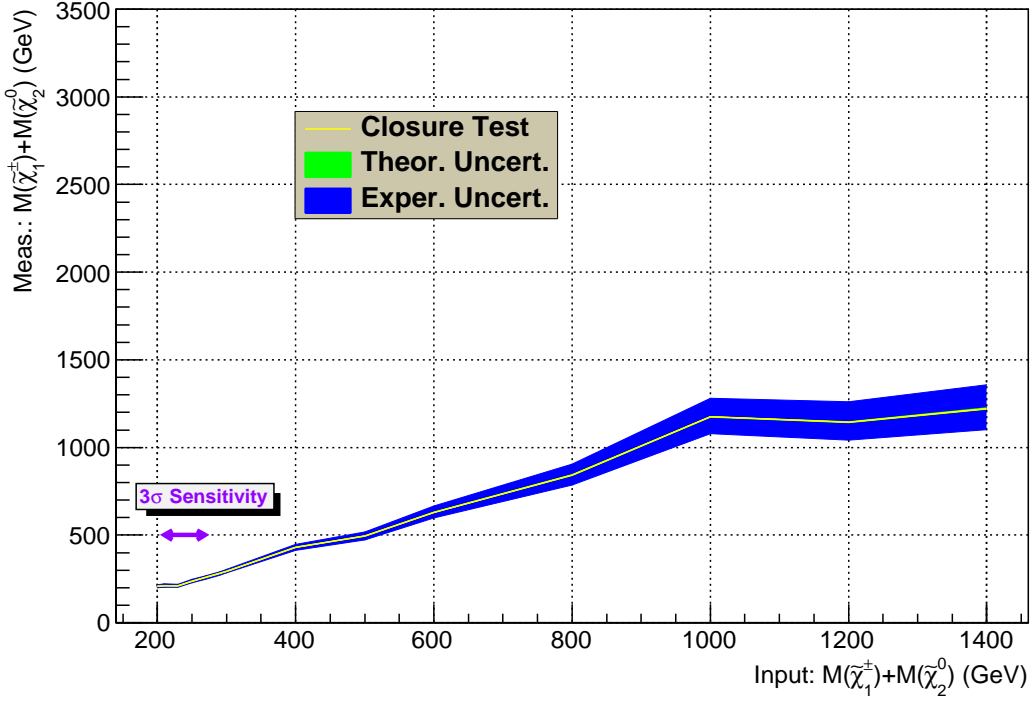}
\end{center}
\caption{\label{sec:Part4:Closure_S2} Closure test of the indirect measurement of $M_{\tilde\chi^{\pm}_{1}}+M_{\tilde\chi^{0}_{2}}$ for the S2 signal samples with both theoretical and experimental uncertainties. The sub-range with a signal sensitivity of $3\sigma$ is highlighted.}
\end{figure}

\noindent
For the S2 sub-samples with a signal significance in excess of $3\sigma$, the indirect measurements of $M_{\tilde\chi^{\pm}_{1}}+M_{\tilde\chi^{0}_{2}}$
are performed with an overall accuracy better than $4.5\%$ for respective input masses $M_{\tilde\chi^{0}_{2}}=M_{\tilde\chi^{\pm}_{1}}$ in the [105,145] GeV interval
and better than $11.1\%$ for considered masses outside this interval. This is reported in table \ref{sec:Part4:Final_Results_S2_Full}
and displayed in figure \ref{sec:Part4:Closure_S2}.

\subsection{Summary of the $M_{\tilde\chi^{\pm}_{1}}+M_{\tilde\chi^{0}_{2}}$ Measurements and their Accuracy}
\label{sec:Part2-PDF-Info}
\noindent
We sum up the indirect mass measurements of $M_{\tilde\chi^{\pm}_{1}}+M_{\tilde\chi^{0}_{2}}$ extracted from the integral charge asymmetry 
of the $\tilde\chi^{\pm}_{1}+\tilde\chi^{0}_{2}\to 3\ell^{\pm}+\rlap{\kern0.25em/}E_{T}$ inclusive process within tables \ref{w1z2-Summary:Tab_S1} (S1 signal) and \ref{w1z2-Summary:Tab_S2} (S2 signal). 

\begin{table}[h]
\begin{center}
\begin{tabular}{|c|c|c|c|}
\hline\hline
S1 Signal                                                                                            & \multicolumn{3}{c|}{Figures of Merit} \\ 
Input $M_{\tilde\chi^{\pm}_{1}}+M_{\tilde\chi^{0}_{2}}$    & 1.  & 2. & 3. \\
  (GeV) &     $\frac{\delta M^{Fit}_{\tiny\tilde\chi^{\pm}_{1}\tilde\chi^{0}_{2}}}{M^{Fit}_{\tiny\tilde\chi^{\pm}_{1}\tilde\chi^{0}_{2}}} $      &   $\frac{(M^{Fit}_{\tiny\tilde\chi^{\pm}_{1}\tilde\chi^{0}_{2}}-M^{True}_{\tiny\tilde\chi^{\pm}_{1}\tilde\chi^{0}_{2}})}{M^{True}_{\tiny\tilde\chi^{\pm}_{1}\tilde\chi^{0}_{2}}}$ &    $\frac{(M^{Fit}_{\tiny\tilde\chi^{\pm}_{1}\tilde\chi^{0}_{2}}-M^{True}_{\tiny\tilde\chi^{\pm}_{1}\tilde\chi^{0}_{2}})}{\delta M^{Fit}_{\tiny\tilde\chi^{\pm}_{1}\tilde\chi^{0}_{2}}}$ \\
\hline
200.    &   $5.8\%$    &  $+0.2\%$   &  $+0.03\sigma$  \\
400.    &   $3.8\%$    &  $-2.5\%$   &  $-0.7\sigma$  \\
600.    &   $4.5\%$    &  $+3.0\%$   &  $+0.7\sigma$  \\
800.    &   $5.7\%$    &  $+3.1\%$   &  $+0.5\sigma$  \\
1000.  &   $7.1\%$    &  $+8.3\%$   &  $+1.1\sigma$  \\
1200.  &   $8.0\%$    &  $-9.8\%$   &  $-1.4\sigma$  \\
1400.  &   $9.1\%$    &  $-6.9\%$   &  $-0.8\sigma$  \\
\hline\hline
\end{tabular}       
\end{center}
\caption{\label{w1z2-Summary:Tab_S1} Summary of the indirect mass measurements of $M_{\tilde\chi^{\pm}_{1}}+M_{\tilde\chi^{0}_{2}}$ extracted from the integral charge asymmetry of the S1 signal samples. Different figures of merit of the accuracy of these measurements are presented.}
\end{table}

\begin{table}[h]
\begin{center}
\begin{tabular}{|c|c|c|c|}
\hline\hline
S2 Signal                                                                                            & \multicolumn{3}{c|}{Figures of Merit} \\ 
Input $M_{\tilde\chi^{\pm}_{1}}+M_{\tilde\chi^{0}_{2}}$   & 1.  & 2. & 3. \\

 (GeV) &     $\frac{\delta M^{Fit}_{\tiny\tilde\chi^{\pm}_{1}\tilde\chi^{0}_{2}}}{M^{Fit}_{\tiny\tilde\chi^{\pm}_{1}\tilde\chi^{0}_{2}}} $      &   $\frac{(M^{Fit}_{\tiny\tilde\chi^{\pm}_{1}\tilde\chi^{0}_{2}}-M^{True}_{\tiny\tilde\chi^{\pm}_{1}\tilde\chi^{0}_{2}})}{M^{True}_{\tiny\tilde\chi^{\pm}_{1}\tilde\chi^{0}_{2}}}$ &    $\frac{(M^{Fit}_{\tiny\tilde\chi^{\pm}_{1}\tilde\chi^{0}_{2}}-M^{True}_{\tiny\tilde\chi^{\pm}_{1}\tilde\chi^{0}_{2}})}{\delta M^{Fit}_{\tiny\tilde\chi^{\pm}_{1}\tilde\chi^{0}_{2}}}$ \\
\hline
200.    &   $4.6\%$    &  $+4.2\%$   &  $+0.9\sigma$  \\
210.    &   $4.4\%$    &  $+1.0\%$   &  $+0.2\sigma$  \\
230.    &   $4.0\%$    &  $-8.7\%$   &  $-2.4\sigma$  \\
250.    &   $3.8\%$    &  $-4.9\%$   &  $-1.4\sigma$  \\
270.    &   $3.7\%$    &  $-4.2\%$   &  $-1.2\sigma$  \\
290.    &   $3.7\%$    &  $-3.0\%$   &  $-0.8\sigma$  \\
300.    &   $3.6\%$    &  $-1.9\%$   &  $-0.5\sigma$  \\  
400.    &   $4.0\%$    &  $+7.7\%$  &  $+1.8\sigma$  \\   
500.    &   $4.7\%$    &  $-0.9\%$   &  $-0.2\sigma$  \\   
600.    &   $5.6\%$    &  $+5.1\%$  &  $+0.9\sigma$  \\   
800.    &   $7.3\%$    &  $+5.4\%$  &  $+0.7\sigma$  \\  
1000.  &   $9.0\%$    &  $+17.5\%$ &  $+1.7\sigma$  \\  
1200.  &   $10.1\%$  &  $-4.6\%$   &  $-0.5\sigma$  \\   
1400.  &   $11.1\%$  &  $-12.7\%$  &  $-1.3\sigma$  \\   
\hline\hline
\end{tabular}       
\end{center}
\caption{\label{w1z2-Summary:Tab_S2} Summary of the indirect mass measurements of $M_{\tilde\chi^{\pm}_{1}}+M_{\tilde\chi^{0}_{2}}$ extracted from the integral charge asymmetry of the S2 signal samples. Different figures of merit of the accuracy of these measurements are presented.}
\end{table}

\noindent
For the S1 signal at LO, this new method enables to get an accuracy better than $6\%$ for the range with $5\sigma$ sensitivity to the signal and better than $10\%$ elsewhere. Whereas for the S2 signal at LO, we get an accuracy better than $4.5\%$ for the range with $3\sigma$ sensitivity to the signal and better than $11.2\%$ elsewhere. All these indirect measurements are statistically compatible with the total uncertainty of the method.
\par\noindent
One should bear in mind however that these results do not account for the dominant theoretical uncertainty ($\delta (A_{C})_{PDF}$).

\clearpage\newpage
\subsection{Comparison with Other Mass Measurement Methods}
\label{sec:Part2-Comp}
\subsubsection{Dilepton Mass Edge}
\vspace*{1.5mm}

In this sub-section, we'll compare the {\it ICA} (Integral Charge Asymmetry) indirect mass measurement technique with two other direct mass measurement techniques.
\par\noindent
But before entering this topic, let us mention the issue of the combinatorics within the trilepton search topology we've chosen. For our signal,
resolving this combinatorics consists in matching the correct dilepton to its parent $\tilde\chi^{0}_{2}$ whilst associating the third lepton
to its parent $\tilde\chi^{\pm}_{1}$. The $\tilde\chi^{0}_{2}$ leptonic decay yields  two leptons with opposite-signs (OS) and same flavours (SF).
In events with mixed flavours ($e^{+}e^{-}\mu^{\pm}$ or $\mu^{+}\mu^{-}e^{\pm}$), the correct assignment is obvious: the dilepton of SF comes
from the $\tilde\chi^{0}_{2}$ and the single lepton with the other flavour comes from the $\tilde\chi^{\pm}_{1}$. However in order to exploit the full
signal statistics, one also needs to resolve this combinatorics in tri-electron and tri-muon events. For each of these event topology involving a single flavour, 
there are always two combinations of OS dileptons and one combination of same-sign (SS) dilepton. Therefore we adopt a statistical solution to lift the combinatorics.
In the calculation of any physical observable, for each $3e^{\pm}$ or $3\mu^{\pm}$ event, we  fill the corresponding histogram with two entries from the two OS dileptons with a weight of +1 and with one entry from the single SS  dilepton with a weight of -1. This systematically subtracts from the observable histogram the wrong combination which associates a lepton from the  $\tilde\chi^{\pm}_{1}$ decay with one of the $\tilde\chi^{0}_{2}$ decay.

\vspace*{0.5cm}
\par\noindent
3.6.1.\ a.\ Experimental\ Observable
\vspace*{0.1cm}
\par\noindent
The fact that the OS-SF dilepton coming from the second neutralino decay has an edge in its invariant mass was noted 
long ago in \cite{Baer:1994nr}. It has been used extensively in the litterature \cite{Muanza:1996fu}\cite{Muanza:1996rk}
\cite{Muanza:1996gha}\cite{Bachacou:1999zb}, including in a few reviews like \cite{Barr:2010zj} and in references therein. 

\par\noindent
For the S1 signal, we have the following mass hierarchy 
$M_{\tilde\chi^{0}_{2}}=M_{\tilde\chi^{\pm}_{1}} > M_{\tilde\ell^{\pm}} > M_{\tilde\chi^{0}_{1}}$ and
we consider $\tilde\chi^{0}_{2}$ and $\tilde\chi^{\pm}_{1}$ two-body decays proceeding through an intermediate slepton.
In this case, the edge is given by:
\begin{equation}
M^{Max}_{\ell^{\pm}\ell^{\mp}} = M_{\tilde\chi^{0}_{2}}
\times\sqrt{ \left ( 1-\frac{M^{2}_{\tilde\ell^{\pm}}}{M^{2}_{\tilde\chi^{0}_{2}}} \right )
                      \left ( 1-\frac{M^{2}_{\tilde\chi^{0}_{1}}}{M^{2}_{\tilde\ell^{\pm}}} \right ) }
\end{equation}

\par\noindent
For the S2 signal, we have the following mass hierarchy 
$M_{\tilde\chi^{0}_{2}}=M_{\tilde\chi^{\pm}_{1}} > M_{\tilde\chi^{0}_{1}}$ and
we consider $\tilde\chi^{0}_{2}$ and $\tilde\chi^{\pm}_{1}$ decays proceeding through $W^{\pm}$ and $Z$ bosons. 
In these cases, the edge is given by:
\begin{equation}
M^{Max}_{\ell^{\pm}\ell^{\mp}} = (M_{\tilde\chi^{0}_{2}} - M_{\tilde\chi^{0}_{1}}) < M_{Z}
\end{equation}
for a $\tilde\chi^{0}_{2}$ three-body decay proceeding through  an off-shell $Z^{*}$ (S2a), and by
\begin{equation}
M^{Max}_{\ell^{\pm}\ell^{\mp}} = (M_{\tilde\chi^{0}_{2}} - M_{\tilde\chi^{0}_{1}})\geq M_{Z}
\end{equation}
for a $\tilde\chi^{0}_{2}$ two-body decay proceeding through  an on-shell $Z$ (S2b).

\par\noindent
In light of these formulae, we see that the mass reconstruction capabilities of this method that we'll call {\it DileME}, for "Dilepton Mass Edge",
regard exclusively the reconsctruction of mass differences.
\par\noindent
The main systematic uncertainties of the {\it DileME}  method come from the lepton energy scales. These are known to a $0.05\%$ accuracy in the ATLAS
experiment at the LHC Run1, both for the electrons \cite{Aad:2014nim} and the muons \cite{Aad:2014rra}. Since the dilepton invariant mass is:
\begin{equation}
M^{2}_{\ell^{\pm}_{1}\ell^{\mp}_{2}} = 2 E_{\ell^{\pm}_{1}} E_{\ell^{\mp}_{2}} (1-cos\alpha_{1,2})
\end{equation}
The index with values 1 or 2 refers to either of the two OS-SF leptons from the $\tilde\chi^{0}_{2}$ decay, and $\alpha_{1,2}$ is the angle in space between
their flight directions. Neglecting the uncertainty on the angle, the relative uncertainty on $M_{\ell^{\pm}\ell^{\mp}}$ writes:
\begin{equation}
\frac{\delta M_{\ell^{\pm}\ell^{\mp}}}{M_{\ell^{\pm}\ell^{\mp}}} = \frac{\delta E_{\ell^{\pm}}}{E_{\ell^{\pm}}} 
\end{equation}

\vspace*{0.5cm}
\par\noindent
3.6.1.\ b.\ Theoretical\ Shape
\vspace*{0.1mm}
\par\noindent
For unpolarized $\tilde\chi^{0}_{2}$  and for their two-body decays, the theoretical shape of the dilepton invariant mass is known \cite{Barr:2004ze} to be:
\begin{equation}
\frac{1}{\Gamma}\frac{d\Gamma}{dM_{\ell^{\pm}\ell^{\mp}}} = 2 M_{\ell^{\pm}\ell^{\mp}}
\end{equation}
%where $x = \frac{ M_{\ell^{\pm}\ell^{\mp}} }{ M^{Max}_{\ell^{\pm}\ell^{\mp}} }$.
%
%%%For N2 from squarks decays, spin correlation implies,
%%%$$ \frac{1}{\Gamma}\frac{d\Gamma}{dM_{\ell^{+}\ell^{-}}} =  4\times M_{\ell^{+}\ell^{-}}\times (1-M_{\ell^{+}\ell^{-}}^{2}),  (\rm %%%Hump\ shape)$$
%%%$$ \frac{1}{\Gamma}\frac{d\Gamma}{dM_{\ell^{+}\ell^{-}}} =  4\times M_{\ell^{+}\ell^{-}}^{3}, (\rm Cusp\ shape) $$
%
\par\noindent
As seen in subsection \ref{sec:Part4:W-Analysis}, the main background process in the $\tilde\chi^{\pm}_{1}+\tilde\chi^{0}_{2}\to 3\ell^{\pm}+\rlap{\kern0.25em/}E_{T}$ analysis
is the $W^{\pm}+\gamma^{*}/Z^{0}\to 3\ell^{\pm}+\rlap{\kern0.25em/}E_{T}$ process, which constitutes an irreducible background. The OS-SF dilepton
coming from the $\gamma^{*}/Z^{0}$ decay forms a peak centered around $M_{Z}$. Therefore, we model the invariant mass distribution
of events surviving our selection using the following 6-parameters functional form:
\begin{equation}
M^{Fit}(x) =   \frac{C_{3}}{2\pi}\times\frac{ C_{5} }{(x-C_{4})^{2} +\frac{C_{5}^{2}}{4}}    +
\begin{cases}
2C_{1}\times\frac{x}{C^{2}_{0}}, & \ for\ x < C_{0};\ and\  \\
0, & \ for\ x > C_{0} 
\label{Functions-Fit-to-DileME}
\end{cases}
\end{equation}

\noindent
In order to account for the detector finite resolution, we convoluted the previous functional form with a gaussian distribution centered on zero and with an RMS
set to $C_{2}$. The other parameters represent:
\begin{itemize}
\item $C_{0}$: $M^{Max}_{\ell^{\pm}\ell^{\mp}}$, i.e. the position of the dilepton edge;
\item $C_{1}$: $N^{Exp}_{S}$, i.e. the number of expected signal events under the triangle;
\item $C_{3}$: $N^{Exp}_{B}$, i.e. the number of expected background events under the Z peak;
\item $C_{4}$: $M_{Z}$, i.e. the position of the Z peak; and,
\item $C_{5}$: $\Gamma_{Z}$, i.e. the width of the Z peak.
\end{itemize}
\par\noindent
For the S2b signal samples, we expect $N^{Exp}_{S}+N^{Exp}_{B}$ events under the Z peak.
\par\noindent
After a few trials we find it is sufficient to use the same triangle distribution to describe both the two-body and the three-body
$\tilde\chi^{0}_{2}$ decay in these fits.
\par
\begin{figure}[hbp]
\begin{center}
\includegraphics[scale=0.35]{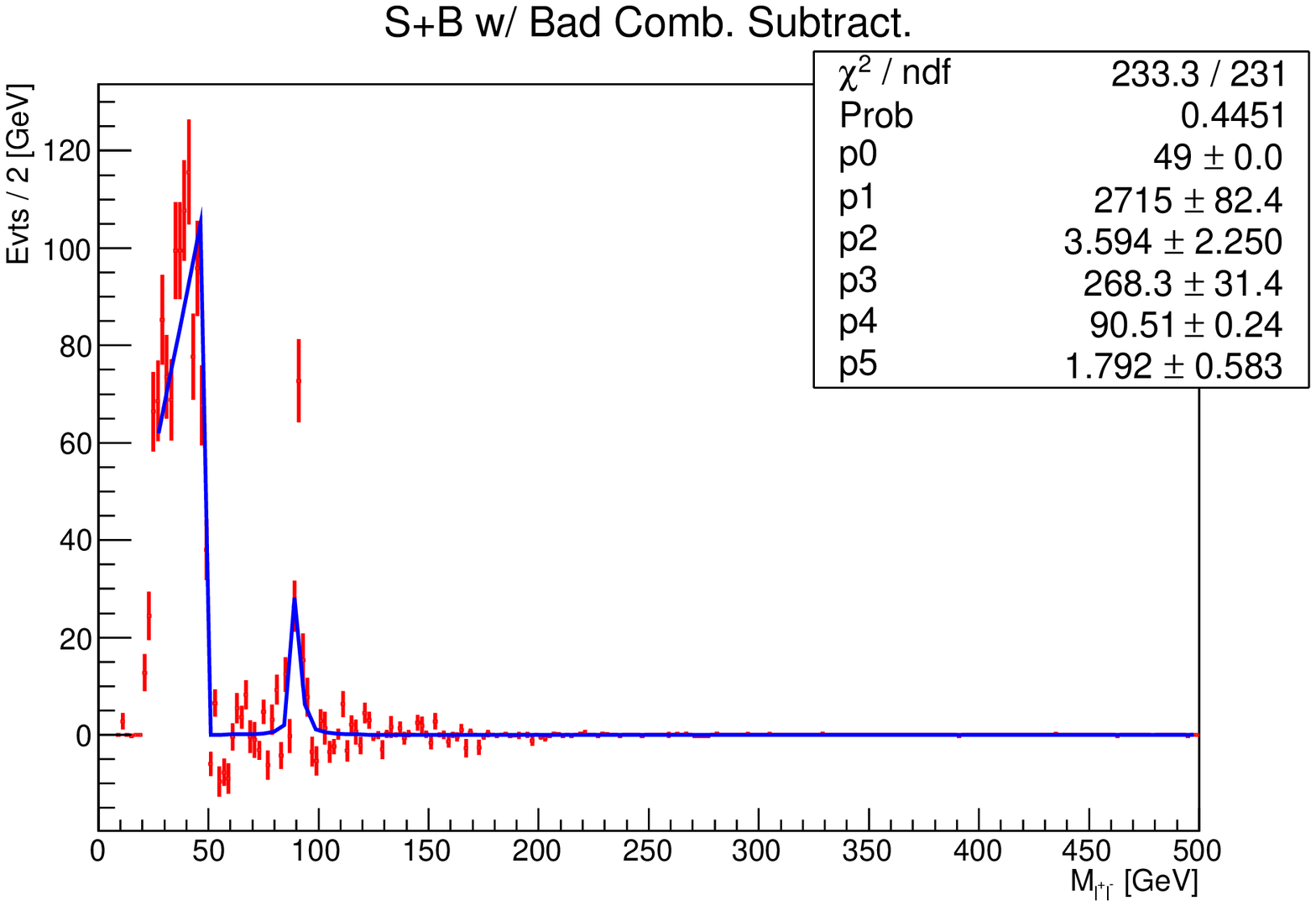}
\includegraphics[scale=0.35]{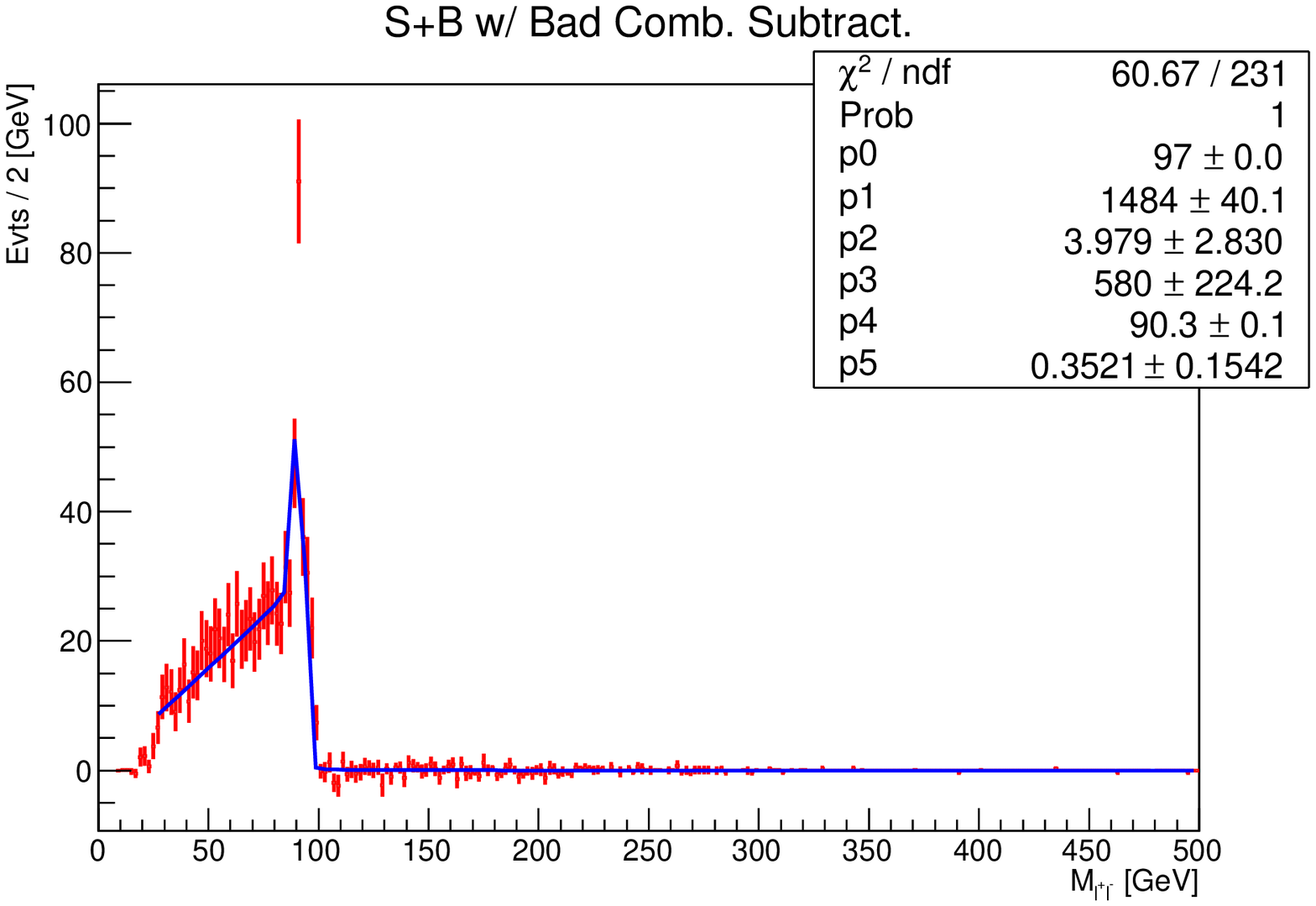}
\includegraphics[scale=0.35]{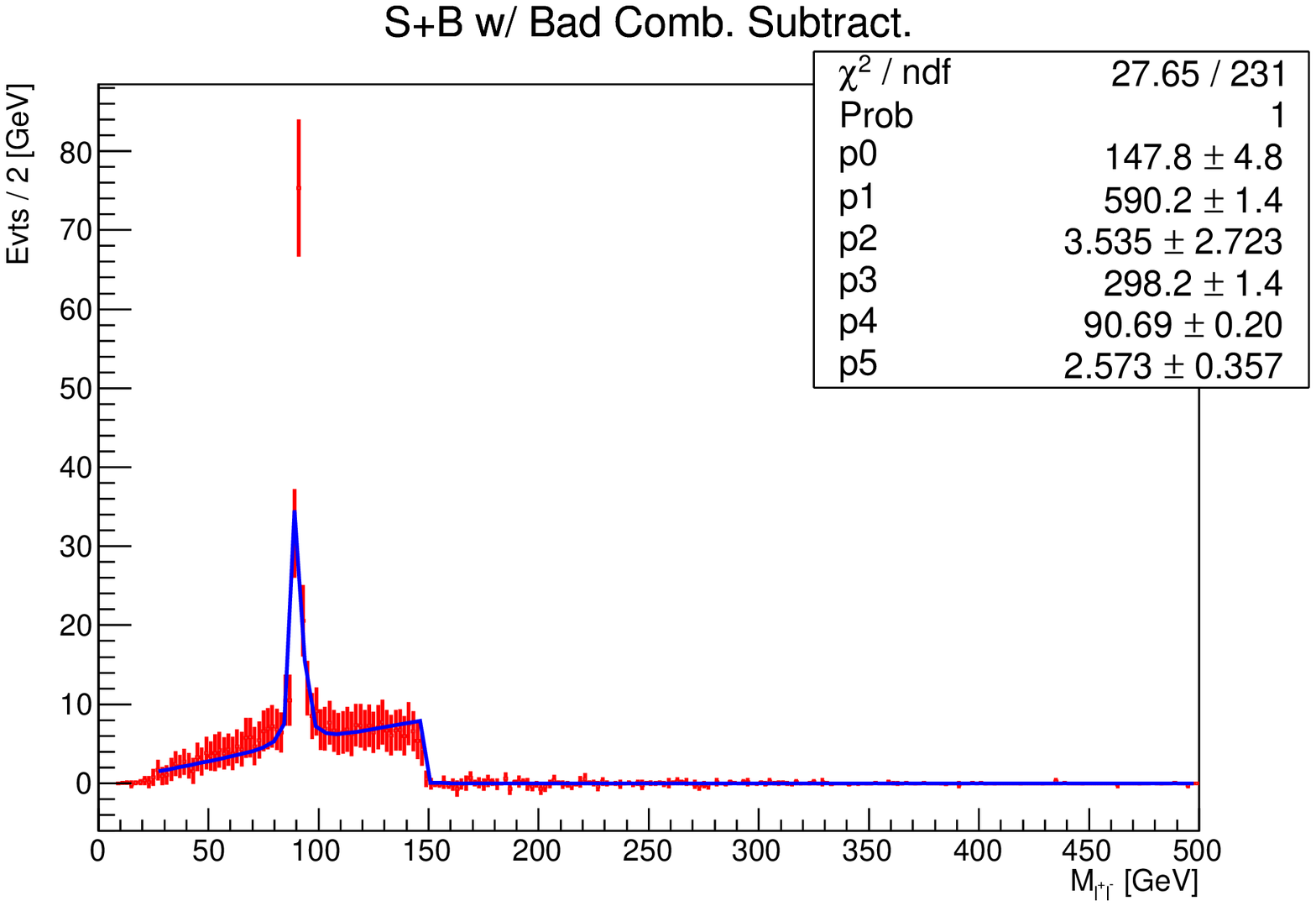}
\includegraphics[scale=0.35]{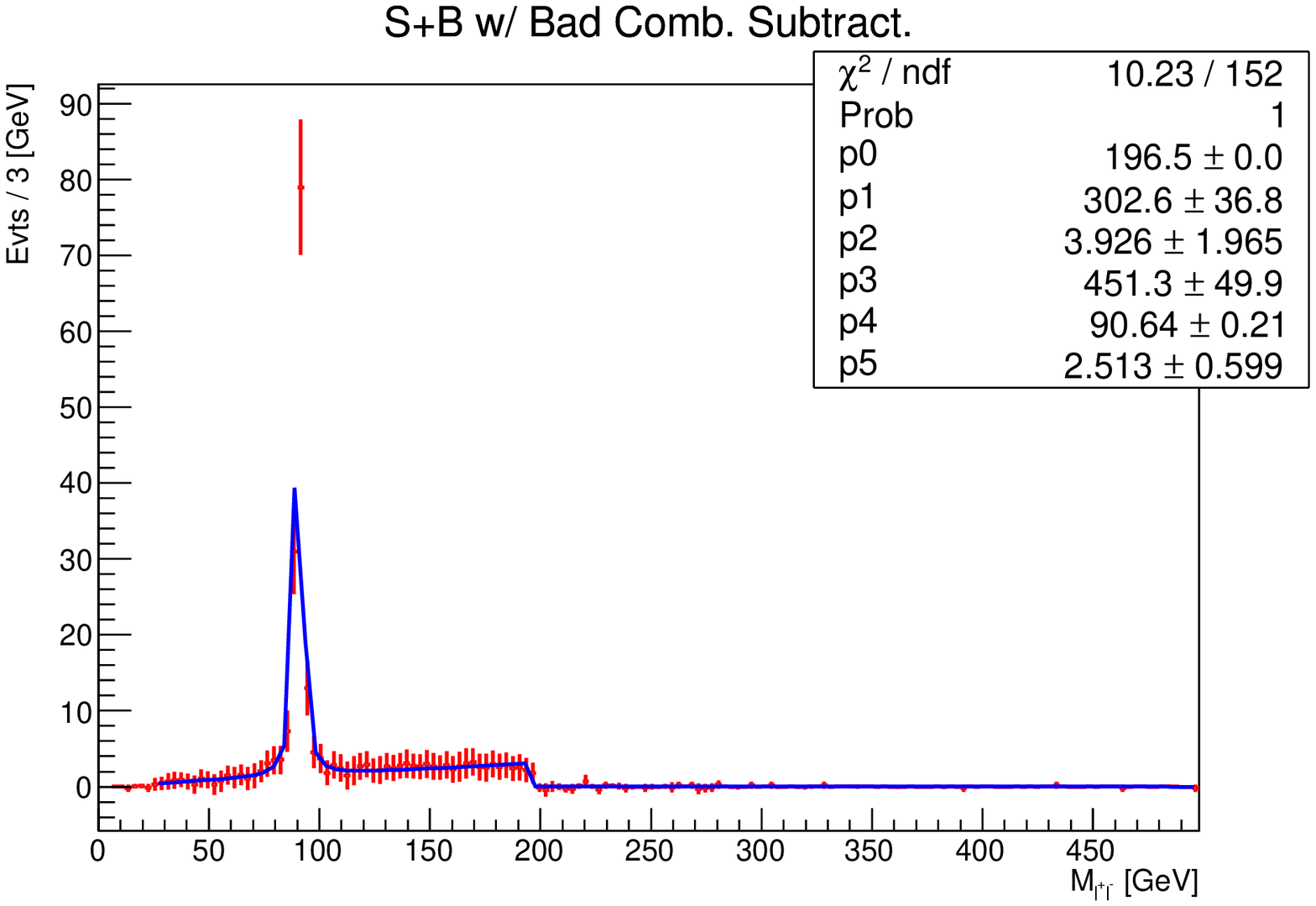}
\caption{\label{DileME_Plots_S1} A few examples of {\it DileME} measurements on the S1 samples for 100 $\leq M_{\tilde\chi^{0}_{2}}\leq$ 400 GeV}
\end{center}
\end{figure}

\begin{figure}[hbp]
\begin{center}
\includegraphics[scale=0.35]{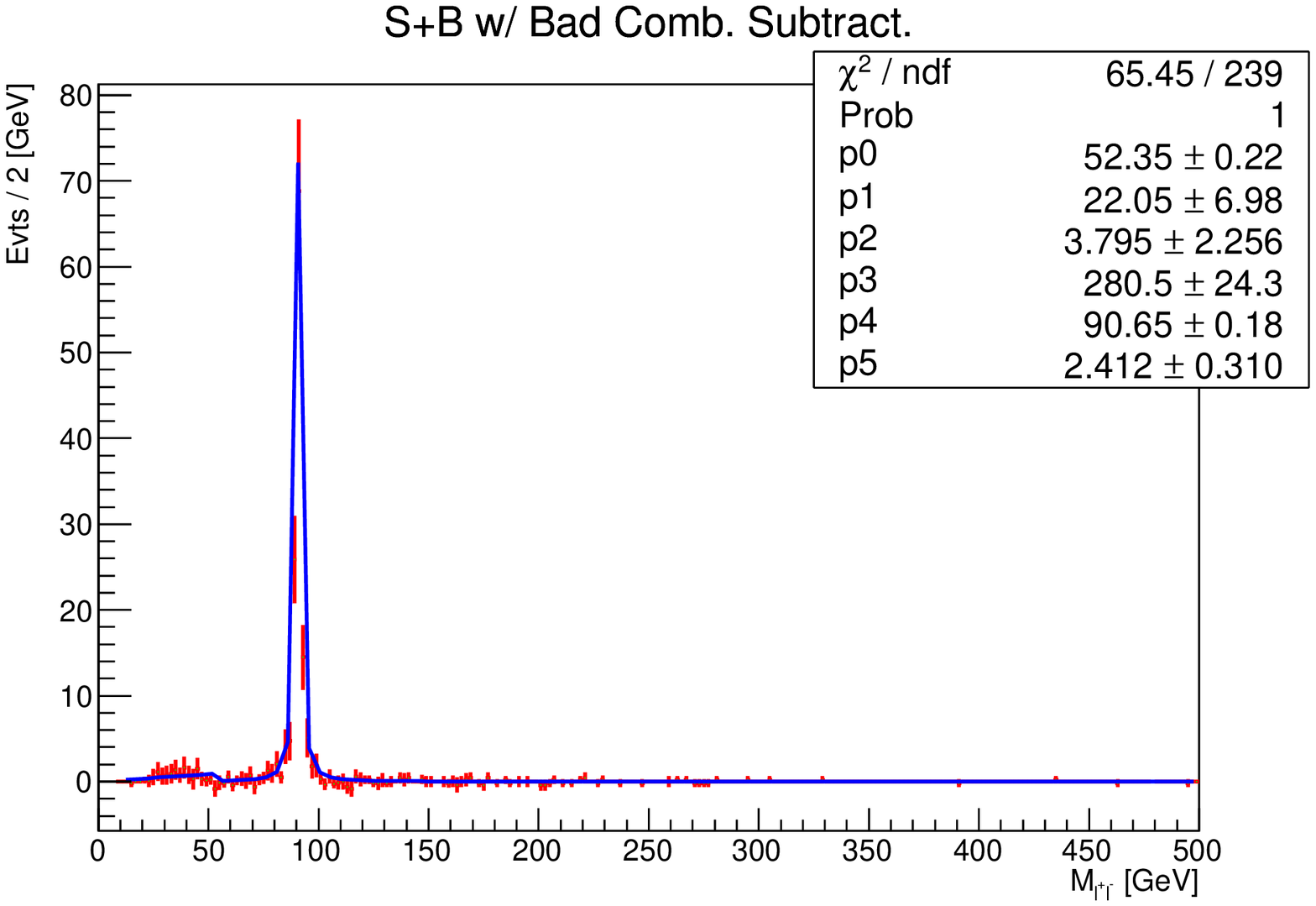}
\includegraphics[scale=0.35]{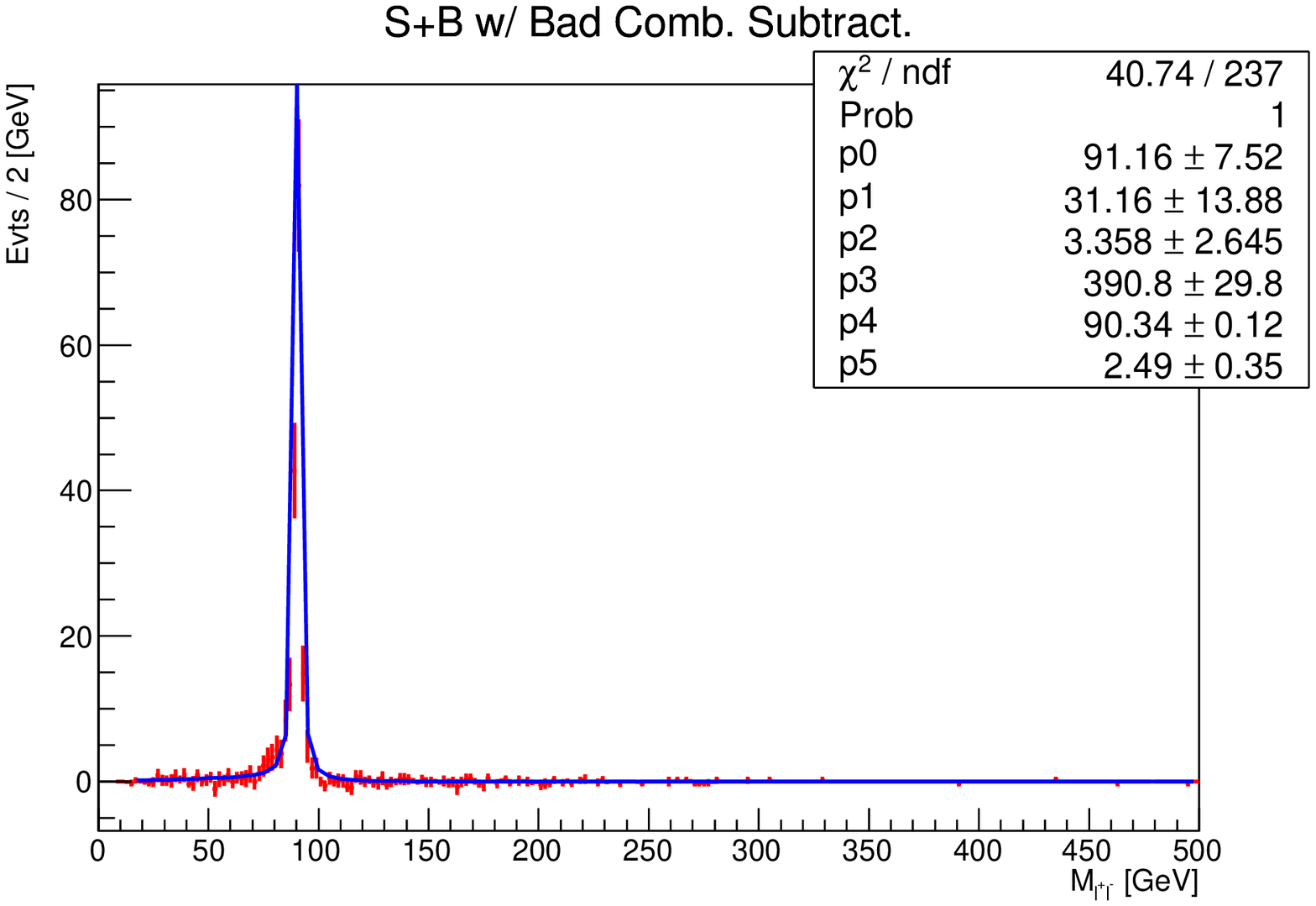}
\includegraphics[scale=0.35]{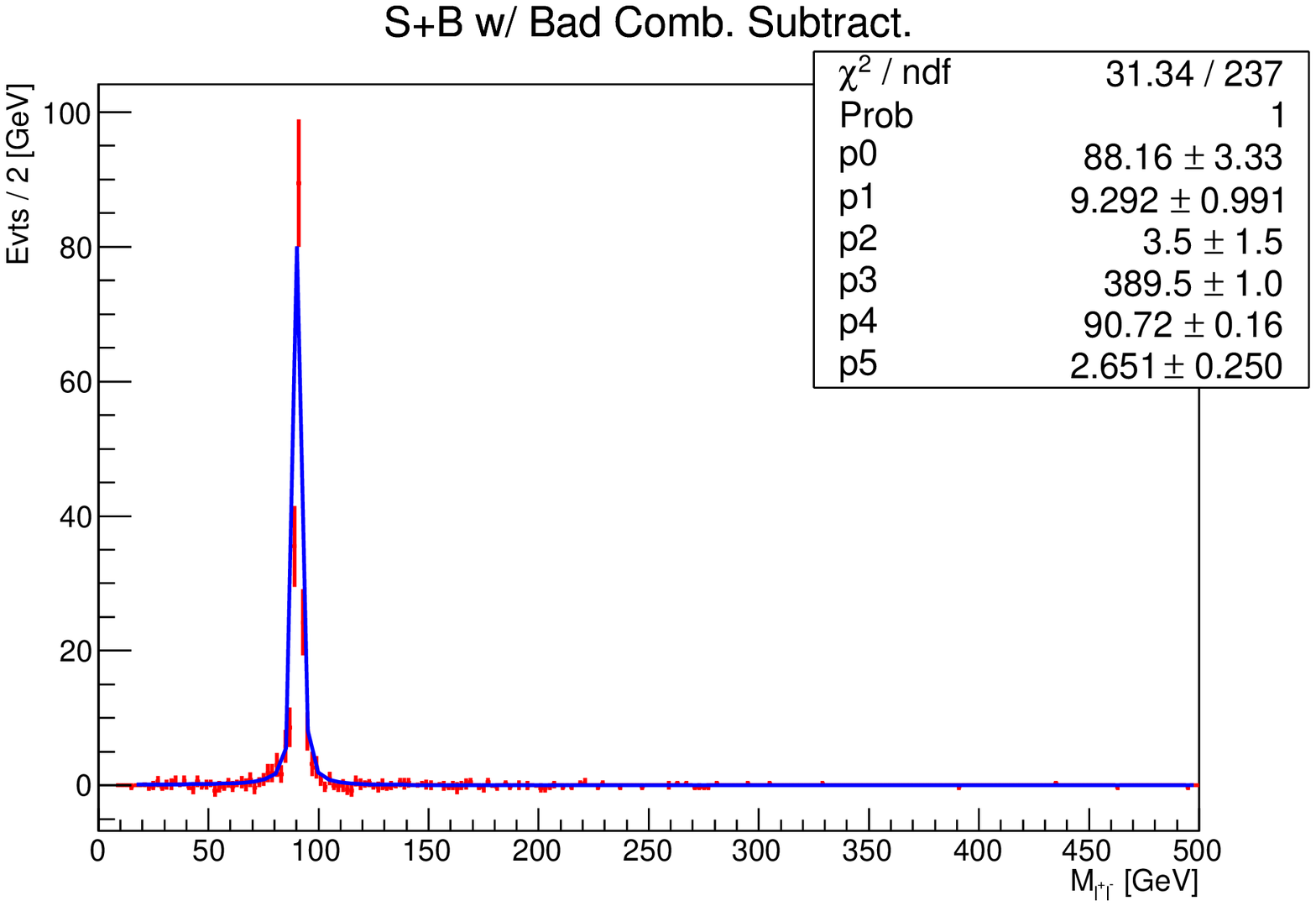}
\includegraphics[scale=0.35]{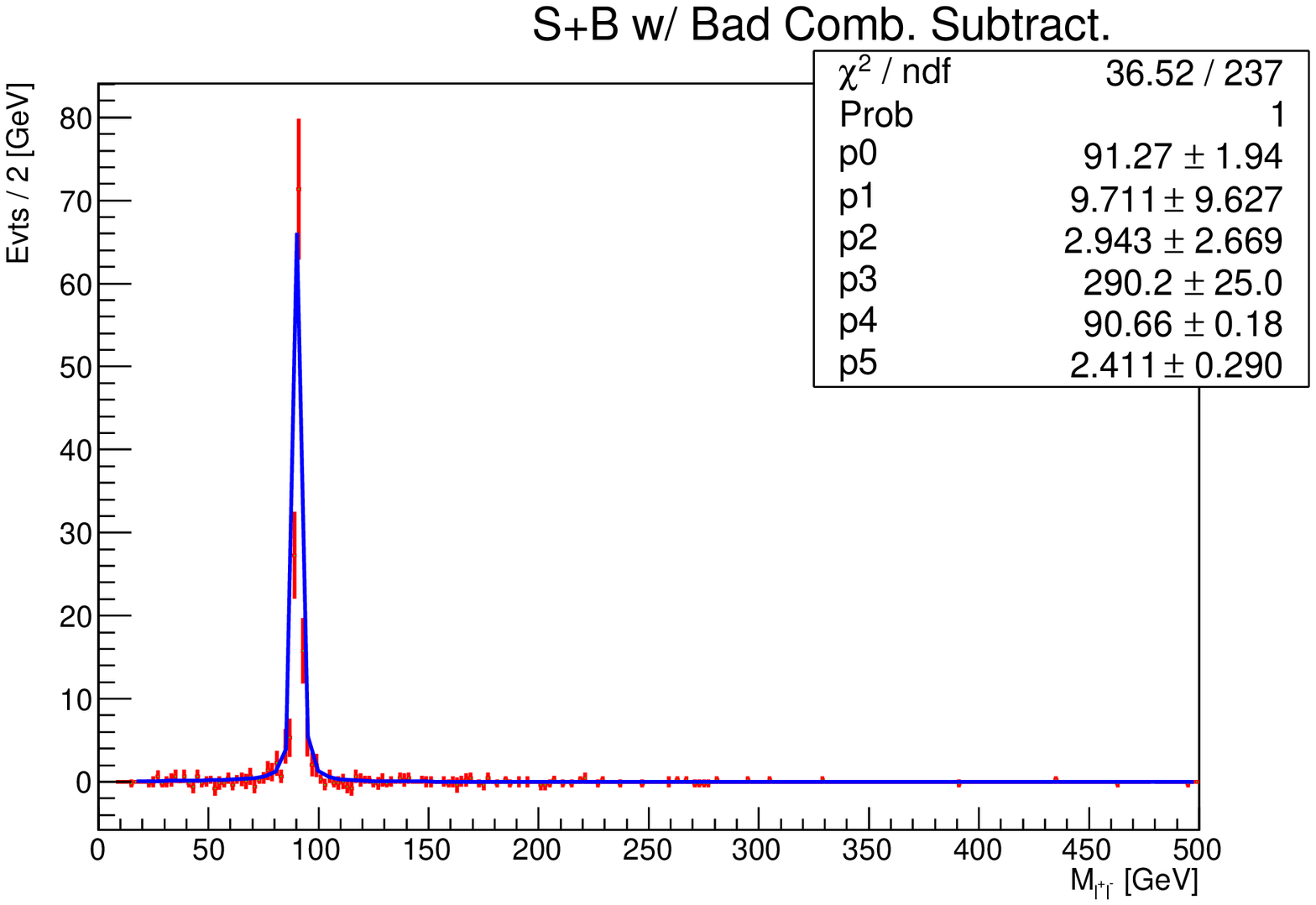}
\caption{\label{DileME_Plots_S2} A few examples of {\it DileME} measurements on the S2 samples for 100 $\leq M_{\tilde\chi^{0}_{2}}\leq$ 250 GeV}
\end{center}
\end{figure}

\noindent
The results of these fits are presented in tables \ref{S1:2l_Edge} and \ref{S2:2l_Edge}. The plots \ref{DileME_Plots_S1} and \ref{DileME_Plots_S2} illustrate a few of these fits.
Obviously the highest $M_{\tilde\chi^{0}_{2}}$ mass hypotheses unable any measurement of the dilepton invariant mass edge because of their
unsufficient signal-to-noise ratio. This situation is met for $M_{\tilde\chi^{0}_{2}}\geq$ 700 GeV for the S1 samples and
$M_{\tilde\chi^{0}_{2}}\geq$ 400 GeV for the S2 samples.

\begin{table}[h]
\begin{center}
\begin{tabular}{|c|c|c|c|}
\hline\hline
Process     & Theor. $M_{\ell^{+}\ell^{-}}^{Edge}$       & Meas. $M_{\ell^{+}\ell^{-}}^{Edge}$ & Fit $\chi^{2}/N_{dof}$   \\ 
                      &                   (GeV)                                          & (GeV)                                                         &    	                                \\
\hline
\underline{Signal S1} & 	           &	 &		     \\
$[M_{\tilde\chi^{0}_{2}},M_{\tilde\ell^{\pm}},M_{\tilde\chi^{0}_{1}}]$  \rm\ GeV	        &	 &		 &	\\ 
\hline
$[100,75,50]$	     &   49.301    &    $49.000\pm 0.000(stat)\pm 0.025(syst)$          &     1.010          \\
$[200,150,100]$    &   98.601	 &    $97.000\pm 0.000(stat)\pm 0.049(syst)$          &     0.263         \\
$[300,225,150]$    & 147.902    &    $147.8\pm   4.8(stat)\pm 0.074(syst)$              &     0.120         \\
$[400,300,200]$    & 197.203    &    $196.500\pm 0.000(stat)\pm 0.098(syst)$        &     0.067         \\
$[500,375,250]$    & 246.503	&    $246.93\pm 0.08(stat)\pm 0.123(syst)$            &     0.093         \\
$[600,450,300]$    & 295.804	&    $300.8\pm 0.7(stat)\pm 0.150(syst)$                &     0.097         \\
$[700,525,350]$    & 345.105	&                                                --                                &          --	          \\
\hline\hline
\end{tabular}       
\end{center}
\caption{\label{S1:2l_Edge} Dilepton mass edge measurements for the S1 samples.}
\end{table}
%
%%%
%
\begin{table}[h]
\begin{center}
\begin{tabular}{|c|c|c|c|}
\hline 
Process     & Theor. $M_{\ell^{+}\ell^{-}}^{Edge}$       & Meas. $M_{\ell^{+}\ell^{-}}^{Edge}$ & Fit $\chi^{2}/N_{dof}$   \\ 
                      &                   (GeV)                                          & (GeV)                                                         &    	                                \\
\hline
\underline{Signal S2}  & 	 &	&	     \\
$[M_{\tilde\chi^{0}_{2}},M_{\tilde\chi^{0}_{1}}]$  \rm\ GeV	        &	&	 &	\\ 
\hline
$[100,50]$	 &   50.0 &   $52.35\pm 0.22(stat)\pm 0.026(syst)$         &	    0.274  \\
$[105,13.8]$    &   91.2 &   $91.16\pm 7.52(stat)\pm 0.046(syst)$	  &	   0.172  \\
$[115,13.8]$    & 101.2 &   $90.28\pm 6.62(stat)\pm 0.045(syst)$	 &	  0.154	 \\
$[125,13.8]$    & 111.2 &   $88.16\pm 3.33(stat)\pm 0.040(syst)$	 &        0.132	 \\
$[135,13.8]$    & 121.2 &   $90.13\pm 6.54(stat)\pm 0.045(syst)$	 &	  0.116	 \\
$[145,13.8]$    & 131.2 &   $88.29\pm 6.03(stat)\pm 0.044(syst)$	 &	  0.125	 \\
$[150,50]$	 & 100.0 &   $99.54\pm 4.16(stat)\pm 0.050(syst)$         &	   0.230  \\
$[200,100]$  	& 100.0 &   $91.92\pm 1.99(stat)\pm 0.046(syst)$         &	  0.125  \\
$[250,125]$	& 125.0 &   $91.27\pm 1.97(stat)\pm 0.046(syst)$         &	  0.154	 \\
$[300,150]$  	& 150.0 &   $91.17\pm 0.94(stat)\pm 0.046(syst)$         &	  0.126  \\
$[400,200]$  	& 200.0 &   --         &	  --	 \\
$[500,250]$  	& 250.0 &   --         &	  --	 \\
$[600,300]$  	& 300.0 &   --         &	  --	 \\
$[700,350]$  	& 350.0 &   --         &	  --	 \\
\hline
\end{tabular}       
\end{center}
\caption{\label{S2:2l_Edge} Dilepton mass edge measurements for the S2 samples.}
\end{table}
\noindent
First of all we notice, that {\it ICA} and {\it DileME} methods do not give access to the same informations: $M_{\tilde\chi^{0}_{2}}+M_{\tilde\chi^{\pm}_{1}}$,
versus  $M_{\tilde\chi^{0}_{2}}-M_{\tilde\chi^{0}_{1}}$ or $M_{\tilde\chi^{0}_{2}} \times\sqrt{ \left ( 1-\frac{M^{2}_{\tilde\ell^{\pm}}}{M^{2}_{\tilde\chi^{0}_{2}}} \right ) \left ( 1-\frac{M^{2}_{\tilde\chi^{0}_{1}}}{M^{2}_{\tilde\ell^{\pm}}} \right ) }$, respectively.
We notice that the {\it DileME} method is very accurate: better than $3.5\%$ (and most often better than $1\%$) for the S1 samples, and better than $0.5\%$ for the S2a sample. However, for the S2b signal samples, it fails to extract any sensible informations about the
mass difference because of the resonant mode of the $\tilde\chi^{0}_{2}$ decay. For the sample $(105,13.8)$ S2b sample, the correct mass difference is found by chance, whereas for the other S2b samples, the {\it DileME} method systematically provides a wrong answer: $M_{\tilde\chi^{0}_{2}}-M_{\tilde\chi^{0}_{1}}=M_{Z}$.
\par\noindent
In regard of these observations, we conclude that the {\it ICA} and {\it DileME} methods complement very well each other.

%
%%%%%%%%%%%%%%%%%%%%%%%%%%%%%%%%%%%%%%%%%%%%%%%%%%%%%%%%%%%%%%%%%%%%%%
%
\newpage\clearpage
\subsubsection{Stransverse Mass End-Point}

\vspace*{1.5mm}
\noindent
3.6.2.\ a.\ Experimental\ Observable
\vspace*{0.5mm}

Let's consider an event where two particles (X) and (Y) are produced. Let's consider they both undergo decay chains, both ending up 
by the same invisible particle, denoted $\chi$, while emitting some visible energy in each hemispheres (A) and (B): $E^{vis_{(A)}}$
and $E^{vis_{(B)}}$. For an hypothesized mass of $\chi$, $M^{trial}_{\chi}$, the event stranverse mass $M_{T2}$ is defined as:
\begin{equation}
M_{T2}(vis^{(A)},vis^{(B)}|M^{trial}_{\chi}) = 
\underset{\rlap{\kern0.25em/}\vec{E}^{(A)}_{T}+\rlap{\kern0.25em/}\vec{E}^{(B)}_{T}=\rlap{\kern0.25em/}\vec{E}_{T}}{Min}
\left\{ Max [M_{T}(\vec{p}^{vis_{(A)}}_{T},\rlap{\kern0.25em/}\vec{E}^{(A)}_{T});
                      M_{T}(\vec{p}^{vis_{(B)}}_{T},\rlap{\kern0.25em/}\vec{E}^{(B)}_{T})] \right\}
\end{equation}
where
\begin{equation}
M^{2(A)}_{T}  = M^{2(A)} + M^{2(\chi_{_{A}})} + 2[E^{(A)}_{T}\cdot E^{(\chi_{_{A}})}_{T}-\vec{p}^{(A)}_{T}\cdot\vec{p}^{(\chi_{_{A}})}_{T}]
\end{equation}
, and
\begin{equation}
E^{2}_{T}         = M^{2} + p^{2}_{T}
\end{equation}

\par\noindent
The stranverse mass has two important properties. On the one hand, it's very effective to discriminitate R-parity conserved SUSY signals from their SM background processes.
On the other hand it enables to measure the mass of the parent particles (X) and (Y) and of children particle ($\chi$) and for this second purpose, we'll denote this method {\it MT2} in the rest of this article.
\par\noindent
Regarding the signal and background discrimination described in section \ref{sec:next-particle-LVL}, we arbitrarily chose the following assignment:
\begin{itemize}
\item $\ell^{\pm}_{1}\leftrightarrow$ visible energy (A),
\item $\ell^{\pm}_{2}\leftrightarrow$ visible energy (B),
\item $\ell^{\pm}_{3}\leftrightarrow$ downstream additional visible particle,
\end{itemize}
\noindent
where the index $i=1,2,3$ refers to the decreasing $p_{T}$ of the leptons, and we set $M^{trial}_{\chi}=0$ GeV. This choice does not accurately reflect the actual kinematics of our signal samples, but it is sufficient to provide a good and simple signal to background discrimination applicable to all of them.

\par\noindent
On the contrary, in the current section, in order to assess the mass measurement capability of the {\it MT2} method we have to properly assign the OS-SF dilepton
to the $\tilde\chi^{0}_{2}$ decay, say into the visible energy (A), and the additional lepton to the $\tilde\chi^{\pm}_{1}$ decay into the visible energy (B).
This precise assignment is done via the solution we adopted to solve the trilepton combinatorics which is presented in the preamble of the current section.

\par\noindent
The main systematic uncertainties for the {\it MT2} method come from the reconstruction of the different objects in our search topology. As inferred from \cite{Aad:2012twa}, we consider as sources of uncertainty: the trigger, the reconstruction, the identification, the energy resolution and the isolation for both the electrons and the muons. The resulting uncertainties are $4.6\%\ (e^{\pm})$ and $1.1\%\ (\mu^{\pm})$, respectively.  These changes in the electrons and muons kinematics are propagated
onto a corrected missing transverse energy $\rlap{\kern0.25em/}E^{Corr}_{T}$. Then, the impact of the uncertainties of the calorimeter cluster energy scale, of the jet energy scale and the jet energy resolution, and of the pile-up on the $\rlap{\kern0.25em/}E_{T}$, are also summed in quadrature, amounting to an uncertainty of $0.8\%$ with which the $\rlap{\kern0.25em/}E^{Corr}_{T}$ is smeared. 
We input the smeared  $\rlap{\kern0.25em/}E^{Corr}_{T}$ and the smeared lepton kinematics into the calculation of a smeared $M_{T2}^{Smear}$. Finally, the systematic uncertainty on $M_{T2}$ is taken as the absolute
value of the relative difference between the nominal $M_{T2}$ and $M_{T2}^{Smear}$:
\begin{equation}
\frac{\delta M_{T2}}{M_{T2}} = \frac{|M_{T2}-M_{T2}^{Smear}|}{M_{T2}}.
\end{equation}
This procedure is re-iterated for each value of $M^{trial}_{\chi}$, as reported in table \ref{MT2_Systematics}.

\begin{table}[h]
\begin{center}
\begin{tabular}{|c|c|}
\hline
$M^{trial}_{\chi}$   (GeV) & $\frac{\delta M_{T2}}{M_{T2}}$     ($\%$) \\
\hline
       0.                                   &  1.86 \\
     13.8                                 &  1.80 \\
     50.                                   &  1.47 \\
   100.                                   &  1.10 \\
   125.                                   &  1.02 \\
   150.                                   &  0.97 \\
   200.                                   &  0.90 \\
   250.                                   &  0.85 \\
   300.                                   &  0.83 \\
   350.                                   &  0.81 \\
\hline
\end{tabular}       
\end{center}
\caption{\label{MT2_Systematics} Systematic uncertainty on $M_{T2}$ for different $M^{trial}_{\chi}$.}
\end{table}
\vspace*{1.5mm}
\noindent
3.6.2.\ b.\ Theoretical\ End-Points
\vspace*{0.5mm}
\par\noindent
In order to measure the end-points ($M_{T2}^{Max}$) of the $M_{T2}$ distributions we use either descending step functions or continuous but not derivable linear functions, depending on the position of this end-points with respect to the remaining background.
\par\noindent
The positions of these end-points depend on the hypothesized value of $M^{trial}_{\chi}$ and have a kink at $M^{trial}_{\chi}=M_{\tilde\chi^{0}_{1}}$
\cite{Cho:2007dh}. Therefore, they are described by continuous functions (yet not derivable at the kink position): one, that we'll denote $f_{down}$ for $M^{trial}_{\chi}<M_{\tilde\chi^{0}_{1}}$ and another one, denoted $f_{up}$ for $M^{trial}_{\chi}>M_{\tilde\chi^{0}_{1}}$.

\par\noindent
For two-body decays, the  $f_{down}$ and $f_{up}$ functions are:

\begin{equation}
\begin{split}
f^{2-body}_{down} = M^{Max}_{T2}\left (\ell^{\pm} _{(A)},\ell^{\pm}\ell^{\mp}_{(B)}|M^{trial}_{\chi} < M_{\tilde\chi^{0}_{1}}\right ) = 
\left ( \frac{M^{2}_{\tilde\chi^{0}_{2}}+M^{2}_{\tilde\ell^{\pm}}-M^{2}_{\tilde\chi^{0}_{1}}}{2M_{\tilde\chi^{0}_{2}}} \right )   +\\
\sqrt{\left ( \frac{M^{2}_{\tilde\chi^{0}_{2}}-M^{2}_{\tilde\ell^{\pm}}+M^{2}_{\tilde\chi^{0}_{1}}}{2M_{\tilde\chi^{0}_{2}}} \right )^{2} + [(M^{trial}_{\chi})^{2}-M^{2}_{\tilde\chi^{0}_{1}}]}
\end{split}
\end{equation}

\noindent
and,

\begin{equation}
\begin{split}
f^{2-body}_{up} = M^{Max}_{T2}\left (\ell^{\pm} _{(A)},\ell^{\pm}\ell^{\mp}_{(B)}|M^{trial}_{\chi} > M_{\tilde\chi^{0}_{1}}\right ) =
\frac{M_{\tilde\chi^{0}_{2}}}{2}\left ( (1-\frac{M^{2}_{\tilde\ell^{\pm}}}{M^{2}_{\tilde\chi^{0}_{2}}}) + (1-\frac{M^{2}_{\tilde\chi^{0}_{1}}}{M^{2}_{\tilde\ell^{\pm}}})\right ) + \\
\sqrt{ \left [ \frac{M_{\tilde\chi^{0}_{2}}}{2}\left ( (1-\frac{M^{2}_{\tilde\ell^{\pm}}}{M^{2}_{\tilde\chi^{0}_{2}}}) -  (1-\frac{M^{2}_{\tilde\chi^{0}_{1}}}{M^{2}_{\tilde\ell^{\pm}}})\right )\right ]^{2} + (M^{trial}_{\chi})^{2} }
\end{split}
\end{equation}

\par\noindent
Whereas, for three-body decays, the  $f_{down}$ and $f_{up}$ functions are:

\begin{equation}
\begin{split}
f^{3-body}_{down} = M^{Max}_{T2}\left (\ell^{\pm} _{(A)},\ell^{\pm}\ell^{\mp}_{(B)}|M^{trial}_{\chi} < M_{\tilde\chi^{0}_{1}}\right ) = 
\left ( \frac{M^{2}_{\tilde\chi^{0}_{2}}-M^{2}_{\tilde\chi^{0}_{1}}}{2M_{\tilde\chi^{0}_{2}}} \right )   +\\
\sqrt{\left ( \frac{M^{2}_{\tilde\chi^{0}_{2}}-M^{2}_{\tilde\chi^{0}_{1}}}{2M_{\tilde\chi^{0}_{2}}} \right )^{2} +
(M^{trial}_{\chi})^{2} }
\end{split}
\end{equation}

\noindent
and,
\begin{equation}
f^{3-body}_{up} = M^{Max}_{T2}\left (\ell^{\pm} _{(A)},\ell^{\pm}\ell^{\mp}_{(B)}|M^{trial}_{\chi} > M_{\tilde\chi^{0}_{1}}\right ) = (M_{\tilde\chi^{0}_{2}}-M_{\tilde\chi^{0}_{1}}) +M^{trial}_{\chi}
\end{equation}

\par\noindent
It's important to note, that for $f^{2-body}_{down}$, small values of $M^{trial}_{\chi}$ are not always permitted. In the particular
for our simplified models, we have the following relations: $M_{\tilde\chi^{0}_{2}}=2M_{\tilde\chi^{0}_{1}}$, and for the S1 samples,
 $M_{\tilde\ell^{\pm}}=\frac{3}{2}M_{\tilde\chi^{0}_{1}}$. Therefore we need to keep $M^{trial}_{\chi}\geq \sqrt{\frac{135}{256}}\times M_{\tilde\chi^{0}_{1}}$  in order for $f^{2-body}_{down}$ to be defined.
\par\noindent
For the {\it MT2} method, we need to perform two series of fits. We start with primary fits to the $M_{T2}$ distributions for each signal sample so as to measure their $M^{Max}_{T2}$. 
Then we proceed with the secondary fits for each signal sample. The latter use as inputs the different $M^{Max}_{T2}$ values obtained for each $M^{trial}_{\chi}$ hypothesis and they enable simultaneoulsy to measure the mass of the parent particle, here $M_{\tilde\chi^{0}_{2}}=M_{\tilde\chi^{\pm}_{1}}$, of the end daughter particle $M_{\tilde\chi^{0}_{1}}$ and, for the S1 samples, the mass of the intermediate particle, $M_{\tilde\ell^{\pm}}$. The 2-body functional forms are utilized to fit the S1 samples and the 3-body ones are utilized to fit the S2 samples. Note that these functional forms also provide the prior knowledge of the $M^{Max}_{T2}$ for each signal hypothesis which serve as starting points in the minimization process of the primary fits.
\par\noindent
Here are a few important observations that justify our strategy for the primary fits:
\begin{itemize}
\item the $M_{T2}$ distribution of the remaining background events cluster into a Z peak which is located at $M_{Z}+M^{trial}_{\chi}$,
\item the $M_{T2}$ distribution of the S2b samples also cluster into a Z peak which is located at $M_{Z}+M^{trial}_{\chi}$ and which may either be truncated or exhibit
an asymmetric shoulder,
\item S1 samples: without an analytical description of the full $M_{T2}$ distribution, we just fit the $M_{T2}$ falling edge.
\end{itemize}

\par\noindent
This leads us  to use similar functional forms as for the dilepton mass distributions for the primary fits, but with 8 parameters:

\begin{equation}
M^{Fit}(x) =   \frac{C_{5}}{2\pi}\times\frac{ C_{7} }{(x-C_{6})^{2} +\frac{C_{7}^{2}}{4}}    +
\begin{cases}
C_{1}\times (x-C_{0} )+C_{2}, & \rm\ for\ x < C_{0};\ and\  \\
C_{3}\times (x-C_{0} )+C_{2}, & \rm\ for\ x > C_{0} 
\label{Functions-Fit-to-MT2-Max}
\end{cases}
\end{equation}

\noindent
In order to account for the detector finite resolution, we convoluted the previous functional form with a gaussian distribution centered on zero and with an RMS
set to $C_{4}$. The other parameters represent:
\begin{itemize}
\item $C_{0}$: $M^{Max}_{T2}$, i.e. the position of the $M_{T2}$ end-point;
\item $C_{1}$: slope of the first line;
\item $C_{2}$: height of the kink between the two lines;
\item $C_{3}$: slope of the second line;
\item $C_{5}$: $N^{Exp}_{B}$, i.e. the number of expected background events under the Z peak;
\item $C_{6}$: $M_{Z}+M^{trial}_{\chi}$, i.e. the position of the (pseudo) Z peak; and,
\item $C_{7}$: the width of the pseudo Z peak.
\end{itemize}

\par
\begin{figure}[hbp]
\begin{center}
\includegraphics[scale=0.35]{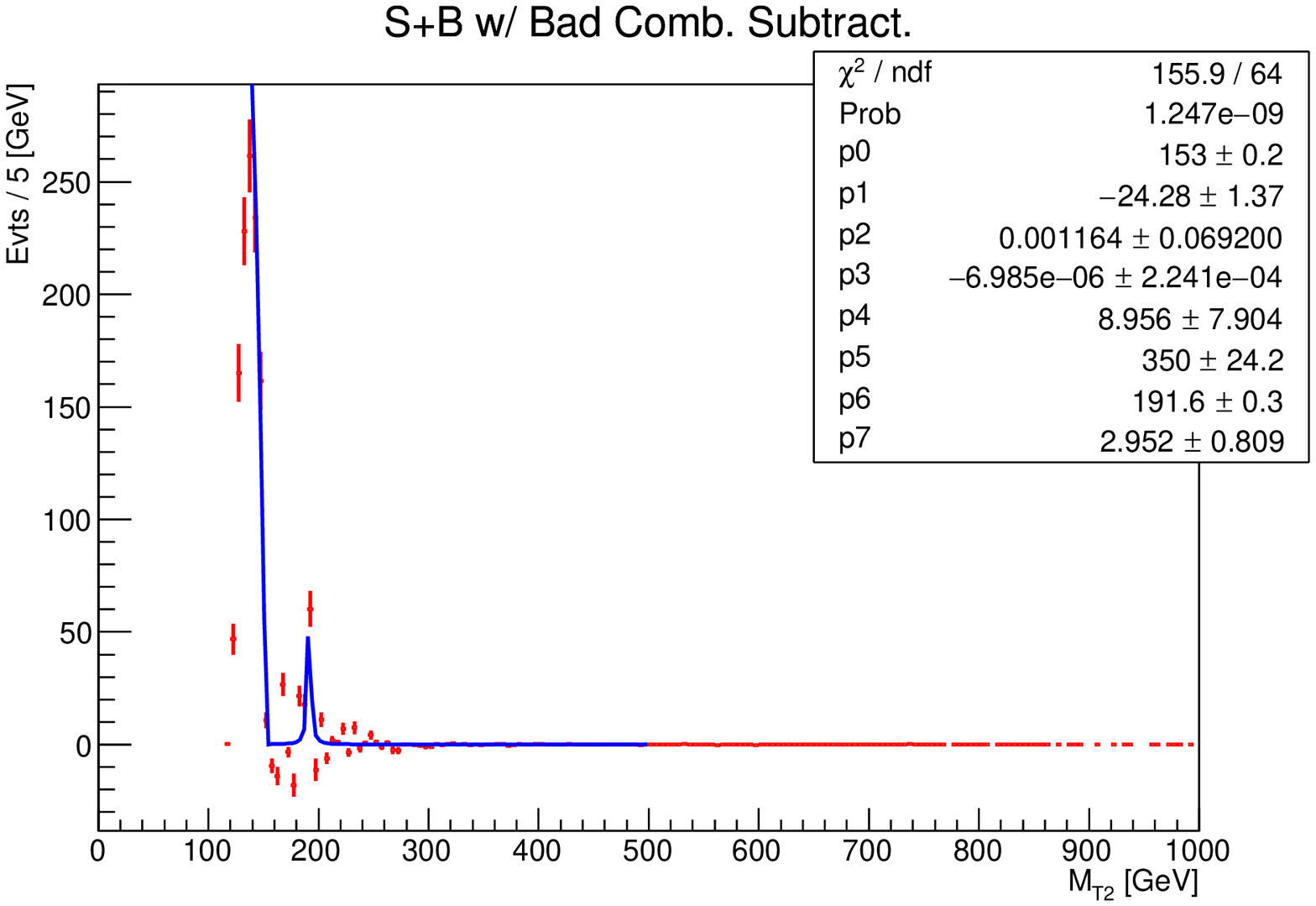}
\includegraphics[scale=0.35]{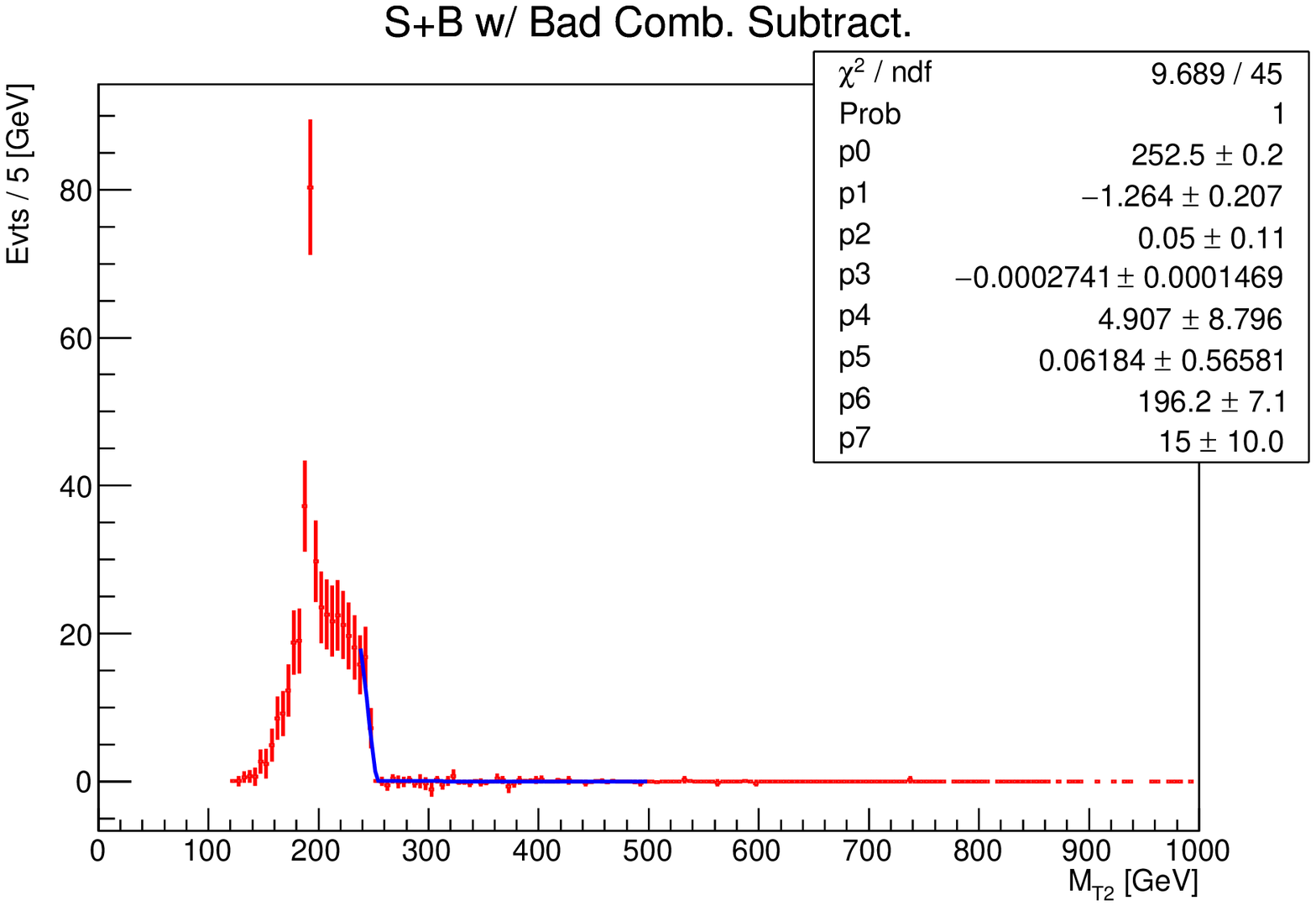}
\includegraphics[scale=0.35]{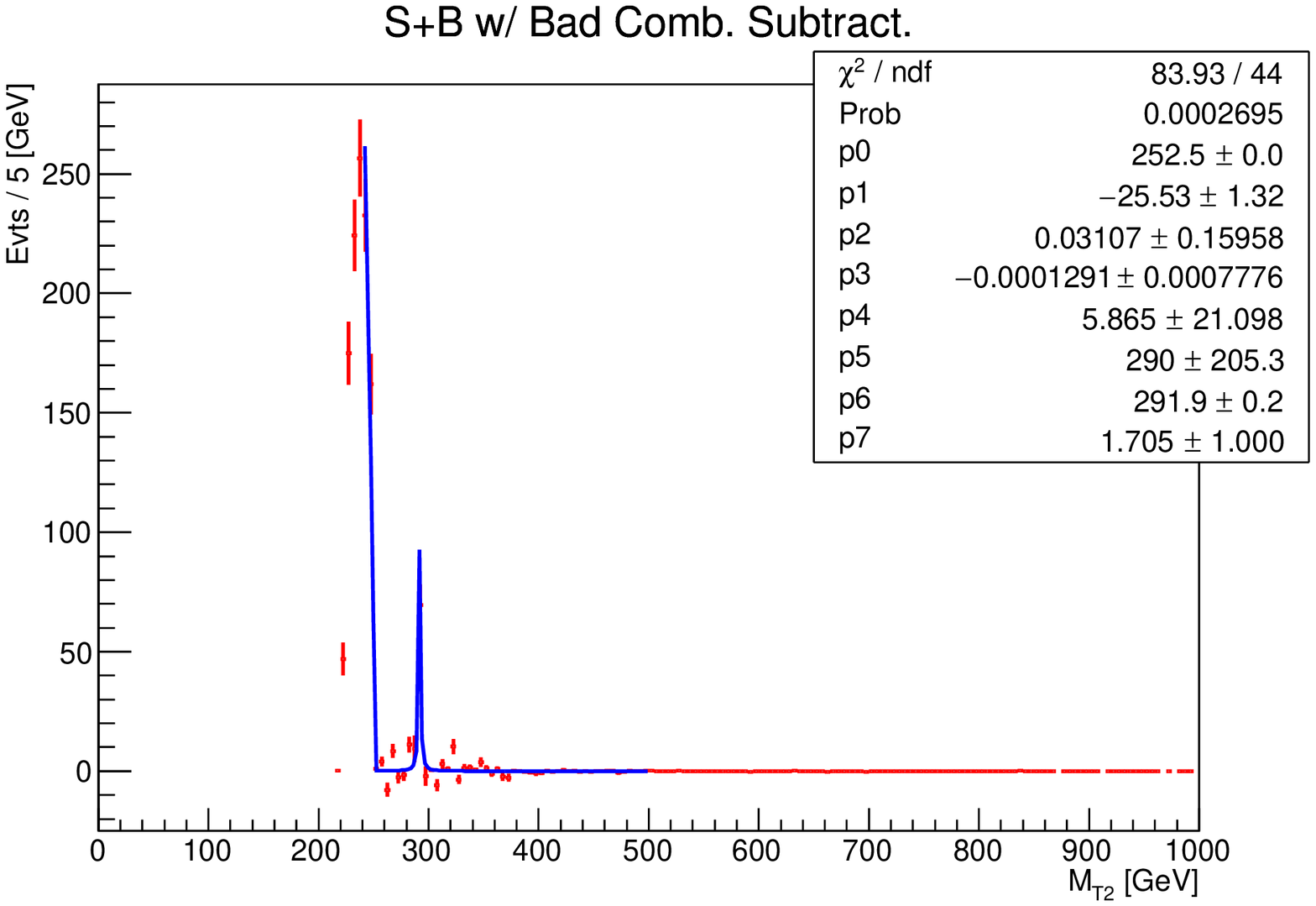}
\includegraphics[scale=0.35]{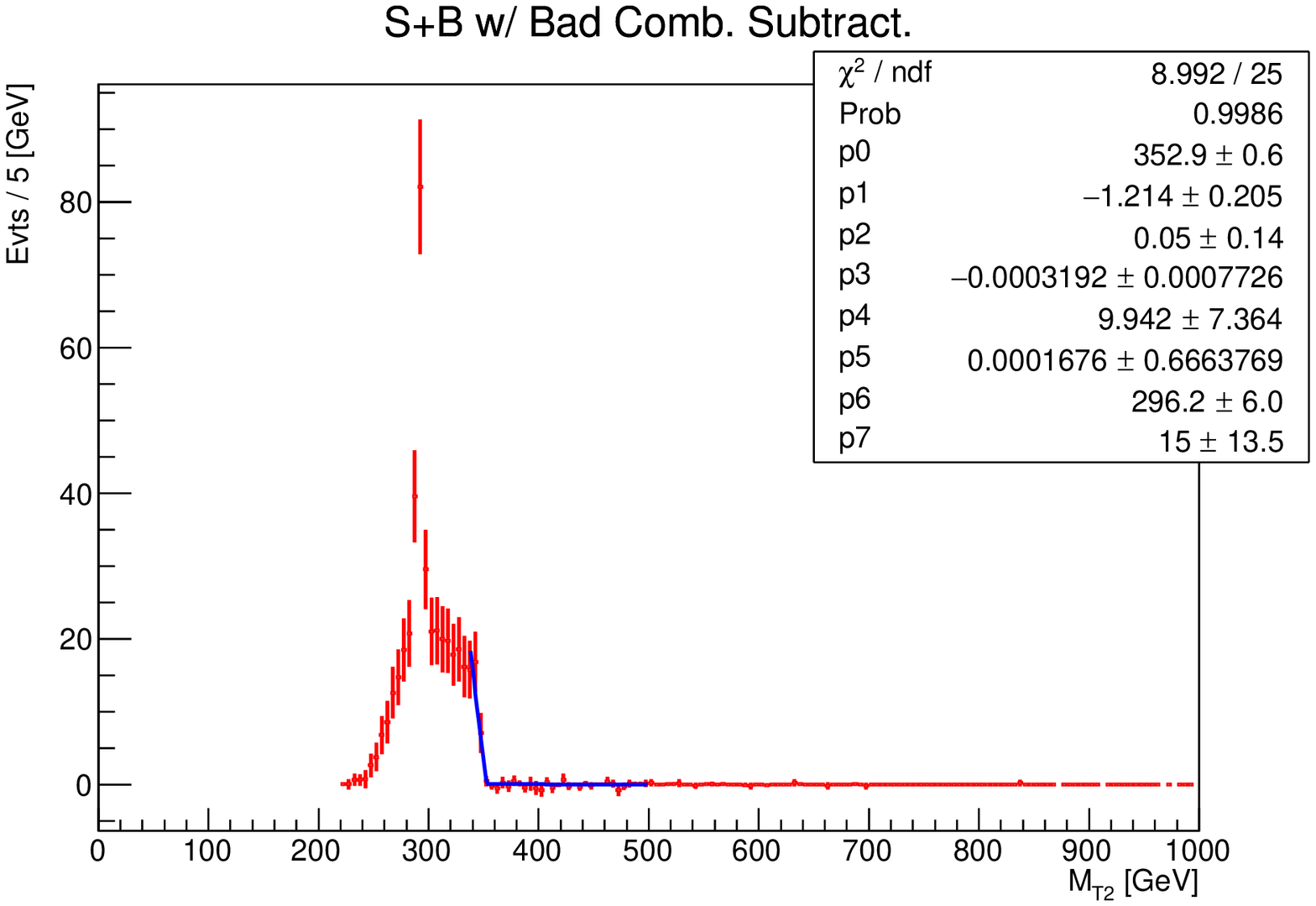}
\caption{\label{MT2-Max-Plots-S1} A few examples of $M_{T2}^{Max}$ measurements on the S1 samples. For the top and the bottom row $M_{trial}^{\chi}=$100 and 200 GeV, respectively. For the left and the right column $M_{\tilde\chi^{0}_{2}}=$ 100 and 300 GeV, respectively.}
\end{center}
\end{figure}

\begin{figure}[hbp]
\begin{center}
\includegraphics[scale=0.35]{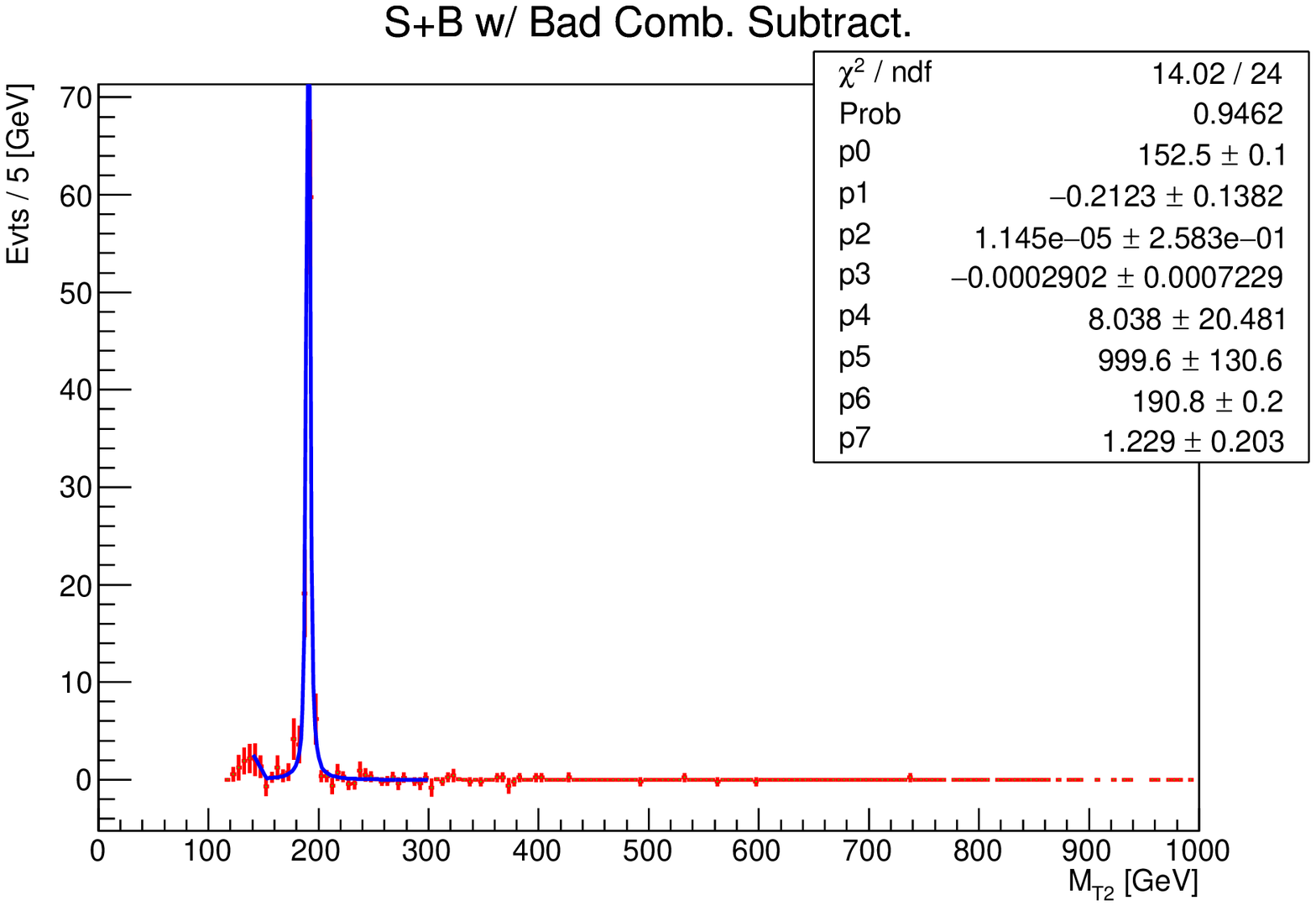}
\includegraphics[scale=0.35]{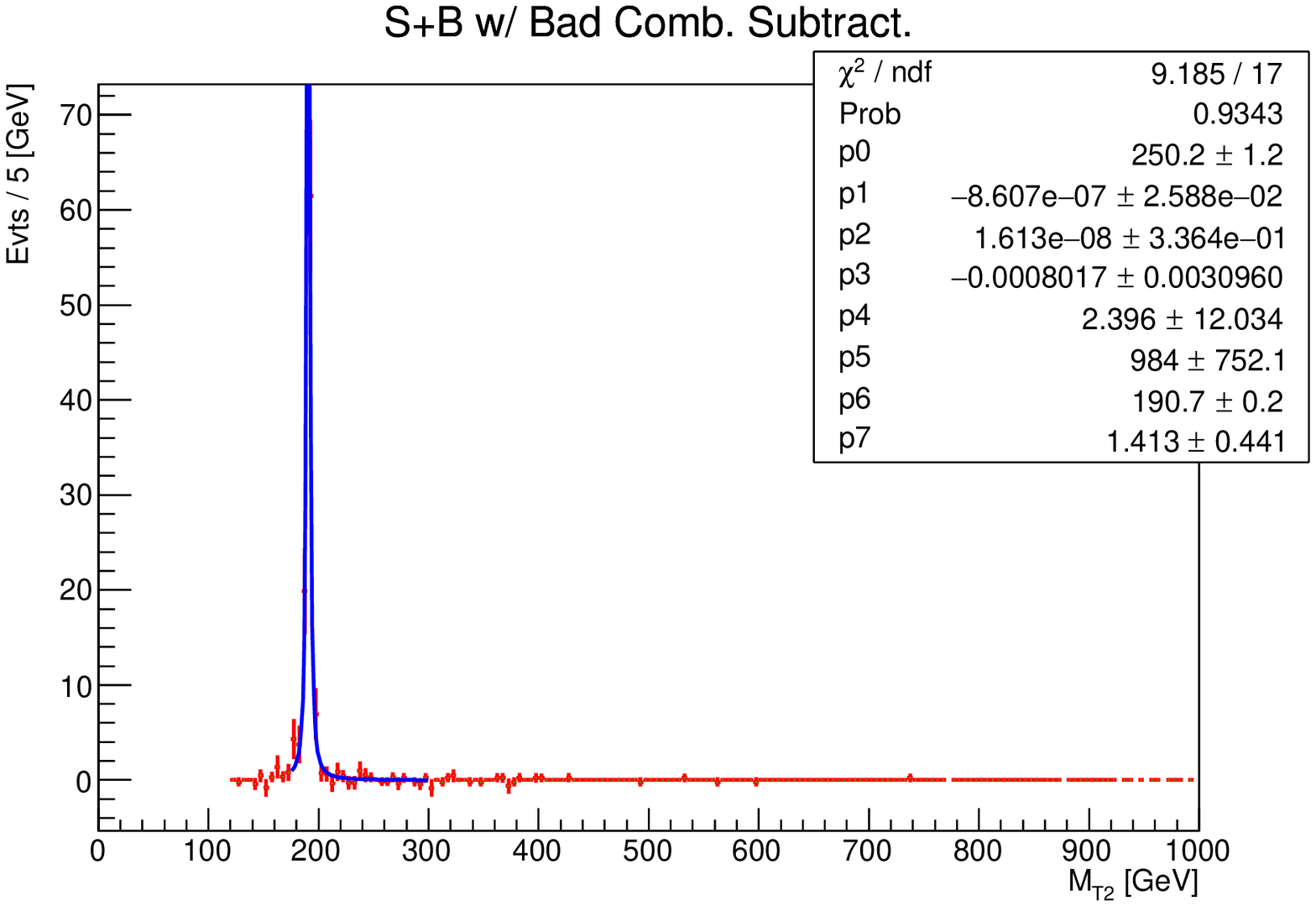}
\includegraphics[scale=0.35]{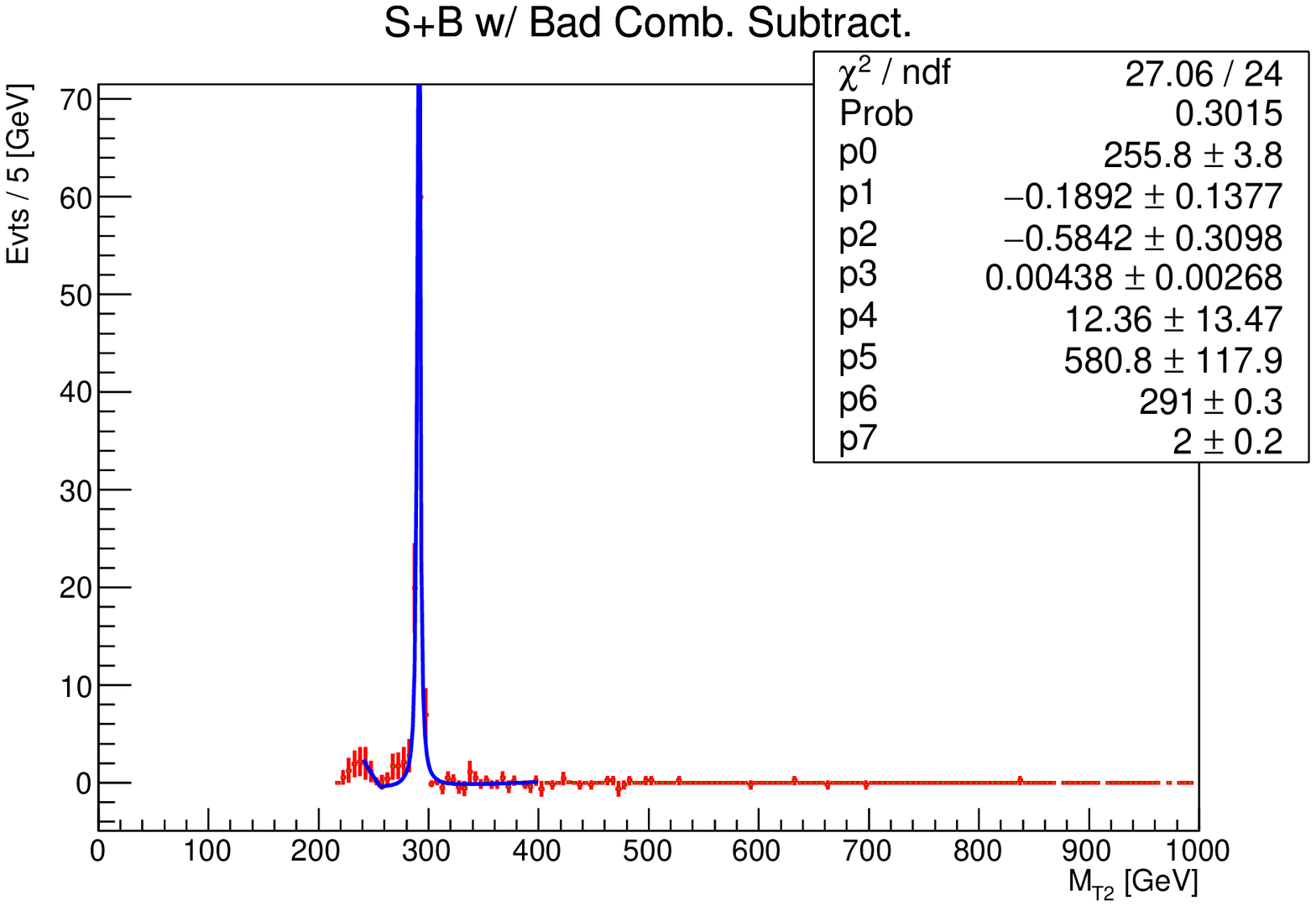}
\includegraphics[scale=0.35]{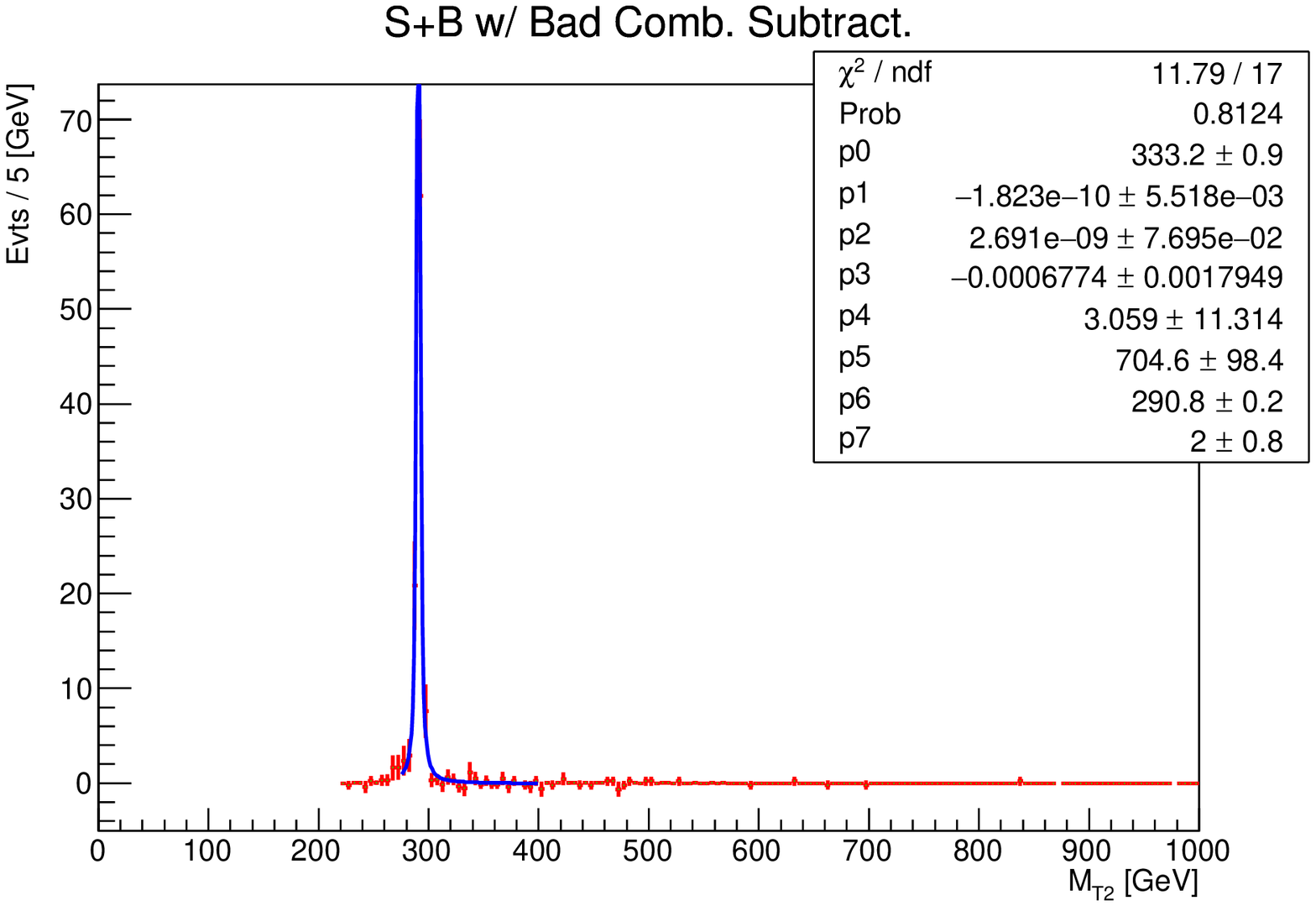}
\caption{\label{MT2-Max-Plots-S2} A few examples of $M_{T2}^{Max}$ measurements on the S2 samples. For the top and the bottom row $M_{trial}^{\chi}=$100 and 200 GeV, respectively. For the left and the right column $M_{\tilde\chi^{0}_{2}}=$ 100 and 300 GeV, respectively.}
\end{center}
\end{figure}

\noindent
The results of the primary fits are presented in tables \ref{MT2-Max-first} to \ref{MT2-Max-last}. Figures \ref{MT2-Max-Plots-S1} and \ref{MT2-Max-Plots-S2}  illustrate a few of them. Again, no $M^{Max}_{T2}$ measurements on our samples are feasible when $M_{\tilde\chi^{0}_{2}}\geq$ 700 GeV for the S1 samples and $M_{\tilde\chi^{0}_{2}}\geq$ 400 GeV for the S2 samples.
\par
\begin{figure}[htbp]
\begin{center}
\includegraphics[scale=0.35]{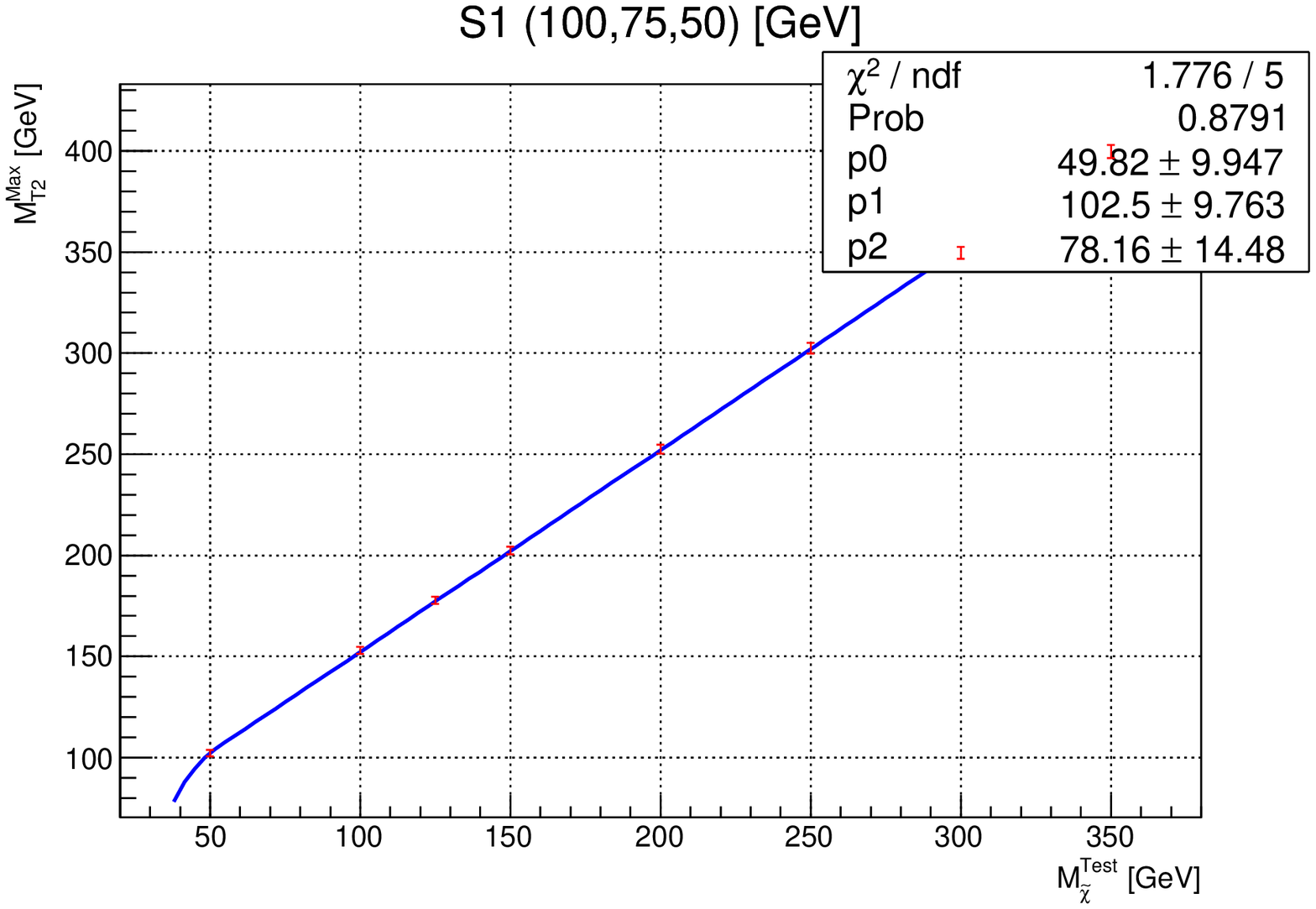}
\includegraphics[scale=0.35]{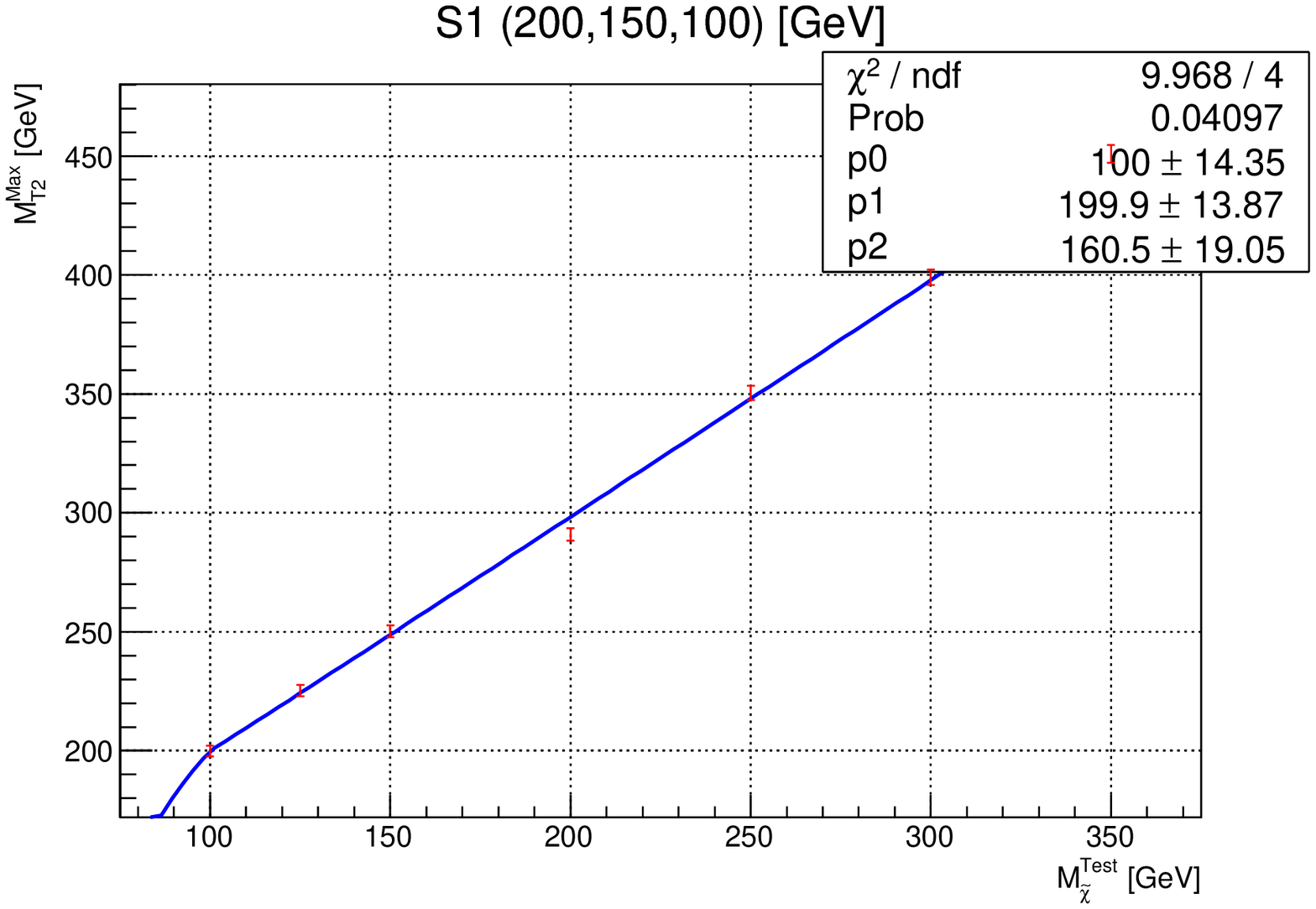}
\includegraphics[scale=0.35]{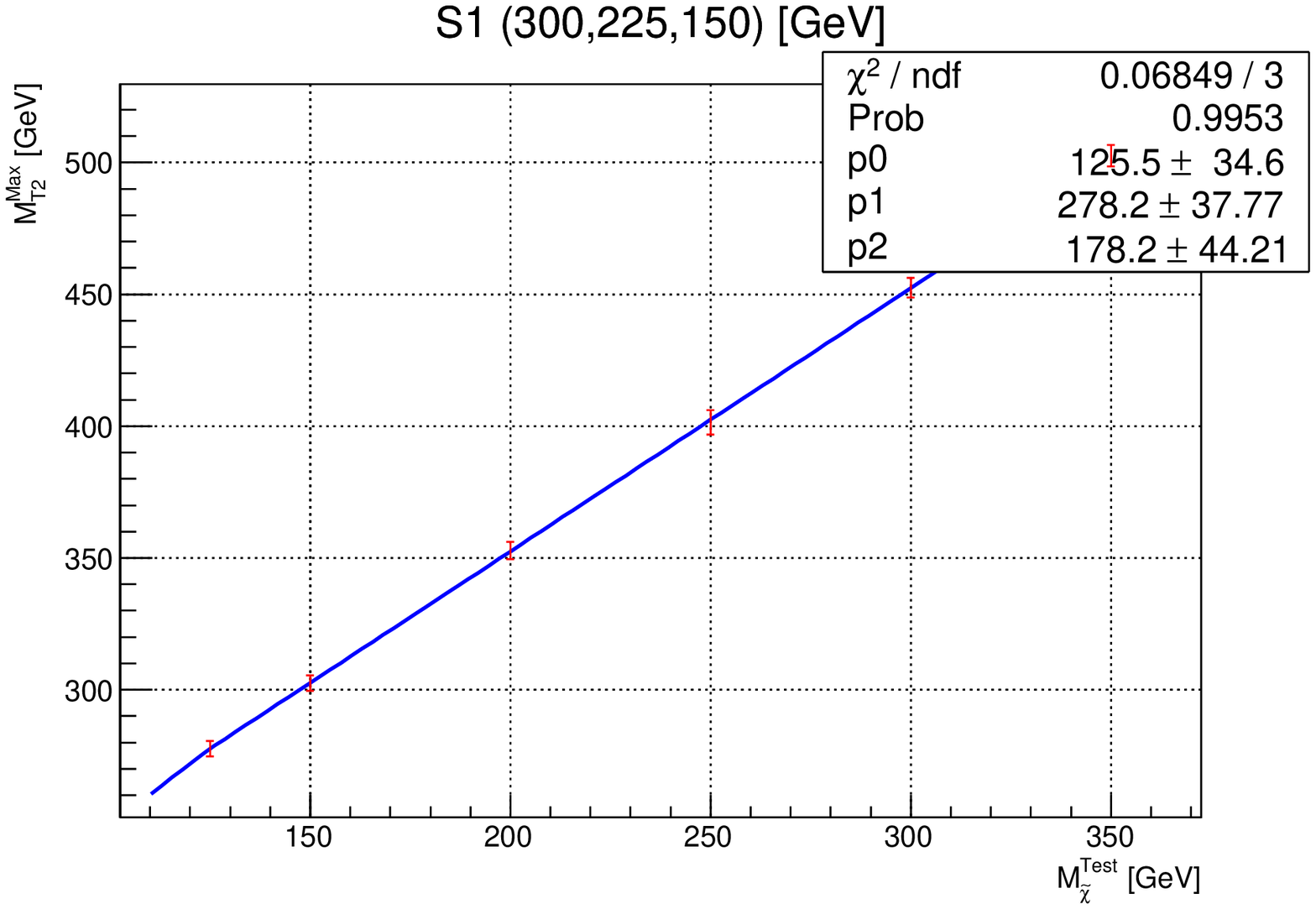}
\includegraphics[scale=0.35]{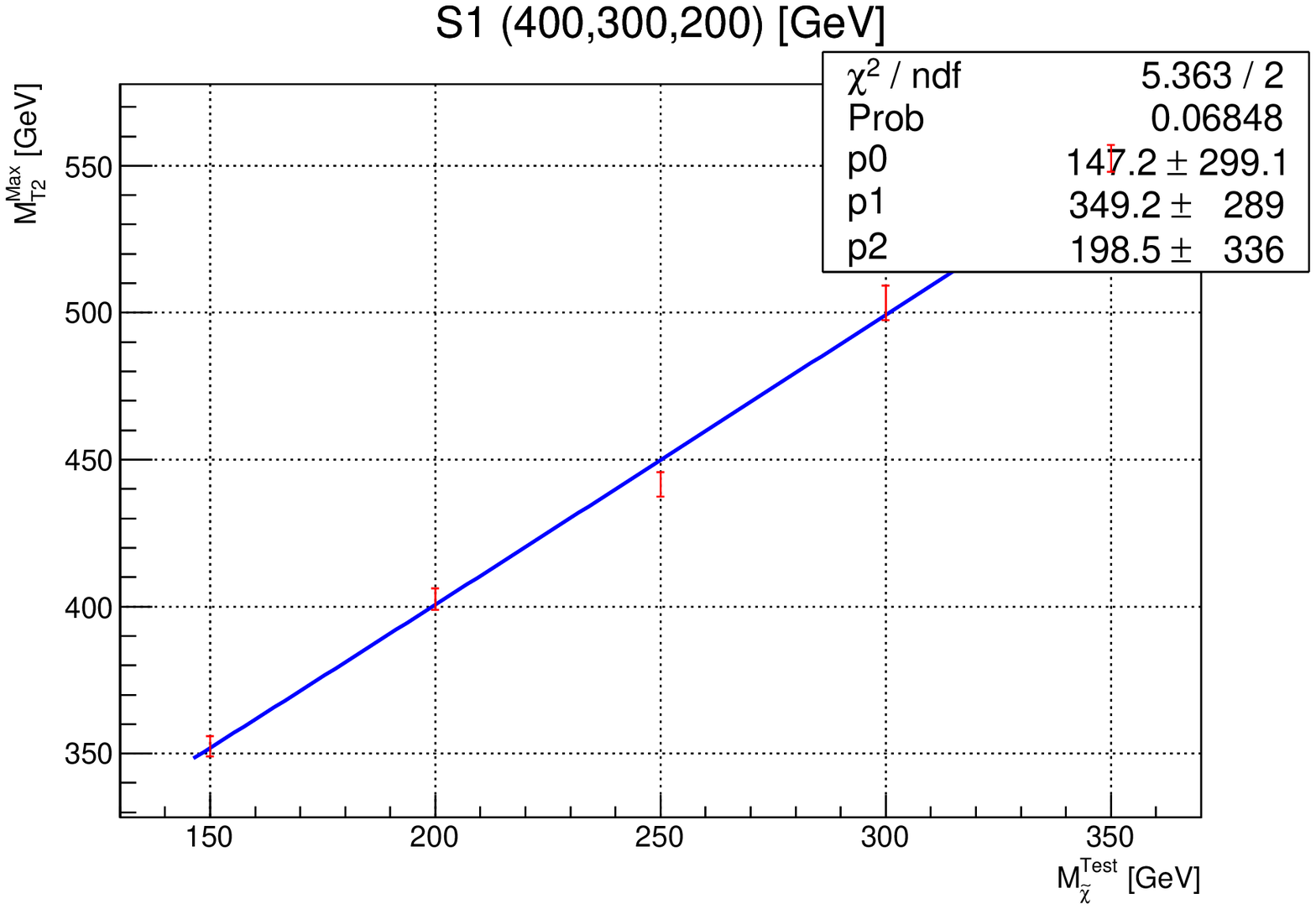}
\caption{\label{Mass-Extract-Plots-S1} Examples of {\it MT2} secondary fits to the S1 samples for $M_{\tilde\chi^{0}_{2}}=$100 (top left), 200 (top right), 300 (bottom left) and 400 (bottom right) GeV.}
\end{center}
\end{figure}

\begin{figure}[htbp]
\begin{center}
\includegraphics[scale=0.35]{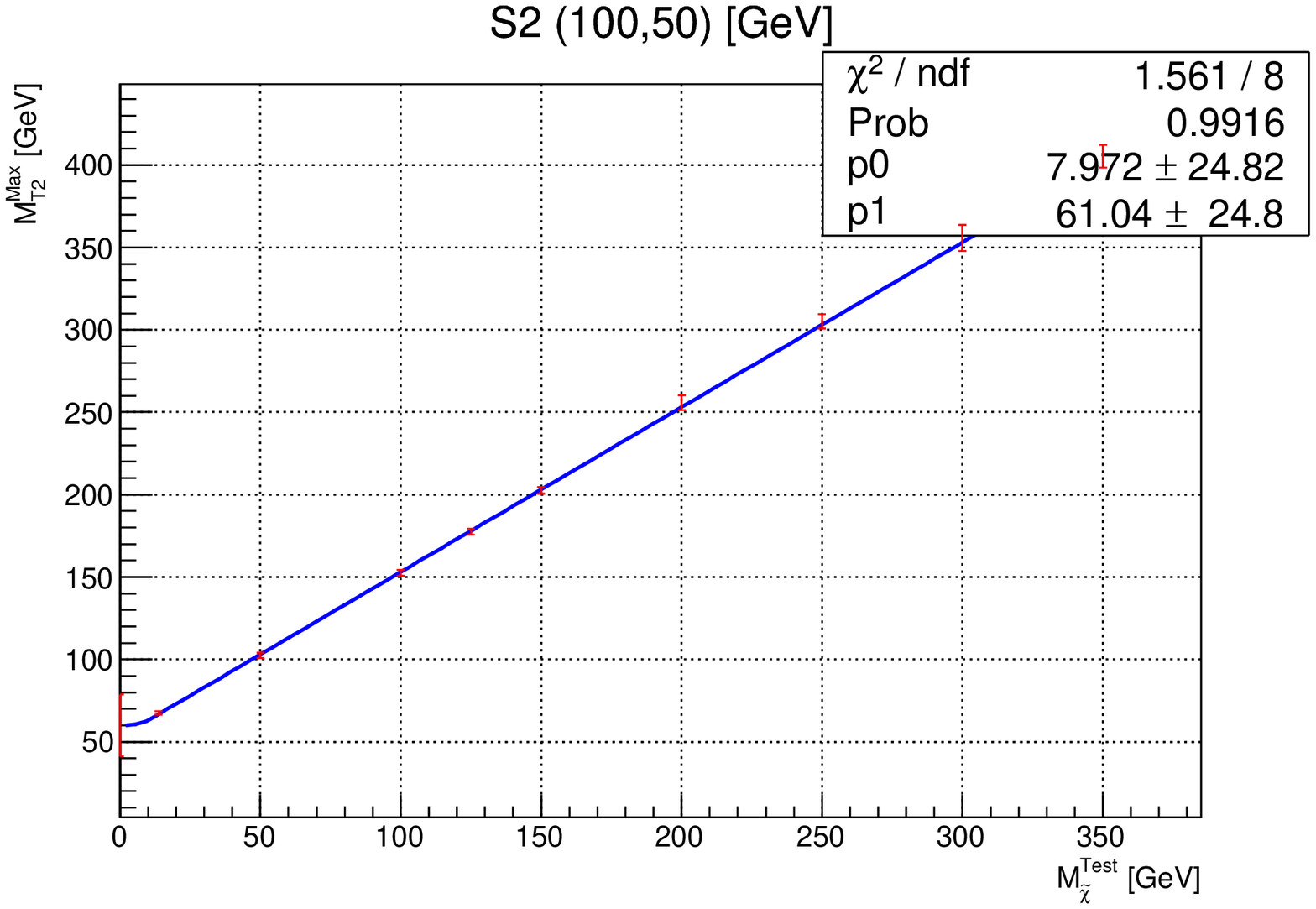}
\includegraphics[scale=0.35]{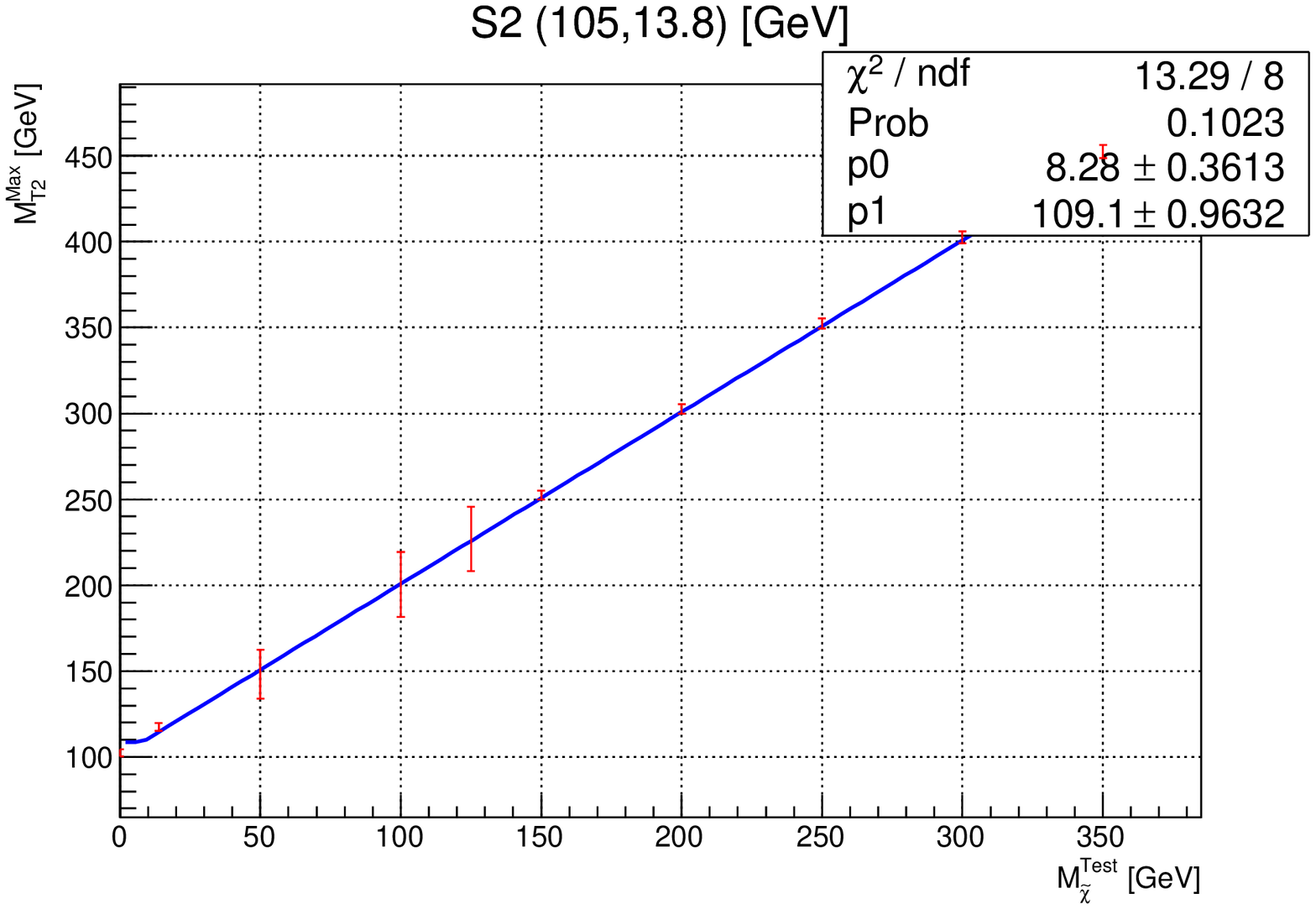}
\includegraphics[scale=0.35]{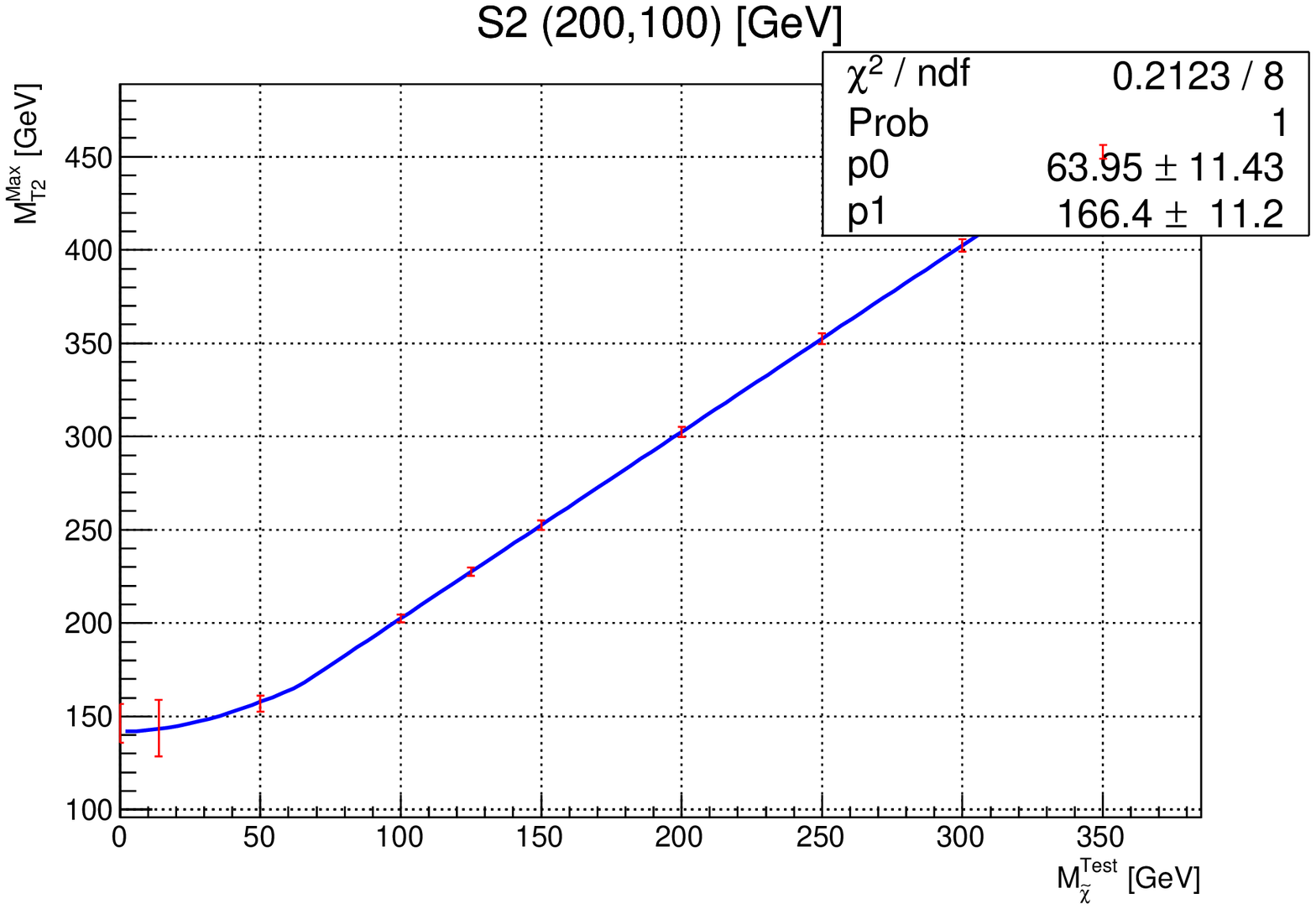}
\includegraphics[scale=0.35]{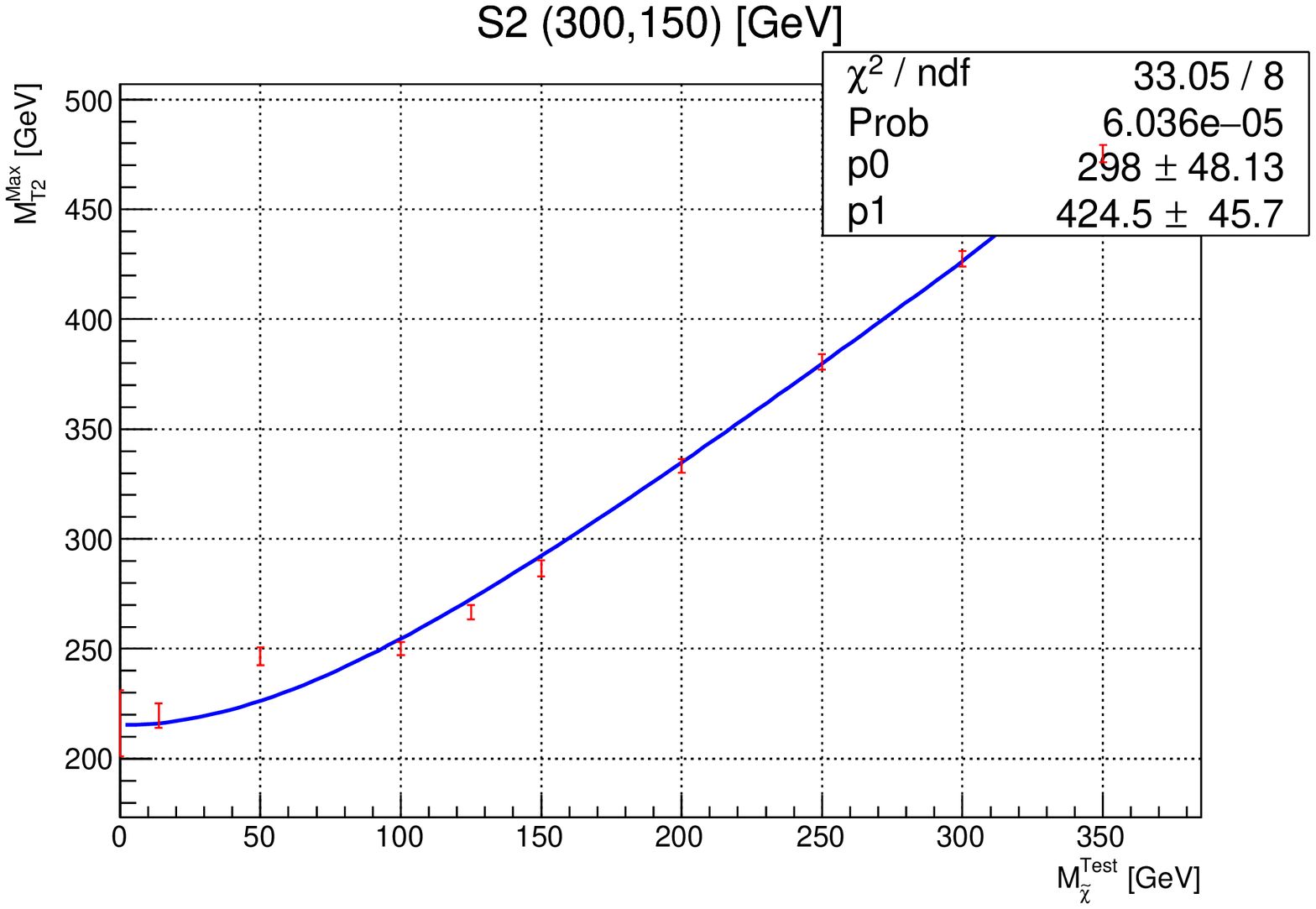}
\caption{\label{Mass-Extract-Plots-S2}  Examples of {\it MT2} secondary fits to the S2 samples for $M_{\tilde\chi^{0}_{2}}=$100 (top left), 105 (top right), 200 (bottom left) and 300 (bottom right) GeV.}
\end{center}
\end{figure}

\noindent
For the secondary fits, the $f_{down}$ and $f_{up}$ functional forms are directly applied onto the $(M^{Max}_{T2},M^{trial}_{\chi})$ two-dimensional plots. The results
of these latter fits, that allow to extract the mass measurements, are presentend in tables \ref{Mass-Extract-first} to \ref{Mass-Extract-last} and a few of them are illustrated in figures \ref{Mass-Extract-Plots-S1} and \ref{Mass-Extract-Plots-S2}.
\begin{table}[h]
\begin{center}
\begin{tabular}{|c|c|c|c|}
\hline 
Process        &  Theor. $M^{Max}_{T2}$   & $M_{T2}$ & Fit $\chi^{2}/N_{dof}$   \\ 
                         & (GeV)                                      & (GeV)         &    	                                   \\
\hline
\underline{Signal S1} & 	 &	&	     \\
$[M_{\tilde\chi^{0}_{2}},M_{\tilde\ell^{\pm}},M_{\tilde\chi^{0}_{1}}]$  \rm\ GeV	        &	&	 &	\\ 
\hline
    $[100,75,50]$	&   Undef.  &   --     &   --    \\  %ok
$[200,150,100]$       &   Undef.  &   --     &   --    \\  %ok
$[300,225,150]$       &   Undef.  &   --     &   --    \\  %ok
$[400,300,200]$       &   Undef.  &   --     &   --    \\  %ok
$[500,375,250]$       &   Undef.  &   --     &   --    \\  %ok
$[600,450,300]$       &   Undef.  &   --     &   --    \\  %ok
$[700,525,350]$       &   Undef.  &   --     &   --    \\  %ok
\hline
\end{tabular}       
\end{center}
\caption{\label{MT2-Max-first} $M^{Max}_{T2}$ measurements of the S1 samples for $M^{trial}_{\chi}=0$ GeV.}
\end{table}
%
%%%
%
\begin{table}[h]
\begin{center}
\begin{tabular}{|c|c|c|c|}
\hline 
Process            & Theor. $M^{Max}_{T2}$ & $M_{T2}$ & Fit $\chi^{2}/N_{dof}$   \\ 
                             &                   (GeV)                 & (GeV)         &                                         \\
\hline
\underline{Signal S2}        &	   &	 &	\\ 
$[M_{\tilde\chi^{0}_{2}},M_{\tilde\chi^{0}_{1}}]$  \rm\ GeV        &   & & \\ 
\hline
   $[100,50]$  &    75.0         &    $  60.00\pm  18.71(stat)\pm  1.12(syst)$   &       1.239     \\  %OK    
$[105,13.8]$  &  103.2         &    $102.39\pm    0.42(stat)\pm  1.90(syst)$   &        2.637    \\  %OK
$[115,13.8]$  &  113.3         &    $102.50\pm    0.12(stat)\pm  1.91(syst)$   &        0.764    \\  %OK
$[125,13.8]$  &  123.5         &    $122.50\pm    0.03(stat)\pm  2.28(syst)$   &        1.006    \\  %OK
$[135,13.8]$  &  133.6         &    $127.80\pm    2.46(stat)\pm  2.38(syst)$   &        0.806    \\  %OK
$[145,13.8]$  &  143.7         &    $136.52\pm  14.31(stat)\pm  2.54(syst)$   &        0.719    \\  %OK
   $[150,50]$  &  133.3         &    $119.99\pm  17.12(stat)\pm  2.23(syst)$   &        1.205    \\  %OK
 $[200,100]$  &  150.0         &    $146.15\pm    9.99(stat)\pm  2.72(syst)$   &        1.210    \\  %OK
 $[250,125]$  &  187.5         &    $188.52\pm  14.57(stat)\pm  3.51(syst)$   &        1.245    \\  %OK
 $[300,150]$  &  225.0         &    $216.17\pm  14.50(stat)\pm  4.02(syst)$   &        1.007    \\  %OK
 $[400,200]$  &             --    &              --                    &          --                \\
 $[500,250]$  &             --    &              --                    &          --                \\
 $[600,300]$  &             --    &              --                    &          --                \\
 $[700,350]$  &             --    &              --                    &          --                \\
\hline
\end{tabular}       
\end{center}
\caption{\label{Mass-Extract-last} $M^{Max}_{T2}$ measurements of the S2 samples for $M^{trial}_{\chi}=0$ GeV.}
\end{table}
%
%%%
%
\begin{table}[h]
\begin{center}
\begin{tabular}{|c|c|c|c|}
\hline 
Process        &  Theor. $M^{Max}_{T2}$   & $M_{T2}$ & Fit $\chi^{2}/N_{dof}$   \\ 
                         & (GeV)                                      & (GeV)         &    	                                   \\
\hline
\underline{Signal S1} & 	 &	&	     \\
$[M_{\tilde\chi^{0}_{2}},M_{\tilde\ell^{\pm}},M_{\tilde\chi^{0}_{1}}]$  \rm\ GeV	        &	&	 &	\\ 
\hline
$[100,75,50]$	   &    Undef.  &   --     &   --    \\  %OK
$[200,150,100]$  &    Undef.  &   --     &   --    \\  %OK
$[300,225,150]$  &    Undef.  &   --     &   --    \\  %OK
$[400,300,200]$  &    Undef.  &   --     &   --    \\  %OK
$[500,375,250]$  &    Undef.  &   --     &   --    \\  %OK
$[600,450,300]$  &    Undef.  &   --     &   --    \\  %OK
$[700,525,350]$  &        --        &   --     &   --    \\
\hline
\end{tabular}       
\end{center}
\caption{$M^{Max}_{T2}$ measurements of the S1 samples for $M^{trial}_{\chi}=13.8$ GeV.}
\end{table}
%
%%%
%
\begin{table}[h]
\begin{center}
\begin{tabular}{|c|c|c|c|}
\hline 
Process         & Theor. $M^{Max}_{T2}$   & $M_{T2}$ & Fit $\chi^{2}/N_{dof}$   \\ 
                         & (GeV)       & (GeV)        &    	                           \\
\hline
\underline{Signal S2} & 	 &	&	     \\
$[M_{\tilde\chi^{0}_{2}},M_{\tilde\chi^{0}_{1}}]$  \rm\ GeV	        &	&	 &	\\ 
\hline
   $[100,50]$	&    77.5  &   $  67.49\pm    0.05(stat)\pm   1.21(syst)$  &    0.976	   \\ % OK
$[105,13.8]$	& 105.0  &   $117.50\pm    0.07(stat)\pm   2.11(syst)$  &    1.423	\\ % OK
$[115,13.8]$	& 115.0  &   $117.50\pm    0.24(stat)\pm   2.11(syst)$  &    1.993 	\\ % OK
$[125,13.8]$	& 125.0  &   $117.50\pm    0.22(stat)\pm   2.11(syst)$  &    0.776	\\ % OK
$[135,13.8]$	& 135.0  &   $128.25\pm    7.48(stat)\pm   2.31(syst)$  &    0.687	\\ % OK
$[145,13.8]$	& 145.0  &   $158.99\pm    1.12(stat)\pm   2.86(syst)$  &    0.478	\\ % OK
   $[150,50]$	 & 134.7  &   $142.67\pm    9.03(stat)\pm   2.57(syst)$  &    0.974	 \\ % OK
 $[200,100]$  	& 151.3  &   $143.74\pm  14.88(stat)\pm   2.59(syst)$  &    0.794      \\ % OK
 $[250,125]$	& 188.5  &   $192.72\pm    4.00(stat)\pm   3.47(syst)$  &    0.590	\\ % OK
 $[300,150]$  	& 225.8  &   $219.64\pm    3.88(stat)\pm   3.95(syst)$  &    0.697	\\ % OK
 $[400,200]$  	& --  &   --          &	  	      --       	     	 \\
 $[500,250]$  	& --  &   --          &	  	      --       	     	 \\
 $[600,300]$  	& --  &   --          &	  	      --       	     	 \\
 $[700,350]$  	& --  &   --          &	  	      --       	     	 \\
\hline
\end{tabular}       
\end{center}
\caption{$M^{Max}_{T2}$ measurements of the S2 samples for $M^{trial}_{\chi}=13.8$ GeV.}
\end{table}
%
%%%
%
\begin{table}[h]
\begin{center}
\begin{tabular}{|c|c|c|c|}
\hline 
Process        & Theor. $M^{Max}_{T2}$       & $M_{T2}$ & Fit $\chi^{2}/N_{dof}$   \\ 
                         & (GeV)         & (GeV)        &    	                           \\
\hline
\underline{Signal S1}        &	&		 &	\\ 
$[M_{\tilde\chi^{0}_{2}},M_{\tilde\ell^{\pm}},M_{\tilde\chi^{0}_{1}}]$  \rm\ GeV	  &	      &		 &	\\ 
\hline
    $[100,75,50]$  &  100.0     &   $102.20\pm   0.31(stat)\pm 1.50(syst)$    &    2.555   \\   % OK
$[200,150,100]$  &  Undef.  &   --     &   --                                                                           \\   % OK
$[300,225,150]$  &  Undef.  &   --     &   --    \\  %OK
$[400,300,200]$  &  Undef.  &   --     &   --    \\  %OK
$[500,375,250]$  &  Undef.  &   --     &   --    \\  %OK
$[600,450,300]$  &  Undef.  &   --     &   --    \\  %OK
$[700,525,350]$  &      --        &   --    &    --    \\
\hline
\end{tabular}  
\end{center}
\caption{$M^{Max}_{T2}$ measurements of the S1 samples for $M^{trial}_{\chi}=50$ GeV.}
\end{table}
%
%%%
%
\begin{table}[h]
\begin{center}
\begin{tabular}{|c|c|c|c|}
\hline 
Process          & Theor. $M^{Max}_{T2}$      & $M_{T2}$ & Fit $\chi^{2}/N_{dof}$   \\ 
                          & (GeV)         & (GeV)        &    	                           \\
\hline
\underline{Signal S2} & 	 &	&	     \\
$[M_{\tilde\chi^{0}_{2}},M_{\tilde\chi^{0}_{1}}]$  \rm\ GeV	        &	&	 &	\\ 
\hline
   $[100,50]$   & 100.0  &    $102.49\pm   0.18(stat)\pm 1.51(syst)$     &    1.096    \\    % OK    
$[105,13.8]$   & 141.2  &    $148.26\pm 14.09(stat)\pm 2.18(syst)$     &    1.371    \\    % OK    
$[115,13.8]$   & 151.2  &    $152.50\pm   0.01(stat)\pm 2.24(syst)$     &    1.366    \\    % OK    
$[125,13.8]$   & 161.2  &    $153.14\pm  3.67(stat)\pm  2.25(syst)$     &    0.759    \\    % OK    
$[135,13.8]$   & 171.2  &    $152.50\pm  0.05(stat)\pm 2.24(syst)$      &    0.493    \\    % OK    
$[145,13.8]$   & 181.2  &    $190.26\pm  9.54(stat)\pm 2.80(syst)$      &    0.602    \\    % OK    
   $[150,50]$   & 150.0  &    $152.50\pm  0.06(stat)\pm 2.24(syst)$      &    1.101    \\    % OK    
 $[200,100]$   & 165.1  &    $156.85\pm  3.68(stat)\pm 2.31(syst)$      &    1.038    \\    % OK    
 $[250,125]$   & 200.0  &    $197.50\pm  2.89(stat)\pm 2.90(syst)$      &    0.630    \\    % OK    
 $[300,150]$   & 235.6  &    $246.67\pm  1.91(stat)\pm 3.63(syst)$      &    0.680    \\    % OK    
 $[400,200]$   & -- &	 -- 	 &	      --			     \\ 
 $[500,250]$   & -- &	 -- 	 &	      --			     \\
 $[600,300]$   & -- &	 -- 	 &	      --			     \\
 $[700,350]$   & -- &	 -- 	 &	      --			     \\
\hline
\end{tabular}       
\end{center}
\caption{$M^{Max}_{T2}$ measurements of the S2 samples for $M^{trial}_{\chi}=50$ GeV.}
\end{table}
%
%%%
%
\begin{table}[h]
\begin{center}
\begin{tabular}{|c|c|c|c|}
\hline 
Process           & Theor. $M^{Max}_{T2}$      & $M_{T2}$ & Fit $\chi^{2}/N_{dof}$   \\ 
                           & (GeV)         & (GeV)        &    	                           \\
\hline
\underline{Signal S1}        &	 &		 &	\\ 
$[M_{\tilde\chi^{0}_{2}},M_{\tilde\ell^{\pm}},M_{\tilde\chi^{0}_{1}}]$  \rm\ GeV	        &	 &		 &	\\ 
\hline
    $[100,75,50]$  &    149.8     &    $152.98\pm 0.15(stat)\pm 1.68(syst)$  &  2.436   \\    % ok   M_SL 
$[200,150,100]$  &    200.0 	&    $199.91\pm 0.35(stat)\pm 2.20(syst)$  &  0.559   \\    % OK     
$[300,225,150]$  &    Undef.  &     --        &      --                                                              \\    % ok   M_SL
$[400,300,200]$  &    Undef.  &     --        &      --                                                              \\    % ok   M_SL
$[500,375,250]$  &    Undef.  &     --        &      --                                                              \\    % ok   M_SL
$[600,450,300]$  &    Undef.  &     --        &      --                                                              \\    % ok   M_SL
$[700,525,350]$  &    --            &     --        &      --              \\
\hline
\end{tabular}  
\end{center}
\caption{$M^{Max}_{T2}$ measurements of the S1 samples for $M^{trial}_{\chi}=100$ GeV.}
\end{table}
%
%%%
%
\begin{table}[h]
\begin{center}
\begin{tabular}{|c|c|c|c|}
\hline 
Process          & Theor. $M^{Max}_{T2}$  & $M_{T2}$ & Fit $\chi^{2}/N_{dof}$   \\ 
                            & (GeV)   & (GeV)        &    	                           \\
\hline
\underline{Signal S2}        &	&	 &	\\ 
$[M_{\tilde\chi^{0}_{2}},M_{\tilde\chi^{0}_{1}}]$  \rm\ GeV	        &&		 &	\\ 
\hline
   $[100,50]$	  & 150.0  & $152.49\pm   0.09(stat)\pm 1.68(syst)$  &  0.584   \\    % OK    
$[105,13.8]$     & 191.2  & $200.44\pm 18.86(stat)\pm 2.20(syst)$  &  1.052   \\    % OK
$[115,13.8]$     & 201.2  & $202.50\pm   0.01(stat)\pm 2.23(syst)$  &  1.138   \\    % OK
$[125,13.8]$     & 211.2  & $202.50\pm   0.13(stat)\pm 2.23(syst)$  &  0.565   \\    % OK
$[135,13.8]$     & 221.2  & $210.14\pm   4.50(stat)\pm 2.31(syst)$  &  0.491   \\    % OK
$[145,13.8]$     & 231.2  & $237.70\pm 12.79(stat)\pm 2.61(syst)$  &  0.558   \\    % OK
   $[150,50]$	  & 200.0  & $202.50\pm   0.10(stat)\pm 2.23(syst)$  &  0.799	\\    % OK
 $[200,100]$     & 200.0  & $202.49\pm   0.01(stat)\pm 2.23(syst)$  &  0.673   \\    % OK
 $[250,125]$	 & 230.8  & $239.16\pm 14.75(stat)\pm 2.63(syst)$  &  0.574   \\    % OK
 $[300,150]$  	 & 263.0  & $250.15\pm   1.24(stat)\pm 2.75(syst)$  &  0.540   \\    % OK
 $[400,200]$  	 & --  &                   --                     &     --          \\
 $[500,250]$  	 & --  &                   --                     &     --          \\
 $[600,300]$  	 & --  &                   --                     &     --          \\
 $[700,350]$  	 & --  &                   --                     &     --          \\
\hline
\end{tabular}       
\end{center}
\caption{$M^{Max}_{T2}$ measurements of the S2 samples for $M^{trial}_{\chi}=100$ GeV.}
\end{table}
%
%%%
%
\begin{table}[h]
\begin{center}
\begin{tabular}{|c|c|c|c|}
\hline 
Process           & Theor. $M^{Max}_{T2}$      & $M_{T2}$ & Fit $\chi^{2}/N_{dof}$   \\ 
                           & (GeV)         & (GeV)        &    	                           \\
\hline
\underline{Signal S1}        &	 &		 &	\\ 
$[M_{\tilde\chi^{0}_{2}},M_{\tilde\ell^{\pm}},M_{\tilde\chi^{0}_{1}}]$  \rm\ GeV        &     &    &   \\ 
\hline
    $[100,75,50]$  &    174.8     &    $177.86\pm 0.13(stat)\pm 1.81(syst)$  &  1.814     \\    % ok
$[200,150,100]$  &    224.9     &    $225.28\pm 0.78(stat)\pm 2.30(syst)$  &  1.284     \\    % ok
$[300,225,150]$  &    258.2     &    $277.64\pm 0.32(stat)\pm 2.83(syst)$  &  0.526    \\    % ok
$[400,300,200]$  &    Undef.  &     --        &      --                                                                \\    % ok
$[500,375,250]$  &    Undef.  &     --        &      --                                                                \\    % ok
$[600,450,300]$  &    Undef.  &     --        &      --                                                                \\    % ok
$[700,525,350]$  &           --     &     --        &      --                                                                \\
\hline
\end{tabular}  
\end{center}
\caption{$M^{Max}_{T2}$ measurements of the S1 samples for $M^{trial}_{\chi}=125$ GeV.}
\end{table}
%
%%%
%
\begin{table}[h]
\begin{center}
\begin{tabular}{|c|c|c|c|}
\hline 
Process       & Theor. $M^{Max}_{T2}$    & $M_{T2}$ & Fit $\chi^{2}/N_{dof}$   \\ 
                        &                       (GeV)                & (GeV)         &                                            \\
\hline
\underline{Signal S2}     &     &     &    \\ 
$[M_{\tilde\chi^{0}_{2}},M_{\tilde\chi^{0}_{1}}]$  \rm\ GeV     &     &     &     \\ 
\hline
   $[100,50]$     & 175.0   &  $177.50\pm    0.06(stat)\pm   1.81(syst)$  &  0.742   \\ %
$[105,13.8]$     & 216.2   &  $227.01\pm  18.62(stat)\pm   2.32(syst)$  &  1.296   \\ %
$[115,13.8]$     & 226.2   &  $227.50\pm    0.01(stat)\pm   2.32(syst)$  &  1.228   \\ %
$[125,13.8]$     & 236.2   &  $227.49\pm    0.03(stat)\pm   2.32(syst)$  &  0.493   \\ %
$[135,13.8]$     & 246.2   &  $227.50\pm    0.04(stat)\pm   2.32(syst)$  &  0.461   \\ %
$[145,13.8]$     & 256.2   &  $246.11\pm    6.54(stat)\pm   2.51(syst)$  &  0.566   \\ %
   $[150,50]$     & 225.0   &  $227.50\pm    0.005(stat)\pm 2.32(syst)$  &  1.167   \\ %
 $[200,100]$     & 225.0   &  $227.50\pm    0.02(stat)\pm   2.32(syst)$  &  0.965   \\ %
 $[250,125]$     & 250.0   &  $250.99\pm  18.17(stat)\pm   2.56(syst)$  &  0.586   \\ %
 $[300,150]$     & 280.7   &  $266.70\pm    1.93(stat)\pm   2.72(syst)$  &  0.566   \\ %
 $[400,200]$     & --  &		   --			 &     --	  \\
 $[500,250]$     & --  &		   --			 &     --	  \\
 $[600,300]$     & --  &		   --			 &     --	  \\
 $[700,350]$     & --  &		   --			 &     --	  \\
\hline
\end{tabular}       
\end{center}
\caption{$M^{Max}_{T2}$ measurements of the S2 samples for $M^{trial}_{\chi}=125$ GeV.}
\end{table}
%
%%%
%
\begin{table}[htpb]
\begin{center}
\begin{tabular}{|c|c|c|c|}
\hline 
Process       &   Theor. $M^{Max}_{T2}$        & $M_{T2}$    & Fit $\chi^{2}/N_{dof}$   \\ 
                        &                    (GeV)                         &    (GeV)         &                                            \\
\hline
\underline{Signal S1}        &   &  &  \\ 
$[M_{\tilde\chi^{0}_{2}},M_{\tilde\ell^{\pm}},M_{\tilde\chi^{0}_{1}}]$  \rm\ GeV      &  &    &     \\
\hline
    $[100,75,50]$     &   199.8     &    $202.50\pm 0.0003(stat)\pm 1.96(syst)$  &   1.857  \\ % OK
$[200,150,100]$     &   249.8     &    $250.29\pm 0.37(stat)\pm     2.43(syst)$  &   0.623  \\ % OK
$[300,225,150]$     &   300.0     &    $302.54\pm 0.52(stat)\pm     2.93(syst)$  &   0.345  \\ % OK
$[400,300,200]$     &   300.0     &    $352.50\pm 0.01(stat)\pm     3.42(syst)$  &   0.239  \\ % OK
$[500,375,250]$     &   Undef.  &     --          &      --                                                                \\    % ok
$[600,450,300]$     &   Undef.  &     --          &      --                                                                \\    % ok
$[700,525,350]$     &        --       &     --          &      --                                                                \\
\hline
\end{tabular}  
\end{center}
\caption{$M^{Max}_{T2}$ measurements of the S1 samples for $M^{trial}_{\chi}=150$ GeV.}
\end{table}
%
%%%
%
\begin{table}[h]
\begin{center}
\begin{tabular}{|c|c|c|c|}
\hline 
Process        &   Theor. $M^{Max}_{T2}$            & $M_{T2}$ & Fit $\chi^{2}/N_{dof}$   \\ 
                        &                           (GeV)                       &    (GeV)      &                                            \\
\hline
\underline{Signal S2}        &	&		 &	\\ 
$[M_{\tilde\chi^{0}_{2}},M_{\tilde\chi^{0}_{1}}]$  \rm\ GeV	        &	&		 &	\\ 
 \hline
   $[100,50]$  & 200.0 & $202.50\pm	0.51(stat)\pm   1.96(syst)$  &  0.920  \\ % OK
$[105,13.8]$  & 241.2 & $252.50\pm     0.003(stat)\pm 2.45(syst)$  &  1.684  \\ % OK
$[115,13.8]$  & 251.2 & $252.50\pm     0.02(stat)\pm   2.45(syst)$  &  1.574  \\ % OK
$[125,13.8]$  & 261.2 & $252.50\pm     0.02(stat)\pm   2.45(syst)$  &  0.716  \\ % OK
$[135,13.8]$  & 271.2 & $252.50\pm     0.03(stat)\pm   2.45(syst)$  &  0.505  \\ % OK
$[145,13.8]$  & 281.2 & $267.16\pm   19.58(stat)\pm   2.59(syst)$  &  0.600  \\ % OK
   $[150,50]$  & 250.0 & $252.50\pm     0.003(stat)\pm 2.45(syst)$  &  1.552  \\ % OK
$[200,100]$   & 250.0 & $252.50\pm     0.07(stat)\pm   2.45(syst)$  &  1.372  \\ % OK
$[250,125]$   & 275.0 & $252.48\pm     0.09(stat)\pm   2.45(syst)$  &  0.701  \\ % OK
$[300,150]$   & 300.0 & $286.68\pm     2.41(stat)\pm   2.78(syst)$  &  0.645  \\ % OK
$[400,200]$   &  -- &    --	&	      -- 		    \\
$[500,250]$   &  -- &    --	&	      -- 		    \\
$[600,300]$   &  -- &    --	&	      -- 		    \\
$[700,350]$   &  -- &    --	&	      -- 		    \\
\hline
\end{tabular}       
\end{center}
\caption{$M^{Max}_{T2}$ measurements of the S2 samples for $M^{trial}_{\chi}=150$ GeV.}
\end{table}
%
%%%
%
\begin{table}[h]
\begin{center}
\begin{tabular}{|c|c|c|c|}
\hline 
Process         &   Theor. $M^{Max}_{T2}$      & $M_{T2}$ & Fit $\chi^{2}/N_{dof}$   \\ 
                          &              (GeV)                             & (GeV)         &    	                           \\
\hline
\underline{Signal S1}        &		& &	\\ 
$[M_{\tilde\chi^{0}_{2}},M_{\tilde\ell^{\pm}},M_{\tilde\chi^{0}_{1}}]$  \rm\ GeV	        &	&	 &	\\ 
\hline
    $[100,75,50]$ & 249.7     &    $252.54\pm  0.04(stat)\pm  2.27(syst)$  &   1.908    \\ % OK
$[200,150,100]$ & 299.7     &    $290.92\pm  0.16(stat)\pm  2.62(syst)$  &   2.085    \\ % OK
$[300,225,150]$ & 349.7     &    $352.89\pm  0.55(stat)\pm  3.18(syst)$  &   0.360    \\ % OK
$[400,300,200]$ & 400.0     &    $402.50\pm  0.01(stat)\pm  3.62(syst)$  &   0.217    \\ % OK
$[500,375,250]$ & 412.0     &    $432.60\pm  0.01(stat)\pm  3.89(syst)$  &   0.008    \\ % OK
$[600,450,300]$ & Undef.  &     --          &      --                                                                \\
$[700,525,350]$ &   --          &     --          &      --                                                                \\
\hline
\end{tabular}  
\end{center}
\caption{$M^{Max}_{T2}$ measurements of the S1 samples for $M^{trial}_{\chi}=200$ GeV.}
\end{table}
%
%%%
%
\begin{table}[h]
\begin{center}
\begin{tabular}{|c|c|c|c|}
\hline 
Process        &   Theor. $M^{Max}_{T2}$           & $M_{T2}$ & Fit $\chi^{2}/N_{dof}$   \\ 
                            & (GeV)         & (GeV)        &    	                           \\
\hline
\underline{Signal S2}   &     &		 &	\\ 
$[M_{\tilde\chi^{0}_{2}},M_{\tilde\chi^{0}_{1}}]$  \rm\ GeV	        &	&	 &	\\ 
\hline
   $[100,50]$  & 250.0  &   $255.83\pm  3.81(stat)\pm   2.30(syst)$  &   1.128    \\ % OK
$[105,13.8]$  & 291.2  &   $302.50\pm  0.05(stat)\pm   2.72(syst)$  &   1.739    \\ % OK
$[115,13.8]$  & 301.2  &   $302.49\pm  0.18(stat)\pm   2.72(syst)$  &   1.733    \\ % OK
$[125,13.8]$  & 311.2  &   $302.50\pm  0.05(stat)\pm   2.72(syst)$  &   0.603    \\ % OK
$[135,13.8]$  & 321.2  &   $302.49\pm  0.05(stat)\pm   2.72(syst)$  &   0.642    \\ % OK
$[145,13.8]$  & 331.2  &   $302.50\pm  0.07(stat)\pm   2.72(syst)$  &   0.592    \\ % OK
   $[150,50]$  & 300.0  &   $302.49\pm  0.07(stat)\pm   2.72(syst)$  &   1.597    \\ % OK
 $[200,100]$  & 300.0  &   $302.50\pm  0.08(stat)\pm   2.72(syst)$  &   1.613    \\ % OK
 $[250,125]$  & 325.0  &   $313.23\pm  3.74(stat)\pm   2.82(syst)$  &   0.844    \\ % OK
 $[300,150]$  & 350.0  &   $333.17\pm  0.86(stat)\pm   3.00(syst)$  &   0.694    \\ % OK
 $[400,200]$  & --  &    --	       &	     --  		   \\
 $[500,250]$  & --  &    --	       &	     --  		   \\
 $[600,300]$  & --  &    --	       &	     --  		   \\
 $[700,350]$  & --  &    --	       &	     --  		   \\
\hline
\end{tabular}       
\end{center}
\caption{$M^{Max}_{T2}$ measurements of the S2 samples for $M^{trial}_{\chi}=200$ GeV.}
\end{table}
%
%%%
%
\begin{table}[h]
\begin{center}
\begin{tabular}{|c|c|c|c|}
\hline 
Process        &   Theor. $M^{Max}_{T2}$           & $M_{T2}$ & Fit $\chi^{2}/N_{dof}$   \\ 
                            & (GeV)        & (GeV)        &    	                           \\
\hline
\underline{Signal S1}    &	    &		 &	\\ 
$[M_{\tilde\chi^{0}_{2}},M_{\tilde\ell^{\pm}},M_{\tilde\chi^{0}_{1}}]$  \rm\ GeV	&	        &		 &	\\ 
\hline
   $[100,75,50]$  &  299.7  &  $302.51\pm  0.02(stat)\pm   2.57(syst)$  &  2.215     \\ % ok
$[200,150,100]$ &  349.6  &  $350.38\pm  0.38(stat)\pm   2.98(syst)$  &  1.160     \\ % ok
$[300,225,150]$ &  399.6  &  $401.39\pm  3.10(stat)\pm   3.41(syst)$  &  0.329     \\ % ok
$[400,300,200]$ &  449.7  &  $441.63\pm  1.70(stat)\pm   3.75(syst)$  &  1.042     \\ % ok
$[500,375,250]$ &  500.0  &  $502.50\pm  0.15(stat)\pm   4.27(syst)$  &  0.212     \\ % OK
$[600,450,300]$ &  516.4  &  $556.34\pm  10.55(stat)\pm 4.73(syst)$  &  0.102     \\ % ok
$[700,525,350]$ &       --    &          --    &      --             \\ 
\hline
\end{tabular}  
\end{center}
\caption{$M^{Max}_{T2}$ measurements of the S1 samples for $M^{trial}_{\chi}=250$ GeV.}
\end{table}
%
%%%
%
\begin{table}[h]
\begin{center}
\begin{tabular}{|c|c|c|c|}
\hline 
Process         &   Theor. $M^{Max}_{T2}$          & $M_{T2}$ & Fit $\chi^{2}/N_{dof}$   \\ 
                            & (GeV)         & (GeV)        &    	                           \\
\hline
\underline{Signal S2}        &	&		 &	\\ 
$[M_{\tilde\chi^{0}_{2}},M_{\tilde\chi^{0}_{1}}]$  \rm\ GeV	        &	&		 &	\\ 
\hline
   $[100,50]$  & 300.0 &    $305.18\pm  3.40(stat)\pm   2.59(syst)$           &    1.084      \\ % OK
$[105,13.8]$  &	341.2 &    $352.49\pm  0.02(stat)\pm   3.00(syst)$           &    1.717      \\ % OK
$[115,13.8]$  &	351.2 &    $352.49\pm  0.21(stat)\pm   3.00(syst)$           &    1.898      \\ % OK
$[125,13.8]$  &	361.2 &    $352.50\pm  0.05(stat)\pm   3.00(syst)$           &    0.796      \\ % OK
$[135,13.8]$  &	371.2 &    $352.50\pm  0.06(stat)\pm   3.00(syst)$           &    0.614      \\ % OK
$[145,13.8]$  &	381.2 &    $362.14\pm  2.43(stat)\pm   3.08(syst)$           &    0.608      \\ % OK
   $[150,50]$  & 350.0 &    $352.50\pm  0.10(stat)\pm   3.00(syst)$           &    1.874      \\ % OK
 $[200,100]$  &	350.0 &    $352.50\pm  0.05(stat)\pm   3.00(syst)$           &    1.551      \\ % OK
 $[250,125]$  &	375.0 &    $362.70\pm  4.50(stat)\pm   3.08(syst)$           &    0.878      \\ % OK
 $[300,150]$  &	400.0 &    $380.55\pm  1.37(stat)\pm   3.23(syst)$           &    0.643      \\ % OK
 $[400,200]$  &	-- &    --   &	       --		     \\
 $[500,250]$  &	-- &    --   &	       --		     \\
 $[600,300]$  &	-- &    --   &	       --		     \\
 $[700,350]$  &	-- &    --   &	       --		     \\
\hline
\end{tabular}       
\end{center}
\caption{$M^{Max}_{T2}$ measurements of the S2 samples for $M^{trial}_{\chi}=250$ GeV.}
\end{table}
%
%%%
%
\begin{table}[h]
\begin{center}
\begin{tabular}{|c|c|c|c|}
\hline 
Process         &   Theor. $M^{Max}_{T2}$          & $M_{T2}$ & Fit $\chi^{2}/N_{dof}$   \\ 
                            & (GeV)         & (GeV)        &    	                           \\
\hline
\underline{Signal S1}      &   &		 &	\\ 
$[M_{\tilde\chi^{0}_{2}},M_{\tilde\ell^{\pm}},M_{\tilde\chi^{0}_{1}}]$  \rm\ GeV	        &	 &	 &	\\ 
\hline
    $[100,75,50]$  &  349.7  &    $349.56\pm    0.12(stat)\pm   2.90(syst)$  &   1.852    \\ % ok
$[200,150,100]$  &  399.5  &    $399.04\pm    0.15(stat)\pm   3.31(syst)$  &   0.746    \\ % ok
$[300,225,150]$  &  449.5  &    $452.50\pm    0.01(stat)\pm   3.76(syst)$  &   0.360    \\ % ok
$[400,300,200]$  &  499.5  &    $503.32\pm    4.22(stat)\pm   4.18(syst)$  &   0.298    \\ % ok
$[500,375,250]$  &  549.7  &    $552.50\pm    0.57(stat)\pm   4.59(syst)$  &   0.253    \\ % ok
$[600,450,300]$  &  600.0  &    $599.40\pm  16.88(stat)\pm   4.97(syst)$  &   0.113    \\ % OK
$[700,525,350]$  &     --      &                        --                                                  &       --        \\
\hline
\end{tabular}  
\end{center}
\caption{$M^{Max}_{T2}$ measurements of the S1 samples for $M^{trial}_{\chi}=300$ GeV.}
\end{table}
\begin{table}[h]
\begin{center}
\begin{tabular}{|c|c|c|c|}
\hline 
Process         &   Theor. $M^{Max}_{T2}$     &        $M_{T2}$   & Fit $\chi^{2}/N_{dof}$   \\ 
                      &                       (GeV)                   &           (GeV)        &                                            \\
\hline
\underline{Signal S2}        &	&		 &	\\ 
$[M_{\tilde\chi^{0}_{2}},M_{\tilde\chi^{0}_{1}}]$  \rm\ GeV	        &	&		 &	\\ 
\hline
   $[100,50]$  &  350.0  &    $355.75\pm    7.34(stat)\pm 2.95(syst)$          &   0.982       \\ % OK
$[105,13.8]$  &  391.2  &    $402.50\pm    0.31(stat)\pm 3.34(syst)$          &   1.465       \\ % OK
$[115,13.8]$  &  401.2  &    $402.50\pm    0.15(stat)\pm 3.34(syst)$          &   1.596       \\ % OK
$[125,13.8]$  &  411.2  &    $402.50\pm    0.18(stat)\pm 3.34(syst)$          &   0.643       \\ % OK
$[135,13.8]$  &  421.2  &    $402.50\pm    0.21(stat)\pm 3.34(syst)$          &   0.545       \\ % OK
$[145,13.8]$  &  431.2  &    $402.50\pm    0.07(stat)\pm 3.34(syst)$          &   0.517       \\ % OK
   $[150,50]$  &  400.0  &    $402.50\pm    0.02(stat)\pm 3.34(syst)$          &   1.755       \\ % OK
 $[200,100]$  &  400.0  &    $402.50\pm    0.26(stat)\pm 3.34(syst)$          &   1.456       \\ % OK
 $[250,125]$  &  425.0  &    $403.75\pm    9.21(stat)\pm 3.35(syst)$          &   0.730       \\ % OK
 $[300,150]$  &  450.0  &    $427.50\pm    0.34(stat)\pm 3.55(syst)$          &   0.635       \\ % OK
 $[400,200]$  &  --  &    --   & 	    --			  \\
 $[500,250]$  &  --  &    --   & 	    --			  \\
 $[600,300]$  &  --  &    --   & 	    --			  \\
 $[700,350]$  &  --  &    --   & 	    --			  \\
\hline
\end{tabular}       
\end{center}
\caption{$M^{Max}_{T2}$ measurements of the S2 samples for $M^{trial}_{\chi}=300$ GeV.}
\end{table}
%
%%%
%
\clearpage
\begin{table}
\begin{center}
\begin{tabular}{|c|c|c|c|}
\hline 
Process        &   Theor. $M^{Max}_{T2}$           & $M_{T2}$ & Fit $\chi^{2}/N_{dof}$   \\ 
                          & (GeV)       & (GeV)        &    	                           \\
\hline
\underline{Signal S1}        &		& &	\\ 
$[M_{\tilde\chi^{0}_{2}},M_{\tilde\ell^{\pm}},M_{\tilde\chi^{0}_{1}}]$  \rm\ GeV	        &	&	 &	\\ 
\hline
    $[100,75,50]$ &  399.7   &	$399.58\pm    0.12(stat)\pm   3.24(syst)$  &   2.513	\\% OK
$[200,150,100]$ &  449.5   &    $450.82\pm 0.69(stat)\pm   3.65(syst)$	 &   1.140    \\% OK
$[300,225,150]$ &  499.4   &    $502.50\pm 0.04(stat)\pm   4.07(syst)$	 &   0.429    \\% OK
$[400,300,200]$ &  549.4   &    $552.50\pm 0.01(stat)\pm   4.48(syst)$	 &   0.384    \\% OK
$[500,375,250]$ &  599.5   &    $586.74\pm  11.04(stat)\pm   4.75(syst)$  &   0.130    \\% OK
$[600,450,300]$ &  649.7   &    $651.60\pm 2.46(stat)\pm   5.28(syst)$	 &   0.108    \\% OK
$[700,525,350]$ &       --     &    --   &	     --  		   \\	
\hline
\end{tabular}  
\caption{$M^{Max}_{T2}$ measurements of the S1 samples for $M^{trial}_{\chi}=350$ GeV.}
\end{center}
\end{table}
%
%%%
%
\clearpage
\begin{table}[h]
\begin{center}
\begin{tabular}{|c|c|c|c|}
\hline 
Process         &   Theor. $M^{Max}_{T2}$      & $M_{T2}$ & Fit $\chi^{2}/N_{dof}$   \\ 
                      &                        (GeV)                   & (GeV)        &                                            \\
\hline
\underline{Signal S2}   &                                   &                    &                                            \\ 
$[M_{\tilde\chi^{0}_{2}},M_{\tilde\chi^{0}_{1}}]$  \rm\ GeV        & &  &                    \\ 
\hline
   $[100,50]$   & 400.0 &     $405.33\pm    5.85(stat)\pm 3.28(syst)$          &     1.156     \\ % OK
$[105,13.8]$   & 441.2 &     $452.50\pm    0.01(stat)\pm 3.67(syst)$          &     1.391     \\ % OK
$[115,13.8]$   & 451.2 &     $452.50\pm    0.29(stat)\pm 3.67(syst)$          &     1.656     \\ % OK
$[125,13.8]$   & 461.2 &     $452.50\pm    0.07(stat)\pm 3.67(syst)$          &     0.748     \\ % OK
$[135,13.8]$   & 471.2 &     $452.48\pm    0.06(stat)\pm 3.67(syst)$          &     0.612     \\ % OK
$[145,13.8]$   & 481.2 &     $452.50\pm    0.19(stat)\pm 3.67(syst)$          &     0.593     \\ % OK
   $[150,50]$   & 450.0 &     $470.85\pm    3.14(stat)\pm 3.81(syst)$          &     1.471     \\ % OK
 $[200,100]$   & 450.0 &     $452.50\pm    0.06(stat)\pm 3.67(syst)$          &     1.147     \\ % OK
 $[250,125]$   & 475.0 &     $480.09\pm  11.87(stat)\pm 3.89(syst)$          &     0.846     \\ % OK
 $[300,150]$   & 500.0 &     $475.29\pm    0.63(stat)\pm 3.85(syst)$          &     0.709     \\ % OK
 $[400,200]$   & -- &    --     &     -- \\
 $[500,250]$   & -- &    --     &     -- \\
 $[600,300]$   & -- &    --     &     -- \\
 $[700,350]$   & -- &    --     &     -- \\
\hline
\end{tabular}       
\caption{\label{MT2-Max-last} $M^{Max}_{T2}$ measurements of the S2 samples for $M^{trial}_{\chi}=350$ GeV.}
\end{center}
\end{table}

%%%%%%%%%%%%%%%%%%%%%%%%%%%%%%%%%%%%%%%%%%%%%%%%%%%%%%%%%%%%%%%%%%%%%

\vspace*{1.5mm}
\noindent
3.6.2.\ c.\ Mass\ Extraction
\vspace*{0.5mm}
\begin{table}[h]
\begin{center}
\begin{tabular}{|c|c|c|c|c|}
\hline 
Process        &   $M^{Fit}_{\tilde\chi^{0}_{2}}$     & $M^{Fit}_{\tilde\ell^{\pm}}$  &  $M^{Fit}_{\tilde\chi^{0}_{1}}$  &  Fit $\chi^{2}/N_{dof}$    \\ 
                     &                           (GeV)                          &                           (GeV)                 &                             (GeV)                    &                                              \\
\hline
\underline{Signal S1}        &                                       &                                                        &                                                           &                                              \\ 
$[M_{\tilde\chi^{0}_{2}},M_{\tilde\ell^{\pm}},M_{\tilde\chi^{0}_{1}}]$  \rm\ GeV & & &                                                  &                                              \\ 
\hline
$[100,  75,  50]$ &   $102.49\pm       9.76$  &    $  78.16\pm    14.48$   &  $  49.82\pm       9.95$   &  0.355  \\
$[200,150,100]$ &   $199.86\pm     13.87$  &    $160.50\pm    19.05$   &  $100.00\pm     14.35$   &  2.492  \\
$[300,225,150]$ &   $278.16\pm     37.77$  &    $178.15\pm    44.21$   &  $125.45\pm     34.60$   &  0.023  \\
$[400,300,200]$ &   $349.20\pm   288.96$  &    $198.48\pm  336.04$   &  $147.22\pm   299.12$   &  2.681  \\
$[500,375,250]$ &   $501.50\pm       2.96$  &     $339.83\pm    15.23$  &  $250.00\pm       0.10$   &  1.576  \\
$[600,450,300]$ &   $555.32\pm 1059.60$  &     $312.33\pm1239.21$  &  $249.66\pm 1125.28$   &      --    \\
$[700,525,350]$ &                   --                   &                          --             &                     --                 &      --    \\   
\hline
\end{tabular} 
\caption{\label{Mass-Extract-first} Mass extraction from $M^{Max}_{T2}$ measurements of the S1 samples.}
\end{center}
\end{table}

\begin{table}[h]
\begin{center}
\begin{tabular}{|c|c|c|c|}
\hline 
Process        &   $M^{Fit}_{\tilde\chi^{0}_{2}}$     &   $M^{Fit}_{\tilde\chi^{0}_{1}}$  &  Fit $\chi^{2}/N_{dof}$    \\ 
                         &                           (GeV)                          &                                       (GeV)           &                                              \\
\hline
\underline{Signal S2}        &    &      &   \\ 
$[M_{\tilde\chi^{0}_{2}},M_{\tilde\chi^{0}_{1}}]$  \rm\ GeV	        &	 & &	\\ 
\hline
$[100,   50]$ &   $   61.04\pm  24.80$   &    $    7.97\pm 24.82$  &  0.195   \\
$[105,13.8]$ &   $ 109.09\pm    0.96$   &    $    8.28\pm   0.36$  &  1.661   \\
$[115,13.8]$ &   $ 109.67\pm    0.78$   &    $    8.28\pm   0.32$  &  1.788   \\
$[125,13.8]$ &   $ 122.14\pm    2.26$   &    $  19.61\pm   2.65$  &  0.561   \\
$[135,13.8]$ &   $ 135.55\pm    5.53$   &    $  32.76\pm   5.76$  &  0.276   \\
$[145,13.8]$ &   $ 217.75\pm 14.22$    &    $112.56\pm 15.09$  &  2.706   \\
$[150,   50]$ &   $152.17\pm  18.13$    &    $  49.01\pm 18.22$  &  1.811   \\
$[200, 100]$ &   $166.44\pm  11.20$    &    $  63.95\pm 11.43$  &  0.027   \\
$[250, 125]$ &   $262.12\pm    1.55$    &    $150.00\pm   0.03$  &  4.118   \\
$[300, 150]$ &   $424.48\pm  45.70$    &    $297.99\pm 48.13$  &  4.131   \\
$[400, 200]$ &   -- &   -- & -- \\
$[500, 250]$ &   -- &   -- & -- \\
$[600, 300]$ &   -- &   -- & -- \\
$[700, 350]$ &   -- &   -- & -- \\
\hline
\end{tabular}  
\end{center}
\caption{\label{Mass-Extract-last} Mass extraction from $M^{Max}_{T2}$ measurements of the S2 samples.}
\end{table}


\par\noindent
Once again, we notice, that {\it ICA} and {\it MT2} methods do not give access to the same informations: $M_{\tilde\chi^{0}_{2}}+M_{\tilde\chi^{\pm}_{1}}$,
versus  $M_{\tilde\chi^{0}_{2}}$, and $M_{\tilde\chi^{0}_{1}}$ (plus possibly  $M_{\tilde\ell^{\pm}}$), respectively. The precision of the {\it MT2} mass measurements are summarized hereafter:
\begin{itemize}
\item S1 signal:
	\begin{itemize}
	\item $\delta M_{\tilde\chi^{0}_{2}} / M_{\tilde\chi^{0}_{2}} <   7-14\%$ for  $M_{\tilde\chi^{0}_{2}} < 400$ GeV
	\item $\delta M_{\tilde\ell^{\pm}} / M_{\tilde\ell^{\pm}}        < 12-25\%$ for  $M_{\tilde\chi^{0}_{2}} < 400$ GeV
	\item $\delta M_{\tilde\chi^{0}_{1}} / M_{\tilde\chi^{0}_{1}} < 14-28\%$ for  $M_{\tilde\chi^{0}_{2}} < 400$ GeV
	\end{itemize}
\item S2a signal:
	\begin{itemize}
	\item $\delta M_{\tilde\chi^{0}_{2}} / M_{\tilde\chi^{0}_{2}} < 41\%$ for  $M_{\tilde\chi^{0}_{2}} < 400$ GeV
	\item bad sensitivity to $M_{\tilde\chi^{0}_{1}}$
	\end{itemize}
\item S2b signal:
	\begin{itemize}
	\item $\delta M_{\tilde\chi^{0}_{2}} / M_{\tilde\chi^{0}_{2}} < 0.6-12\%$ for  $M_{\tilde\chi^{0}_{2}} < 400$ GeV
	\item $\delta M_{\tilde\chi^{0}_{1}} / M_{\tilde\chi^{0}_{1}} < 4-13\%$    for  $M_{\tilde\chi^{0}_{2}} < 150$ GeV
	\end{itemize}
\end{itemize}

\noindent
Even though the {\it MT2} method, appears to be slightly less accurate than {\it ICA} (itself being much less accurate than {\it DileME}), it provides much more informations on different individual particles mass than {\it ICA}, or {\it DileME}, or even a combination of {\it ICA} and {\it DileME}. 
However $M_{T2}$ end-points are known to be sometimes difficult to measure \cite{Curtin:2011ng}, especially for small signals in the presence of some background. 
\par\noindent
The last remark, is that {\it ICA} appears to have  a higher mass reach than {\it DileME} and {\it MT2}. This is mostly due to the {\it ICA}  reduced systematic uncertainty in its background subtraction.
\par\noindent
So, we see that the three methods have quite different advantages and drawbacks, they also have different systematic uncertainties. They are therefore complementary and the best SUSY mass informations can be extracted by combining them.

%%%%%%%%%%%%%%%%%%%%%%%%%%%%%%%%%%%%%%%%%%%%%%%%%%%%%%%%%%%%%%%%%%%%%%%%%%%%%%%%%%%%%%%%%%%%%%%% 
% CONCLUSIONS & PROSPECTS
%%%%%%%%%%%%%%%%%%%%%%%%%%%%%%%%%%%%%%%%%%%%%%%%%%%%%%%%%%%%%%%%%%%%%%%%%%%%%%%%%%%%%%%%%%%%%%%%

\vfill\eject\clearpage\newpage
\vspace*{5mm}
\section{\label{Concl} Conclusions}
\vspace*{2.5mm}
\par
We propose a new method to measure the mass of charged final states using the integral charge asymmetry $A_{C}$ at the LHC.
\par\noindent
At first we detail and test this method on the $p+p\to W^{\pm}\to\ell^{\pm}\nu$ inclusive process. Then we apply it on a SUSY search of interest, namely the 
$p+p\to\tilde\chi^{\pm}_{1}+\tilde\chi^{0}_{2}\to 3\ell^{\pm}+\rlap{\kern0.25em/}E_{T}$ inclusive process.
\noindent
For each process, we start by calculating the central values of $A_{C}$ using cross section integrators with LO MEs and with three different LO PDFs. MCFM is used for the SM process and Resummino is used for the SUSY process. The same tools are also used to estimate the theoretical unceratinties on $A_{C}$. These calculations are repeated varying the mass of the charged final state. Over the studied mass ranges we find that $A_{C}$ is a monotically increasing function of $M(FS^{\pm})$. This function is well described by a polynomial of logarithms of logarithms of $M(FS^{\pm})$. The PDF uncertainty turns out to be the dominant source of the theoretical uncertainty. 
\par\noindent
The experimental extraction of $A_{C}$ requires a quantitative estimate of the biases caused by the event selection and by the residual background. To this end MC samples
are generated for the considered signal and its related background processes. These samples are passed through a fast simulation of the ATLAS detector response. Realistic values for the systematic uncertainties are taken from publications of LHC data analyses. The full experimetal uncertainties as well as the effect of the residual background are consistently propagated through a central value and uncertainties of the measured $A_{C}$. This way the measured $A_{C}$ of each signal sample can be translated into a central value and uncertainties of an indirect measurement of the corresponding $M(FS^{\pm})$. The theoretical uncertainties of each measured  $M(FS^{\pm})$ is summed in quadrature with the experimental uncertainties so as to provide the full uncertainty for this new method.
\par\noindent
For the $p+p\to W^{\pm}\to\ell^{\pm}\nu$ inclusive process, $M_{W^{\pm}}$ can be indirectly measured with an overall accuracy better than $1.2\%$. We note that the dispersion of the central values of  $M_{W^{\pm}}$ indirectly measured with the three PDFs are compatible with the total uncertainty of the MSTW2008lo68cl prediction.
\par\noindent
For the $p+p\to\tilde\chi^{\pm}_{1}+\tilde\chi^{0}_{2}\to 3\ell^{\pm}+\rlap{\kern0.25em/}E_{T}$ inclusive process, without accounting for $\delta (A_{C})_{PDF}$, $M_{\tilde\chi^{\pm}_{1}}+M_{\tilde\chi^{0}_{2}}$ can be measured with an overall accuracy better than $6\%$ for a sensitivity to the signal in excess of $5\sigma$ and with an accuracy better than $4.5\%$ for a sensitivity to the signal in excess of $3\sigma$. These indirect mass measurements are independent of the details of the decay chains of the signal samples. For the considered SUSY process, basic closure tests indicate the indirect mass estimate does not need any linearity nor offset corrections.
\par\noindent
We recommend to apply this method using at least NLO $A_{C}$ templates both for the theoretical and the experimental parts. Indeed, the most precise cross sections and event generations constitute more reliable theory predictions and are in better agreement with the data than LO predictions. NLO or NLL theoretical templates reduce
the theoretical uncertainty, as shown in table \ref{II-parton-LVL:Tab:AC_MSTW} for example. Besides, the measurements of $\frac{dA_{C}(W^{\pm}\to\ell^{\pm}\nu)}{d\eta(\ell^{\pm})}$ by the LHC experiments \cite{Aad:2011yna} \cite{Chatrchyan:2012xt} \cite{Chatrchyan:2013mza} 
 \cite{LHCb:2011xha} \cite{ATLAS:2011pha} were found to agree well with NLO theory predictions. Even if our asymmetry ratios of the $A_{C}$ theoretical templates: $\frac{A_{C}^{NLO}}{A_{C}^{LO}}$ in Fig. \ref{I-parton-LVL:Fig4} and  $\frac{A_{C}^{NLL}}{A_{C}^{LO}}$ in Fig. \ref{AC_w1z2_Compare}, reveal important shape difference of the higher orders with respect to LO, the size of the corrections remain nevertheless quite modest.
\par\noindent
Finally, the comparison of the {\it ICA} (Integral Charge Asymmetry) method for SUSY mass measurements, to the {\it DileME} (Dilepton Mass Edge) and to the {\it MT2} (stransverse mass), shows that these three methods are quite complementary.
\begin{itemize}
\item the {\it DileME} method is the most precise one, but it can only access  a mass difference and it has a strong bias in certain situations (S2b signal);
\item the {\it MT2} method is the least precise one, it may be difficult to exploit in certain cases, but it provides constraints on individual mass (parent, possibly intermediate and end daughter particle);
\item the {\it MT2} method is slightly more precise than {\it MT2}, it has the largest mass reach, but it can only access a mass sum.
\end{itemize}

%\newpage
\vspace*{5mm}
\section{\label{Prospects} Prospects}
\vspace*{2.5mm}
\par\noindent
In this article we have envisaged two production processes for which the mass measurement from the integral charge asymmetry is applicable. One SM inclusive process 
$p+p\to W^{\pm}\to 1\ell^{\pm}+\rlap{\kern0.25em/}E_{T}$ and one SUSY inclusive process $p+p\to\tilde\chi^{\pm}_{1}+\tilde\chi^{0}_{2}\to 3\ell^{\pm}+\rlap{\kern0.25em/}E_{T}$. Here are the typical physics cases where we think the indirect mass measurement is applicable and complementary with respect to usual mass reconstruction techniques: 
\begin{itemize}
\item Initial state (IS): processes induced by $q+\bar q$, or $q+g$
\item Final state (FS): situations where the clasiscal reconstruction techniques are degraded because of
	\begin{itemize}
	\item bad energy resolution for some objects ($\tau^{\pm}_{had}$, jets, b-jets,...)  combined with a limited statistical significance\\
                  (i.e. channels with $\tau^{\pm}_{had}$ compared to channels with $e^{\pm}$ or $\mu^{\pm}$)
	\item and especially where many particles are undetected 
	\end{itemize}
\end{itemize}

\par\noindent
For models with an extended Higgs sector:
the $H^{\pm\pm}(\to W^{\pm}W^{\pm})+H^{\mp}(\to\ell^{\mp}\nu)\to \ell^{\pm}\ell^{\pm}+\ell^{\mp}+\rlap{\kern0.25em/}E_{T}$ channel could be a good physics case because there are 3 undetected neutrinos. On the contrary, for $H^{\pm\pm}+H^{\mp}\to \ell^{\pm}\ell^{\pm}+\ell^{\mp}+\rlap{\kern0.25em/}E_{T}$, $M_{T}$ templates should be more accurate.

\par\noindent
Other physics cases could be searches for $W'^{\pm}\to\mu^{\pm}\nu$ and for $W'^{\pm}\to t\bar b$. 

\par\noindent
In SUSY models, here's a non-exhaustive list of processes of interest:
\begin{itemize}
\item For "semi-weak" processes: 
	\begin{itemize}
	\item $\tilde\chi^{\pm}_{1}+\tilde q$, for which $M_{\tilde\chi^{\pm}_{1}}+M_{\tilde q}$ could be measured
	\item $\tilde\chi^{\pm}_{1}+\tilde g$, for which $M_{\tilde\chi^{\pm}_{1}}+M_{\tilde g}$ could be measured
	\end{itemize}
\item For "weak" processes:
	\begin{itemize}
	\item Slepton sector: $\tilde\ell^{\pm}+\tilde\nu$, for which  $M_{\tilde\ell^{\pm}}+M_{\tilde\nu}$ could be measured
	\item Chargino-neutralino sector: $\tilde\chi^{\pm}_{1}+\tilde\chi^{0}_{1,2,3}$, to measure $M_{\tilde\chi^{\pm}_{1}}+M_{\tilde\chi^{0}_{1,2,3}}$
	\end{itemize}
\end{itemize}
Note, that with the increasing center-of-mass energies and the increasing integrated luminosities of the LHC runs in the years to come, all the vector boson fusion production
modes of the above cited processes could also become testable.
\par\noindent
This new method only applies after a given event selection and it is indicative of the mass of the final state produced by a charged current process, only when
the event selection provides a good statistical significance for that process. 
Further studies should determine wether a differential charge asymmetry can be used to improve the separation between a given signal and its related background processes and therefore improve the sensitivity to some of this signal properties.
\par\noindent
Differential charge asymmetries have been extensively used in other search contexts. For example, in attempts to explain the large forward-backward asymmetries of the $t\bar t$ production measured at the TEVATRON by both the CDF \cite{Aaltonen:2011kc} and the D0 \cite{Abazov:2011rq} experiments, some studies were carried out at the LHC to constrain possible contributions from an extra $W'^{\pm}$ boson. See for example \cite{Knapen:2011hu}\cite{Chatrchyan:2012su}, using a differential charge asymmetry with respect to a three-body invariant mass, and also \cite{Craig:2011an}, using an integral charge asymmetry, and the references therein. Such analyses, using charge asymmetries with respect to the $t\bar t$ system rapidity, invariant mass and transverse momentum, have also been performed by the ATLAS and CMS collaborations, see \cite{Aad:2013cea} and \cite{Chatrchyan:2014yta}, respectively. 
We should also mention the differential charge asymmetry with respect to a two-body invariant mass which served as a discriminant between 
some BSM underlying models \cite{Barr:2004ze}\cite{Smillie:2005ar}, namely SUSY versus Universal Extra Dimension \cite{Appelquist:2000nn} models, in the study of some specific decay chains.
\par\noindent
For what concerns the current article, a first look at the differential charge asymmetry versus the pseudo-rapidity of the charged lepton coming from the chargino decay, reveals promising shape differences between the SM background and the $p+p\to\tilde\chi^{\pm}_{1}+\tilde\chi^{0}_{2}$ SUSY signals. However detailed results are awaiting further studies.

%%%%%%%%%%%%%%%%%%%%%%%%%%%%%%%%%%%%%%%%%%%%%%%%%%%%%%%%%%%%%%%%%
\acknowledgments

We would like to thank the CCIN2P3 computing facilty in Lyon where we produced, stored and analyzed our MC samples.
\noindent
The corresponding author thanks Ben O'Leary, Abdelhak Djouadi, and Gordon Watts for useful discussions. 
He also adresses a special word of thanks to the authors of Resummino, of MCFM, and of Delphes for their help and availability.

%%%%%%%%%%%%%%%%%%%%%%%%%%%%%%%%%%%%%%%%%%%%%%%%%%%%%%%%%%%%%%%%%
\newpage
\appendix
\section{Appendix: Toy Models for the Evolution of $A_{C}$} % Main appendix title
\label{AppendixA} % For referencing this appendix elsewhere, use \ref{AppendixA}
This section is by no mean a formal proof of the properties of the functional forms utilized to fit the different $A_{C}$ template curves.
It's rather a numerical illsutration that render these properties plausible.

\subsection{Numerical Example of Evolution of the PDFs, the Quark Currents and $A_{C}$}
In this paragraph, we describe in a simplified scheme, the choice of these functional forms aimed at fitting:
\begin{enumerate}
\item the proton u and d quarks and anti-quarks density functions,
\item the quark currents in the initial state,
\item the dominant flavour contribution to the LO expression of $A_{C}$ which is recalled in Eq. \ref{AC_LO_2FS}.
\end{enumerate}

\begin{equation}
A_{C} \approx \frac{u(x_{1,2},Q^{2})\bar{d}(x_{2,1},Q^{2})-\bar{u}(x_{1,2},Q^{2})d(x_{2,1},Q^{2})}{u(x_{1,2},Q^{2})\bar{d}(x_{2,1},Q^{2})+\bar{u}(x_{1,2},Q^{2})d(x_{2,1},Q^{2})}
\label{AC_LO_2FS}
\end{equation}

\par
\begin{figure}[h]
\begin{center}
\includegraphics[scale=0.375]{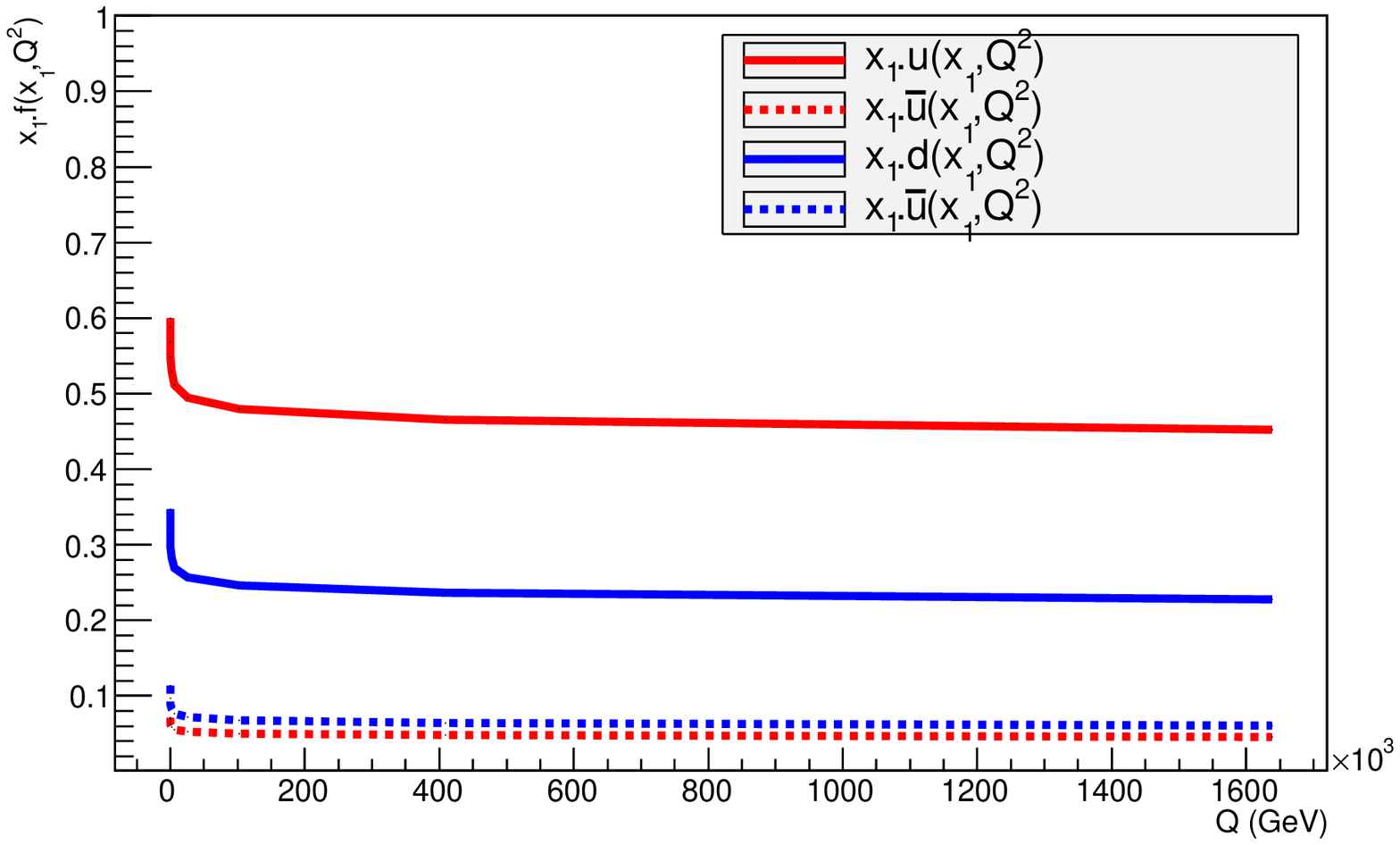}
\includegraphics[scale=0.375]{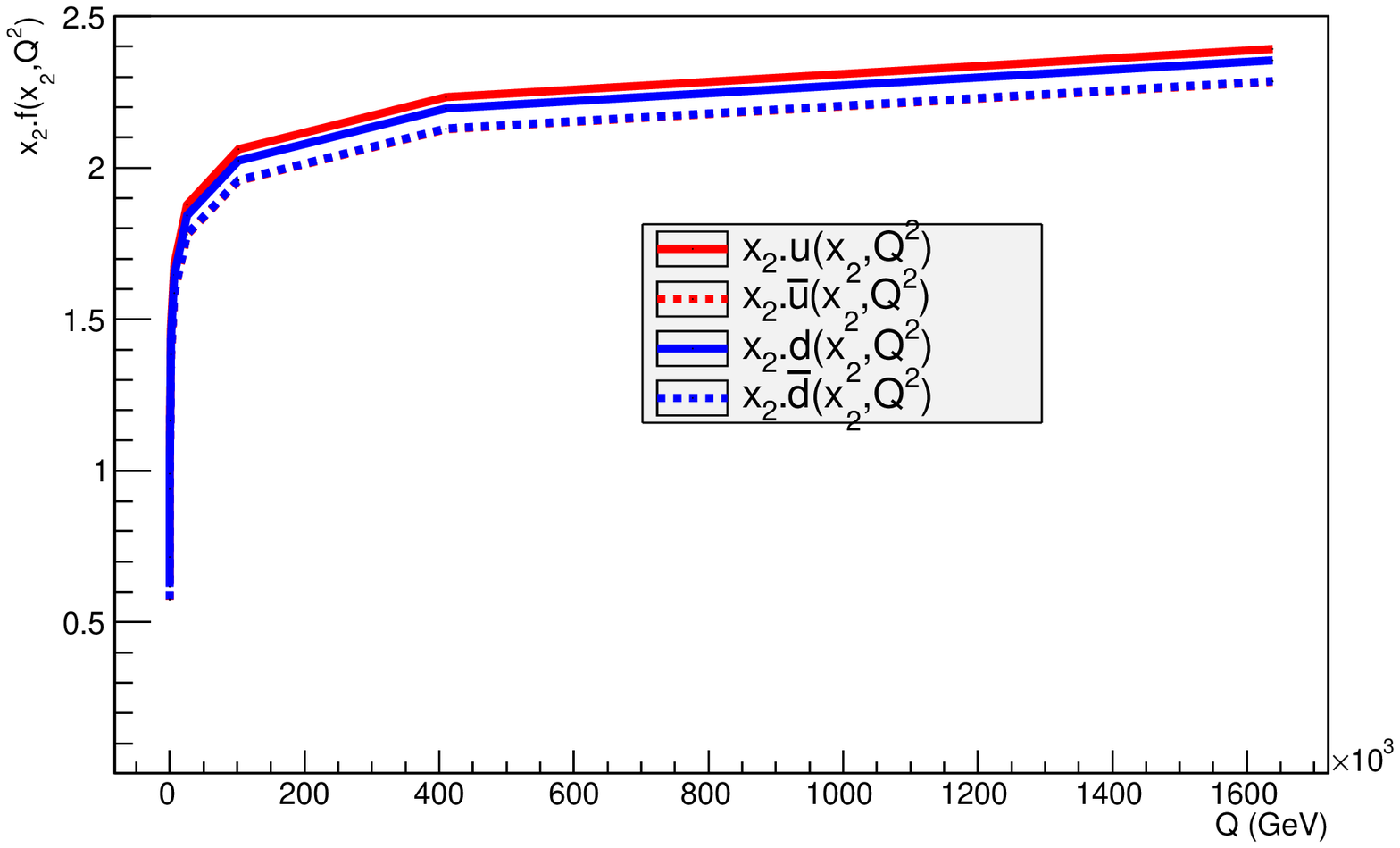}\\
\includegraphics[scale=0.375]{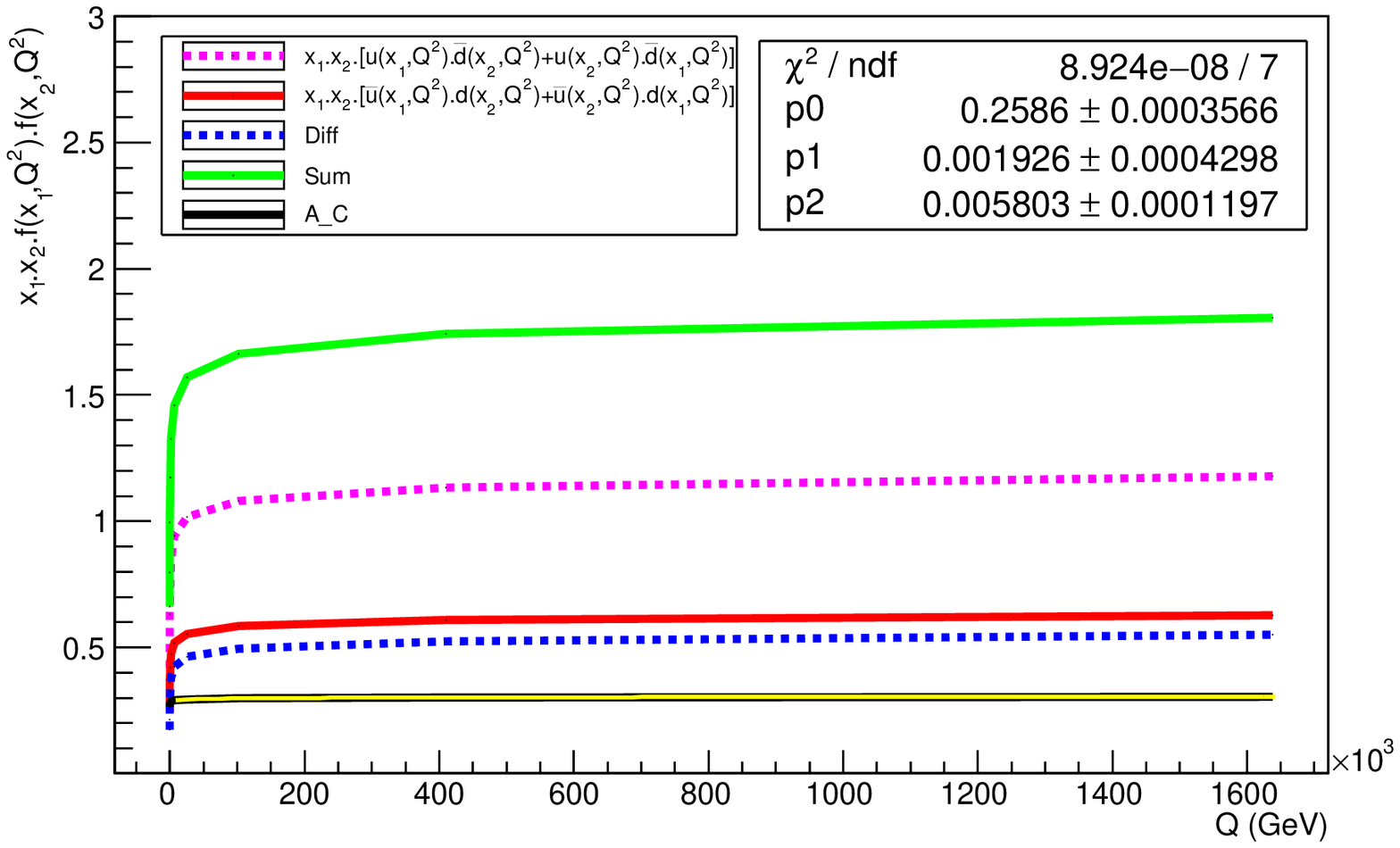}
\includegraphics[scale=0.375]{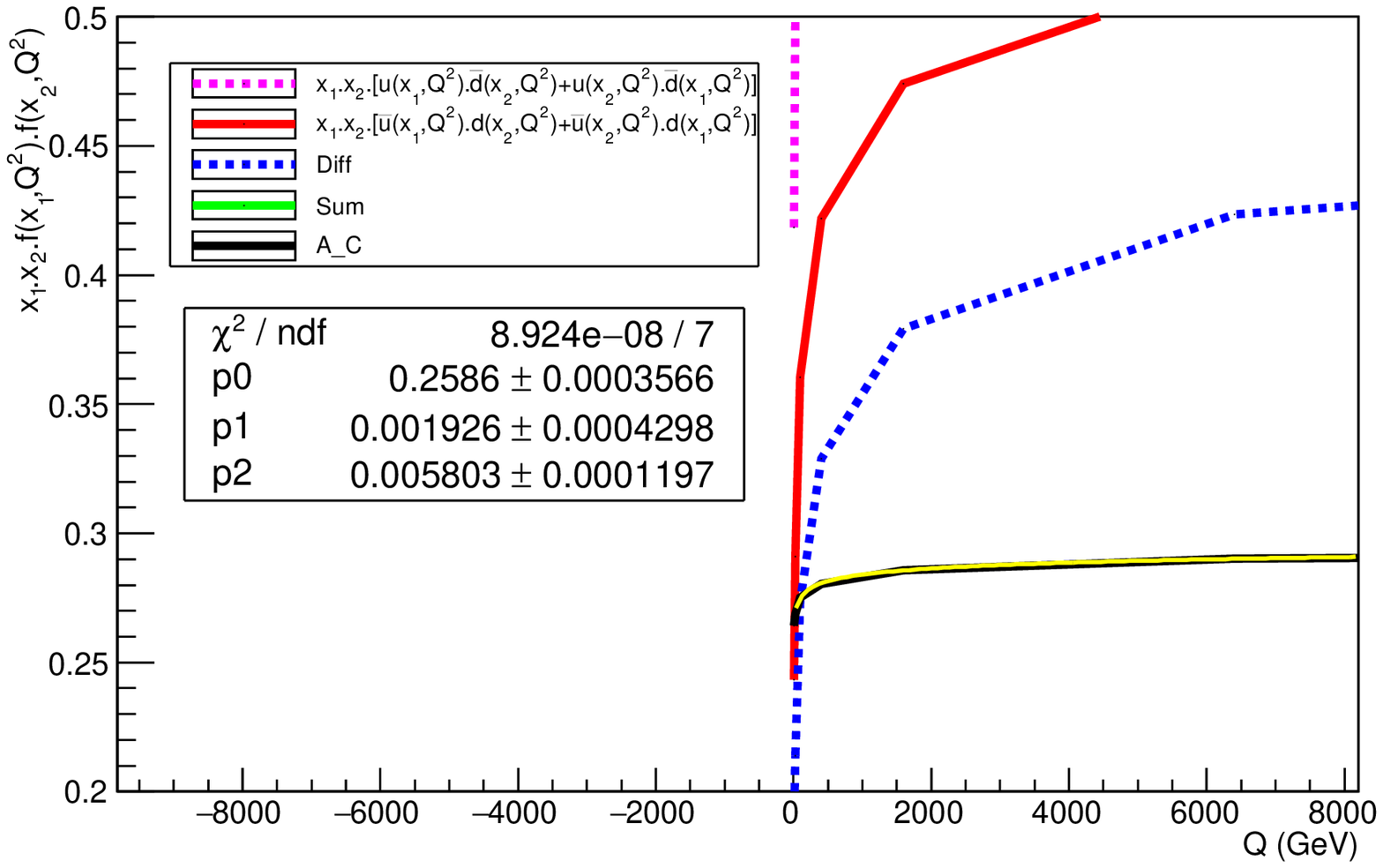}
\caption{Evolutions of the quark PDFs (top), of the quark currents in the IS and of $A_{C}$ (bottom) calculated with QCDNUM using the MSTW2008nlo68cl parametrization.}
\label{QCDNUM}
\end{center}
\end{figure}

\noindent
In order to illustrate numerically the Q evolution of the different quantities listed above, we used QCDNUM and the MSTW2008nlo68cl PDF. We set the Bj\" orken momentum fractions to arbitray values (compatible with the $W^{\pm}$ production in p+p collisions at $\sqrt{s}=7$ TeV), $x_{1}=0.15$ and $x_{2}=8.79\times 10^{-4}$, and varied $Q$. The quark density functions $x_{1}\cdot u(x_{1},Q^{2})$, $x_{1}\cdot \bar{u}(x_{1},Q^{2})$, $x_{1}\cdot d(x_{1},Q^{2})$, $x_{1}\cdot \bar{d}(x_{1},Q^{2})$, and $x_{2}\cdot u(x_{2},Q^{2})$, $x_{2}\cdot \bar{u}(x_{2},Q^{2})$, $x_{2}\cdot d(x_{2},Q^{2})$, $x_{2}\cdot \bar{d}(x_{2},Q^{2})$ are shown in the top RHS and LHS of  Fig. \ref{QCDNUM}, respectively. At the bottom row of the same figure the positively and negatively charged currents 
$x_{1,2}\cdot x_{2,1}\cdot u(x_{1,2},Q^{2})\cdot\bar{d}(x_{2,1},Q^{2})$, and $x_{1,2}\cdot x_{2,1}\cdot \bar{u}(x_{1,2},Q^{2})\cdot d(x_{2,1},Q^{2})$ as well
as $A_{C}$ are displayed on the LHS, with a zoom on the low $Q$ end on the RHS. 
\par\noindent
In sub-section \ref{I-parton-LVL:Fit_Functions} we consider different polynomials of functions of Q as fit functions to describe the Q evolution of the PDFs.
Let's consider here a polynomial of $Log(Log(Q))$ , in this example, the momentum fractions carried by the incoming quarks: $x_{i}\cdot f(x_{i},Q^{2})$ can be fitted by first degree polynomials of $Log(Log(Q))$ (though $x_{2}\cdot f(x_{2},Q^{2})$ fits are actually improved by using a second degree polynomial). First degree
polynomials of $Log(Log(Q))$ give very good fits of the evolution of the "quark currents": $x_{1}\cdot x_{2}\cdot f_{flav1}(x_{1},Q^{2})\cdot f_{flav2}(x_{2},Q^{2})$ c, and, given the hierarchy of the coefficients of these quark currents polynomials, of the $A_C$ as well.

\noindent

%%%%%%%%%%%%%%%%%%%%%%%%%%%%%%%%%%%%%%%%%%%%%%%%%%%%%%%%%%%%%%%%%%%%%

\subsection{Toy Models for the Main Properties of $A_{C}^{Fit}$}
\noindent
Hereafter, we make the hypothesis that quark currents and $A_{C}$ can be fitted by the different polynomials of functions of $Q$ evoked above. We want to figure out how the coefficients of such polynomials arrange so as to give the $A_{C}$ template curves presented in sub-section ~\ref{sec:Part1-parton-LVL}, i.e. monotonically increasing functions of $Q$ with a monotonically decreasing slope.
\par\noindent
Again, let's consider the simplest case where the first degree polynomials are sufficient. If we denote $x=Q$, and $f(x)$ the fit function, we can write the charged cross sections:

\begin{equation}
\begin{cases}
\sigma^{+}(x) = P_{0}  + P_{1}  \cdot f(x)\\
\sigma^{-}(x)  = M_{0} + M_{1}\cdot f(x)\\
\end{cases}
\end{equation}
 
\noindent
therefore

\begin{equation}
A_{C}(x) = \frac{(P_{0}-M_{0})+(P_{1}-M_{1})\cdot f(x)}{(P_{0}+M_{0})+(P_{1}+M_{1})\cdot f(x)}
\end{equation}

\noindent
Provided that $\lim\limits_{x \to+\infty}|f(x)| = +\infty$ (which holds for all the fit functions we considered), it appears that $A_{C}$ has an asymptote given by: 
\begin{equation}
\lim\limits_{x \to+\infty}A_ {C}(x)= \frac{(P_{1}-M_{1})}{(P_{1}+M_{1})}
\end{equation}

\noindent
The derivative of $A_{C}(x)$ can be expressed as:
\begin{equation}
\frac{dA_{C}(x)}{dx} = \frac{2\cdot(P_{1}M_{0}-P_{0}M_{1})\cdot f'(x)}{\lbrack (P_{0}+M_{0})+(P_{1}+M_{1})\cdot f(x)\rbrack^{2}}
\end{equation}

\noindent
Hence the condition to get a monotonically increasing $A_{C}(x)$ writes: 
\begin{equation}
\frac{dA_{C}(x)}{dx}\geq 0\Longleftrightarrow (P_{1}M_{0}-P_{0}M_{1})\cdot f'(x)\geq 0
\end{equation}

\noindent
And finally, that fact that $A_{C}$ can be fitted with the same functional form as $\sigma^{+}(x)$ and $\sigma^{-}(x)$ relies on the (approximate) fullfilment
of the following second degree functional equation:
\begin{equation}
(A_{1}M_{1})\cdot\left(f(x)\right)^{2} + (A_{0}M_{1}+A_{1}M_{0}-P_{1})\cdot f(x) + (A_{0}M_{0}-P_{0}) = 0
\end{equation}
\noindent
This equation has an analyitical solution if it's determinant is positive or null:
$$\Delta=\sqrt{(A_{0}M_{1}+A_{1}M_{0}-P_{1})^{2}-4\cdot (A_{1}M_{1})\cdot (A_{0}M_{0}-P_{0})}\geq 0.$$

\noindent
The fits of $\sigma^{+}(x)$, $\sigma^{-}(x)$ and $A_{C}$ with the 3 considered functional forms are performed and the corresponding values
of the fit parameters are presented in table \ref{Fit_Results}.

\begin{table}[h]
\begin{center}
\begin{tabular}{|c|c|c|c|}
\hline\hline
Fit Parameter & Polynomial                  & Polynomial                               & Laguerre          \\ 
                         & of $Log(Q)$                 & of $Log\left(Log(Q)\right)$   & Polynomials	\\
\hline		
$P_{0}$	          &  $0.33\pm 0.03$           &  $0.01\pm 0.03$                         &       $0.79\pm  0.08$    \\
$P_{1}$	          &  $0.064\pm 0.004$       &  $0.43\pm 0.02$                         &    $(-2.9\pm  1.5)\times 10^{-7}$       \\
\hline		
$M_{0}$	        &  $0.21\pm 0.02$           &   $0.04\pm 0.01$                         &       $0.44\pm 0.04$    \\
$M_{1}$	        &  $0.032\pm 0.002$       &   $0.220\pm 0.006$                     &    $(-1.4\pm 0.8)\times 10^{-7}$       \\
\hline		
$A_{0}$	          &  $0.258\pm 0.002$      &  $0.242\pm 0.002$                      &       $0.283\pm 0.004$    \\
$A_{1}$	          &  $0.0036\pm 0.0002$  &  $0.023\pm 0.001$                      &    $(-1.6\pm 0.8)\times 10^{-8}$       \\
\hline\hline
\end{tabular}       
\end{center}
\caption{\label{Fit_Results} Values of the fits parameters.}
\end{table}

\subsubsection{Polynomials of $Log(x)$}
\noindent
In this case, our toy model writes:
\begin{equation}
A_{C}(x) = \frac{(P_{0}-M_{0})+(P_{1}-M_{1})\cdot Log(x)}{(P_{0}+M_{0})+(P_{1}+M_{1})\cdot Log(x)}
\end{equation}

\noindent
with
\begin{equation}
\frac{dA_{C}(x)}{dx} = \frac{2\cdot(P_{1}M_{0}-P_{0}M_{1})}{x\cdot \lbrack (P_{0}+M_{0})+(P_{1}+M_{1})\cdot Log(x)\rbrack^{2}}
\end{equation}

\noindent
and, since x > 0,
\begin{equation}
\frac{dA_{C}(x)}{dx}\geq 0\Longleftrightarrow (P_{1}M_{0}-P_{0}M_{1})\geq 0
\end{equation}

\noindent
Given the values of the fits parameters:
\begin{itemize}
\item the asymptoteic $A_{C}$ is $33.0\%$
\item $P_{1}M_{0}-P_{0}M_{1}=2.51\times 10^{-3} \geq 0$
\item $\Delta=3.12\times 10^{-3} \geq 0$
\end{itemize}
Therefore $A_{C}(x)$ can be fitted by a first order polynomial of $Log(x)$, it's a monotonically increasing function, yet its has an asymptote.

\subsubsection{Polynomials of $Log\left( Log(x)\right)$}
\noindent
In this case, our toy model writes:
\begin{equation}
A_{C}(x) = \frac{(P_{0}-M_{0})+(P_{1}-M_{1})\cdot Log\left( Log(x)\right)}{(P_{0}+M_{0})+(P_{1}+M_{1})\cdot Log\left( Log(x)\right)}
\end{equation}

\noindent
with
\begin{equation}
\frac{dA_{C}(x)}{dx} = \frac{2\cdot(P_{1}M_{0}-P_{0}M_{1})}{x\cdot Log(x)\cdot \lbrack (P_{0}+M_{0})+(P_{1}+M_{1})\cdot Log\left( Log(x)\right)\rbrack^{2}}
\end{equation}

\noindent
and, since x > 0 (in practice x > 10 GeV) and Log(x) > 0, 
\begin{equation}
\frac{dA_{C}(x)}{dx}\geq 0\Longleftrightarrow (P_{1}M_{0}-P_{0}M_{1})\geq 0
\end{equation}

\noindent
Given the values of the fits parameters:
\begin{itemize}
\item the asymptotic $A_{C}$ is $32.6\%$
\item $P_{1}M_{0}-P_{0}M_{1}=1.57\times 10^{-2} \geq 0$
\item $\Delta=0.144 \geq 0$
\end{itemize}
Therefore $A_{C}(x)$ can be fitted by a first order polynomial of $Log\left(Log(x)\right)$, it's a monotonically increasing function, yet its has an asymptote.

\subsubsection{Laguerre Polynomials $L_{n}(x)$}
\noindent
The toy model writes:
\begin{equation}
A_{C}(x) = \frac{(P_{0}-M_{0})+(P_{1}-M_{1})\cdot (1-x)}{(P_{0}+M_{0})+(P_{1}+M_{1})\cdot (1-x)}
\end{equation}

\noindent
with
\begin{equation}
\frac{dA_{C}(x)}{dx} = \frac{-2\cdot(P_{1}M_{0}-P_{0}M_{1})}{\lbrack (P_{0}+M_{0})+(P_{1}+M_{1})\cdot (1-x)\rbrack^{2}}
\end{equation}

\noindent
and, 
\begin{equation}
\frac{dA_{C}(x)}{dx}\geq 0\Longleftrightarrow (P_{1}M_{0}-P_{0}M_{1})\leq 0
\end{equation}

\noindent
Given the values of the fits parameters:
\begin{itemize}
\item the asymptoteic $A_{C}$ is $34.2\%$
\item $P_{1}M_{0}-P_{0}M_{1}=-1.46\times 10^{-3} \leq 0$
\item $\Delta=6.3\times 10^{-14} \geq 0$
\end{itemize}
Therefore $A_{C}(x)$ can be fitted by a first order polynomial of $(1-x)$, it's a monotonically increasing function, yet its has an asymptote.

\par\noindent
We verified that for the case without longitudinal boost: $x_{1}=x_{2}=1.15\times 10^{-2}$, the conclusions listed above remain valid.

%%%%%%%%%%%%%%%%%%%%%%%%%%%%%%%%%%%%%%%%%%%%%%%%%%%%%%%%%%%%%%%%%

\end{document}